%% file: main.tex
\documentclass[11pt,a4paper]{article}
\usepackage[T1]{fontenc}
\usepackage{lmodern}
\usepackage{microtype}
\DisableLigatures[-]{family=tt*}
\usepackage[utf8]{inputenc}
\usepackage{amsmath,amssymb}
\usepackage{fancyvrb}
\usepackage{booktabs}
\usepackage{longtable}
\usepackage[margin=2.6cm]{geometry}
\usepackage[hidelinks]{hyperref}
\newcommand{\lean}{\textsf{Lean\,4}}
\newcommand{\isa}{\textsf{Isabelle/HOL}}
\newcommand{\tc}[1]{\texttt{#1}}

\title{G\"odel's and Scott's Variants of the Ontological Argument\\ in \lean{} and TPTP THF}

\author{Christoph Benzm\"uller\\[1ex]
  {\normalsize AI Systems Engineering, Otto-Friedrich-Universit\"at Bamberg, Germany}\\
  {\normalsize Faculty of Mathematics and Computer Science, Freie Universit\"at Berlin, Germany}\\
  {\normalsize\texttt{christoph.benzmueller@uni-bamberg.de}}}
\date{October 2026}

\input{thf-numbers}

\begin{document}
\maketitle

\begin{abstract}
The \isa{} dataset of Benzm\"uller and Scott's study of G\"odel's ontological argument and
Scott's variant (\emph{Monatshefte f\"ur Mathematik}, 2025) is carried to \lean{} and from
there back to the automated provers, as a benchmark independent of either proof
assistant.  The port covers all thirty theories, structure and names preserved: 548
statements compare identical as parsed, every named result is proved again, and five results the
original reports without replaying them are proved here.  For every theorem,
\tc{\#print axioms} gives the postulates its proof consumes: Scott's necessary existence and
modal collapse need only a symmetric frame, confirming that \textsf{KB} suffices.

The benchmark, in TPTP THF and SMT-LIB, turns the steps of an argument debated in
philosophy into \thfProblems{} theorems, alongside \thfNonTheorems{} statements the original
refutes or leaves open, ten left open there.  Five THF
provers, and cvc5 on SMT-LIB, prove \thfSolvedAny{} theorems within ten
seconds on one core and \thfSolvedAnySixty{} within sixty, and none proves any of the
\thfNonTheorems{}.  E and Leo-II solve the most, although Leo-II's calculus has been unchanged for
about a decade and was only repaired and modernised here, as release 2.2.  Vampire, whose later version
won the higher-order division of CASC-30, solves the most in no configuration.  Only E and Leo-II
are measured in their own automatic mode: Zipperposition proves \thfPlainZipperposition{} in a single
mode and \thfTenZipperposition{} with its developers' portfolio, Vampire \thfPlainVampire{}
without options and \thfTenVampire{} with a higher-order schedule that its CASC mode does not select,
and Leo-III \thfPlainLeoIII{} alone and \thfTenLeoIII{} with E as partner.
\end{abstract}

\tableofcontents
\addtocontents{toc}{\protect\setcounter{tocdepth}{1}}
\newpage

\section{Introduction}

G\"odel's ontological argument~\cite{GoedelNotes}, a derivation in higher-order modal logic
of the necessary existence of a God-like being from its possibility, stands in a tradition that
reaches from Anselm through Descartes and Leibniz, and it draws attention well beyond logic:
G\"odel's philosophical notebooks, now being edited in full~\cite{GoedelNotebooks}, show how
much of his thought was theological, and whose existence he believed he had proved is debated to this
day.\footnote{Most recently at the conference \emph{Kurt G\"odel, `Gott' und `Teufel'} at the
Berlin-Brandenburg Academy, 29--31 July 2026, reported in the
\emph{Frankfurter Allgemeine Zeitung} of 30 September 2026, p.~12.}  Formal verification bears on that debate directly: which axioms
an argument consumes, whether they are consistent, and whether they force modal collapse are
questions a prover can answer.  Scott's variant of the argument~\cite{ScottNotes} has been a recurring
benchmark for computational metaphysics since its first fully automated treatment, in which it
was first encoded in the TPTP THF syntax, explored with higher-order automated theorem provers
such as Leo-II~\cite{J29} and formalised in \isa{}~\cite{IsabelleHOL}, archived in the Archive
of Formal Proofs in 2013~\cite{GoedelGod-AFP} and presented at ECAI~2014~\cite{C40}; a second
study, presented at IJCAI~2016~\cite{C55}, found the inconsistency of G\"odel's own 1970
axioms.  That work was followed by an encoding of Fitting's
reconstruction of the argument~\cite{Fitting2002}, with intensional and extensional positive
properties, and of Anderson's variant~\cite{Types_Tableaus_and_Goedels_God-AFP}, by a computer-supported study of
positive properties, ultrafilters and modal collapse~\cite{J50}, and by simplified
variants of the argument~\cite{C82,SimplifiedOntologicalArgument-AFP}; the underlying
methodology was later generalised into the LogiKEy framework~\cite{J48}.  That line of work
is computer-supported throughout.  Kanckos and Woltzenlogel Paleo worked in an explicit
calculus instead of over an embedding: four proofs by hand in a natural deduction calculus
for a rigid higher-order modal logic~\cite{KanckosPaleo2017}, two of them of
Scott's variant of G\"odel's argument, the second needing only \textsf{KB} and no equality, and one each of Anderson's
and Bj{\o}rdal's variants.  Derivations in that style, they note, have been checked in
Coq~\cite{C44}.
In~\cite{J75}, the \emph{Notes} for short, Benzm\"uller and Scott extended the computer-supported experiments considerably.  For the first
time, G\"odel's original manuscript~\cite{GoedelNotes} was formalised as closely as possible;
the inconsistency of its postulates was explained in detail and removed in two ways, one of
them new, and both repairs take the manuscript's conjunction of ``any number of summands'' at
its word: the conjunction axiom then quantifies over sets of properties, and the argument
becomes one of third-order modal logic, where Scott's variant needs only second order; Scott's variant was compared with it; and variants were studied in \textsf{S4}
and \textsf{K} as well as in \textsf{S5}, and, systematically for the first time, under
combinations of possibilist and actualist quantification; and concepts of evil were
examined, and whether evil is derivable was put to the test.  The study answered a number of
questions that had accumulated over the years, and its dataset is an entry in the Archive of
Formal Proofs~\cite{NotesAFP}.

This paper carries the dataset of~\cite{J75} from \isa{} to \lean{}~\cite{Lean4}
and from there back to the automated provers, as a benchmark that can be checked
independently of either proof assistant, and uses the port to account for the postulates each
proof consumes and to measure how much the provers' results depend on the configurations they are
run in.

The embedding has been ported before to a proof assistant without a tool like
\isa{}'s \tc{sledgehammer}~\cite{Sledgehammer}, which hands a goal to external automated provers and turns a
found proof into a checked one: Benzm\"uller and
Woltzenlogel Paleo carried Scott's version of the argument to Coq~\cite{C44}, with
interactive proofs that follow Scott's notes step by step.  That
transfer was done entirely by hand, with no mechanical check against its source; here the
correspondence to the \isa{} sources is checked mechanically, statement by statement
(Section~\ref{sec:method}).  What is new here is the dataset --- the thirty
theories of~\cite{J75} in full: the variants, the test theories, and the 45 statements the
original refutes or leaves open --- and the
per-theorem dependency accounting of Section~\ref{sec:axioms}.  The dataset is also
rendered in TPTP THF~\cite{TPTP} and in SMT-LIB~\cite{SMTLIB} (Section~\ref{sec:thf}), as a benchmark of
\thfProblems{} theorems and \thfNonTheorems{} statements of known status, ten of them left open
by the original;
measuring six higher-order provers on it showed how much their results depend on the configuration
they are run in.  Unlike the exports of mathematical libraries that form the largest families
of the higher-order part of TPTP, these problems are the steps of an argument whose answers
matter outside automated reasoning, and the library analysis of
Section~\ref{sec:library} shows that they are structurally representative nonetheless.

The motivation is threefold.  First, the technique the dataset rests on --- a shallow
semantical embedding of a higher-order modal logic in classical higher-order logic --- is
deliberately system-independent, and it is worth exhibiting that independence concretely.
The same holds for the wider methodology it belongs to, LogiKEy~\cite{J48},
\url{https://logikey.org}, which is tied neither to one logic nor to one way of embedding
it~\cite{B31,B30}.  The port required no change to any statement and no idea that is not
already in the \isa{} sources; the embedding itself, the modal library and all thirty
theories went across by the same routine translation.  What carried over is the
representation: the statements, found identical as parsed, and their proofs.  What did not
carry over is the way the dataset was produced.  LogiKEy is a method of exploration as much as
a representation: a conjecture is put to automated provers and to a model finder, and each
answer shapes the next question.  \lean{}'s core has neither a sledgehammer nor a model finder, so
the port proves again what was already known and records the countermodels instead of
finding them; exploring an open question there would be a different kind of work.  The
rendering in TPTP THF and SMT-LIB returns the statements to the automated provers that this
loop needs.  This one case study thus supports the portability of the LogiKEy framework, its
library of embedded logics and the case studies built on them, to \lean{} as far as the
representation goes, without establishing it.

Second, \lean{}'s user community does not
overlap greatly with \isa{}'s, and the same holds for teaching.  The LogiKEy methodology has been
taught for more than a decade, with G\"odel's argument as a highlight, among others in lecture courses in
Bamberg, Berlin and Luxembourg; in invited courses and tutorials at Zhejiang University and East
China Normal University in Shanghai, at the Berkeley--Stanford Circle in Logic and Philosophy, at
the University of Campinas and PUCRS in Porto Alegre, and in Toulouse; and at summer schools,
from the Reasoning Web and ANU Logic Summer Schools in 2015 to ESSLLI 2026 in
Prague,\footnote{A record of these courses and tutorials is at
\url{https://christoph-benzmueller.de/teaching.html}; the materials of the ESSLLI course are
at \url{https://logikey.org/CoursesAndTutorials/2026-ESSLLI/}.} so far always in \isa{}.
With the port and the benchmark, the same material can now be taught in \lean{}, or with
automated provers alone.

Third, the port was an occasion to
make the dependency structure of the arguments explicit: which postulates, and which frame
conditions --- hence which modal logic --- each theorem actually uses.  In \lean{} this is
read off with the \tc{\#print axioms} command.  No advantage of \lean{} is claimed or has
been detected in this respect: \isa{} offers the same information, through its
\tc{thm\_deps} command, which lists the facts a theorem depends on, and through its proof
automation, which reports the facts used by each proof it finds; the same account could
have been produced there.  The port merely did it systematically, for every theorem of the
dataset, as a by-product of writing the proofs by hand, and reports the result
(Section~\ref{sec:axioms}).

What this paper is \emph{not} should be said as well.  It is not a new philosophical or logical
analysis of the arguments; for that, see~\cite{J75}, of which the present development
is a faithful re-implementation.  Nor does it change what the original leaves
\emph{open}: a statement that~\cite{J75} neither proves nor refutes stays open here.  A companion article~\cite{NominalsArXiv}, drawing on this port, has since
answered two further questions.  It checks the port's hand-written proofs,
and the original's own, for nominals --- terms of hybrid logic that the embedding admits but
the modal object language lacks --- and finds none, so that every result the original proves
has a proof in the modal object language; and it settles the three questions the original
leaves open, and with them all ten open statements of the dataset (Section~\ref{sec:formalised}), again without nominals.  Each of its results
is verified in \isa{} and independently in \lean{}.  This paper
does close five statements that the original leaves unreplayed after an automated prover
had found a proof, and it records one dependency as dispensable; see
Sections~\ref{sec:formalised} and~\ref{sec:axioms}.

\section{The embedding of higher-order modal logic in \texorpdfstring{\lean{}}{Lean 4}}
\label{sec:embedding}

The embedding is that of Benzm\"uller and Paulson~\cite{J26}, and follows~\cite{J75}
exactly.  Two uninterpreted types are postulated,
\tc{i} for possible worlds and \tc{e} for individuals; world-lifted propositions are
\tc{$\sigma$ := i $\rightarrow$ Prop} and modal properties are
\tc{$\tau$ := e $\rightarrow$ $\sigma$}.  An accessibility relation \tc{R} is
postulated, together with the frame conditions of the logic under consideration --- the
properties of \tc{R}, such as reflexivity, symmetry and transitivity, that determine which
modal logic holds --- and the modal connectives are defined as operations on truth sets.  The base module
\tc{HOMLinHOL} (Figure~3 of~\cite{J75}) postulates \tc{Rrefl}, \tc{Rsymm} and
\tc{Rtrans}, which makes the full strength of \textsf{S5} available to the theories
built on it.  The two modules that the sources label
``slight variations'' of that figure, \tc{HOMLinHOLonlyK} and \tc{HOMLinHOLonlyS4},
differ from it in exactly this axiom block and otherwise only in comments: the former postulates no
frame condition at all (logic \textsf{K}), the latter only \tc{Rrefl} and \tc{Rtrans}
(logic \textsf{S4}).  Every other declaration --- types, connectives, quantifiers,
\tc{existsAt}, validity --- is identical in the three modules, so a theory can be moved
between the logics by changing one \tc{import}.  Quantifiers come in a
polymorphic possibilist form and, for individuals, an actualist form guarded by a
postulated existence predicate \tc{existsAt}.  Global validity is
\tc{$\lfloor\psi\rfloor$ := $\forall$ w, $\psi$ w}.

The one substantive adaptation concerns logical strength.  \isa{}'s \tc{HOL} is
classical, whereas \lean{}'s \tc{Prop} is intuitionistic; classical reasoning is
therefore enabled globally.  Since \isa{}'s \tc{typedecl} introduces a non-empty type whereas a
\lean{} type may be empty, \tc{Nonempty i} and \tc{Nonempty e} are postulated in
addition; this is used, for instance, by the existential-import tests.

Table~\ref{tab:notation} relates the notation of the two developments.  The bold
operators of the \isa{} sources are reproduced with a superscript-\tc{m} convention, which
keeps the formulas close to the figures of~\cite{J75} while avoiding clashes with
\lean{} core notation.

\begin{table}[t]
\centering\small
\begin{tabular}{llll@{\qquad}llll}
\toprule
\isa{} & \lean{} & \isa{} & \lean{} & \isa{} & \lean{} & \isa{} & \lean{}\\
\midrule
$\bot$ & \tc{$\bot$\textsuperscript{m}}       & $\top$ & \tc{$\top$\textsuperscript{m}}
  & $\forall$ & \tc{$\forall$\textsuperscript{m}} & $\exists$ & \tc{$\exists$\textsuperscript{m}}\\
$\neg$ & \tc{$\neg$\textsuperscript{m}}       & $\wedge$ & \tc{$\wedge$\textsuperscript{m}}
  & $\forall^{E}$ & \tc{$\forall$\textsuperscript{E}} & $\exists^{E}$ & \tc{$\exists$\textsuperscript{E}}\\
$\vee$ & \tc{$\vee$\textsuperscript{m}}       & $\supset$ & \tc{$\rightarrow$\textsuperscript{m}}
  & $\leftrightarrow$ & \tc{$\leftrightarrow$\textsuperscript{m}} & $\mathbf{r}$ & \tc{$\mathsf{r}$}\\
$\Box$ & \tc{$\Box$}                    & $\Diamond$ & \tc{$\Diamond$}
  & $=$ & \tc{$=$\textsuperscript{m}}         & $\neq$ & \tc{$\neq$\textsuperscript{m}}\\
$\equiv$ & \tc{$\equiv$\textsuperscript{m}}   & $\lfloor\cdot\rfloor$ & \tc{$\lfloor\cdot\rfloor$}
  & $\supset_{N}$ & \tc{$\supset$\textsuperscript{N}} & $@$ & \tc{$@$\textsuperscript{m}}\\
\bottomrule
\end{tabular}
\caption{Notation of the \isa{} sources and of the \lean{} port.}
\label{tab:notation}
\end{table}

\section{Method of the port}
\label{sec:method}

The port is structure-preserving: there is exactly one \lean{} module per \isa{} theory,
carrying the same name, the same sections, the same declaration order and the same names
for all axioms, definitions and theorems, up to the renamings listed at the end of this
section, and the same statements as parsed.  Nothing beyond that is claimed; in particular
the two developments are not asserted to have the same semantics.  This was checked
mechanically, at two levels.  First, names: each of the 569 distinct names
introduced by \tc{lemma}, \tc{theorem}, \tc{definition}, \tc{abbreviation} and
\tc{axiomatization} across the thirty theories occurs in the corresponding \lean{} module,
modulo the renamings listed at the end of this section and the twelve binder abbreviations
discussed below, which the port realises as macros.  The raw declaration count is 572:
\tc{Th4} is declared twice in each of the three variants of Section~4.5 of~\cite{J75},
once abandoned and once postulated.  Second, propositions: a comparison tool shipped with the development parses the
statement of every item on both sides, with the operator precedences of the two
developments (which agree), into abstract syntax trees over a common vocabulary, up to the
names of bound and free variables, and compares the trees.  All 548 pairs of statements
are identical under this comparison.  The comparison is sensitive to what matters:
redundant parentheses disappear, while a changed connective, a moved parenthesis or an
exchange of two variables is reported, as was confirmed by mutating the \lean{} sources.

The tool re-parses the \isa{} source text, and that deserves a word.  Several of the
\isa{} mixfix declarations carry no priorities (\tc{Mprimeq}, \tc{Mprimneg}, \tc{R},
\tc{existsAt}), the base theory switches off \tc{syntax\_ambiguity\_warning}, and so
some readings are fixed by \isa{} through type inference rather than by precedence.  The
tool assigns these operators the priorities of the corresponding \lean{} declarations.
That this yields \isa{}'s reading is checked by \isa{} itself: the tool writes, for each
of the 417 bracketed statements, its reading back as a fully parenthesised \isa{} formula,
and a generated theory per source theory (\tc{tools/isabelle/check/}; 25 of the 30, the other five
being the three base theories, the modal filter theory and the Cantor illustration) parses the original
and the reparenthesised formula in the theory's own context and fails unless both yield
the same term.  All 417 agree; the check is live, since moving one operator in a generated
theory makes it fail.  The same is done on the \lean{} side: for each of the 399
declarations that carry those 417 formulas, some of them several, the tool writes \tc{example : (statement) = (reparenthesised statement) :=
by with\_reducible rfl}, closed under the declaration's binders, into a file per module
(\tc{tools/lean-check/}); \tc{with\_reducible rfl} accepts the two sides only if they
agree up to reducible unfolding, which for these statements means the same term, and all
25 files compile.  The generator rejects a line whose two sides coincide textually, which
would be vacuous.  Definitions are compared by the tool but are not covered by these two
checks.  Twelve Isabelle items have no \lean{}
counterpart by construction: the four binder abbreviations \tc{Mallpossb}, \tc{Mexipossb},
\tc{Mallactb} and \tc{Mexiactb}, declared in each of the three base modules, which the port
realises as \lean{} \tc{macro}s; nor do the satisfiability probes \tc{lemma True nitpick}, which
the port records as comments.  The declared types, the polymorphic
versus entity-specific split of the quantifiers and equalities, and the operator
precedences were checked by hand, as were the twelve statements that contain no modal
formula: the three inconsistency lemmas, the four of the Cantor illustration, two of them kept as comments as \isa{} leaves them at
\tc{oops}, and the five
frame axioms --- \tc{Rrefl}, \tc{Rsymm} and \tc{Rtrans} of the \textsf{S5} theory,
\tc{Rrefl} and \tc{Rtrans} of the \textsf{S4} one.  Table~\ref{tab:corr} lists the constructions used.

\begin{table}[t]
\centering\footnotesize
\begin{tabular}{ll}
\toprule
\isa{} & \lean{}\\
\midrule
\tc{typedecl i}, \tc{typedecl e} & \tc{axiom i : Type}, \tc{axiom e : Type} (with \tc{Nonempty})\\
\tc{type\_synonym} & \tc{abbrev}\\
\tc{consts}, \tc{axiomatization where} & \tc{axiom} (free variables become explicit binders)\\
\tc{abbreviation} (unfolded at parse time) & \tc{@[simp, grind] def} (unfolded by \tc{simp}/\tc{grind})\\
\tc{definition f} with \tc{f\_def} & \tc{def f}, unfolded definitionally or by \tc{simp [f]}\\
\tc{lemma \dots{} oops} after \tc{nitpick} & \tc{example \dots{} := by countermodel}\\
\tc{lemma \dots{} oops} (open problem) & \tc{example \dots{} := by openproblem}\\
\tc{lemma \dots{} oops} (proof not replayed) & \tc{theorem}, proved here\\
\tc{lemma True nitpick[satisfy] oops} & a comment recording the model-finder result\\
\tc{ROOT} & \tc{lakefile.toml}\\
\bottomrule
\end{tabular}
\caption{Correspondence of \isa{} and \lean{} constructions.}
\label{tab:corr}
\end{table}

Since the layout of the sources is mirrored as well --- one item per line group, one blank
line between items --- their sizes can be compared.  Table~\ref{tab:loc} counts the
non-blank lines by kind.  The declaration lines agree to within a few per cent once the \lean{} notation
declarations are taken into account (each of the three base modules carries some twenty
lines of \tc{notation}, \tc{infix} and binder \tc{macro}s for the bold operators, which
\isa{} folds into its mixfix annotations; that accounts for 64 of the 89 lines of
difference, the rest being spread thinly over the other modules); the two developments have
the same shape.  The
\lean{} sources are longer in two places, for two reasons.  Prose grows because the
\lean{} modules carry a doc-string per declaration and record the 65 Nitpick~\cite{Nitpick}
results as comments; proof lines more than double because a one-line \tc{by smt} or
\tc{by blast} becomes an explicit proof term or tactic script, the price of writing every proof by hand.  The
per-module table is
\tc{tools/loc-report.txt} in the development; both it and Table~\ref{tab:loc} are
generated from the sources by \tc{tools/loc.py}.

\begin{table}[t]
\centering\small
\input{tab-loc}
\caption{Non-blank lines of source by kind.  A one-line statement with an inline proof
  (\tc{by blast}, \tc{:= term}) counts as one declaration line on either side; only lines
  of proof below the statement count as proof.  Declarations include notation and module
  headers.}
\label{tab:loc}
\end{table}

\paragraph{Abandoned statements.}
\isa{}'s \tc{oops} states a goal, abandons it, and adds nothing to the theory.  The port
reproduces this with two tactics, \tc{countermodel} and \tc{openproblem}, which are
\tc{sorry} under a descriptive name, applied \emph{only} to anonymous \tc{example}s.
The statement is therefore elaborated and type-checked, but no name is introduced and
nothing downstream can depend on it.  There are 45 such statements, 35 \tc{countermodel}
and 10 \tc{openproblem} (Section~\ref{sec:formalised}); \tc{\#print axioms} confirms
that no named result of the development depends on \tc{sorryAx}.

\paragraph{Automation.}
\lean{}'s core has no counterpart of \isa{}'s \tc{sledgehammer}, the port uses no external
hammer, and the automation available in the core (\tc{simp}, \tc{grind}) does not reach the higher-order instantiations
these proofs need --- typically the choice of a witness property such as
$\lambda z.\,z \equiv x$ or the constant property $\lambda \_ .\,\varphi$.  Almost all
proofs are therefore explicit terms.  Where a step recurs across the variants a
small named auxiliary lemma is introduced and marked as such in the source: \tc{Ax2a'}
(the $\neg P\varphi \leftrightarrow P\,{\sim}\varphi$ reading of G\"odel's \tc{Ax2a}),
\tc{PosOfGod} and \tc{NegOfEvil} (a God-like being has only positive properties, an
Evil-like being only non-positive ones),
\tc{PosTop} (the universal property is positive), \tc{GNonempty} and \tc{PosIncl}
(\tc{Ax4} applied to $G$, in the third variant), and \tc{Pos} and \tc{NegP}
(everything true of every God-like being at the successor worlds is positive),
\tc{EmptyEssL} in \tc{GoedelVariantHOML1inS4} (the empty property is an essence of anything,
which the \textsf{S5} theory states as a lemma and the \textsf{S4} theory leaves to
\tc{fastforce}), and \tc{noFixpoint} in \tc{SurjectiveCantor} (no proposition is equivalent
to its own negation).  Apart from the two nonemptiness axioms of
Section~\ref{sec:embedding}, these are the only declarations in the development that have no
counterpart in~\cite{J75}.

The postulates each proof consumes were then reduced, with \tc{\#print axioms} as the
instrument (Section~\ref{sec:axioms}): to what the corresponding \isa{} proof cites, and
further where a shorter argument was found.

\paragraph{Model finding.}
\lean{} has no analogue of \tc{nitpick}, so the countermodels and the satisfiability
(consistency) checks of the original cannot be re-run.  All 65 \tc{nitpick}
invocations are recorded verbatim --- the parameters and the outcome each reported --- beside the
corresponding statement, so that no experiment of the original is silently dropped.
This is documentation, not verification, and is flagged as such: a reader who wants the
countermodels must consult the \isa{} sources.  The THF rendering of
Section~\ref{sec:thf} restores the experiment outside \lean{}.

\paragraph{Renamings.}
Definitions that the \isa{} sources declare under a long name with a short notation are
named after the notation: \tc{God} becomes \tc{G}, \tc{Essence} \tc{Ess}, \tc{NecExist}
\tc{E} (\tc{NE} in Scott's variant), \tc{PropertyInclusion} \tc{PInc}, and the constant
\tc{PositiveProperty} \tc{P}; the comparison tool applies the same map.  Further,
\isa{} distinguishes the namespace of terms from that of facts, \lean{} does not.
Three groups of names therefore change: the lemmas \tc{Filter} and \tc{UltraFilter}
become \tc{PisFilter}/\tc{PisUFilter} (respectively \tc{PisFilterP}/\tc{PisUFilterP}
and \tc{NisFilter}/\tc{NisUFilter}), and the lemma \tc{PosProps} becomes \tc{PosProps'}
in those theories where the abbreviation \tc{PosProps} already claims the name.

\paragraph{Tool support.}
The port, including the proof terms, was produced interactively with an AI coding assistant
(Anthropic Claude), working from the \isa{} sources of~\cite{J75}; the assistant was also
used to support the drafting of this paper.  No claim rests on its reliability: every statement is fixed by
the original development and mechanically compared against it, as this section describes,
every proof is checked by \lean{}'s kernel, and every count reported about the development or
the experiments is produced by the scripts distributed with the paper --- three of the tables
and the THF counts in the text are generated from their output.  The author checked and revised
every claim and is responsible for the content.

\section{What is formalised}
\label{sec:formalised}

The 30 modules fall into five groups, following the \tc{ROOT} file of the original: the
base embeddings (\tc{HOMLinHOL} for \textsf{S5}, \tc{HOMLinHOLonlyS4}, \tc{HOMLinHOLonlyK};
Section~\ref{sec:embedding}) with
their test theories; the modal filter theory; G\"odel's 1970 argument in three variants
(the original inconsistent axioms, and two repairs, each in an actualist, a possibilist
and a mixed ``Anderson'' quantifier setting, restricted to \textsf{S4} in the actualist
setting for the original and in the actualist and possibilist settings for the repairs);
Scott's variant in the same three settings, restricted to \textsf{S4} in the actualist and
possibilist ones and to \textsf{K} in the actualist one; and the three ``evil'' theories.
The thirtieth module is the Cantor illustration of Figure~2 of~\cite{J75}.

Every named result that~\cite{J75} proves is proved again.  In particular:
\begin{itemize}
\item \tc{GoedelVariantHOML1}, \tc{...poss} and \tc{...AndersonQuant} derive
  \tc{Inconsistency~:~False}: G\"odel's axioms as stated in the 1970
  manuscript~\cite{GoedelNotes} are inconsistent, in all three quantifier settings.
\item \tc{GoedelVariantHOML2} and \tc{GoedelVariantHOML3} (repairing, respectively, the
  definition of essence and the notion of necessary property inclusion) establish
  \tc{Th1}--\tc{Th5}, modal collapse \tc{MC}, \tc{Monotheism}, and that the positive
  properties form an ultrafilter.
\item \tc{ScottVariantHOML} establishes \tc{T1}--\tc{T3}, \tc{MC}, \tc{Monotheism} and
  the ultrafilter property.
\item \tc{ThereIsNoEvil1} proves that necessarily no entity possesses all non-positive
  properties; \tc{EvilDerivable} derives the necessary existence of an Evil-like entity
  from the correspondingly negated assumptions.
\end{itemize}
Where the original does not establish a statement, \cite{J75} distinguishes three
reasons, and the port treats them differently.  A statement may be \emph{refuted}: \tc{nitpick}
produced a countermodel, as for the Barcan formulas with actualist quantifiers, for
\tc{T3} and \tc{MC} of Scott's variant in \textsf{K} and in \textsf{S4}, and for modal
collapse in the bare embedding.  There are 35 of these; they are left unproved,
marked \tc{countermodel}.  It may be \emph{genuinely open}: \tc{Th3} of the \textsf{S4}
restrictions, where~\cite{J75} reports that neither a proof nor a counterexample has been
obtained, and likewise \tc{UniqueEss1} and \tc{UniqueEss2} of the third G\"odel variant,
which the authors ``expect \dots{} to be provable''.  There are 10 of these, marked
\tc{openproblem}; they are left open too.  A companion
article~\cite{NominalsArXiv} settles all ten: under the
conjunction axiom as it stands, though only through its empty instance, and again once the
axiom is restricted, under side conditions and, for the third variant, with the non-emptiness
clause of its repair confined to the essence.  For Scott's variant as it stands the source answers the question negatively, with a Nitpick countermodel to \tc{T3} in \textsf{S4} that the dataset reproduces, and leaves it open only once Scott's \tc{A3} is replaced by the conjunction axiom~\cite[Sect.~5.3]{J75}, a variant the dataset does not contain; in the inconsistent 1970 variant \tc{Th3} is provable from reflexivity alone, as Section~\ref{sec:axioms} shows.  For that countermodel~\cite[p.~586]{J75} points at its Appendix~A, which is about the inconsistency of the 1970 axioms under possibilist quantifiers and says nothing about \textsf{S4}; the experiment is in the sources~\cite{NotesAFP}, in \tc{ScottVariantHOMLinS4}, where the \tc{nitpick} call sits on \tc{T3} itself.  Both markers are those of Section~\ref{sec:method}, so
the 45 statements establish nothing.  Or, third, an automated prover may have reported a proof that, in the experiments
of~\cite{J75}, which used Isabelle2024, no trusted \isa{} tactic could replay --- \tc{Th4} of the third variant in its three quantifier
settings (for the actualist setting \cite{J75} names Vampire~\cite{HOVampire2024},
Zipperposition~\cite{Zipperposition2023}, Leo-II~\cite{J29} and Leo-III~\cite{J53}), \tc{MC} of
\tc{EvilDerivable}, and \tc{NecNoEvil} of \tc{ThereIsNoEvil2}, whose derivation~\cite{J75}
presents as succeeding in its Figure~11.  The AFP sources~\cite{NotesAFP} close these five with \tc{oops}
and, for \tc{Th4}, postulate the statement afterwards, so that \tc{Th5}, \tc{MC} and the
ultrafilter results of the third variant rest there on an axiom.  These statements are not
open, and the port proves them.  Two further \tc{oops} lie outside the modal
development, in the Cantor illustration of Figure~2 of~\cite{J75}.  That theory states the
surjective Cantor theorem four times.  Twice it proves it, once interactively and once
avoiding proof by contradiction, and the port proves both.  The third statement is the same
proposition again, stated to show what the automated provers make of it and abandoned with
\tc{oops} after the remote Leo provers reported a proof.  The fourth is a deliberately wrong
formalisation, differing in the types, which Nitpick refutes.  Neither of the last two asserts
anything the port could check --- one repeats a proposition already proved, the other is
false --- so the port records them as comments rather than as \tc{example}s, and they are not
among the 45.  The five modal statements are provable in \isa{} as well: Isar proofs of all
five, closed by \tc{simp} and \tc{blast}, are in the \tc{Crosscheck} session of
\tc{tools/isabelle/} (\tc{Replay2026*.thy}) and build with Isabelle2025-2; what the
experiments lacked was a sledgehammer-based automated reconstruction of the proofs, not the
proofs, and each of them is short.
\tc{Ax1Gen} is the formalisation, following Anderson and Gettings~\cite{AndersonGettings1996},
of G\"odel's footnote that the conjunction axiom should cover ``any number of summands''.
\tc{Th4} --- possibly there is a God-like being --- follows from \tc{Ax1Gen} and \tc{Ax2a}
alone, with no frame condition, by three instances of \tc{Ax1Gen}.  For the set of all
positive properties it yields \tc{L}, that $G$ is positive, since $G$ is by definition their
conjunction; this is the \isa{} lemma.  For the empty set it makes the universal property
positive, the conjunction of nothing.  And for the singleton $\{G\}$ it makes every property
that is necessarily coextensive with $G$ positive --- in particular, if no God-like being
were possible, the empty property; the last two violate the exclusivity in \tc{Ax2a}.
\tc{NecNoEvil} needs only the empty-set instance: at no world is the empty property
positive, and an Evil-like being would have to possess it.  \tc{MC} of \tc{EvilDerivable}
uses no \tc{Ax1Gen} at all; it is the modal-collapse proof of the God-like variants with
\tc{Evil} for $G$.

Two remarks.  First, what the proofs draw from \tc{Ax1Gen}, beyond \tc{L}, is not its reach
to arbitrarily many, even infinitely many, summands but its two degenerate instances, which
the binary \tc{Ax1} cannot express: for the empty set, that the universal property is
positive; for a singleton, that positivity is closed under necessary coextension.  In the
second variant \tc{Ax4}
provides both; in the third, where \tc{Ax4} is weakened so as not to apply to an empty
$G$, the two instances of \tc{Ax1Gen} take its place.  Second, the proofs do not make
\tc{Th4} frame-free --- the \isa{} proof in the second variant already is --- but they show
that it needs neither \tc{Ax4} nor, in the third variant, the postulate.  That the
possible existence of a God-like being follows from the conjunction axiom and \tc{Ax2a} alone is arguably
more than G\"odel's footnote intends; the derivation disappears once the axiom is
restricted to non-empty sets of conjuncts, in the spirit of the third variant's non-empty
property inclusion.  The point is taken up in the companion article~\cite{NominalsArXiv}, which restricts the axiom to at
least two different conjuncts, the reading G\"odel's footnote suggests.  The
mixed-quantifier setting, where the argument falls short, is discussed in
Section~\ref{sec:axioms}.

\section{Which postulates each result uses}
\label{sec:axioms}

\lean{} records, for every theorem, the set of axioms its proof term depends on, and
\tc{\#print axioms} reports it, as \tc{thm\_deps} does in \isa{}.  Because the embedding postulates the frame conditions
as named axioms, this yields a per-theorem account of the frame conditions --- and hence
of the modal logic --- consumed by each proof.  Table~\ref{tab:frames} collects the data for the principal theorems; it is
generated from the sources by the script \tc{tools/print\_axioms.py}, which runs
\tc{\#print axioms} on every named theorem of every module (the complete output is
\tc{tools/axioms-report.txt}); the development's \tc{Makefile} regenerates it whenever
a module changes, and the script and its output are part of the ancillary files.

\begin{table}[t]
\centering\scriptsize
\input{tab-frames}
\caption{Postulates consumed by the proofs of the principal theorems, as reported by
  \tc{\#print axioms}; upper bounds for what the theorems require.  Omitted are only the three standard
  \lean{} axioms (\tc{propext}, \tc{Classical.choice}, \tc{Quot.sound}) and the
  embedding's signature constants (\tc{i}, \tc{e}, \tc{R}, \tc{existsAt}, \tc{P});
  everything else a proof depends on is listed, including the non-emptiness postulates
  \tc{iNonempty}.  The possibilist and mixed-quantifier variants agree with the entries
  of their actualist counterparts except where a row of their own is shown
  (\tc{GoedelVariantHOML2AndersonQuant}, Section~\ref{sec:axioms}).}
\label{tab:frames}
\end{table}

Three comments are in order.

First, these are \emph{upper} bounds, and they are the dependencies of one particular
proof.  \tc{\#print axioms} reports what that proof consumed, not what the theorem
requires, and the number depends on the proof: the first proofs of the ultrafilter results
and of the positivity lemmas of the third variant consumed reflexivity, where the \isa{}
proofs cite only symmetry, because they obtained a God-like being at the world under
consideration by \tc{Rrefl}; the first proof of the inconsistency consumed
\tc{eNonempty}, the non-emptiness of the type of individuals, only because a \tc{simp}
call happened to use it, and a first replay of it discharged one of its six steps by
reflexivity.  All three were found with \tc{\#print axioms}, the first two by comparing against
the \isa{} citations, and removed.  That comparison is the calibration used throughout: every entry of the table is at or
below what the corresponding \isa{} proof cites, once the postulated \tc{Th4} of the third
variant is replaced by its proof from \tc{Ax1Gen} and \tc{Ax2a}; and it is below in one
place, \tc{Th4} of the second variant in its actualist and possibilist settings, where
\tc{Ax4} is dispensable and \tc{Th5} and \tc{MC} inherit the saving.  The calibration has a
blind spot: it cannot flag a dependency that \isa{} has no name for, as \tc{eNonempty}
showed, and it would inherit any slack the \isa{} citation itself carries.  Each entry
should be read accordingly.  One more distinction matters for reading upper against lower
bounds.  The countermodels of~\cite{J75} are models of the embedding, in which the
property quantifiers range over all functions \tc{e $\rightarrow$ i $\rightarrow$ Prop} --- among them properties
such as ``being at world $w$'', which the modal object language cannot express.  The
proofs of this port instantiate property and proposition variables only with terms of the
object language, so the bounds of the table hold not only in that full semantics but for
every model whose properties are closed under the object-language operations.  An upper
bound of the table and a countermodel of~\cite{J75} therefore need not meet, and where
they do not, neither is at fault.  Establishing that a
frame condition is \emph{necessary} needs a countermodel.  \lean{} cannot find one, but
it can check one, as \isa{} can: a two-world Kripke frame refuting \tc{T3} in \textsf{K} is a finite
structure whose properties \tc{decide} can verify.  What prevents this here is a design
decision, not a limitation of the system: to keep the one-to-one correspondence with the
\isa{} sources that Section~\ref{sec:method} verifies, \tc{i}, \tc{e}, \tc{R} and \tc{P}
are global signature constants, declared with \tc{axiom}, and cannot be instantiated.  The alternative --- parameters in place of the
signature constants and hypotheses in place of the axioms, so that a countermodel is a term
of the right type --- is
demonstrated for two of the results in the \tc{palomar/} directory of the ancillary files
(the inconsistency in \textsf{K}, and Scott's \tc{T3}, each depending on no axiom beyond
\lean{}'s three), and is the natural next step for the 35 \tc{countermodel} statements.  In the present port
sharpness is settled by~\cite{J75}, not here: \tc{nitpick} refutes \tc{T3} and
\tc{MC} in \textsf{K} (its Figures~13 and~14) and \tc{T3} in \textsf{S4} (its
Section~5.3), so symmetry really is needed there.  The table corroborates the positive
half of that analysis and contributes nothing to the negative half.

Second, with that caveat, the data is informative.  Scott's \tc{T3} and modal collapse,
and the corresponding \tc{Th5} and \tc{MC} of the repaired G\"odel variants, consume
symmetry alone: the arguments go through in \textsf{KB}, and neither reflexivity nor
transitivity is touched.  This is the observation of~\cite{J75}, recorded there for these
variants and, for Scott's argument, already in the 2013 AFP entry~\cite{GoedelGod-AFP};
Kanckos and Woltzenlogel Paleo prove it in a natural deduction
calculus~\cite{KanckosPaleo2017}.  What the dependency account adds is that it holds
theorem by theorem.  The ultrafilter results consume nothing beyond that.  \tc{Th1},
\tc{T1}, \tc{T2} and \tc{Monotheism} consume no frame condition at all and hold already
in \textsf{K}; so does \tc{Th4}, the possible existence of a God-like being, in the
actualist and possibilist settings of both repaired variants, from \tc{Ax1Gen} and
\tc{Ax2a} alone (Section~\ref{sec:formalised}).

Third, the inconsistency of G\"odel's 1970 axioms consumes no frame condition.  This
confirms mechanically what~\cite{J75} already states in so many words about that
derivation: it ``does not require any specific properties of modal logic \textsf{S5}''
and ``is instead valid already in base modal logic \textsf{K}''.  The argument runs by
cases on whether some world is a dead end.  If $v$ has no successor, the premise of
\tc{Ax4} is vacuously true at $v$, so every property is positive at $v$, contradicting
the exclusivity in \tc{Ax2a}.  If no world is a dead end, seriality is forced --- not
assumed --- and under seriality the property $\lambda x.\,\Box\bot$, which \tc{Ax3} and
\tc{Ax4} show to be positive, \emph{is} the empty property, which \tc{Ax2a} and \tc{Ax4}
show not to be.  The derivation does consume \tc{iNonempty}, the existence of at least
one world, which \isa{} builds into every type and which the embedding has to postulate:
with no world at all every $\lfloor\cdot\rfloor$ holds vacuously and the 1970 axioms are
consistent.  It does not consume \tc{eNonempty}: an empty type of individuals leaves a
single property, and \tc{Ax2a} fails on it at once.

As a cross-check in the other direction the \isa{}
theory was copied with its import changed to \tc{HOMLinHOLonlyK}, which postulates nothing
about the accessibility relation, and with \tc{Th3} dropped, which needs the symmetry that is
not there: the proof of \tc{Inconsistency} replays unchanged, \tc{smt} steps included.  The theory (\tc{GoedelVariantHOML1inK}) is in the
\tc{Crosscheck} session of \tc{tools/isabelle/}, which builds with Isabelle2025-2 (log
included).

Relatedly, \tc{Th3} of \tc{GoedelVariantHOML1inS4} is proved from \tc{Ax3} and
reflexivity alone, without \tc{Ax4}.  This too is an observation of~\cite{J75}, which
reports that in the inconsistent variant ``a proof for theorem \tc{Th3} based on
reflexivity alone is indeed found by the theorem provers''; an explicit proof --- the
\lean{} one in the appendix, or the Isar proof \tc{Th3\_refl} in \tc{Replay2026S4.thy} of
the \tc{Crosscheck} session --- shows why, since the antecedent $\Diamond\exists^{E}x.\,G\,x$ is already refutable there, necessary
existence applied to the empty essence forcing the world to have no successor.

The one divergence from the pattern is \tc{Th4} in the mixed-quantifier setting of the
second variant, module \tc{GoedelVariantHOML2AndersonQuant}.  In the actualist and
possibilist settings \tc{Th4} follows from \tc{Ax1Gen} and \tc{Ax2a}; in the mixed
setting \tc{Ax1Gen} speaks of all beings at the successor worlds and \tc{Th4} only of the
actual ones, so the two-line argument yields only a \emph{possibilist} God-like being at
a successor, which has to be turned into an actual one --- the proof does this through
\tc{Th2} and symmetry, consuming \tc{Ax2b}, \tc{Ax3}, \tc{Ax4} and \tc{Rsymm} in
addition, and \tc{Th5} and \tc{MC} inherit \tc{Ax4}.  This is the \isa{} proof
(\tc{Ax2a Ax4 L Rsymm Th2}), and it is what \cite[Footnote~25]{J75} records for the
mixed-quantifier variant of its Figure~7: the proof ``requires symmetry of the
accessibility relation (or modal axiom B), as an additional dependency (reflexivity would
actually also do the job)''.  The footnote attributes the extra dependency to \tc{Th2};
in the AFP sources~\cite{NotesAFP}, as in the port, \tc{Th2} is frame-free and the dependency sits on
\tc{Th4}.  A separate footnote of~\cite{J75} notes that the corresponding variant of Figure~8 needs no
such addition \cite[Footnote~29]{J75}, and indeed in \tc{GoedelVariantHOML3AndersonQuant}
the conjunction in \tc{Ax1Gen} quantifies actualistically, the two-line argument applies,
and \tc{Th4} is frame-free there.  The companion article~\cite{NominalsArXiv} shows that
\tc{Th4} holds in \textsf{K} in the mixed setting of the second variant too, by a different
proof that uses no frame condition, so the symmetry is a property of this proof, not of
the theorem.

\section{The dataset in TPTP THF}
\label{sec:thf}

\subsection{The rendering}

The embedding used here~\cite{J26} was taken up by higher-order provers reading TPTP
THF~\cite{J22}, and that is the form in which Scott's variant of G\"odel's argument was
first mechanised~\cite{C40}.  Returning the dataset to that format does not depend on the port
--- it could equally be generated from the \isa{} sources --- and here it is done by a
metaprogram that prints each statement, a \lean{} term, as
THF.  For each of the
30 theories, \tc{tools/thf/export.py} writes the embedding together with that theory's
definitions and axioms as one THF axiom file, and each of its theorems as a problem of its
own --- 294 problems in TH0, the monomorphic fragment.  Worlds have type \tc{mu}
and individuals type \tc{ent}, a modal proposition is a term of type \tc{mu > \$o}, and the
connectives are definitions, so that the box reads

{\small\begin{verbatim}
thf(mbox_def, definition,
    mbox = (^[A: mu > $o, W: mu]: ![V: mu]: ((r @ W @ V) => (A @ V)))).
\end{verbatim}}

\noindent Why from the port and not from the \isa{} sources?  Only because it was at hand.
The \isa{} sources declare the connectives as \tc{abbreviation}s, which are unfolded in the
internal term, so an exporter has to fold the object-level structure back before printing it;
in the port they are definitions and the structure is already there.  That is a property of
the sources, not of the system.  Nothing hinges on the choice: the comparison of
Section~\ref{sec:method} finds all 548 statements identical as parsed, so what is generated are the
statements of the \emph{Notes}, and the model finder refutes here what it refutes there.

\noindent Three conventions.  TH0 has no type variables, so the few statements that are
polymorphic in the sources are instantiated at the type of entities.  A problem carries the
axioms of its theory and nothing else: the theorems proved earlier in the same theory are not
included, so a prover must find the whole chain, which makes these problems harder than the
corresponding step in the \emph{Notes}.  And the frame conditions are axioms of the theory
they belong to, so the \textsf{S4} and \textsf{K} theories yield problems in those logics.
Statements the \emph{Notes} refute are not among these problems; a countermodel is not a proof
task, and they form a second set of their own, discussed below.

\subsection{The provers and their results}
\label{sec:thf-results}

\begin{table}[ht]
\centering
\input{tab-thf.tex}
\caption{The 294 THF problems at 10 and at 60 seconds on one core and on the whole machine,
  against the four higher-order provers Isabelle2025-2 bundles, E~\cite{Eprover},
  Vampire~\cite{HOVampire2024}, Zipperposition~\cite{Zipperposition2023} and cvc5~\cite{cvc5},
  and against Leo-II~\cite{J29,LeoII20} and Leo-III~\cite{J53}; the versions are in the rows.
  Each prover's first row is its recommended configuration (Section~\ref{sec:thf-results}), the
  rows beneath it are further configurations, named by their options; a dash marks a column a
  configuration does not have.  Provers are ordered by their first rows, summed over the three
  columns.  ``Alone'' counts the problems a prover proves in its first row at 10 seconds that no
  other first row proves; the last row is the union of the first rows, with the one-core figures
  of Leo-III and cvc5 in the machine column.  cvc5 runs on the SMT-LIB rendering of the same
  statements.}
\label{tab:thf}
\end{table}

The table measures what each prover proves in a stated configuration; it is not a controlled
ranking of speed.  E and Leo-II run under a limit on processor time, the others under one on
elapsed time, so ten seconds is not quite the same resource in every row, and a prover close to
its limit may fall on either side of it from one run to the next: of Leo-II's problems, two were
proved in three and in two of ten runs.  A difference of one or two problems between two
provers is therefore not meaningful.
Of the \thfProblems{} problems, \thfSolvedAny{} are proved by at least one prover within ten
seconds on one core; \thfInconsistentSolved{} of those come from the three theories that
formalise G\"odel's 1970 axioms, where everything is provable because the axioms are
inconsistent, so the figure over the theories with consistent axioms is
\thfConsistentSolved{} of \thfConsistentTotal{}.  \thfIncBest{} establishes that inconsistency
in \thfIncBestCount{} of the \thfInconsistentTotal{}, the first of them in \thfIncBestSeconds{}
seconds --- the result automated provers found first~\cite{C55}; Leo-II~2.2 establishes it in one
of the three at ten seconds and in all three at sixty.  What resists is what needs a
higher-order instantiation of a set variable: \tc{L} above all, where the conjunction axiom
has to be applied to the positive properties themselves, and everything downstream of it ---
the third-order quantification that G\"odel's ``any number of summands'' brings in.

The two provers of the \textsc{Leo} family are not part of the \isa{} distribution and were
obtained separately.  Leo-III~1.7.0 installs from its jar, which calls E through a helper
the jar carries for Linux on x86-64 only; on the machine of these measurements no call to E
starts until that helper is replaced by a native one (\tc{anc/tools/thf/treelimitedrun/}).
Leo-II, last released about a decade ago,
no longer built with a current toolchain nor cooperated with a current E, so part of this
work was to modernise and fix it; the result is Leo-II~2.2~\cite{LeoII20}, and that is the
version the table measures (Section~\ref{sec:leo2}).  Its calculus and its main loop are unchanged:
the work fixed errors and made individual algorithms more efficient, among them how often it
calls its first-order partner and in what order it spends its time slices.

The 45 statements the \emph{Notes} do not prove travel with the set as well, as problems for
which a proof is not the expected answer.  Nitpick, which the \emph{Notes} use and
which Isabelle's TPTP interface runs on these very files, finds a countermodel for 35 of them
and gives up on the remaining ten --- exactly the ten the \emph{Notes} record as open.  Since cvc5 no longer
reads TPTP, the dataset is rendered a second time in SMT-LIB with higher-order sorts; there
cvc5 proves \thfCvcProved{} of the theorems, which is its row in the table, and finds
\thfCvcCountermodels{} countermodels.  In its recommended configuration at ten seconds, cvc5
there with finite-model finding, no prover claims a proof of any of the 45, which for the 35 the
\emph{Notes} refute would signal an error and for the ten they leave open would agree with the
companion article~\cite{NominalsArXiv}, which proves all ten.

Each prover is called in its \emph{recommended configuration}, the one used for it in CASC,
on SystemOnTPTP or in Sledgehammer,\footnote{What counts as a prover's automatic mode is evident from the prover
itself only for E and Leo-II, whose documented automatic modes are the ones measured here.
Vampire's portfolio runs a first-order schedule unless a higher-order one is named, and its
CASC mode in version~4.8 runs that first-order schedule too;
Zipperposition's schedules are scripts distributed with its sources rather than modes of the
executable, and several of their configurations require a later version than 2.1; Leo-III
works without a first-order partner unless one is given, although its developers give it E in
CASC and Sledgehammer does the same; and cvc5 proves more with \tc{--full-saturate-quant},
which Sledgehammer passes in every slice it uses for proving.  A single
documented automatic mode that selects such configurations by itself would make measurements like these
easier to reproduce.} with its time limit and nothing else tuned to the dataset:

{\footnotesize\begin{verbatim}
eprover   --auto-schedule --tptp3-format --cpu-limit=T -s P
vampire   --mode portfolio --schedule snake_tptp_hol --cores 1 --input_syntax tptp -t T P
python3   portfolio.sh.sequential.py P T DIR      (Zipperposition, cut off at T)
cvc5      --full-saturate-quant --tlimit 1000*T P.smt2   (--finite-model-find for those 45)
leo       --atp e=eprover -t T P
java      -Xss32m -Xmx1g -jar leo3.jar P -t T -a e=eprover
\end{verbatim}}

\noindent Zipperposition's schedule is its developers' sequential portfolio as SystemOnTPTP
runs it, a script of 15 configurations, of which 2.1 rejects 8 because they need options of
later versions.  Every one it accepts passes \tc{--tptp-def-as-rewrite}, without which
Zipperposition does not unfold the definitions of the embedding and proves almost nothing.
The script does not stop at the budget by itself and is cut off there.  Leo-II and Leo-III were built to explore how a
higher-order prover can cooperate with a first-order one~\cite{C16}, and they are given E as that
partner; Leo-II runs on one core (\tc{--cores}~1, its default).  E then receives only
first-order clauses, not the THF problem it is given in its own row: the higher-order
reasoning, the instantiation of set variables included, is Leo-II's and Leo-III's own, so
their rows measure that cooperation.  Nothing else is set: no strategy is chosen per problem, and
no prover is given a list of the axioms it should use.  Leo-III's figures without E were
measured before this invocation was adopted, with the JVM's default stack: with the 32\,MB of
Leo-III's own launcher it proves one problem more at ten seconds, 160 rather than
\thfPlainLeoIII{}, and its column in Table~\ref{tab:strata} was measured the same way.  One prover is not independent of the
dataset, though: Leo-II~2.2 was repaired and improved on these very problems
(Section~\ref{sec:leo2}), so that its row is not a measurement on unseen problems.  On the
problems of the library that share their profile, on which it was not tuned, though a sample
across the library that served as a check of its changes overlaps them in part, it proves 262 of 438 where E proves 267 (Table~\ref{tab:strata}).

Six times the budget moves little.  At 60 seconds the six recommended configurations together prove
\thfSolvedAnySixty{} of the problems rather than \thfSolvedAny{}, and only \thfSixtyNew{} of
the problems are proved that were not proved at 10 seconds; \thfUnsolvedSixty{} resist every
prover at either budget.  What changes is which prover the extra time helps: Vampire gains
\thfGainVampire{} problems, Leo-III \thfGainLeoIII{}, cvc5 \thfGainCvc{}, Leo-II
\thfGainLeoII{}, E \thfGainE{} and Zipperposition \thfGainZipperposition{}.  For the two
provers at the head of the table the budget is thus not the binding constraint; for Vampire's
schedule it is.  Ten cores instead of one move as little: the six together
prove \thfSolvedAnyMachine{} rather than \thfSolvedAny{}.  What the whole machine means is each
prover's own: E a processor count for its schedule, Vampire its portfolio over all cores,
Zipperposition the parallel portfolio its developers entered in CASC, Leo-II six configurations
at once, and cvc5 its portfolio over ten jobs without options; Leo-III has no such
configuration.  Offered all ten cores, Vampire's portfolio uses fewer: on six problems it does
not solve, it draws between seven and a half and eight cores' worth of processor time in its ten
seconds (\tc{anc/thf/results-machine-cores.csv}).

Four choices a reader may question.  The provers are the versions Isabelle2025-2 bundles, so
that a single installation reproduces the table; later versions exist, among them Vampire~5.0,
the CASC-30 winner in higher-order logic~\cite{CASC30}, and were not measured.  On the machine
of these measurements Isabelle brings Vampire and Zipperposition as x86-64 builds, which run
under Rosetta~2.  Vampire's figures are those of that binary; a native build of the same
version proves \thfNativeVampire{} rather than \thfTenVampire{} at ten seconds and as many,
\thfSixtyVampire{}, at sixty, which leaves the order of the table unchanged.  Zipperposition's
are those of a native build of the same version, 2.1 from opam, the emulated one proving
\thfEmulatedZipperposition{} rather than \thfTenZipperposition{}
(\tc{anc/thf/native-vampire/}, \tc{anc/thf/emulated-zipperposition/}).  The library of
Section~\ref{sec:library} is TPTP v7.5.0, the release installed when that run was made; a later
one would change the counts, not the method.  Satallax, another higher-order prover of 2013,
is not among the systems Isabelle bundles and appears only in that section's comparison with
the provers of the time.  And the machine is a laptop with performance and efficiency cores
rather than a cluster node: the figures are counts at a limit, compared across provers on one
machine, not timings that transfer to another.

The set is offered as a resource rather than as a measurement of proof automation, whose
results depend on versions, time limits and hardware.  It makes the dataset usable without a
proof assistant, and it is included with this paper as ancillary material (\tc{anc/thf/} and
\tc{anc/smt/}, with the generator and the runners under \tc{anc/tools/thf/}).

\subsection{The problems in the library}
\label{sec:library}

\begin{table}[t]
\centering\small
\begin{tabular}{lrrrrr}
\toprule
stratum & problems & E & Leo-II & Vampire & Leo-III\\
\midrule
a variable applied to a lambda & 872 & 510 & 214 & 265 & 203\\
\quad no such application & 2765 & 1907 & 1675 & 1539 & 1428\\
100 formulas or more & 1033 & 614 & 282 & 318 & 174\\
\quad fewer than 100 & 2604 & 1803 & 1607 & 1486 & 1457\\
\midrule
the profile of these problems & 438 & 267 & 262 & 207 & 193\\
these problems & \thfProblems{} & 218 & 218 & 174 & 159\\
\bottomrule
\end{tabular}
\caption{The 3637 monomorphic TH0 problems of TPTP v7.5.0 by stratum, at ten seconds on
  one core, in the versions of Table~\ref{tab:thf} but with Vampire without options and Leo-III
  without E, and the same figures for the problems of this paper, which are those of the
  corresponding rows of Table~\ref{tab:thf}.  The profile is: a lambda somewhere, no variable applied to a lambda, no Boolean
  argument, highest type order three, between 50 and 100 formulas.  Zipperposition and cvc5
  were not run over the library.}
\label{tab:strata}
\end{table}

Are these problems representative?  The library answers, and not as its totals suggest.  Over
its 3637 monomorphic TH0 problems Leo-II proves 1889 where E proves 2417, but that total is a
weighted average of family sizes, and the largest families are exports from mathematical
libraries.  Leo-II was built to explore small problems in cooperation with a first-order
prover, not to search a large library.  What separates the provers is less the height of the
types than what is done with them.  Table~\ref{tab:strata} gives the two strata that separate
them most.  A variable applied to a lambda --- the application that forces higher-order
unification --- costs Leo-II more than half of what it otherwise reaches relative to E, and size
nearly as much.  The highest type order barely separates anything: at orders two and three Leo-II
stands at 0.73 and 0.74 of E, at order one at 0.96.  Every one of the \thfProblems{} problems
of this paper contains lambdas and applied variables, and not one applies a variable to a
lambda or quantifies over anything with a Boolean argument; the highest type order is three in
291 of them, and the median problem has 82 formulas.  The library holds 438 problems of that
profile, and there the four provers stand as they stand here.  These problems are therefore a
delimited region of the library rather than an outlier, and that E and Leo-II prove as many on
them is a property of the region.

The argument itself is in that region.  The \tc{PHI} family of TPTP v7.5.0, the version measured throughout this section, is the argument as it
was first mechanised~\cite{C40,C55}, mostly in Scott's variant: 19 problems, 14 of them of the profile above, three of them
asserting the inconsistency of the 1970 axioms.  Two higher-order provers of that time
prove 7 of the 19 (Satallax~2.7~\cite{Satallax}, of April 2013) and 9 (Leo-II~1.6.0, of
August 2013), against 12 for Leo-II~2.2 and 14 for E; and of the three inconsistency problems
Satallax proves none at this budget while Leo-II proves all three, in the code of 2013 as in
2.2, E answering \tc{ContradictoryAxioms} to them.  That the inconsistency was first reported
by Leo-II is, of the provers measured here, not surprising.  Four cautions belong with that.
Both old provers talk to a current E, which is an anachronism; ten seconds is not the budget
under which the original experiments ran; E, Vampire and cvc5 acquired their higher-order
reasoning only in recent years, so they say nothing about that time; and Leo-II~1.6.0 ran past
its own limit on two problems and had to be stopped from outside, which is the missing
deadline that 2.2 repairs.  It is built as it stands, on a compiler configured for the string
semantics of its day and with the C++ repair its bundled MiniSAT needs under a current
compiler, and it reaches E through the wrapper in \tc{anc/tools/thf/}.  The library run, the
stratification and every figure of this subsection are under \tc{anc/thf/tptp-library/}.

\section{Leo-II 2.2}
\label{sec:leo2}

Leo-II is a prover of this dataset's own history, and the version reachable when this work
began could neither be built with a current toolchain nor call a current E.  Repairing it
was a precondition of its row in Table~\ref{tab:thf}, and it is a side result of this work
that Leo-II, winner of the THF division at CASC-J5 in 2010~\cite{CASCJ5}, is available to
the community again, as release 2.2.  What follows documents that release, so that the row
can be reproduced.  Repair work turned into improvement.  The calculus is untouched --- not one inference was added, removed
or changed.  Most of what the improvement reflects is arithmetic and bookkeeping.  The time budget
was cut into slices too short to hold the first-order calls they scheduled.  Indexed terms,
which are equal exactly when their integer identifiers are, were compared structurally, so
every comparison cost the size of everything proved so far.  And the grammar could not read a
negated quantification, though that is ordinary THF: on this dataset it never occurs, but of
584 TH0 problems sampled across the TPTP library 197 could not be read at all, and none of them
resists now.  What that repair left is a second count: over the whole TH0 part of
the library, 28 of 3637 problems still defeated its parser or its choice rule, and 2.2 reads all 28 and proves five of them.  Twenty-four are the type at
which a quantifier combinator is applied, which THF writes as \tc{?? @ \$o @ (P @ I)} and the
grammar did not know.  Three are a guard in the choice rule that rejects a term carrying a
variable the clause does not have free, and the literal was built before the guard could speak,
which is what raised the exception.  The failure on the twenty-eighth could not be
reproduced.  It is worth noting that neither defect shows on a curated benchmark: both were
found only by running the whole library.  That run travels with this paper, under
\tc{anc/thf/tptp-library/}: the script, the class of each of the 4515 problems whose name
carries a \tc{\textasciicircum}, and one row per problem and prover, so that the figures of
this paragraph can be recounted.  Every figure of this section is in the
\tc{CHANGES} file of the Leo-II release, beside the code it describes and with what was
measured and rejected alongside, or in the results the tables are computed from; the release is at the page named in
Section~\ref{sec:availability}.

Four changes are of substance, and none is in the calculus.  The first is when Leo-II calls
its first-order partner: after every iteration of its main loop, which is what it had always
done.  Each call costs the whole second a time slice allows it, so ten calls exhaust a ten
second limit, and on some problems the loop then ran no iterations at all --- every call
answering that it had saturated, so the clauses a refutation needs had not been derived,
because deriving them is what the calls were preventing.  Leo-II now calls it every fifth
iteration instead, which was worth 25 problems when it was made, 205 against 180.  The second is the encoding: Leo-II
tags every subterm it hands that prover with its type, and a type was itself a term, so a
single atom arrived as some 150 characters of which most was type.  A ground function type is
now one constant --- the same encoding up to a renaming, so that it neither adds nor removes a
refutation, but without the weight a first-order prover would carry through every ordering
comparison and every index lookup.  The third is the order of the time slices: Leo-II's
relevance filter, which keeps only axioms sharing symbols with the conjecture, loses as a
default because it drops axioms a proof needs, but what it finds it finds in about a second,
so a short filtered slice now comes first and the unfiltered search gets the rest, which is
worth twelve, 217 against 205.  Given the whole machine it runs six configurations at once and proves 226 of the problems,
which is its whole-machine entry in the table.  On one
core it proves as many as E, and no order between the two should be read into the table
either way (Section~\ref{sec:thf-results}).  A fourth change is what the
budget is measured in.  Ten seconds used to mean ten seconds of wall clock, so a figure
depended on what else the machine was doing; it is now ten seconds of processor time, of
Leo-II and of the first-order prover it waits for together, which is the kind of second
that prover is given anyway.  That buys a problem and, more to the point, makes the figure
independent of the load; it does not make the two problems of
Section~\ref{sec:thf-results} reproducible.  The patches that make 1.7.0 build
and call a current E travel with the dataset; what came after them is in 2.2 itself.

\section{Availability and reproduction}
\label{sec:availability}

The development is a \tc{lake} package requiring \lean{} 4.33.1 and no further
dependencies; \tc{lake build} type-checks all 30 modules.  A separate document in the
style of the Archive of Formal Proofs, containing the complete sources with their
prose, is generated from the sources themselves.  Appendix~\ref{app:sources}
reproduces it, and it is among the ancillary files as \tc{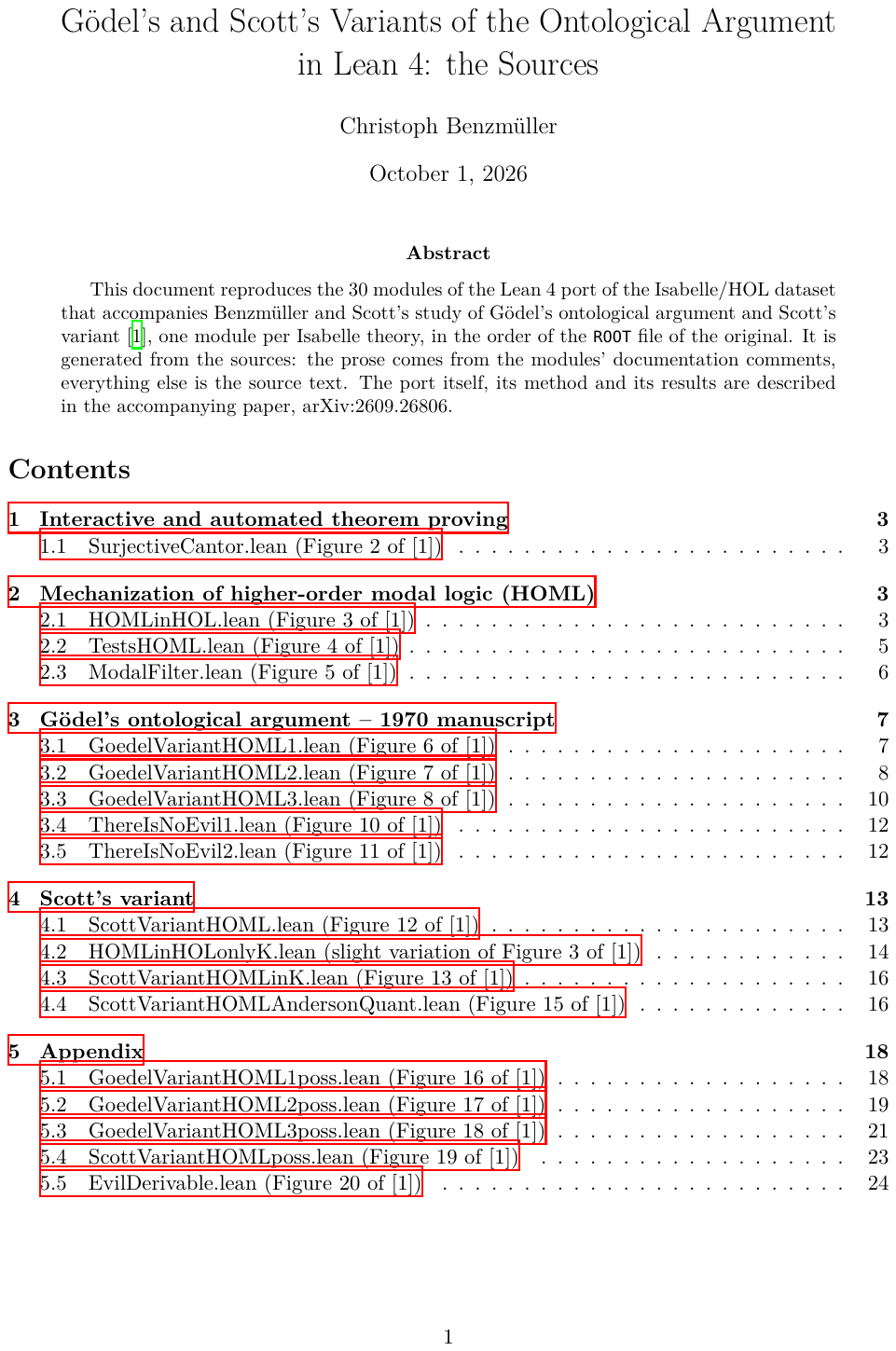}.

\paragraph{Sources.} The \lean{} sources are distributed with this paper as ancillary
files (directory \tc{anc/}): the 30 modules, \tc{lakefile.toml}, \tc{lake-manifest.json}
and \tc{lean-toolchain}, together with a \tc{README} --- a self-contained \tc{lake}
package.  Unpack it and run \tc{lake build}.  Also included, under \tc{anc/tools/}, are
the comparison tool of Section~\ref{sec:method} (\tc{compare\_statements.py}), the script
that generates Table~\ref{tab:frames} (\tc{print\_axioms.py}) with its complete output,
the \lean{} parse-agreement files (\tc{lean-check/}), and the two \isa{} check sessions
(\tc{isabelle/}) together with the log of their build (\tc{isabelle/BUILD-LOG.txt}); the
comparison tool and the \isa{} sessions expect the \isa{} sources of the Archive of Formal
Proofs entry~\cite{NotesAFP} alongside, in the release of 6~February 2026, which is the one
that accompanies Isabelle2025-2 and is archived at
\url{https://www.isa-afp.org/release/afp-Notes_On_Goedels_Ontological_Argument-2026-02-06.tar.gz}.
The sources the comparison was made against are pinned by their checksums in
\tc{anc/tools/afp-sources.sha256}, and the tool warns when a reader's copy differs.

\paragraph{The THF and SMT-LIB renderings.}  Both travel as ancillary files as well,
\tc{anc/thf/} and \tc{anc/smt/}, each with the \thfProblems{} theorems, the
\thfNonTheorems{} statements the \emph{Notes} do not prove, and the results of every
prover as CSV.  The metaprogram that prints them from the \lean{} sources is
\tc{anc/tools/thf/export.py}, and the runners beside it reproduce the measurements:
\tc{run\_provers.py} for E, Vampire and Zipperposition and for the two of the \textsc{Leo}
family, \tc{run\_cvc5.py} for cvc5 on the SMT-LIB rendering, \tc{run\_nitpick.py} for the
countermodels.  The runners need Isabelle2025-2, which brings E, Vampire,
Zipperposition and cvc5 with it; Leo-II~2.2 and Leo-III~1.7.0 are installed separately and
are used when found.  On macOS, Isabelle brings Vampire and Zipperposition as x86-64 builds
only; \tc{anc/tools/thf/build\_native.sh} builds both natively, and
\tc{anc/tools/thf/treelimitedrun/build.sh} the Leo-III jar whose calls to E start there.  Every prover can be pointed at a different binary through an
environment variable, and the invocation of each is printed by the runner and listed in
\tc{anc/thf/README.md}.  Leo-II~2.2, and every earlier release, is at
\url{https://christoph-benzmueller.de/leo/download.html}; it is BSD-licensed and builds with
\tc{./configure} and \tc{make opt} in the \tc{src} directory of its archive.

\paragraph{The machine.}  Table~\ref{tab:thf} was measured on a MacBookPro~18,1 with an
Apple M1 Pro processor and 16\,GB memory under macOS~15, one core per prover and at most
two problems at a time, except the whole-machine column, which runs one problem at a time.  Its ten cores are not alike, eight performance and two efficiency,
so a prover on one of the latter gets less of a second than the limit suggests; and the
timings of a cooperative prover move with whatever else the machine is doing, by how much
Section~\ref{sec:leo2} says.  Isabelle2025-2 brings Vampire and Zipperposition for this
machine as x86-64 builds only, which run under Rosetta~2 (Section~\ref{sec:thf-results}).

\section*{Acknowledgements}

The \lean{} port and this paper were produced with the tool support described in
Section~\ref{sec:method}.

\bibliographystyle{abbrv}
\bibliography{root}

\appendix
\addtocontents{toc}{\protect\setcounter{tocdepth}{3}}
\section{The \texorpdfstring{\lean{}}{Lean 4} sources}
\label{app:sources}

The following is generated directly from the \lean{} modules, in the order of the
\tc{ROOT} file of the \isa{} dataset.  Prose comes from the modules' documentation
comments; everything else is the source text.

\input{lean}

\end{document}

%% file: thf-numbers.tex
\newcommand{\thfProblems}{294}
\newcommand{\thfNonTheorems}{45}
\newcommand{\thfSolvedAny}{227}
\newcommand{\thfInconsistentSolved}{24}
\newcommand{\thfConsistentSolved}{203}
\newcommand{\thfConsistentTotal}{270}
\newcommand{\thfInconsistentTotal}{3}
\newcommand{\thfIncBest}{E}
\newcommand{\thfIncBestCount}{3}
\newcommand{\thfIncBestSeconds}{8}
\newcommand{\thfCvcProved}{160}
\newcommand{\thfCvcCountermodels}{22}
\newcommand{\thfSolvedAnySixty}{232}
\newcommand{\thfSixtyNew}{5}
\newcommand{\thfUnsolvedSixty}{62}
\newcommand{\thfSolvedAnyMachine}{232}
\newcommand{\thfPlainVampire}{174}
\newcommand{\thfPlainZipperposition}{101}

\newcommand{\thfPlainLeoIII}{159}
\newcommand{\thfGainE}{2}
\newcommand{\thfGainVampire}{10}
\newcommand{\thfGainZipperposition}{2}
\newcommand{\thfGainCvc}{4}
\newcommand{\thfGainLeoII}{3}
\newcommand{\thfGainLeoIII}{8}

\newcommand{\thfTenVampire}{209}
\newcommand{\thfTenZipperposition}{213}

\newcommand{\thfTenLeoIII}{177}
\newcommand{\thfNativeVampire}{212}

\newcommand{\thfSixtyVampire}{219}

\newcommand{\thfEmulatedZipperposition}{203}

%% file: tab-loc.tex
\begin{tabular}{lrrrr}
\toprule
 & total & prose & declarations & proof\\
\midrule
\isa{} (30 theories) & 1157 & 180 & 765 & 212\\
\lean{} (30 modules) & 1868 & 517 & 854 & 497\\
\bottomrule
\end{tabular}

%% file: tab-frames.tex
\setlength{\tabcolsep}{4pt}%
\begin{tabular}{llll}
\toprule
Module & Result & Frame conditions & Other postulates used\\
\midrule
\tc{GoedelVariantHOML1} & \tc{Inconsistency} & --- & \tc{Ax2a Ax3 Ax4 iNonempty}\\
\tc{GoedelVariantHOML1inS4} & \tc{Th3} & \tc{Rrefl} & \tc{Ax3}\\
\tc{GoedelVariantHOML2} & \tc{Th1} & --- & \tc{Ax2a Ax2b}\\
\tc{GoedelVariantHOML2} & \tc{Th4} & --- & \tc{Ax1Gen Ax2a}\\
\tc{GoedelVariantHOML2} & \tc{Th5}, \tc{MC} & \tc{Rsymm} & \tc{Ax1Gen Ax2a Ax2b Ax3}\\
\tc{GoedelVariantHOML2} & \tc{Monotheism} & --- & \tc{Ax2a}\\
\tc{GoedelVariantHOML2} & \tc{PisUFilter} & \tc{Rsymm} & \tc{Ax1 Ax1Gen Ax2a Ax2b Ax3 Ax4}\\
\tc{GoedelVariantHOML2AndersonQuant} & \tc{Th4} & \tc{Rsymm} & \tc{Ax1Gen Ax2a Ax2b Ax3 Ax4}\\
\tc{GoedelVariantHOML2AndersonQuant} & \tc{Th5}, \tc{MC} & \tc{Rsymm} & \tc{Ax1Gen Ax2a Ax2b Ax3 Ax4}\\
\tc{GoedelVariantHOML3} & \tc{Th4} & --- & \tc{Ax1Gen Ax2a}\\
\tc{GoedelVariantHOML3} & \tc{Th5}, \tc{MC} & \tc{Rsymm} & \tc{Ax1Gen Ax2a Ax2b Ax3}\\
\tc{GoedelVariantHOML3} & \tc{PisUFilter} & \tc{Rsymm} & \tc{Ax1 Ax1Gen Ax2a Ax2b Ax3 Ax4}\\
\tc{ScottVariantHOML} & \tc{T1} & --- & \tc{A1 A2}\\
\tc{ScottVariantHOML} & \tc{T2} & --- & \tc{A1 A4}\\
\tc{ScottVariantHOML} & \tc{T3}, \tc{MC} & \tc{Rsymm} & \tc{A1 A2 A3 A4 A5}\\
\tc{ScottVariantHOML} & \tc{Monotheism} & --- & \tc{A1}\\
\tc{ScottVariantHOML} & \tc{PisUFilter} & \tc{Rsymm} & \tc{A1 A2 A3 A4 A5}\\
\tc{ThereIsNoEvil1} & \tc{NecNoEvil} & --- & \tc{Ax2a Ax4}\\
\tc{ThereIsNoEvil2} & \tc{NecNoEvil} & --- & \tc{Ax1Gen Ax2a}\\
\tc{EvilDerivable} & \tc{T5}, \tc{MC} & \tc{Rsymm} & \tc{A1 A2 A3 A4 A5}\\
\tc{EvilDerivable} & \tc{NisUFilter} & \tc{Rsymm} & \tc{A1 A2 A3 A4 A5}\\
\bottomrule
\end{tabular}%

%% file: tab-thf.tex
\begin{tabular}{llrrrr}
\toprule
prover & configuration & 10\,s & 60\,s & machine & alone\\
\midrule
E~3.2.5-ho & \tc{--auto-schedule} & 218 & 220 & 227 & 8\\
 & \quad \tc{--auto} & 209 & 212 & -- & \\
Leo-II~2.2 & with first-order E & 218 & 221 & 226 & 1\\
Vampire~4.8~HO & \tc{snake\_tptp\_hol} & 209 & 219 & 218 & 0\\
 & \quad no options & 174 & 174 & -- & \\
 & \quad \tc{--mode casc} & 167 & 186 & 185 & \\
Zipperposition~2.1 & developers' portfolio & 213 & 215 & 217 & 0\\
 & \quad \tc{ho-competitive} & 101 & 109 & -- & \\
Leo-III~1.7.0 & with first-order E & 177 & 185 & -- & 0\\
 & \quad without E & 159 & 166 & -- & \\
cvc5~1.2.0 (SMT-LIB) & \tc{--full-saturate-quant} & 160 & 164 & -- & 0\\
 & \quad no options & 148 & 150 & 148 & \\
\midrule
at least one & & 227 & 232 & 232 & \\
\bottomrule
\end{tabular}

%% file: lean.tex
\subsection{Interactive and automated theorem proving}
\subsubsection{SurjectiveCantor.lean (Figure 2 of \texorpdfstring{\cite{J75}}{the Notes})}
The surjective Cantor theorem is used in \cite{J75} to illustrate some aspects of interactive and automated theorem proving as relevant for the paper.

\begin{Verbatim}[commandchars=\\\{\},fontsize=\scriptsize,xleftmargin=1em]
/-- Auxiliary: no proposition is equivalent to its own negation. -/
theorem noFixpoint \{p : Prop\} (h : p = \ensuremath{\neg}p) : False :=
  have k : \ensuremath{\neg}p := fun x => (iff_of_eq h).1 x x
  k ((iff_of_eq h).2 k)
\end{Verbatim}
Surjective Cantor theorem: traditional interactive proof

\begin{Verbatim}[commandchars=\\\{\},fontsize=\scriptsize,xleftmargin=1em]
theorem SurjectiveCantor \{\ensuremath{\alpha} : Type\} : \ensuremath{\neg} \ensuremath{\exists} G : \ensuremath{\alpha} \ensuremath{\rightarrow} (\ensuremath{\alpha} \ensuremath{\rightarrow} Prop), \ensuremath{\forall} F : \ensuremath{\alpha} \ensuremath{\rightarrow} Prop, \ensuremath{\exists} X : \ensuremath{\alpha}, G
    X = F := by
  intro h1
  let \ensuremath{\langle}g, h2\ensuremath{\rangle} := h1                                   -- `g` is assumed surjective
  let F := fun X => \ensuremath{\neg} g X X                           -- the diagonal property
  let \ensuremath{\langle}a, h4\ensuremath{\rangle} := h2 F                                 -- `g a = F`
  have h5 : g a a = F a := congrFun h4 a
  exact noFixpoint h5
\end{Verbatim}
Avoiding proof by contradiction (Fuenmayor \& Benzm\"uller, 2021)

\begin{Verbatim}[commandchars=\\\{\},fontsize=\scriptsize,xleftmargin=1em]
theorem SurjectiveCantor' \{\ensuremath{\alpha} : Type\} : \ensuremath{\neg} \ensuremath{\exists} G : \ensuremath{\alpha} \ensuremath{\rightarrow} (\ensuremath{\alpha} \ensuremath{\rightarrow} Prop), \ensuremath{\forall} F : \ensuremath{\alpha} \ensuremath{\rightarrow} Prop, \ensuremath{\exists} X : \ensuremath{\alpha},
    G X = F :=
  have h5 : \ensuremath{\forall} G : \ensuremath{\alpha} \ensuremath{\rightarrow} (\ensuremath{\alpha} \ensuremath{\rightarrow} Prop), \ensuremath{\exists} F : \ensuremath{\alpha} \ensuremath{\rightarrow} Prop, \ensuremath{\forall} X : \ensuremath{\alpha}, \ensuremath{\neg} (G X = F) :=
    fun g => \ensuremath{\langle}fun Z => \ensuremath{\neg} g Z Z, fun X h => noFixpoint (congrFun h X)\ensuremath{\rangle}
  fun \ensuremath{\langle}G, h\ensuremath{\rangle} => let \ensuremath{\langle}F, hF\ensuremath{\rangle} := h5 G; let \ensuremath{\langle}X, hX\ensuremath{\rangle} := h F; hF X hX
\end{Verbatim}
Surjective Cantor theorem: automated proof by some internal/external theorem provers

\begin{Verbatim}[commandchars=\\\{\},fontsize=\scriptsize,xleftmargin=1em]
-- SurjectiveCantor'' : \ensuremath{\neg} \ensuremath{\exists} G : \ensuremath{\alpha} \ensuremath{\rightarrow} (\ensuremath{\alpha} \ensuremath{\rightarrow} Prop), \ensuremath{\forall} F : \ensuremath{\alpha} \ensuremath{\rightarrow} Prop, \ensuremath{\exists} X : \ensuremath{\alpha}, G X = F
--   `nitpick[expect=none]`: no counterexample found.
--   `sledgehammer`: most internal provers give up; `remote_leo2`/`remote_leo3` find a proof.
--   Abandoned in Isabelle with `oops`.  In Lean, `grind` likewise gives up: the proof needs the
--   diagonal witness supplied explicitly, as above.
\end{Verbatim}
Surjective Cantor theorem (wrong formalization attempt): the types are crucial

\begin{Verbatim}[commandchars=\\\{\},fontsize=\scriptsize,xleftmargin=1em]
-- SurjectiveCantor''' : \ensuremath{\neg} \ensuremath{\exists} G : \ensuremath{\alpha} \ensuremath{\rightarrow} \ensuremath{\beta}, \ensuremath{\forall} F : \ensuremath{\beta}, \ensuremath{\exists} X : \ensuremath{\alpha}, G X = F
--   `nitpick`: countermodel found for card \ensuremath{\alpha} = 1 and card \ensuremath{\beta} = 1 (G = \ensuremath{\lambda}x. (a1 := b1)).
--   `nitpick[satisfy, expect=genuine]`: model found for card \ensuremath{\alpha} = 1 and card \ensuremath{\beta} = 2.
--   `nitpick[card \ensuremath{\alpha}=2, card \ensuremath{\beta}=3, expect=none]`: no counterexample found.
--   Abandoned in Isabelle with `oops`.
\end{Verbatim}
\subsection{Mechanization of higher-order modal logic (HOML)}
\subsubsection{HOMLinHOL.lean (Figure 3 of \texorpdfstring{\cite{J75}}{the Notes})}
Shallow embedding of higher-order modal logic (HOML) in the classical higher-order logic (HOL) of Lean 4, utilizing the logic-pluralistic LogiKEy methodology.  Here logic S5 is introduced.

The global parameter settings of the Isabelle sources configure the model finder \texttt{nitpick} and the parser; they have no Lean counterpart, as Lean has no model finder.

\begin{Verbatim}[commandchars=\\\{\},fontsize=\scriptsize,xleftmargin=1em]
-- nitpick_params[user_axioms,expect=genuine,show_all,format=2,max_genuine=3]
-- declare[[syntax_ambiguity_warning=false]]
\end{Verbatim}
Lean's \texttt{Prop} is intuitionistic, HOL is classical; classical reasoning is enabled globally.

\begin{Verbatim}[commandchars=\\\{\},fontsize=\scriptsize,xleftmargin=1em]
attribute [instance] Classical.propDecidable

/-- Counterpart of Isabelle's `oops` after `nitpick` found a countermodel.  Lean has no model
finder, so such statements are only type-checked, never proved (and never added to the context). -/
macro "countermodel" : tactic => `(tactic| sorry)
/-- Counterpart of Isabelle's `oops` on a statement left genuinely open: neither a proof
nor a countermodel was obtained. -/
macro "openproblem" : tactic => `(tactic| sorry)
\end{Verbatim}
Type \texttt{i} is associated with possible worlds and type \texttt{e} with entities

\begin{Verbatim}[commandchars=\\\{\},fontsize=\scriptsize,xleftmargin=1em]
axiom i : Type                       -- Possible worlds
axiom e : Type                       -- Individuals/entities
@[instance] axiom iNonempty : Nonempty i
@[instance] axiom eNonempty : Nonempty e
abbrev \ensuremath{\sigma} := i \ensuremath{\rightarrow} Prop                 -- World-lifted propositions
abbrev \ensuremath{\tau} := e \ensuremath{\rightarrow} \ensuremath{\sigma}                    -- Modal properties

/-- Accessibility relation between worlds -/
axiom R : i \ensuremath{\rightarrow} i \ensuremath{\rightarrow} Prop
@[inherit_doc] infix:60 " \textsf{\bfseries r} " => R

axiom Rrefl  : \ensuremath{\forall} x, x \textsf{\bfseries r} x
axiom Rsymm  : \ensuremath{\forall} x y, x \textsf{\bfseries r} y \ensuremath{\rightarrow} y \textsf{\bfseries r} x
axiom Rtrans : \ensuremath{\forall} x y z, x \textsf{\bfseries r} y \ensuremath{\wedge} y \textsf{\bfseries r} z \ensuremath{\rightarrow} x \textsf{\bfseries r} z
\end{Verbatim}
Logical connectives (operating on truth-sets)

\begin{Verbatim}[commandchars=\\\{\},fontsize=\scriptsize,xleftmargin=1em]
@[simp, grind] def Mbot : \ensuremath{\sigma} := fun _ => False
@[simp, grind] def Mtop : \ensuremath{\sigma} := fun _ => True
@[simp, grind] def Mneg (\ensuremath{\varphi} : \ensuremath{\sigma}) : \ensuremath{\sigma} := fun w => \ensuremath{\neg} \ensuremath{\varphi} w
@[simp, grind] def Mand (\ensuremath{\varphi} \ensuremath{\psi} : \ensuremath{\sigma}) : \ensuremath{\sigma} := fun w => \ensuremath{\varphi} w \ensuremath{\wedge} \ensuremath{\psi} w
@[simp, grind] def Mor (\ensuremath{\varphi} \ensuremath{\psi} : \ensuremath{\sigma}) : \ensuremath{\sigma} := fun w => \ensuremath{\varphi} w \ensuremath{\vee} \ensuremath{\psi} w
@[simp, grind] def Mimp (\ensuremath{\varphi} \ensuremath{\psi} : \ensuremath{\sigma}) : \ensuremath{\sigma} := fun w => \ensuremath{\varphi} w \ensuremath{\rightarrow} \ensuremath{\psi} w
@[simp, grind] def Mequiv (\ensuremath{\varphi} \ensuremath{\psi} : \ensuremath{\sigma}) : \ensuremath{\sigma} := fun w => \ensuremath{\varphi} w \ensuremath{\leftrightarrow} \ensuremath{\psi} w
@[simp, grind] def Mbox (\ensuremath{\varphi} : \ensuremath{\sigma}) : \ensuremath{\sigma} := fun w => \ensuremath{\forall} v, w \textsf{\bfseries r} v \ensuremath{\rightarrow} \ensuremath{\varphi} v
@[simp, grind] def Mdia (\ensuremath{\varphi} : \ensuremath{\sigma}) : \ensuremath{\sigma} := fun w => \ensuremath{\exists} v, w \textsf{\bfseries r} v \ensuremath{\wedge} \ensuremath{\varphi} v
@[simp, grind] def Mprimeq \{\ensuremath{\alpha}\} (x y : \ensuremath{\alpha}) : \ensuremath{\sigma} := fun _ => x = y
@[simp, grind] def Mprimneg \{\ensuremath{\alpha}\} (x y : \ensuremath{\alpha}) : \ensuremath{\sigma} := fun _ => x \ensuremath{\neq} y
@[simp, grind] def Mnegpred (\ensuremath{\Phi} : \ensuremath{\tau}) : \ensuremath{\tau} := fun x w => \ensuremath{\neg} \ensuremath{\Phi} x w
@[simp, grind] def Mconpred (\ensuremath{\Phi} \ensuremath{\Psi} : \ensuremath{\tau}) : \ensuremath{\tau} := fun x w => \ensuremath{\Phi} x w \ensuremath{\wedge} \ensuremath{\Psi} x w

notation:max "\ensuremath{\bot}\textsuperscript{m}" => Mbot
notation:max "\ensuremath{\top}\textsuperscript{m}" => Mtop
prefix:53 "\ensuremath{\neg}\textsuperscript{m}" => Mneg
infixl:50 " \ensuremath{\wedge}\textsuperscript{m} " => Mand
infixl:49 " \ensuremath{\vee}\textsuperscript{m} " => Mor
infixr:48 " \ensuremath{\rightarrow}\textsuperscript{m} " => Mimp
infixl:47 " \ensuremath{\leftrightarrow}\textsuperscript{m} " => Mequiv
prefix:55 "\ensuremath{\Box}" => Mbox
prefix:55 "\ensuremath{\Diamond}" => Mdia
infix:50 " =\textsuperscript{m} " => Mprimeq
infix:50 " \ensuremath{\neq}\textsuperscript{m} " => Mprimneg
prefix:max "~\textsuperscript{m}" => Mnegpred
infixl:50 " \ensuremath{\centerdot}\textsuperscript{m} " => Mconpred

@[simp, grind] def Mexclor (\ensuremath{\varphi} \ensuremath{\psi} : \ensuremath{\sigma}) : \ensuremath{\sigma} := (\ensuremath{\varphi} \ensuremath{\vee}\textsuperscript{m} \ensuremath{\psi}) \ensuremath{\wedge}\textsuperscript{m} \ensuremath{\neg}\textsuperscript{m}(\ensuremath{\varphi} \ensuremath{\wedge}\textsuperscript{m} \ensuremath{\psi})
infixl:49 " \ensuremath{\vee}\textsuperscript{e} " => Mexclor
\end{Verbatim}
Possibilist quantifiers (polymorphic)

\begin{Verbatim}[commandchars=\\\{\},fontsize=\scriptsize,xleftmargin=1em]
@[simp, grind] def Mallposs \{\ensuremath{\alpha}\} (\ensuremath{\Phi} : \ensuremath{\alpha} \ensuremath{\rightarrow} \ensuremath{\sigma}) : \ensuremath{\sigma} := fun w => \ensuremath{\forall} x, \ensuremath{\Phi} x w
@[simp, grind] def Mexiposs \{\ensuremath{\alpha}\} (\ensuremath{\Phi} : \ensuremath{\alpha} \ensuremath{\rightarrow} \ensuremath{\sigma}) : \ensuremath{\sigma} := fun w => \ensuremath{\exists} x, \ensuremath{\Phi} x w
\end{Verbatim}
Actualist quantifiers (for individuals/entities)

\begin{Verbatim}[commandchars=\\\{\},fontsize=\scriptsize,xleftmargin=1em]
/-- Existence (actuality) of an entity at a world -/
axiom existsAt : e \ensuremath{\rightarrow} \ensuremath{\sigma}
@[inherit_doc] infix:60 " @\textsuperscript{m} " => existsAt
@[simp, grind] def Mallact (\ensuremath{\Phi} : e \ensuremath{\rightarrow} \ensuremath{\sigma}) : \ensuremath{\sigma} := fun w => \ensuremath{\forall} x, x @\textsuperscript{m} w \ensuremath{\rightarrow} \ensuremath{\Phi} x w
@[simp, grind] def Mexiact (\ensuremath{\Phi} : e \ensuremath{\rightarrow} \ensuremath{\sigma}) : \ensuremath{\sigma} := fun w => \ensuremath{\exists} x, x @\textsuperscript{m} w \ensuremath{\wedge} \ensuremath{\Phi} x w

open Lean TSyntax.Compat in
macro "\ensuremath{\forall}\textsuperscript{m}" xs:explicitBinders ", " b:term : term => expandExplicitBinders ``Mallposs xs b
open Lean TSyntax.Compat in
macro "\ensuremath{\exists}\textsuperscript{m}" xs:explicitBinders ", " b:term : term => expandExplicitBinders ``Mexiposs xs b
open Lean TSyntax.Compat in
macro "\ensuremath{\forall}\textsuperscript{E}" xs:explicitBinders ", " b:term : term => expandExplicitBinders ``Mallact xs b
open Lean TSyntax.Compat in
macro "\ensuremath{\exists}\textsuperscript{E}" xs:explicitBinders ", " b:term : term => expandExplicitBinders ``Mexiact xs b
\end{Verbatim}
Leibniz equality (polymorphic)

\begin{Verbatim}[commandchars=\\\{\},fontsize=\scriptsize,xleftmargin=1em]
@[simp, grind] def Mleibeq \{\ensuremath{\alpha}\} (x y : \ensuremath{\alpha}) : \ensuremath{\sigma} := \ensuremath{\forall}\textsuperscript{m} (P : \ensuremath{\alpha} \ensuremath{\rightarrow} \ensuremath{\sigma}), P x \ensuremath{\rightarrow}\textsuperscript{m} P y
infix:50 " \ensuremath{\equiv}\textsuperscript{m} " => Mleibeq
\end{Verbatim}
Meta-logical predicate for global validity

\begin{Verbatim}[commandchars=\\\{\},fontsize=\scriptsize,xleftmargin=1em]
@[simp, grind] def Mvalid (\ensuremath{\psi} : \ensuremath{\sigma}) : Prop := \ensuremath{\forall} w, \ensuremath{\psi} w
notation:max "\ensuremath{\lfloor}" \ensuremath{\psi} "\ensuremath{\rfloor}" => Mvalid \ensuremath{\psi}
\end{Verbatim}
\subsubsection{TestsHOML.lean (Figure 4 of \texorpdfstring{\cite{J75}}{the Notes})}
Tests and verifications of properties for the embedding of HOML (S5) in HOL.
\begin{Verbatim}[commandchars=\\\{\},fontsize=\scriptsize,xleftmargin=1em]
import Notes.HOMLinHOL
\end{Verbatim}

\begin{Verbatim}[commandchars=\\\{\},fontsize=\scriptsize,xleftmargin=1em]
variable \{\ensuremath{\alpha} : Type\} \{A B C \ensuremath{\varphi} \ensuremath{\psi} : \ensuremath{\sigma}\} \{p q : \ensuremath{\alpha} \ensuremath{\rightarrow} \ensuremath{\sigma}\} \{P Q : e \ensuremath{\rightarrow} \ensuremath{\sigma}\} \{x y z t : \ensuremath{\alpha}\}
\end{Verbatim}
Test for S5 modal logic

\begin{Verbatim}[commandchars=\\\{\},fontsize=\scriptsize,xleftmargin=1em]
theorem axM : \ensuremath{\lfloor}\ensuremath{\Box}\ensuremath{\varphi} \ensuremath{\rightarrow}\textsuperscript{m} \ensuremath{\varphi}\ensuremath{\rfloor} := fun w h => h w (Rrefl w)
theorem axD : \ensuremath{\lfloor}\ensuremath{\Box}\ensuremath{\varphi} \ensuremath{\rightarrow}\textsuperscript{m} \ensuremath{\Diamond}\ensuremath{\varphi}\ensuremath{\rfloor} := fun w h => \ensuremath{\langle}w, Rrefl w, h w (Rrefl w)\ensuremath{\rangle}
theorem axB : \ensuremath{\lfloor}\ensuremath{\varphi} \ensuremath{\rightarrow}\textsuperscript{m} \ensuremath{\Box}\ensuremath{\Diamond}\ensuremath{\varphi}\ensuremath{\rfloor} := fun w h v hv => \ensuremath{\langle}w, Rsymm w v hv, h\ensuremath{\rangle}
theorem ax4 : \ensuremath{\lfloor}\ensuremath{\Box}\ensuremath{\varphi} \ensuremath{\rightarrow}\textsuperscript{m} \ensuremath{\Box}\ensuremath{\Box}\ensuremath{\varphi}\ensuremath{\rfloor} := fun w h v hv u hu => h u (Rtrans w v u \ensuremath{\langle}hv, hu\ensuremath{\rangle})
theorem ax5 : \ensuremath{\lfloor}\ensuremath{\Diamond}\ensuremath{\varphi} \ensuremath{\rightarrow}\textsuperscript{m} \ensuremath{\Box}\ensuremath{\Diamond}\ensuremath{\varphi}\ensuremath{\rfloor} :=
  fun w \ensuremath{\langle}u, hu, h\ensuremath{\rangle} v hv => \ensuremath{\langle}u, Rtrans v w u \ensuremath{\langle}Rsymm w v hv, hu\ensuremath{\rangle}, h\ensuremath{\rangle}
\end{Verbatim}
Test for Barcan and converse Barcan formula

\begin{Verbatim}[commandchars=\\\{\},fontsize=\scriptsize,xleftmargin=1em]
-- BarcanAct   -- `nitpick[expect=genuine]`: countermodel found
example : \ensuremath{\lfloor}(\ensuremath{\forall}\textsuperscript{E} x, \ensuremath{\Box}(P x)) \ensuremath{\rightarrow}\textsuperscript{m} \ensuremath{\Box}(\ensuremath{\forall}\textsuperscript{E} x, P x)\ensuremath{\rfloor} := by countermodel
-- ConvBarcanAct   -- `nitpick[expect=genuine]`: countermodel found
example : \ensuremath{\lfloor}\ensuremath{\Box}(\ensuremath{\forall}\textsuperscript{E} x, P x) \ensuremath{\rightarrow}\textsuperscript{m} (\ensuremath{\forall}\textsuperscript{E} x, \ensuremath{\Box}(P x))\ensuremath{\rfloor} := by countermodel
theorem BarcanPoss : \ensuremath{\lfloor}(\ensuremath{\forall}\textsuperscript{m} x, \ensuremath{\Box}(p x)) \ensuremath{\rightarrow}\textsuperscript{m} \ensuremath{\Box}(\ensuremath{\forall}\textsuperscript{m} x, p x)\ensuremath{\rfloor} := fun _ h v hv x => h x v hv
theorem ConvBarcanPoss : \ensuremath{\lfloor}\ensuremath{\Box}(\ensuremath{\forall}\textsuperscript{m} x, p x) \ensuremath{\rightarrow}\textsuperscript{m} (\ensuremath{\forall}\textsuperscript{m} x, \ensuremath{\Box}(p x))\ensuremath{\rfloor} := fun _ h x v hv => h v hv x
\end{Verbatim}
A simple Hilbert system for classical propositional logic is derived

\begin{Verbatim}[commandchars=\\\{\},fontsize=\scriptsize,xleftmargin=1em]
theorem Hilbert_A1 : \ensuremath{\lfloor}A \ensuremath{\rightarrow}\textsuperscript{m} (B \ensuremath{\rightarrow}\textsuperscript{m} A)\ensuremath{\rfloor} := by grind
theorem Hilbert_A2 : \ensuremath{\lfloor}(A \ensuremath{\rightarrow}\textsuperscript{m} (B \ensuremath{\rightarrow}\textsuperscript{m} C)) \ensuremath{\rightarrow}\textsuperscript{m} ((A \ensuremath{\rightarrow}\textsuperscript{m} B) \ensuremath{\rightarrow}\textsuperscript{m} (A \ensuremath{\rightarrow}\textsuperscript{m} C))\ensuremath{\rfloor} := by grind
theorem Hilbert_MP (h1 : \ensuremath{\lfloor}A\ensuremath{\rfloor}) (h2 : \ensuremath{\lfloor}A \ensuremath{\rightarrow}\textsuperscript{m} B\ensuremath{\rfloor}) : \ensuremath{\lfloor}B\ensuremath{\rfloor} := fun w => h2 w (h1 w)
\end{Verbatim}
We have a polymorphic possibilist quantifier for which existential import holds

\begin{Verbatim}[commandchars=\\\{\},fontsize=\scriptsize,xleftmargin=1em]
theorem Quant_1 (h : \ensuremath{\lfloor}A\ensuremath{\rfloor}) : \ensuremath{\lfloor}\ensuremath{\forall}\textsuperscript{m} (_ : \ensuremath{\alpha}), A\ensuremath{\rfloor} := by grind
\end{Verbatim}
Existential import holds for possibilist quantifiers

\begin{Verbatim}[commandchars=\\\{\},fontsize=\scriptsize,xleftmargin=1em]
theorem ExImPossibilist1 : \ensuremath{\lfloor}\ensuremath{\exists}\textsuperscript{m} (x : e), x =\textsuperscript{m} x\ensuremath{\rfloor} := fun _ => \ensuremath{\langle}Classical.ofNonempty, rfl\ensuremath{\rangle}
theorem ExImPossibilist2 : \ensuremath{\lfloor}\ensuremath{\exists}\textsuperscript{m} (x : e), x \ensuremath{\equiv}\textsuperscript{m} x\ensuremath{\rfloor} := fun _ => \ensuremath{\langle}Classical.ofNonempty, fun _ h => h\ensuremath{\rangle}
theorem ExImPossibilist3 \{t : e\} : \ensuremath{\lfloor}\ensuremath{\exists}\textsuperscript{m} (x : e), x =\textsuperscript{m} t\ensuremath{\rfloor} := fun _ => \ensuremath{\langle}t, rfl\ensuremath{\rangle}
theorem ExImPossibilist4 : \ensuremath{\lfloor}\ensuremath{\exists}\textsuperscript{m} (x : \ensuremath{\alpha}), x \ensuremath{\equiv}\textsuperscript{m} t\ensuremath{\rfloor} := fun _ => \ensuremath{\langle}t, fun _ h => h\ensuremath{\rangle}
theorem ExImPossibilist [Nonempty \ensuremath{\alpha}] : \ensuremath{\lfloor}\ensuremath{\exists}\textsuperscript{m} (_ : \ensuremath{\alpha}), \ensuremath{\top}\textsuperscript{m}\ensuremath{\rfloor} := fun _ => \ensuremath{\langle}Classical.ofNonempty, trivial\ensuremath{\rangle}
\end{Verbatim}
We have an actualist quantifier for individuals for which existential import does not hold

\begin{Verbatim}[commandchars=\\\{\},fontsize=\scriptsize,xleftmargin=1em]
theorem Quant_2 (h : \ensuremath{\lfloor}A\ensuremath{\rfloor}) : \ensuremath{\lfloor}\ensuremath{\forall}\textsuperscript{E} (_ : e), A\ensuremath{\rfloor} := by grind
\end{Verbatim}
Existential import does not hold for our actualist quantifiers (for individuals)

\begin{Verbatim}[commandchars=\\\{\},fontsize=\scriptsize,xleftmargin=1em]
-- ExImActualist1   -- `nitpick[card=1,expect=genuine]`: countermodel found
example : \ensuremath{\lfloor}\ensuremath{\exists}\textsuperscript{E} (x : e), x =\textsuperscript{m} x\ensuremath{\rfloor} := by countermodel
-- ExImActualist2   -- `nitpick[card=1,expect=genuine]`: countermodel found
example : \ensuremath{\lfloor}\ensuremath{\exists}\textsuperscript{E} (x : e), x \ensuremath{\equiv}\textsuperscript{m} x\ensuremath{\rfloor} := by countermodel
-- ExImActualist3   -- `nitpick[card=1,expect=genuine]`: countermodel found
example \{t : e\} : \ensuremath{\lfloor}\ensuremath{\exists}\textsuperscript{E} (x : e), x =\textsuperscript{m} t\ensuremath{\rfloor} := by countermodel
-- ExImActualist   -- `nitpick[card=1,expect=genuine]`: countermodel found
example : \ensuremath{\lfloor}\ensuremath{\exists}\textsuperscript{E} (_ : e), \ensuremath{\top}\textsuperscript{m}\ensuremath{\rfloor} := by countermodel
\end{Verbatim}
Properties of the embedded primitive equality, which coincides with Leibniz equality

\begin{Verbatim}[commandchars=\\\{\},fontsize=\scriptsize,xleftmargin=1em]
theorem EqRefl : \ensuremath{\lfloor}x =\textsuperscript{m} x\ensuremath{\rfloor} := by grind
theorem EqSym : \ensuremath{\lfloor}(x =\textsuperscript{m} y) \ensuremath{\leftrightarrow}\textsuperscript{m} (y =\textsuperscript{m} x)\ensuremath{\rfloor} := by grind
theorem EqTrans : \ensuremath{\lfloor}((x =\textsuperscript{m} y) \ensuremath{\wedge}\textsuperscript{m} (y =\textsuperscript{m} z)) \ensuremath{\rightarrow}\textsuperscript{m} (x =\textsuperscript{m} z)\ensuremath{\rfloor} := by grind
theorem EQCong \{\ensuremath{\beta}\} \{f : \ensuremath{\alpha} \ensuremath{\rightarrow} \ensuremath{\beta}\} : \ensuremath{\lfloor}(x =\textsuperscript{m} y) \ensuremath{\rightarrow}\textsuperscript{m} ((f x) =\textsuperscript{m} (f y))\ensuremath{\rfloor} := by grind
theorem EQFuncExt : \ensuremath{\lfloor}(p =\textsuperscript{m} q) \ensuremath{\rightarrow}\textsuperscript{m} (\ensuremath{\forall}\textsuperscript{m} x, ((p x) =\textsuperscript{m} (q x)))\ensuremath{\rfloor} := by grind
theorem EQBoolExt1 : \ensuremath{\lfloor}(\ensuremath{\varphi} =\textsuperscript{m} \ensuremath{\psi}) \ensuremath{\rightarrow}\textsuperscript{m} (\ensuremath{\varphi} \ensuremath{\leftrightarrow}\textsuperscript{m} \ensuremath{\psi})\ensuremath{\rfloor} := by grind
-- EQBoolExt2   -- `nitpick[card=2]`: countermodel found
example : \ensuremath{\lfloor}(\ensuremath{\varphi} \ensuremath{\leftrightarrow}\textsuperscript{m} \ensuremath{\psi}) \ensuremath{\rightarrow}\textsuperscript{m} (\ensuremath{\varphi} =\textsuperscript{m} \ensuremath{\psi})\ensuremath{\rfloor} := by countermodel
theorem EQBoolExt3 : \ensuremath{\lfloor}\ensuremath{\varphi} \ensuremath{\leftrightarrow}\textsuperscript{m} \ensuremath{\psi}\ensuremath{\rfloor} \ensuremath{\rightarrow} \ensuremath{\lfloor}\ensuremath{\varphi} =\textsuperscript{m} \ensuremath{\psi}\ensuremath{\rfloor} := fun h _ => funext fun v => propext (h v)
theorem EqPrimLeib : \ensuremath{\lfloor}(x =\textsuperscript{m} y) \ensuremath{\leftrightarrow}\textsuperscript{m} (x \ensuremath{\equiv}\textsuperscript{m} y)\ensuremath{\rfloor} :=
  fun _ => \ensuremath{\langle}fun h _ hp => h \ensuremath{\blacktriangleright} hp, fun h => h (fun z _ => x = z) rfl\ensuremath{\rangle}
\end{Verbatim}
Comprehension is natively supported in HOL (due to lambda-abstraction)

\begin{Verbatim}[commandchars=\\\{\},fontsize=\scriptsize,xleftmargin=1em]
theorem Comprehension1 : \ensuremath{\lfloor}\ensuremath{\exists}\textsuperscript{m} (f : \ensuremath{\alpha} \ensuremath{\rightarrow} \ensuremath{\sigma}), \ensuremath{\forall}\textsuperscript{m} x, (f x) \ensuremath{\leftrightarrow}\textsuperscript{m} A\ensuremath{\rfloor} :=
  fun _ => \ensuremath{\langle}fun _ => A, fun _ => Iff.rfl\ensuremath{\rangle}
theorem Comprehension2 \{A1 : \ensuremath{\alpha} \ensuremath{\rightarrow} \ensuremath{\sigma}\} : \ensuremath{\lfloor}\ensuremath{\exists}\textsuperscript{m} (f : \ensuremath{\alpha} \ensuremath{\rightarrow} \ensuremath{\sigma}), \ensuremath{\forall}\textsuperscript{m} x, (f x) \ensuremath{\leftrightarrow}\textsuperscript{m} (A1 x)\ensuremath{\rfloor} :=
  fun _ => \ensuremath{\langle}A1, fun _ => Iff.rfl\ensuremath{\rangle}
theorem Comprehension3 \{\ensuremath{\beta}\} \{A2 : \ensuremath{\alpha} \ensuremath{\rightarrow} \ensuremath{\beta} \ensuremath{\rightarrow} \ensuremath{\sigma}\} :
    \ensuremath{\lfloor}\ensuremath{\exists}\textsuperscript{m} (f : \ensuremath{\alpha} \ensuremath{\rightarrow} \ensuremath{\beta} \ensuremath{\rightarrow} \ensuremath{\sigma}), \ensuremath{\forall}\textsuperscript{m} x, \ensuremath{\forall}\textsuperscript{m} y, (f x y) \ensuremath{\leftrightarrow}\textsuperscript{m} (A2 x y)\ensuremath{\rfloor} := fun _ => \ensuremath{\langle}A2, fun _ _ => Iff.rfl\ensuremath{\rangle}
\end{Verbatim}
Modal collapse does not hold

\begin{Verbatim}[commandchars=\\\{\},fontsize=\scriptsize,xleftmargin=1em]
-- ModalCollapse   -- `nitpick[card=2,expect=genuine]`: countermodel found
example : \ensuremath{\lfloor}\ensuremath{\forall}\textsuperscript{m} (f : \ensuremath{\sigma}), f \ensuremath{\rightarrow}\textsuperscript{m} \ensuremath{\Box}f\ensuremath{\rfloor} := by countermodel
\end{Verbatim}
Empty property and self-difference

\begin{Verbatim}[commandchars=\\\{\},fontsize=\scriptsize,xleftmargin=1em]
theorem TruePropertyAndSelfIdentity : \ensuremath{\lfloor}(fun _ : e => \ensuremath{\top}\textsuperscript{m}) =\textsuperscript{m} (fun x : e => x =\textsuperscript{m} x)\ensuremath{\rfloor} := by
  intro _; funext x v; simp
theorem EmptyPropertyAndSelfDifference : \ensuremath{\lfloor}(fun _ : e => \ensuremath{\bot}\textsuperscript{m}) =\textsuperscript{m} (fun x : e => x \ensuremath{\neq}\textsuperscript{m} x)\ensuremath{\rfloor} := by
  intro _; funext x v; simp
theorem EmptyProperty2 : \ensuremath{\lfloor}\ensuremath{\exists}\textsuperscript{m} x, P x\ensuremath{\rfloor} \ensuremath{\rightarrow} \ensuremath{\lfloor}P \ensuremath{\neq}\textsuperscript{m} (fun _ : e => \ensuremath{\bot}\textsuperscript{m})\ensuremath{\rfloor} := by
  intro h w hEq; have \ensuremath{\langle}x, hx\ensuremath{\rangle} := h w; rw [hEq] at hx; exact hx
theorem EmptyProperty3 : \ensuremath{\lfloor}\ensuremath{\exists}\textsuperscript{E} x, P x\ensuremath{\rfloor} \ensuremath{\rightarrow} \ensuremath{\lfloor}P \ensuremath{\neq}\textsuperscript{m} (fun _ : e => \ensuremath{\bot}\textsuperscript{m})\ensuremath{\rfloor} := by
  intro h w hEq; have \ensuremath{\langle}x, _, hx\ensuremath{\rangle} := h w; rw [hEq] at hx; exact hx
-- EmptyProperty4   -- `nitpick[expect=genuine]`: countermodel found
example : \ensuremath{\lfloor}P \ensuremath{\neq}\textsuperscript{m} (fun _ : e => \ensuremath{\bot}\textsuperscript{m})\ensuremath{\rfloor} \ensuremath{\rightarrow} \ensuremath{\lfloor}\ensuremath{\exists}\textsuperscript{m} x, P x\ensuremath{\rfloor} := by countermodel
-- EmptyProperty5   -- `nitpick[expect=genuine]`: countermodel found
example : \ensuremath{\lfloor}P \ensuremath{\neq}\textsuperscript{m} (fun _ : e => \ensuremath{\bot}\textsuperscript{m})\ensuremath{\rfloor} \ensuremath{\rightarrow} \ensuremath{\lfloor}\ensuremath{\exists}\textsuperscript{E} x, P x\ensuremath{\rfloor} := by countermodel
\end{Verbatim}
\subsubsection{ModalFilter.lean (Figure 5 of \texorpdfstring{\cite{J75}}{the Notes})}
Set filter and ultrafilter formalized for our modal logic setting.
\begin{Verbatim}[commandchars=\\\{\},fontsize=\scriptsize,xleftmargin=1em]
import Notes.HOMLinHOL
\end{Verbatim}

\begin{Verbatim}[commandchars=\\\{\},fontsize=\scriptsize,xleftmargin=1em]
@[simp, grind] def Element (\ensuremath{\varphi} : \ensuremath{\tau}) (S : \ensuremath{\tau} \ensuremath{\rightarrow} \ensuremath{\sigma}) : \ensuremath{\sigma} := S \ensuremath{\varphi}
@[simp, grind] def EmptySet : \ensuremath{\tau} := fun _ => \ensuremath{\bot}\textsuperscript{m}
@[simp, grind] def UniversalSet : \ensuremath{\tau} := fun _ => \ensuremath{\top}\textsuperscript{m}
@[simp, grind] def Subset (\ensuremath{\varphi} \ensuremath{\psi} : \ensuremath{\tau}) : \ensuremath{\sigma} := \ensuremath{\forall}\textsuperscript{m} x, (\ensuremath{\varphi} x) \ensuremath{\rightarrow}\textsuperscript{m} (\ensuremath{\psi} x)
@[simp, grind] def SubsetE (\ensuremath{\varphi} \ensuremath{\psi} : \ensuremath{\tau}) : \ensuremath{\sigma} := \ensuremath{\forall}\textsuperscript{E} x, (\ensuremath{\varphi} x) \ensuremath{\rightarrow}\textsuperscript{m} (\ensuremath{\psi} x)
@[simp, grind] def Intersection (\ensuremath{\varphi} \ensuremath{\psi} : \ensuremath{\tau}) : \ensuremath{\tau} := fun x => (\ensuremath{\varphi} x) \ensuremath{\wedge}\textsuperscript{m} (\ensuremath{\psi} x)
@[simp, grind] def Inverse (\ensuremath{\psi} : \ensuremath{\tau}) : \ensuremath{\tau} := fun x => \ensuremath{\neg}\textsuperscript{m}(\ensuremath{\psi} x)

infix:90 " \ensuremath{\in}\textsuperscript{m} " => Element
notation:max "\ensuremath{\emptyset}\textsuperscript{m}" => EmptySet
notation:max "U\textsuperscript{m}" => UniversalSet
infix:80 " \ensuremath{\subseteq}\textsuperscript{m} " => Subset
infix:80 " \ensuremath{\subseteq}\textsuperscript{E} " => SubsetE
infix:91 " \ensuremath{\sqcap}\textsuperscript{m} " => Intersection
prefix:max "\textsuperscript{-}" => Inverse

@[simp, grind] def Filter (\ensuremath{\Phi} : \ensuremath{\tau} \ensuremath{\rightarrow} \ensuremath{\sigma}) : \ensuremath{\sigma} :=
  U\textsuperscript{m} \ensuremath{\in}\textsuperscript{m} \ensuremath{\Phi} \ensuremath{\wedge}\textsuperscript{m} \ensuremath{\neg}\textsuperscript{m}(\ensuremath{\emptyset}\textsuperscript{m} \ensuremath{\in}\textsuperscript{m} \ensuremath{\Phi}) \ensuremath{\wedge}\textsuperscript{m}
  (\ensuremath{\forall}\textsuperscript{m} \ensuremath{\varphi}, \ensuremath{\forall}\textsuperscript{m} \ensuremath{\psi}, \ensuremath{\varphi} \ensuremath{\in}\textsuperscript{m} \ensuremath{\Phi} \ensuremath{\wedge}\textsuperscript{m} \ensuremath{\varphi} \ensuremath{\subseteq}\textsuperscript{E} \ensuremath{\psi} \ensuremath{\rightarrow}\textsuperscript{m} \ensuremath{\psi} \ensuremath{\in}\textsuperscript{m} \ensuremath{\Phi}) \ensuremath{\wedge}\textsuperscript{m}
  (\ensuremath{\forall}\textsuperscript{m} \ensuremath{\varphi}, \ensuremath{\forall}\textsuperscript{m} \ensuremath{\psi}, \ensuremath{\varphi} \ensuremath{\in}\textsuperscript{m} \ensuremath{\Phi} \ensuremath{\wedge}\textsuperscript{m} \ensuremath{\psi} \ensuremath{\in}\textsuperscript{m} \ensuremath{\Phi} \ensuremath{\rightarrow}\textsuperscript{m} \ensuremath{\varphi} \ensuremath{\sqcap}\textsuperscript{m} \ensuremath{\psi} \ensuremath{\in}\textsuperscript{m} \ensuremath{\Phi})
@[simp, grind] def UFilter (\ensuremath{\Phi} : \ensuremath{\tau} \ensuremath{\rightarrow} \ensuremath{\sigma}) : \ensuremath{\sigma} := Filter \ensuremath{\Phi} \ensuremath{\wedge}\textsuperscript{m} (\ensuremath{\forall}\textsuperscript{m} \ensuremath{\varphi}, \ensuremath{\varphi} \ensuremath{\in}\textsuperscript{m} \ensuremath{\Phi} \ensuremath{\vee}\textsuperscript{m} (\textsuperscript{-}\ensuremath{\varphi}) \ensuremath{\in}\textsuperscript{m} \ensuremath{\Phi})
@[simp, grind] def FilterP (\ensuremath{\Phi} : \ensuremath{\tau} \ensuremath{\rightarrow} \ensuremath{\sigma}) : \ensuremath{\sigma} :=
  U\textsuperscript{m} \ensuremath{\in}\textsuperscript{m} \ensuremath{\Phi} \ensuremath{\wedge}\textsuperscript{m} \ensuremath{\neg}\textsuperscript{m}(\ensuremath{\emptyset}\textsuperscript{m} \ensuremath{\in}\textsuperscript{m} \ensuremath{\Phi}) \ensuremath{\wedge}\textsuperscript{m}
  (\ensuremath{\forall}\textsuperscript{m} \ensuremath{\varphi}, \ensuremath{\forall}\textsuperscript{m} \ensuremath{\psi}, \ensuremath{\varphi} \ensuremath{\in}\textsuperscript{m} \ensuremath{\Phi} \ensuremath{\wedge}\textsuperscript{m} \ensuremath{\varphi} \ensuremath{\subseteq}\textsuperscript{m} \ensuremath{\psi} \ensuremath{\rightarrow}\textsuperscript{m} \ensuremath{\psi} \ensuremath{\in}\textsuperscript{m} \ensuremath{\Phi}) \ensuremath{\wedge}\textsuperscript{m}
  (\ensuremath{\forall}\textsuperscript{m} \ensuremath{\varphi}, \ensuremath{\forall}\textsuperscript{m} \ensuremath{\psi}, \ensuremath{\varphi} \ensuremath{\in}\textsuperscript{m} \ensuremath{\Phi} \ensuremath{\wedge}\textsuperscript{m} \ensuremath{\psi} \ensuremath{\in}\textsuperscript{m} \ensuremath{\Phi} \ensuremath{\rightarrow}\textsuperscript{m} \ensuremath{\varphi} \ensuremath{\sqcap}\textsuperscript{m} \ensuremath{\psi} \ensuremath{\in}\textsuperscript{m} \ensuremath{\Phi})
@[simp, grind] def UFilterP (\ensuremath{\Phi} : \ensuremath{\tau} \ensuremath{\rightarrow} \ensuremath{\sigma}) : \ensuremath{\sigma} := FilterP \ensuremath{\Phi} \ensuremath{\wedge}\textsuperscript{m} (\ensuremath{\forall}\textsuperscript{m} \ensuremath{\varphi}, \ensuremath{\varphi} \ensuremath{\in}\textsuperscript{m} \ensuremath{\Phi} \ensuremath{\vee}\textsuperscript{m} (\textsuperscript{-}\ensuremath{\varphi}) \ensuremath{\in}\textsuperscript{m} \ensuremath{\Phi})
\end{Verbatim}
\subsection{G\"odel's ontological argument -- 1970 manuscript}
\subsubsection{GoedelVariantHOML1.lean (Figure 6 of \texorpdfstring{\cite{J75}}{the Notes})}
G\"odel's axioms and definitions, as presented in the 1970 manuscript, are inconsistent.  Actualist quantifiers (avoiding existential import) are used for quantification over entities, otherwise possibilist quantifiers are used.
\begin{Verbatim}[commandchars=\\\{\},fontsize=\scriptsize,xleftmargin=1em]
import Notes.HOMLinHOL
\end{Verbatim}

\begin{Verbatim}[commandchars=\\\{\},fontsize=\scriptsize,xleftmargin=1em]
/-- Positive property -/
axiom P : (e \ensuremath{\rightarrow} \ensuremath{\sigma}) \ensuremath{\rightarrow} \ensuremath{\sigma}

axiom Ax1 (\ensuremath{\varphi} \ensuremath{\psi} : e \ensuremath{\rightarrow} \ensuremath{\sigma}) : \ensuremath{\lfloor}P \ensuremath{\varphi} \ensuremath{\wedge}\textsuperscript{m} P \ensuremath{\psi} \ensuremath{\rightarrow}\textsuperscript{m} P (\ensuremath{\varphi} \ensuremath{\centerdot}\textsuperscript{m} \ensuremath{\psi})\ensuremath{\rfloor}

axiom Ax2a (\ensuremath{\varphi} : e \ensuremath{\rightarrow} \ensuremath{\sigma}) : \ensuremath{\lfloor}P \ensuremath{\varphi} \ensuremath{\vee}\textsuperscript{e} P ~\textsuperscript{m}\ensuremath{\varphi}\ensuremath{\rfloor}

/-- Auxiliary reformulation of `Ax2a` (Lean has no `sledgehammer`). -/
theorem Ax2a' (\ensuremath{\varphi} : e \ensuremath{\rightarrow} \ensuremath{\sigma}) : \ensuremath{\lfloor}\ensuremath{\neg}\textsuperscript{m}(P \ensuremath{\varphi}) \ensuremath{\leftrightarrow}\textsuperscript{m} P ~\textsuperscript{m}\ensuremath{\varphi}\ensuremath{\rfloor} :=
  fun w => \ensuremath{\langle}(Ax2a \ensuremath{\varphi} w).1.resolve_left, fun hp hq => (Ax2a \ensuremath{\varphi} w).2 \ensuremath{\langle}hq, hp\ensuremath{\rangle}\ensuremath{\rangle}

/-- God-like -/
def G (x : e) : \ensuremath{\sigma} := \ensuremath{\forall}\textsuperscript{m} (\ensuremath{\varphi} : e \ensuremath{\rightarrow} \ensuremath{\sigma}), P \ensuremath{\varphi} \ensuremath{\rightarrow}\textsuperscript{m} \ensuremath{\varphi} x

/-- Necessary property inclusion -/
@[simp, grind] def PInc (\ensuremath{\varphi} \ensuremath{\psi} : e \ensuremath{\rightarrow} \ensuremath{\sigma}) : \ensuremath{\sigma} := \ensuremath{\Box}(\ensuremath{\forall}\textsuperscript{E} y, \ensuremath{\varphi} y \ensuremath{\rightarrow}\textsuperscript{m} \ensuremath{\psi} y)
infixr:48 " \ensuremath{\supset}\textsuperscript{N} " => PInc

/-- `Ess \ensuremath{\varphi} x`: \ensuremath{\varphi} is an essence of x -/
def Ess (\ensuremath{\varphi} : e \ensuremath{\rightarrow} \ensuremath{\sigma}) (x : e) : \ensuremath{\sigma} := \ensuremath{\forall}\textsuperscript{m} (\ensuremath{\psi} : e \ensuremath{\rightarrow} \ensuremath{\sigma}), \ensuremath{\psi} x \ensuremath{\rightarrow}\textsuperscript{m} (\ensuremath{\varphi} \ensuremath{\supset}\textsuperscript{N} \ensuremath{\psi})

axiom Ax2b (\ensuremath{\varphi} : e \ensuremath{\rightarrow} \ensuremath{\sigma}) : \ensuremath{\lfloor}P \ensuremath{\varphi} \ensuremath{\rightarrow}\textsuperscript{m} \ensuremath{\Box} P \ensuremath{\varphi}\ensuremath{\rfloor}

theorem Ax2b' (\ensuremath{\varphi} : e \ensuremath{\rightarrow} \ensuremath{\sigma}) : \ensuremath{\lfloor}\ensuremath{\neg}\textsuperscript{m}(P \ensuremath{\varphi}) \ensuremath{\rightarrow}\textsuperscript{m} \ensuremath{\Box}(\ensuremath{\neg}\textsuperscript{m}(P \ensuremath{\varphi}))\ensuremath{\rfloor} :=
  fun w hn v hv => (Ax2a' \ensuremath{\varphi} v).2 (Ax2b ~\textsuperscript{m}\ensuremath{\varphi} w ((Ax2a' \ensuremath{\varphi} w).1 hn) v hv)

/-- A property exemplified by a God-like being is positive (a consequence of `Ax2a`). -/
theorem PosOfGod \{x : e\} \{\ensuremath{\psi} : e \ensuremath{\rightarrow} \ensuremath{\sigma}\} \{w : i\} (hG : G x w) (h : \ensuremath{\psi} x w) : P \ensuremath{\psi} w :=
  Classical.byContradiction fun hn => hG ~\textsuperscript{m}\ensuremath{\psi} ((Ax2a' \ensuremath{\psi} w).1 hn) h

theorem Th1 (x : e) : \ensuremath{\lfloor}G x \ensuremath{\rightarrow}\textsuperscript{m} Ess G x\ensuremath{\rfloor} :=
  fun w hG \ensuremath{\psi} h\ensuremath{\psi} v hv _y _ hGy => hGy \ensuremath{\psi} (Ax2b \ensuremath{\psi} w (PosOfGod hG h\ensuremath{\psi}) v hv)

/-- Necessary existence -/
def E (x : e) : \ensuremath{\sigma} := \ensuremath{\forall}\textsuperscript{m} (\ensuremath{\varphi} : e \ensuremath{\rightarrow} \ensuremath{\sigma}), Ess \ensuremath{\varphi} x \ensuremath{\rightarrow}\textsuperscript{m} \ensuremath{\Box}(\ensuremath{\exists}\textsuperscript{E} x, \ensuremath{\varphi} x)

axiom Ax3 : \ensuremath{\lfloor}P E\ensuremath{\rfloor}

theorem Th2 (x : e) : \ensuremath{\lfloor}G x \ensuremath{\rightarrow}\textsuperscript{m} \ensuremath{\Box}(\ensuremath{\exists}\textsuperscript{E} y, G y)\ensuremath{\rfloor} := fun w hG => hG E (Ax3 w) G (Th1 x w hG)

theorem Th3 : \ensuremath{\lfloor}\ensuremath{\Diamond}(\ensuremath{\exists}\textsuperscript{E} x, G x) \ensuremath{\rightarrow}\textsuperscript{m} \ensuremath{\Box}(\ensuremath{\exists}\textsuperscript{E} y, G y)\ensuremath{\rfloor} := by
  have h1 : \ensuremath{\lfloor}(\ensuremath{\exists}\textsuperscript{E} x, G x) \ensuremath{\rightarrow}\textsuperscript{m} \ensuremath{\Box}(\ensuremath{\exists}\textsuperscript{E} y, G y)\ensuremath{\rfloor} := fun _ \ensuremath{\langle}x, _, hG\ensuremath{\rangle} => Th2 x _ hG
  intro w \ensuremath{\langle}v, hv, hex\ensuremath{\rangle}
  exact h1 w (h1 v hex w (Rsymm w v hv))               -- only symmetry of `\textsf{\bfseries r}` is needed

axiom Ax4 (\ensuremath{\varphi} \ensuremath{\psi} : e \ensuremath{\rightarrow} \ensuremath{\sigma}) : \ensuremath{\lfloor}P \ensuremath{\varphi} \ensuremath{\wedge}\textsuperscript{m} (\ensuremath{\varphi} \ensuremath{\supset}\textsuperscript{N} \ensuremath{\psi}) \ensuremath{\rightarrow}\textsuperscript{m} P \ensuremath{\psi}\ensuremath{\rfloor}

-- `lemma True nitpick[satisfy,expect=unknown] oops`
--   No model found (consistency check)

theorem EmptyEssL (x : e) : \ensuremath{\lfloor}Ess (fun _ : e => \ensuremath{\bot}\textsuperscript{m}) x\ensuremath{\rfloor} := fun _ _ _ _ _ _ _ h => h.elim

-- Like the Isabelle original, this proof uses no frame condition at all: the axioms are
-- inconsistent already in K.  (Verified by replaying the Isabelle proof over `HOMLinHOLonlyK`.)
theorem Inconsistency : False := by
  have h1 : \ensuremath{\lfloor}\ensuremath{\neg}\textsuperscript{m}(P (fun _ : e => \ensuremath{\bot}\textsuperscript{m}))\ensuremath{\rfloor} := fun w hp =>
    (Ax2a _ w).2 \ensuremath{\langle}hp, Ax4 (fun _ : e => \ensuremath{\bot}\textsuperscript{m}) _ w \ensuremath{\langle}hp, fun _ _ _ _ h => h.elim\ensuremath{\rangle}\ensuremath{\rangle}
  have h2 : \ensuremath{\lfloor}P (fun x : e => Ess (fun _ : e => \ensuremath{\bot}\textsuperscript{m}) x \ensuremath{\rightarrow}\textsuperscript{m} \ensuremath{\Box}(\ensuremath{\exists}\textsuperscript{E} z, (fun _ : e => \ensuremath{\bot}\textsuperscript{m}) z))\ensuremath{\rfloor} :=
    fun w => Ax4 E _ w \ensuremath{\langle}Ax3 w, fun _ _ _ _ hE => hE (fun _ => \ensuremath{\bot}\textsuperscript{m})\ensuremath{\rangle}
  have h3 : \ensuremath{\lfloor}P (fun _ : e => \ensuremath{\Box}(\ensuremath{\exists}\textsuperscript{E} z, (fun _ : e => \ensuremath{\bot}\textsuperscript{m}) z))\ensuremath{\rfloor} :=
    fun w => Ax4 _ _ w \ensuremath{\langle}h2 w, fun v _ y _ h => h (EmptyEssL y v)\ensuremath{\rangle}
  have h4 : \ensuremath{\lfloor}P (fun _ : e => \ensuremath{\Box} \ensuremath{\bot}\textsuperscript{m})\ensuremath{\rfloor} := by
    have heq : (fun _ : e => \ensuremath{\Box}(\ensuremath{\exists}\textsuperscript{E} z, (fun _ : e => \ensuremath{\bot}\textsuperscript{m}) z)) = (fun _ : e => \ensuremath{\Box} \ensuremath{\bot}\textsuperscript{m}) := by
      -- an explicit term rather than `simp`: `simp` would draw in `eNonempty`, which the
      -- argument does not need (`#print axioms`)
      funext _ w; exact propext \ensuremath{\langle}fun h v hv => match h v hv with | \ensuremath{\langle}_, _, hf\ensuremath{\rangle} => hf, fun h v hv =>
          (h v hv).elim\ensuremath{\rangle}
    exact heq \ensuremath{\blacktriangleright} h3
  -- Isabelle discharges this with `smt` from `4 Ax2a Ax4`; no frame condition is needed.
  have h5 : \ensuremath{\lfloor}P (fun _ : e => \ensuremath{\bot}\textsuperscript{m})\ensuremath{\rfloor} := by
    intro w
    by_cases hdead : \ensuremath{\exists} v : i, \ensuremath{\forall} u, \ensuremath{\neg} v \textsf{\bfseries r} u
    \textperiodcentered{} -- some world is a dead end: there `Ax4`'s premise is vacuous, so `Ax2a` is violated
      have \ensuremath{\langle}v, hv\ensuremath{\rangle} := hdead
      exact ((Ax2a (fun _ : e => \ensuremath{\Box} \ensuremath{\bot}\textsuperscript{m}) v).2
        \ensuremath{\langle}h4 v, Ax4 _ _ v \ensuremath{\langle}h4 v, fun u hu => absurd hu (hv u)\ensuremath{\rangle}\ensuremath{\rangle}).elim
    \textperiodcentered{} -- otherwise every world has a successor, so `\ensuremath{\Box}\ensuremath{\bot}` is false and the properties coincide
      have ser : \ensuremath{\forall} v : i, \ensuremath{\exists} u, v \textsf{\bfseries r} u := fun v =>
        Classical.byContradiction fun h => hdead \ensuremath{\langle}v, fun u hu => h \ensuremath{\langle}u, hu\ensuremath{\rangle}\ensuremath{\rangle}
      have heq : (fun _ : e => \ensuremath{\Box} \ensuremath{\bot}\textsuperscript{m}) = (fun _ : e => \ensuremath{\bot}\textsuperscript{m}) := by
        funext _ v
        have \ensuremath{\langle}u, hu\ensuremath{\rangle} := ser v
        exact propext \ensuremath{\langle}fun h => h u hu, fun h => h.elim\ensuremath{\rangle}
      exact heq \ensuremath{\blacktriangleright} h4 w
  have w : i := Classical.ofNonempty
  exact h1 w (h5 w)
\end{Verbatim}
\subsubsection{GoedelVariantHOML2.lean (Figure 7 of \texorpdfstring{\cite{J75}}{the Notes})}
After an appropriate modification of the definition of essence in G\"odel's 1970 ontological proof, the inconsistency revealed in Figure 6 is avoided, and the argument can be successfully verified in modal logic S5 (indeed, as shown, only symmetry of the accessibility relation is actually needed). Actualist quantifiers (avoiding existential import) are used for quantification over entities, otherwise possibilist quantifiers are used.
\begin{Verbatim}[commandchars=\\\{\},fontsize=\scriptsize,xleftmargin=1em]
import Notes.HOMLinHOL
import Notes.ModalFilter
\end{Verbatim}

\begin{Verbatim}[commandchars=\\\{\},fontsize=\scriptsize,xleftmargin=1em]
/-- Positive property -/
axiom P : (e \ensuremath{\rightarrow} \ensuremath{\sigma}) \ensuremath{\rightarrow} \ensuremath{\sigma}

axiom Ax1 (\ensuremath{\varphi} \ensuremath{\psi} : e \ensuremath{\rightarrow} \ensuremath{\sigma}) : \ensuremath{\lfloor}P \ensuremath{\varphi} \ensuremath{\wedge}\textsuperscript{m} P \ensuremath{\psi} \ensuremath{\rightarrow}\textsuperscript{m} P (\ensuremath{\varphi} \ensuremath{\centerdot}\textsuperscript{m} \ensuremath{\psi})\ensuremath{\rfloor}

axiom Ax2a (\ensuremath{\varphi} : e \ensuremath{\rightarrow} \ensuremath{\sigma}) : \ensuremath{\lfloor}P \ensuremath{\varphi} \ensuremath{\vee}\textsuperscript{e} P ~\textsuperscript{m}\ensuremath{\varphi}\ensuremath{\rfloor}

/-- Auxiliary reformulation of `Ax2a` (Lean has no `sledgehammer`). -/
theorem Ax2a' (\ensuremath{\varphi} : e \ensuremath{\rightarrow} \ensuremath{\sigma}) : \ensuremath{\lfloor}\ensuremath{\neg}\textsuperscript{m}(P \ensuremath{\varphi}) \ensuremath{\leftrightarrow}\textsuperscript{m} P ~\textsuperscript{m}\ensuremath{\varphi}\ensuremath{\rfloor} :=
  fun w => \ensuremath{\langle}(Ax2a \ensuremath{\varphi} w).1.resolve_left, fun hp hq => (Ax2a \ensuremath{\varphi} w).2 \ensuremath{\langle}hq, hp\ensuremath{\rangle}\ensuremath{\rangle}

/-- God-like -/
def G (x : e) : \ensuremath{\sigma} := \ensuremath{\forall}\textsuperscript{m} (\ensuremath{\varphi} : e \ensuremath{\rightarrow} \ensuremath{\sigma}), P \ensuremath{\varphi} \ensuremath{\rightarrow}\textsuperscript{m} \ensuremath{\varphi} x

/-- Necessary property inclusion -/
@[simp, grind] def PInc (\ensuremath{\varphi} \ensuremath{\psi} : e \ensuremath{\rightarrow} \ensuremath{\sigma}) : \ensuremath{\sigma} := \ensuremath{\Box}(\ensuremath{\forall}\textsuperscript{E} y, \ensuremath{\varphi} y \ensuremath{\rightarrow}\textsuperscript{m} \ensuremath{\psi} y)
infixr:48 " \ensuremath{\supset}\textsuperscript{N} " => PInc

/-- `Ess \ensuremath{\varphi} x`: \ensuremath{\varphi} is an essence of x (modified: \ensuremath{\varphi} must be exemplified by x) -/
def Ess (\ensuremath{\varphi} : e \ensuremath{\rightarrow} \ensuremath{\sigma}) (x : e) : \ensuremath{\sigma} := \ensuremath{\varphi} x \ensuremath{\wedge}\textsuperscript{m} (\ensuremath{\forall}\textsuperscript{m} (\ensuremath{\psi} : e \ensuremath{\rightarrow} \ensuremath{\sigma}), \ensuremath{\psi} x \ensuremath{\rightarrow}\textsuperscript{m} (\ensuremath{\varphi} \ensuremath{\supset}\textsuperscript{N} \ensuremath{\psi}))

axiom Ax2b (\ensuremath{\varphi} : e \ensuremath{\rightarrow} \ensuremath{\sigma}) : \ensuremath{\lfloor}P \ensuremath{\varphi} \ensuremath{\rightarrow}\textsuperscript{m} \ensuremath{\Box} P \ensuremath{\varphi}\ensuremath{\rfloor}

theorem Ax2b' (\ensuremath{\varphi} : e \ensuremath{\rightarrow} \ensuremath{\sigma}) : \ensuremath{\lfloor}\ensuremath{\neg}\textsuperscript{m}(P \ensuremath{\varphi}) \ensuremath{\rightarrow}\textsuperscript{m} \ensuremath{\Box}(\ensuremath{\neg}\textsuperscript{m}(P \ensuremath{\varphi}))\ensuremath{\rfloor} :=
  fun w hn v hv => (Ax2a' \ensuremath{\varphi} v).2 (Ax2b ~\textsuperscript{m}\ensuremath{\varphi} w ((Ax2a' \ensuremath{\varphi} w).1 hn) v hv)

/-- A property exemplified by a God-like being is positive (a consequence of `Ax2a`). -/
theorem PosOfGod \{x : e\} \{\ensuremath{\psi} : e \ensuremath{\rightarrow} \ensuremath{\sigma}\} \{w : i\} (hG : G x w) (h : \ensuremath{\psi} x w) : P \ensuremath{\psi} w :=
  Classical.byContradiction fun hn => hG ~\textsuperscript{m}\ensuremath{\psi} ((Ax2a' \ensuremath{\psi} w).1 hn) h

theorem Th1 (x : e) : \ensuremath{\lfloor}G x \ensuremath{\rightarrow}\textsuperscript{m} Ess G x\ensuremath{\rfloor} := fun w hG =>
  \ensuremath{\langle}hG, fun \ensuremath{\psi} h\ensuremath{\psi} v hv _y _ hGy => hGy \ensuremath{\psi} (Ax2b \ensuremath{\psi} w (PosOfGod hG h\ensuremath{\psi}) v hv)\ensuremath{\rangle}

/-- Necessary existence -/
def E (x : e) : \ensuremath{\sigma} := \ensuremath{\forall}\textsuperscript{m} (\ensuremath{\varphi} : e \ensuremath{\rightarrow} \ensuremath{\sigma}), Ess \ensuremath{\varphi} x \ensuremath{\rightarrow}\textsuperscript{m} \ensuremath{\Box}(\ensuremath{\exists}\textsuperscript{E} x, \ensuremath{\varphi} x)

axiom Ax3 : \ensuremath{\lfloor}P E\ensuremath{\rfloor}

theorem Th2 (x : e) : \ensuremath{\lfloor}G x \ensuremath{\rightarrow}\textsuperscript{m} \ensuremath{\Box}(\ensuremath{\exists}\textsuperscript{E} y, G y)\ensuremath{\rfloor} := fun w hG => hG E (Ax3 w) G (Th1 x w hG)

theorem Th3 : \ensuremath{\lfloor}\ensuremath{\Diamond}(\ensuremath{\exists}\textsuperscript{E} x, G x) \ensuremath{\rightarrow}\textsuperscript{m} \ensuremath{\Box}(\ensuremath{\exists}\textsuperscript{E} y, G y)\ensuremath{\rfloor} := by
  have h1 : \ensuremath{\lfloor}(\ensuremath{\exists}\textsuperscript{E} x, G x) \ensuremath{\rightarrow}\textsuperscript{m} \ensuremath{\Box}(\ensuremath{\exists}\textsuperscript{E} y, G y)\ensuremath{\rfloor} := fun _ \ensuremath{\langle}x, _, hG\ensuremath{\rangle} => Th2 x _ hG
  intro w \ensuremath{\langle}v, hv, hex\ensuremath{\rangle}
  exact h1 w (h1 v hex w (Rsymm w v hv))               -- only symmetry of `\textsf{\bfseries r}` is needed

axiom Ax4 (\ensuremath{\varphi} \ensuremath{\psi} : e \ensuremath{\rightarrow} \ensuremath{\sigma}) : \ensuremath{\lfloor}P \ensuremath{\varphi} \ensuremath{\wedge}\textsuperscript{m} (\ensuremath{\varphi} \ensuremath{\supset}\textsuperscript{N} \ensuremath{\psi}) \ensuremath{\rightarrow}\textsuperscript{m} P \ensuremath{\psi}\ensuremath{\rfloor}

-- `lemma True nitpick[satisfy,card=1,eval="\ensuremath{\lfloor}P (\ensuremath{\lambda}x.\ensuremath{\bot})\ensuremath{\rfloor}"] oops`
--   One model found of cardinality one (consistency check)

@[simp, grind] def PosProps (\ensuremath{\Phi} : (e \ensuremath{\rightarrow} \ensuremath{\sigma}) \ensuremath{\rightarrow} \ensuremath{\sigma}) : \ensuremath{\sigma} := \ensuremath{\forall}\textsuperscript{m} \ensuremath{\varphi}, \ensuremath{\Phi} \ensuremath{\varphi} \ensuremath{\rightarrow}\textsuperscript{m} P \ensuremath{\varphi}
@[simp, grind] def ConjOfPropsFrom (\ensuremath{\varphi} : e \ensuremath{\rightarrow} \ensuremath{\sigma}) (\ensuremath{\Phi} : (e \ensuremath{\rightarrow} \ensuremath{\sigma}) \ensuremath{\rightarrow} \ensuremath{\sigma}) : \ensuremath{\sigma} :=
  \ensuremath{\Box}(\ensuremath{\forall}\textsuperscript{E} z, \ensuremath{\varphi} z \ensuremath{\leftrightarrow}\textsuperscript{m} (\ensuremath{\forall}\textsuperscript{m} \ensuremath{\psi}, \ensuremath{\Phi} \ensuremath{\psi} \ensuremath{\rightarrow}\textsuperscript{m} \ensuremath{\psi} z))
axiom Ax1Gen (\ensuremath{\Phi} : (e \ensuremath{\rightarrow} \ensuremath{\sigma}) \ensuremath{\rightarrow} \ensuremath{\sigma}) (\ensuremath{\varphi} : e \ensuremath{\rightarrow} \ensuremath{\sigma}) :
  \ensuremath{\lfloor}(PosProps \ensuremath{\Phi} \ensuremath{\wedge}\textsuperscript{m} ConjOfPropsFrom \ensuremath{\varphi} \ensuremath{\Phi}) \ensuremath{\rightarrow}\textsuperscript{m} P \ensuremath{\varphi}\ensuremath{\rfloor}

theorem L : \ensuremath{\lfloor}P G\ensuremath{\rfloor} := fun w => Ax1Gen P G w \ensuremath{\langle}fun _ h => h, fun _ _ _ _ => Iff.rfl\ensuremath{\rangle}

/-- Possibly there is a God-like being.  The Isabelle proof cites `Ax2a`, `Ax4` and `L`; `Ax4` is
dispensable: if no God-like being were possible, `Ax1Gen` would make the empty property
positive, as the conjunction of the positive properties `\{G\}`, while it makes the universal
property positive as the conjunction of `\ensuremath{\emptyset}`, against the exclusivity in `Ax2a`.  Only `Ax1Gen`
and `Ax2a` are used, and no frame condition. -/
theorem Th4 : \ensuremath{\lfloor}\ensuremath{\Diamond}(\ensuremath{\exists}\textsuperscript{E} x, G x)\ensuremath{\rfloor} := fun w =>
  Classical.byContradiction fun hn =>
    have hbot : P (fun _ : e => \ensuremath{\bot}\textsuperscript{m}) w :=
      Ax1Gen (fun \ensuremath{\psi} => \ensuremath{\psi} =\textsuperscript{m} G) (fun _ => \ensuremath{\bot}\textsuperscript{m}) w
        \ensuremath{\langle}fun \ensuremath{\psi} h\ensuremath{\psi} => by simp only [Mprimeq] at h\ensuremath{\psi}; subst h\ensuremath{\psi}; exact L w,
         fun v hv z hz => \ensuremath{\langle}fun h => h.elim, fun h => hn \ensuremath{\langle}v, hv, z, hz, h G rfl\ensuremath{\rangle}\ensuremath{\rangle}\ensuremath{\rangle}
    have htop : P (fun _ : e => \ensuremath{\top}\textsuperscript{m}) w :=
      Ax1Gen (fun _ => \ensuremath{\bot}\textsuperscript{m}) (fun _ => \ensuremath{\top}\textsuperscript{m}) w
        \ensuremath{\langle}fun _ h => h.elim, fun _ _ _ _ => \ensuremath{\langle}fun _ _ h => h.elim, fun _ => trivial\ensuremath{\rangle}\ensuremath{\rangle}
    have heq : (~\textsuperscript{m}(fun _ : e => \ensuremath{\bot}\textsuperscript{m})) = (fun _ : e => \ensuremath{\top}\textsuperscript{m}) := by funext x v; simp
    (Ax2a (fun _ => \ensuremath{\bot}\textsuperscript{m}) w).2 \ensuremath{\langle}hbot, heq \ensuremath{\blacktriangleright} htop\ensuremath{\rangle}

theorem Th5 : \ensuremath{\lfloor}\ensuremath{\Box}(\ensuremath{\exists}\textsuperscript{E} x, G x)\ensuremath{\rfloor} := fun w => Th3 w (Th4 w)

theorem MC (\ensuremath{\varphi} : \ensuremath{\sigma}) : \ensuremath{\lfloor}\ensuremath{\varphi} \ensuremath{\rightarrow}\textsuperscript{m} \ensuremath{\Box}\ensuremath{\varphi}\ensuremath{\rfloor} := by                   -- modal collapse
  intro w h\ensuremath{\varphi} v hv
  have \ensuremath{\langle}x, _, hGx\ensuremath{\rangle} := Th5 v w (Rsymm w v hv)
  have \ensuremath{\langle}z, hzv, hGz\ensuremath{\rangle} := Th5 w v hv
  exact (Th1 x w hGx).2 (fun _ => \ensuremath{\varphi}) h\ensuremath{\varphi} v hv z hzv hGz

/-- The universal property is positive, from `Ax2a` and `Ax4` alone: were its complement
positive, `Ax4` would make every property positive, against the exclusivity in `Ax2a`. -/
theorem PosTop (w : i) : P (fun _ : e => \ensuremath{\top}\textsuperscript{m}) w :=
  (Ax2a (fun _ : e => \ensuremath{\top}\textsuperscript{m}) w).1.elim id fun hB =>
    ((Ax2a (fun _ : e => \ensuremath{\top}\textsuperscript{m}) w).2 \ensuremath{\langle}Ax4 _ _ w \ensuremath{\langle}hB, fun _ _ _ _ _ => trivial\ensuremath{\rangle}, hB\ensuremath{\rangle}).elim

theorem PosProps' : \ensuremath{\lfloor}P (fun _ : e => \ensuremath{\top}\textsuperscript{m}) \ensuremath{\wedge}\textsuperscript{m} P (fun x : e => x =\textsuperscript{m} x)\ensuremath{\rfloor} := fun w =>
  \ensuremath{\langle}PosTop w, Ax4 _ _ w \ensuremath{\langle}PosTop w, fun _ _ _ _ _ => rfl\ensuremath{\rangle}\ensuremath{\rangle}

theorem NegProps : \ensuremath{\lfloor}\ensuremath{\neg}\textsuperscript{m}P (fun _ : e => \ensuremath{\bot}\textsuperscript{m}) \ensuremath{\wedge}\textsuperscript{m} \ensuremath{\neg}\textsuperscript{m}P (fun x : e => x \ensuremath{\neq}\textsuperscript{m} x)\ensuremath{\rfloor} := fun w =>
  \ensuremath{\langle}fun hp => (Ax2a _ w).2 \ensuremath{\langle}hp, Ax4 _ _ w \ensuremath{\langle}hp, fun _ _ _ _ h => h.elim\ensuremath{\rangle}\ensuremath{\rangle},
   fun hp => (Ax2a _ w).2 \ensuremath{\langle}hp, Ax4 _ _ w \ensuremath{\langle}hp, fun _ _ _ _ h => (h rfl).elim\ensuremath{\rangle}\ensuremath{\rangle}\ensuremath{\rangle}

theorem UniqueEss1 (\ensuremath{\varphi} \ensuremath{\psi} : e \ensuremath{\rightarrow} \ensuremath{\sigma}) (x : e) : \ensuremath{\lfloor}Ess \ensuremath{\varphi} x \ensuremath{\wedge}\textsuperscript{m} Ess \ensuremath{\psi} x \ensuremath{\rightarrow}\textsuperscript{m} \ensuremath{\Box}(\ensuremath{\forall}\textsuperscript{E} y, \ensuremath{\varphi} y \ensuremath{\leftrightarrow}\textsuperscript{m} \ensuremath{\psi} y)\ensuremath{\rfloor} :=
  fun _ \ensuremath{\langle}\ensuremath{\langle}h\ensuremath{\varphi}x, h\ensuremath{\varphi}\ensuremath{\rangle}, \ensuremath{\langle}h\ensuremath{\psi}x, h\ensuremath{\psi}\ensuremath{\rangle}\ensuremath{\rangle} v hv y hy => \ensuremath{\langle}h\ensuremath{\varphi} \ensuremath{\psi} h\ensuremath{\psi}x v hv y hy, h\ensuremath{\psi} \ensuremath{\varphi} h\ensuremath{\varphi}x v hv y hy\ensuremath{\rangle}

-- UniqueEss2 : \ensuremath{\lfloor}Ess \ensuremath{\varphi} x \ensuremath{\wedge}\textsuperscript{m} Ess \ensuremath{\psi} x \ensuremath{\rightarrow}\textsuperscript{m} \ensuremath{\Box}(\ensuremath{\varphi} \ensuremath{\equiv}\textsuperscript{m} \ensuremath{\psi})\ensuremath{\rfloor}   -- `nitpick[card i=1]`: countermodel found
example (\ensuremath{\varphi} \ensuremath{\psi} : e \ensuremath{\rightarrow} \ensuremath{\sigma}) (x : e) : \ensuremath{\lfloor}Ess \ensuremath{\varphi} x \ensuremath{\wedge}\textsuperscript{m} Ess \ensuremath{\psi} x \ensuremath{\rightarrow}\textsuperscript{m} \ensuremath{\Box}(\ensuremath{\varphi} \ensuremath{\equiv}\textsuperscript{m} \ensuremath{\psi})\ensuremath{\rfloor} := by countermodel

theorem UniqueEss3 (\ensuremath{\varphi} : e \ensuremath{\rightarrow} \ensuremath{\sigma}) (x : e) : \ensuremath{\lfloor}Ess \ensuremath{\varphi} x \ensuremath{\rightarrow}\textsuperscript{m} \ensuremath{\Box}(\ensuremath{\forall}\textsuperscript{E} y, \ensuremath{\varphi} y \ensuremath{\rightarrow}\textsuperscript{m} y \ensuremath{\equiv}\textsuperscript{m} x)\ensuremath{\rfloor} :=
  fun _ \ensuremath{\langle}_, h\ensuremath{\rangle} => h (fun z => z \ensuremath{\equiv}\textsuperscript{m} x) (fun _ hp => hp)

theorem Monotheism (x y : e) : \ensuremath{\lfloor}G x \ensuremath{\wedge}\textsuperscript{m} G y \ensuremath{\rightarrow}\textsuperscript{m} x \ensuremath{\equiv}\textsuperscript{m} y\ensuremath{\rfloor} :=
  fun _ \ensuremath{\langle}hx, hy\ensuremath{\rangle} Q hQ => hy Q (PosOfGod hx hQ)

theorem PisFilter : \ensuremath{\lfloor}Filter P\ensuremath{\rfloor} := fun w =>
  \ensuremath{\langle}\ensuremath{\langle}\ensuremath{\langle}(PosProps' w).1, (NegProps w).1\ensuremath{\rangle},
    -- the inclusion holds at `w`; `MC` carries it to the successor worlds, then `Ax4` applies
    fun \ensuremath{\varphi} \ensuremath{\psi} \ensuremath{\langle}h\ensuremath{\varphi}, hsub\ensuremath{\rangle} => Ax4 \ensuremath{\varphi} \ensuremath{\psi} w \ensuremath{\langle}h\ensuremath{\varphi}, MC (\ensuremath{\forall}\textsuperscript{E} x, \ensuremath{\varphi} x \ensuremath{\rightarrow}\textsuperscript{m} \ensuremath{\psi} x) w hsub\ensuremath{\rangle}\ensuremath{\rangle},
   fun \ensuremath{\varphi} \ensuremath{\psi} \ensuremath{\langle}h\ensuremath{\varphi}, h\ensuremath{\psi}\ensuremath{\rangle} => Ax1 \ensuremath{\varphi} \ensuremath{\psi} w \ensuremath{\langle}h\ensuremath{\varphi}, h\ensuremath{\psi}\ensuremath{\rangle}\ensuremath{\rangle}

theorem PisUFilter : \ensuremath{\lfloor}UFilter P\ensuremath{\rfloor} := fun w =>
  \ensuremath{\langle}PisFilter w, fun \ensuremath{\varphi} => (Classical.em (P \ensuremath{\varphi} w)).imp id (Ax2a' \ensuremath{\varphi} w).1\ensuremath{\rangle}

-- `lemma True nitpick[satisfy,card=1,eval="\ensuremath{\lfloor}P (\ensuremath{\lambda}x.\ensuremath{\bot})\ensuremath{\rfloor}"] oops`
--   One model found of cardinality one (consistency check)
\end{Verbatim}
\subsubsection{GoedelVariantHOML3.lean (Figure 8 of \texorpdfstring{\cite{J75}}{the Notes})}
After an appropriate modification of the notion of necessary property inclusion in G\"odel's 1970 ontological proof, the inconsistency revealed in Figure 6 is avoided, and the argument can be successfully verified in modal logic S5 (indeed, as shown, only symmetry of the accessibility relation is actually needed).  Actualist quantifiers (avoiding existential import) are used for quantification over entities, otherwise possibilist quantifiers are used.
\begin{Verbatim}[commandchars=\\\{\},fontsize=\scriptsize,xleftmargin=1em]
import Notes.HOMLinHOL
import Notes.ModalFilter
\end{Verbatim}

\begin{Verbatim}[commandchars=\\\{\},fontsize=\scriptsize,xleftmargin=1em]
/-- Positive property -/
axiom P : (e \ensuremath{\rightarrow} \ensuremath{\sigma}) \ensuremath{\rightarrow} \ensuremath{\sigma}

axiom Ax1 (\ensuremath{\varphi} \ensuremath{\psi} : e \ensuremath{\rightarrow} \ensuremath{\sigma}) : \ensuremath{\lfloor}P \ensuremath{\varphi} \ensuremath{\wedge}\textsuperscript{m} P \ensuremath{\psi} \ensuremath{\rightarrow}\textsuperscript{m} P (\ensuremath{\varphi} \ensuremath{\centerdot}\textsuperscript{m} \ensuremath{\psi})\ensuremath{\rfloor}

axiom Ax2a (\ensuremath{\varphi} : e \ensuremath{\rightarrow} \ensuremath{\sigma}) : \ensuremath{\lfloor}P \ensuremath{\varphi} \ensuremath{\vee}\textsuperscript{e} P ~\textsuperscript{m}\ensuremath{\varphi}\ensuremath{\rfloor}

/-- Auxiliary reformulation of `Ax2a` (Lean has no `sledgehammer`). -/
theorem Ax2a' (\ensuremath{\varphi} : e \ensuremath{\rightarrow} \ensuremath{\sigma}) : \ensuremath{\lfloor}\ensuremath{\neg}\textsuperscript{m}(P \ensuremath{\varphi}) \ensuremath{\leftrightarrow}\textsuperscript{m} P ~\textsuperscript{m}\ensuremath{\varphi}\ensuremath{\rfloor} :=
  fun w => \ensuremath{\langle}(Ax2a \ensuremath{\varphi} w).1.resolve_left, fun hp hq => (Ax2a \ensuremath{\varphi} w).2 \ensuremath{\langle}hq, hp\ensuremath{\rangle}\ensuremath{\rangle}

/-- God-like -/
def G (x : e) : \ensuremath{\sigma} := \ensuremath{\forall}\textsuperscript{m} (\ensuremath{\varphi} : e \ensuremath{\rightarrow} \ensuremath{\sigma}), P \ensuremath{\varphi} \ensuremath{\rightarrow}\textsuperscript{m} \ensuremath{\varphi} x

/-- Necessary property inclusion (modified: \ensuremath{\varphi} must be non-empty) -/
@[simp, grind] def PInc (\ensuremath{\varphi} \ensuremath{\psi} : e \ensuremath{\rightarrow} \ensuremath{\sigma}) : \ensuremath{\sigma} :=
  \ensuremath{\Box}((\ensuremath{\varphi} \ensuremath{\neq}\textsuperscript{m} (fun _ : e => \ensuremath{\bot}\textsuperscript{m})) \ensuremath{\wedge}\textsuperscript{m} (\ensuremath{\forall}\textsuperscript{E} y, \ensuremath{\varphi} y \ensuremath{\rightarrow}\textsuperscript{m} \ensuremath{\psi} y))
infixr:48 " \ensuremath{\supset}\textsuperscript{N} " => PInc

/-- `Ess \ensuremath{\varphi} x`: \ensuremath{\varphi} is an essence of x -/
def Ess (\ensuremath{\varphi} : e \ensuremath{\rightarrow} \ensuremath{\sigma}) (x : e) : \ensuremath{\sigma} := \ensuremath{\forall}\textsuperscript{m} (\ensuremath{\psi} : e \ensuremath{\rightarrow} \ensuremath{\sigma}), \ensuremath{\psi} x \ensuremath{\rightarrow}\textsuperscript{m} (\ensuremath{\varphi} \ensuremath{\supset}\textsuperscript{N} \ensuremath{\psi})

axiom Ax2b (\ensuremath{\varphi} : e \ensuremath{\rightarrow} \ensuremath{\sigma}) : \ensuremath{\lfloor}P \ensuremath{\varphi} \ensuremath{\rightarrow}\textsuperscript{m} \ensuremath{\Box} P \ensuremath{\varphi}\ensuremath{\rfloor}

theorem Ax2b' (\ensuremath{\varphi} : e \ensuremath{\rightarrow} \ensuremath{\sigma}) : \ensuremath{\lfloor}\ensuremath{\neg}\textsuperscript{m}(P \ensuremath{\varphi}) \ensuremath{\rightarrow}\textsuperscript{m} \ensuremath{\Box}(\ensuremath{\neg}\textsuperscript{m}(P \ensuremath{\varphi}))\ensuremath{\rfloor} :=
  fun w hn v hv => (Ax2a' \ensuremath{\varphi} v).2 (Ax2b ~\textsuperscript{m}\ensuremath{\varphi} w ((Ax2a' \ensuremath{\varphi} w).1 hn) v hv)

/-- A property exemplified by a God-like being is positive (a consequence of `Ax2a`). -/
theorem PosOfGod \{x : e\} \{\ensuremath{\psi} : e \ensuremath{\rightarrow} \ensuremath{\sigma}\} \{w : i\} (hG : G x w) (h : \ensuremath{\psi} x w) : P \ensuremath{\psi} w :=
  Classical.byContradiction fun hn => hG ~\textsuperscript{m}\ensuremath{\psi} ((Ax2a' \ensuremath{\psi} w).1 hn) h

theorem Th1 (x : e) : \ensuremath{\lfloor}G x \ensuremath{\rightarrow}\textsuperscript{m} Ess G x\ensuremath{\rfloor} := by
  intro w hG \ensuremath{\psi} h\ensuremath{\psi} v hv
  refine \ensuremath{\langle}fun heq => ?_, fun y _ hGy => hGy \ensuremath{\psi} (Ax2b \ensuremath{\psi} w (PosOfGod hG h\ensuremath{\psi}) v hv)\ensuremath{\rangle}
  rw [heq] at hG; exact hG                             -- `G` is non-empty since `G x` holds

/-- Necessary existence -/
def E (x : e) : \ensuremath{\sigma} := \ensuremath{\forall}\textsuperscript{m} (\ensuremath{\varphi} : e \ensuremath{\rightarrow} \ensuremath{\sigma}), Ess \ensuremath{\varphi} x \ensuremath{\rightarrow}\textsuperscript{m} \ensuremath{\Box}(\ensuremath{\exists}\textsuperscript{E} x, \ensuremath{\varphi} x)

axiom Ax3 : \ensuremath{\lfloor}P E\ensuremath{\rfloor}

theorem Th2 (x : e) : \ensuremath{\lfloor}G x \ensuremath{\rightarrow}\textsuperscript{m} \ensuremath{\Box}(\ensuremath{\exists}\textsuperscript{E} y, G y)\ensuremath{\rfloor} := fun w hG => hG E (Ax3 w) G (Th1 x w hG)

theorem Th3 : \ensuremath{\lfloor}\ensuremath{\Diamond}(\ensuremath{\exists}\textsuperscript{E} x, G x) \ensuremath{\rightarrow}\textsuperscript{m} \ensuremath{\Box}(\ensuremath{\exists}\textsuperscript{E} y, G y)\ensuremath{\rfloor} := by
  have h1 : \ensuremath{\lfloor}(\ensuremath{\exists}\textsuperscript{E} x, G x) \ensuremath{\rightarrow}\textsuperscript{m} \ensuremath{\Box}(\ensuremath{\exists}\textsuperscript{E} y, G y)\ensuremath{\rfloor} := fun _ \ensuremath{\langle}x, _, hG\ensuremath{\rangle} => Th2 x _ hG
  intro w \ensuremath{\langle}v, hv, hex\ensuremath{\rangle}
  exact h1 w (h1 v hex w (Rsymm w v hv))               -- only symmetry of `\textsf{\bfseries r}` is needed

axiom Ax4 (\ensuremath{\varphi} \ensuremath{\psi} : e \ensuremath{\rightarrow} \ensuremath{\sigma}) : \ensuremath{\lfloor}P \ensuremath{\varphi} \ensuremath{\wedge}\textsuperscript{m} (\ensuremath{\varphi} \ensuremath{\supset}\textsuperscript{N} \ensuremath{\psi}) \ensuremath{\rightarrow}\textsuperscript{m} P \ensuremath{\psi}\ensuremath{\rfloor}

-- `lemma True nitpick[satisfy,card=1,eval="\ensuremath{\lfloor}P (\ensuremath{\lambda}x.\ensuremath{\bot})\ensuremath{\rfloor}"] oops`
--   Two models found of cardinality one (consistency check)

@[simp, grind] def PosProps (\ensuremath{\Phi} : (e \ensuremath{\rightarrow} \ensuremath{\sigma}) \ensuremath{\rightarrow} \ensuremath{\sigma}) : \ensuremath{\sigma} := \ensuremath{\forall}\textsuperscript{m} \ensuremath{\varphi}, \ensuremath{\Phi} \ensuremath{\varphi} \ensuremath{\rightarrow}\textsuperscript{m} P \ensuremath{\varphi}
@[simp, grind] def ConjOfPropsFrom (\ensuremath{\varphi} : e \ensuremath{\rightarrow} \ensuremath{\sigma}) (\ensuremath{\Phi} : (e \ensuremath{\rightarrow} \ensuremath{\sigma}) \ensuremath{\rightarrow} \ensuremath{\sigma}) : \ensuremath{\sigma} :=
  \ensuremath{\Box}(\ensuremath{\forall}\textsuperscript{E} z, \ensuremath{\varphi} z \ensuremath{\leftrightarrow}\textsuperscript{m} (\ensuremath{\forall}\textsuperscript{m} \ensuremath{\psi}, \ensuremath{\Phi} \ensuremath{\psi} \ensuremath{\rightarrow}\textsuperscript{m} \ensuremath{\psi} z))
axiom Ax1Gen (\ensuremath{\Phi} : (e \ensuremath{\rightarrow} \ensuremath{\sigma}) \ensuremath{\rightarrow} \ensuremath{\sigma}) (\ensuremath{\varphi} : e \ensuremath{\rightarrow} \ensuremath{\sigma}) :
  \ensuremath{\lfloor}(PosProps \ensuremath{\Phi} \ensuremath{\wedge}\textsuperscript{m} ConjOfPropsFrom \ensuremath{\varphi} \ensuremath{\Phi}) \ensuremath{\rightarrow}\textsuperscript{m} P \ensuremath{\varphi}\ensuremath{\rfloor}

theorem L : \ensuremath{\lfloor}P G\ensuremath{\rfloor} := fun w => Ax1Gen P G w \ensuremath{\langle}fun _ h => h, fun _ _ _ _ => Iff.rfl\ensuremath{\rangle}

/-- The AFP sources leave `Th4` unreplayed (`oops`, then a postulate), although Vampire,
Zipperposition, Leo-II and Leo-III report proofs from `Ax2a`, `L` and `Ax1Gen`.  Here is one.
If no God-like being were possible, `Ax1Gen` would make the empty property positive, as the
conjunction of the positive properties `\{G\}`; it also makes the universal property positive,
as the conjunction of `\ensuremath{\emptyset}`; and the two together contradict the exclusivity in `Ax2a`.  Only
`Ax1Gen` and `Ax2a` are used, and no frame condition: the theorem holds in logic K. -/
theorem Th4 : \ensuremath{\lfloor}\ensuremath{\Diamond}(\ensuremath{\exists}\textsuperscript{E} x, G x)\ensuremath{\rfloor} := fun w =>
  Classical.byContradiction fun hn =>
    have hbot : P (fun _ : e => \ensuremath{\bot}\textsuperscript{m}) w :=
      Ax1Gen (fun \ensuremath{\psi} => \ensuremath{\psi} =\textsuperscript{m} G) (fun _ => \ensuremath{\bot}\textsuperscript{m}) w
        \ensuremath{\langle}fun \ensuremath{\psi} h\ensuremath{\psi} => by simp only [Mprimeq] at h\ensuremath{\psi}; subst h\ensuremath{\psi}; exact L w,
         fun v hv z hz => \ensuremath{\langle}fun h => h.elim, fun h => hn \ensuremath{\langle}v, hv, z, hz, h G rfl\ensuremath{\rangle}\ensuremath{\rangle}\ensuremath{\rangle}
    have htop : P (fun _ : e => \ensuremath{\top}\textsuperscript{m}) w :=
      Ax1Gen (fun _ => \ensuremath{\bot}\textsuperscript{m}) (fun _ => \ensuremath{\top}\textsuperscript{m}) w
        \ensuremath{\langle}fun _ h => h.elim, fun _ _ _ _ => \ensuremath{\langle}fun _ _ h => h.elim, fun _ => trivial\ensuremath{\rangle}\ensuremath{\rangle}
    have heq : (~\textsuperscript{m}(fun _ : e => \ensuremath{\bot}\textsuperscript{m})) = (fun _ : e => \ensuremath{\top}\textsuperscript{m}) := by funext x v; simp
    (Ax2a (fun _ => \ensuremath{\bot}\textsuperscript{m}) w).2 \ensuremath{\langle}hbot, heq \ensuremath{\blacktriangleright} htop\ensuremath{\rangle}

theorem Th5 : \ensuremath{\lfloor}\ensuremath{\Box}(\ensuremath{\exists}\textsuperscript{E} x, G x)\ensuremath{\rfloor} := fun w => Th3 w (Th4 w)

theorem MC (\ensuremath{\varphi} : \ensuremath{\sigma}) : \ensuremath{\lfloor}\ensuremath{\varphi} \ensuremath{\rightarrow}\textsuperscript{m} \ensuremath{\Box}\ensuremath{\varphi}\ensuremath{\rfloor} := by                   -- modal collapse
  intro w h\ensuremath{\varphi} v hv
  have \ensuremath{\langle}x, _, hGx\ensuremath{\rangle} := Th5 v w (Rsymm w v hv)
  have \ensuremath{\langle}z, hzv, hGz\ensuremath{\rangle} := Th5 w v hv
  exact (Th1 x w hGx) (fun _ => \ensuremath{\varphi}) h\ensuremath{\varphi} v hv |>.2 z hzv hGz

/-- `G` is not the empty property: by `Th4` it is exemplified at some successor of `w`. -/
theorem GNonempty (w : i) : G \ensuremath{\neq} (fun _ : e => \ensuremath{\bot}\textsuperscript{m}) := fun h =>
  let \ensuremath{\langle}_, _, _, _, hG\ensuremath{\rangle} := Th4 w; (h \ensuremath{\blacktriangleright} hG : \ensuremath{\bot}\textsuperscript{m} _)

/-- `Ax4` applied to `G`: everything true of every actual God-like being at every successor
world is positive.  Uses `L`, `Th4` and `Ax4`, hence `Ax1Gen`, `Ax2a`, `Ax4`; no frame condition. -/
theorem PosIncl (\ensuremath{\chi} : e \ensuremath{\rightarrow} \ensuremath{\sigma}) (w : i) (h : \ensuremath{\forall} v, w \textsf{\bfseries r} v \ensuremath{\rightarrow} \ensuremath{\forall} x, x @\textsuperscript{m} v \ensuremath{\rightarrow} G x v \ensuremath{\rightarrow} \ensuremath{\chi} x v) : P \ensuremath{\chi} w :=
  Ax4 G \ensuremath{\chi} w \ensuremath{\langle}L w, fun v hv => \ensuremath{\langle}GNonempty w, h v hv\ensuremath{\rangle}\ensuremath{\rangle}

theorem PosProps' : \ensuremath{\lfloor}P (fun _ : e => \ensuremath{\top}\textsuperscript{m}) \ensuremath{\wedge}\textsuperscript{m} P (fun x : e => x =\textsuperscript{m} x)\ensuremath{\rfloor} := fun w =>
  \ensuremath{\langle}PosIncl _ w (fun _ _ _ _ _ => trivial), PosIncl _ w (fun _ _ _ _ _ => rfl)\ensuremath{\rangle}

theorem NegProps : \ensuremath{\lfloor}\ensuremath{\neg}\textsuperscript{m}P (fun _ : e => \ensuremath{\bot}\textsuperscript{m}) \ensuremath{\wedge}\textsuperscript{m} \ensuremath{\neg}\textsuperscript{m}P (fun x : e => x \ensuremath{\neq}\textsuperscript{m} x)\ensuremath{\rfloor} := fun w =>
  \ensuremath{\langle}fun hB => (Ax2a _ w).2 \ensuremath{\langle}hB, (show (~\textsuperscript{m}(fun _ : e => \ensuremath{\bot}\textsuperscript{m})) = (fun _ : e => \ensuremath{\top}\textsuperscript{m}) by
      funext x v; simp) \ensuremath{\blacktriangleright} (PosProps' w).1\ensuremath{\rangle},
   fun hN => (Ax2a _ w).2 \ensuremath{\langle}hN, (show (~\textsuperscript{m}(fun x : e => x \ensuremath{\neq}\textsuperscript{m} x)) = (fun _ : e => \ensuremath{\top}\textsuperscript{m}) by
      funext x v; simp) \ensuremath{\blacktriangleright} (PosProps' w).1\ensuremath{\rangle}\ensuremath{\rangle}

-- UniqueEss1 : \ensuremath{\lfloor}Ess \ensuremath{\varphi} x \ensuremath{\wedge}\textsuperscript{m} Ess \ensuremath{\psi} x \ensuremath{\rightarrow}\textsuperscript{m} \ensuremath{\Box}(\ensuremath{\forall}\textsuperscript{E} y, \ensuremath{\varphi} y \ensuremath{\leftrightarrow}\textsuperscript{m} \ensuremath{\psi} y)\ensuremath{\rfloor}      -- Unclear, open question
example (\ensuremath{\varphi} \ensuremath{\psi} : e \ensuremath{\rightarrow} \ensuremath{\sigma}) (x : e) : \ensuremath{\lfloor}Ess \ensuremath{\varphi} x \ensuremath{\wedge}\textsuperscript{m} Ess \ensuremath{\psi} x \ensuremath{\rightarrow}\textsuperscript{m} \ensuremath{\Box}(\ensuremath{\forall}\textsuperscript{E} y, \ensuremath{\varphi} y \ensuremath{\leftrightarrow}\textsuperscript{m} \ensuremath{\psi} y)\ensuremath{\rfloor} := by openproblem
-- UniqueEss2 : \ensuremath{\lfloor}Ess \ensuremath{\varphi} x \ensuremath{\wedge}\textsuperscript{m} Ess \ensuremath{\psi} x \ensuremath{\rightarrow}\textsuperscript{m} \ensuremath{\Box}(\ensuremath{\varphi} \ensuremath{\equiv}\textsuperscript{m} \ensuremath{\psi})\ensuremath{\rfloor}                -- Unclear, open question
example (\ensuremath{\varphi} \ensuremath{\psi} : e \ensuremath{\rightarrow} \ensuremath{\sigma}) (x : e) : \ensuremath{\lfloor}Ess \ensuremath{\varphi} x \ensuremath{\wedge}\textsuperscript{m} Ess \ensuremath{\psi} x \ensuremath{\rightarrow}\textsuperscript{m} \ensuremath{\Box}(\ensuremath{\varphi} \ensuremath{\equiv}\textsuperscript{m} \ensuremath{\psi})\ensuremath{\rfloor} := by openproblem

theorem UniqueEss3 (\ensuremath{\varphi} : e \ensuremath{\rightarrow} \ensuremath{\sigma}) (x : e) : \ensuremath{\lfloor}Ess \ensuremath{\varphi} x \ensuremath{\rightarrow}\textsuperscript{m} \ensuremath{\Box}(\ensuremath{\forall}\textsuperscript{E} y, \ensuremath{\varphi} y \ensuremath{\rightarrow}\textsuperscript{m} y \ensuremath{\equiv}\textsuperscript{m} x)\ensuremath{\rfloor} :=
  fun _ h v hv => (h (fun z => z \ensuremath{\equiv}\textsuperscript{m} x) (fun _ hp => hp) v hv).2

theorem Monotheism (x y : e) : \ensuremath{\lfloor}G x \ensuremath{\wedge}\textsuperscript{m} G y \ensuremath{\rightarrow}\textsuperscript{m} x \ensuremath{\equiv}\textsuperscript{m} y\ensuremath{\rfloor} :=
  fun _ \ensuremath{\langle}hx, hy\ensuremath{\rangle} Q hQ => hy Q (PosOfGod hx hQ)

theorem PisFilter : \ensuremath{\lfloor}Filter P\ensuremath{\rfloor} := fun w =>
  \ensuremath{\langle}\ensuremath{\langle}\ensuremath{\langle}(PosProps' w).1, (NegProps w).1\ensuremath{\rangle},
    -- the inclusion holds at `w`; `MC` carries it to the successor worlds, then `Ax4` applies
    fun \ensuremath{\varphi} \ensuremath{\psi} \ensuremath{\langle}h\ensuremath{\varphi}, hsub\ensuremath{\rangle} => Ax4 \ensuremath{\varphi} \ensuremath{\psi} w \ensuremath{\langle}h\ensuremath{\varphi}, fun v hv =>
      \ensuremath{\langle}fun h => (NegProps w).1 (h \ensuremath{\blacktriangleright} h\ensuremath{\varphi}), MC (\ensuremath{\forall}\textsuperscript{E} x, \ensuremath{\varphi} x \ensuremath{\rightarrow}\textsuperscript{m} \ensuremath{\psi} x) w hsub v hv\ensuremath{\rangle}\ensuremath{\rangle}\ensuremath{\rangle},
   fun \ensuremath{\varphi} \ensuremath{\psi} \ensuremath{\langle}h\ensuremath{\varphi}, h\ensuremath{\psi}\ensuremath{\rangle} => Ax1 \ensuremath{\varphi} \ensuremath{\psi} w \ensuremath{\langle}h\ensuremath{\varphi}, h\ensuremath{\psi}\ensuremath{\rangle}\ensuremath{\rangle}

theorem PisUFilter : \ensuremath{\lfloor}UFilter P\ensuremath{\rfloor} := fun w =>
  \ensuremath{\langle}PisFilter w, fun \ensuremath{\varphi} => (Classical.em (P \ensuremath{\varphi} w)).imp id (Ax2a' \ensuremath{\varphi} w).1\ensuremath{\rangle}

-- `lemma True nitpick[satisfy,card=1,eval="\ensuremath{\lfloor}P (\ensuremath{\lambda}x.\ensuremath{\top})\ensuremath{\rfloor}"] oops`
--   One model found of cardinality one (consistency check)
\end{Verbatim}
\subsubsection{ThereIsNoEvil1.lean (Figure 10 of \texorpdfstring{\cite{J75}}{the Notes})}
Importing G\"odel's modified axioms from Figure 7 we can prove that necessarily there exists no entity that possesses all non-positive (= negative) properties.
\begin{Verbatim}[commandchars=\\\{\},fontsize=\scriptsize,xleftmargin=1em]
import Notes.GoedelVariantHOML2
\end{Verbatim}

\begin{Verbatim}[commandchars=\\\{\},fontsize=\scriptsize,xleftmargin=1em]
/-- An entity possessing all non-positive properties -/
def Evil (x : e) : \ensuremath{\sigma} := \ensuremath{\forall}\textsuperscript{m} (\ensuremath{\varphi} : e \ensuremath{\rightarrow} \ensuremath{\sigma}), \ensuremath{\neg}\textsuperscript{m}(P \ensuremath{\varphi}) \ensuremath{\rightarrow}\textsuperscript{m} \ensuremath{\varphi} x

theorem NecNoEvil : \ensuremath{\lfloor}\ensuremath{\Box}(\ensuremath{\neg}\textsuperscript{m}(\ensuremath{\exists}\textsuperscript{E} x, Evil x))\ensuremath{\rfloor} :=
  fun _ v _ \ensuremath{\langle}_, _, hE\ensuremath{\rangle} => hE (fun _ : e => \ensuremath{\bot}\textsuperscript{m}) (NegProps v).1
\end{Verbatim}
\subsubsection{ThereIsNoEvil2.lean (Figure 11 of \texorpdfstring{\cite{J75}}{the Notes})}
Importing G\"odel's modified axioms from Figure 8 we can prove that necessarily there exists no entity that possesses all non-positive (= negative) properties.
\begin{Verbatim}[commandchars=\\\{\},fontsize=\scriptsize,xleftmargin=1em]
import Notes.GoedelVariantHOML3
\end{Verbatim}

\begin{Verbatim}[commandchars=\\\{\},fontsize=\scriptsize,xleftmargin=1em]
/-- An entity possessing all non-positive properties -/
def Evil (x : e) : \ensuremath{\sigma} := \ensuremath{\forall}\textsuperscript{m} (\ensuremath{\varphi} : e \ensuremath{\rightarrow} \ensuremath{\sigma}), \ensuremath{\neg}\textsuperscript{m}(P \ensuremath{\varphi}) \ensuremath{\rightarrow}\textsuperscript{m} \ensuremath{\varphi} x

/-- Necessarily there is no actual Evil-like being.  The article presents this derivation
(Figure 11) as succeeding from `Ax1Gen` and `Ax2a`; the AFP source closes it with `oops`, no
trusted tactic having replayed the sledgehammer proof.  Here is one: at any world `v`,
`Ax1Gen` makes the universal property positive (as the conjunction of `\ensuremath{\emptyset}`), so by the
exclusivity in `Ax2a` the empty property is not positive, and an Evil-like being would have
to possess it. -/
theorem NecNoEvil : \ensuremath{\lfloor}\ensuremath{\Box}(\ensuremath{\neg}\textsuperscript{m}(\ensuremath{\exists}\textsuperscript{E} x, Evil x))\ensuremath{\rfloor} := fun _ v _ \ensuremath{\langle}_, _, hE\ensuremath{\rangle} =>
  have htop : P (fun _ : e => \ensuremath{\top}\textsuperscript{m}) v :=
    Ax1Gen (fun _ => \ensuremath{\bot}\textsuperscript{m}) (fun _ => \ensuremath{\top}\textsuperscript{m}) v
      \ensuremath{\langle}fun _ h => h.elim, fun _ _ _ _ => \ensuremath{\langle}fun _ _ h => h.elim, fun _ => trivial\ensuremath{\rangle}\ensuremath{\rangle}
  have hnb : \ensuremath{\neg} P (fun _ : e => \ensuremath{\bot}\textsuperscript{m}) v := fun hb =>
    (Ax2a (fun _ => \ensuremath{\bot}\textsuperscript{m}) v).2 \ensuremath{\langle}hb, (show (~\textsuperscript{m}(fun _ : e => \ensuremath{\bot}\textsuperscript{m})) = (fun _ : e => \ensuremath{\top}\textsuperscript{m}) by
      funext x v; simp) \ensuremath{\blacktriangleright} htop\ensuremath{\rangle}
  hE (fun _ => \ensuremath{\bot}\textsuperscript{m}) hnb
\end{Verbatim}
\subsection{Scott's variant}
\subsubsection{ScottVariantHOML.lean (Figure 12 of \texorpdfstring{\cite{J75}}{the Notes})}
Verification of Scott's variant of G\"odel's ontological argument.  Actualist quantifiers (avoiding existential import) are used for quantification over entities, otherwise possibilist quantifiers.
\begin{Verbatim}[commandchars=\\\{\},fontsize=\scriptsize,xleftmargin=1em]
import Notes.HOMLinHOL
import Notes.ModalFilter
\end{Verbatim}

\begin{Verbatim}[commandchars=\\\{\},fontsize=\scriptsize,xleftmargin=1em]
/-- Positive property -/
axiom P : (e \ensuremath{\rightarrow} \ensuremath{\sigma}) \ensuremath{\rightarrow} \ensuremath{\sigma}

axiom A1 (\ensuremath{\varphi} : e \ensuremath{\rightarrow} \ensuremath{\sigma}) : \ensuremath{\lfloor}\ensuremath{\neg}\textsuperscript{m}(P \ensuremath{\varphi}) \ensuremath{\leftrightarrow}\textsuperscript{m} P ~\textsuperscript{m}\ensuremath{\varphi}\ensuremath{\rfloor}

axiom A2 (\ensuremath{\varphi} \ensuremath{\psi} : e \ensuremath{\rightarrow} \ensuremath{\sigma}) : \ensuremath{\lfloor}P \ensuremath{\varphi} \ensuremath{\wedge}\textsuperscript{m} \ensuremath{\Box}(\ensuremath{\forall}\textsuperscript{E} y, \ensuremath{\varphi} y \ensuremath{\rightarrow}\textsuperscript{m} \ensuremath{\psi} y) \ensuremath{\rightarrow}\textsuperscript{m} P \ensuremath{\psi}\ensuremath{\rfloor}

theorem T1 (\ensuremath{\varphi} : e \ensuremath{\rightarrow} \ensuremath{\sigma}) : \ensuremath{\lfloor}P \ensuremath{\varphi} \ensuremath{\rightarrow}\textsuperscript{m} \ensuremath{\Diamond}(\ensuremath{\exists}\textsuperscript{E} x, \ensuremath{\varphi} x)\ensuremath{\rfloor} := fun w h =>
  Classical.byContradiction fun hn =>
    (A1 \ensuremath{\varphi} w).2 (A2 \ensuremath{\varphi} ~\textsuperscript{m}\ensuremath{\varphi} w \ensuremath{\langle}h, fun v hv y hy h\ensuremath{\varphi} _ => hn \ensuremath{\langle}v, hv, y, hy, h\ensuremath{\varphi}\ensuremath{\rangle}\ensuremath{\rangle}) h

/-- God-like -/
def G (x : e) : \ensuremath{\sigma} := \ensuremath{\forall}\textsuperscript{m} (\ensuremath{\varphi} : e \ensuremath{\rightarrow} \ensuremath{\sigma}), P \ensuremath{\varphi} \ensuremath{\rightarrow}\textsuperscript{m} \ensuremath{\varphi} x

axiom A3 : \ensuremath{\lfloor}P G\ensuremath{\rfloor}

theorem Coro : \ensuremath{\lfloor}\ensuremath{\Diamond}(\ensuremath{\exists}\textsuperscript{E} x, G x)\ensuremath{\rfloor} := fun w => T1 G w (A3 w)

axiom A4 (\ensuremath{\varphi} : e \ensuremath{\rightarrow} \ensuremath{\sigma}) : \ensuremath{\lfloor}P \ensuremath{\varphi} \ensuremath{\rightarrow}\textsuperscript{m} \ensuremath{\Box} P \ensuremath{\varphi}\ensuremath{\rfloor}

/-- `Ess \ensuremath{\varphi} x`: \ensuremath{\varphi} is an essence of x -/
def Ess (\ensuremath{\varphi} : e \ensuremath{\rightarrow} \ensuremath{\sigma}) (x : e) : \ensuremath{\sigma} := \ensuremath{\varphi} x \ensuremath{\wedge}\textsuperscript{m} (\ensuremath{\forall}\textsuperscript{m} (\ensuremath{\psi} : e \ensuremath{\rightarrow} \ensuremath{\sigma}), \ensuremath{\psi} x \ensuremath{\rightarrow}\textsuperscript{m} \ensuremath{\Box}(\ensuremath{\forall}\textsuperscript{E} y, \ensuremath{\varphi} y \ensuremath{\rightarrow}\textsuperscript{m} \ensuremath{\psi} y))

/-- A property exemplified by a God-like being is positive (a consequence of `A1`). -/
theorem PosOfGod \{x : e\} \{\ensuremath{\psi} : e \ensuremath{\rightarrow} \ensuremath{\sigma}\} \{w : i\} (hG : G x w) (h : \ensuremath{\psi} x w) : P \ensuremath{\psi} w :=
  Classical.byContradiction fun hn => hG ~\textsuperscript{m}\ensuremath{\psi} ((A1 \ensuremath{\psi} w).1 hn) h

theorem T2 (x : e) : \ensuremath{\lfloor}G x \ensuremath{\rightarrow}\textsuperscript{m} Ess G x\ensuremath{\rfloor} := fun w hG =>
  \ensuremath{\langle}hG, fun \ensuremath{\psi} h\ensuremath{\psi} v hv _y _ hGy => hGy \ensuremath{\psi} (A4 \ensuremath{\psi} w (PosOfGod hG h\ensuremath{\psi}) v hv)\ensuremath{\rangle}

/-- Necessary existence -/
def NE (x : e) : \ensuremath{\sigma} := \ensuremath{\forall}\textsuperscript{m} (\ensuremath{\varphi} : e \ensuremath{\rightarrow} \ensuremath{\sigma}), Ess \ensuremath{\varphi} x \ensuremath{\rightarrow}\textsuperscript{m} \ensuremath{\Box}(\ensuremath{\exists}\textsuperscript{E} x, \ensuremath{\varphi} x)

axiom A5 : \ensuremath{\lfloor}P NE\ensuremath{\rfloor}

-- `lemma True nitpick[satisfy,card=1,eval="\ensuremath{\lfloor}P (\ensuremath{\lambda}x.\ensuremath{\top})\ensuremath{\rfloor}"] oops`
--   One model found of cardinality one (consistency check)

theorem T3 : \ensuremath{\lfloor}\ensuremath{\Box}(\ensuremath{\exists}\textsuperscript{E} x, G x)\ensuremath{\rfloor} := by
  have h2 : \ensuremath{\lfloor}(\ensuremath{\exists}\textsuperscript{E} x, G x) \ensuremath{\rightarrow}\textsuperscript{m} \ensuremath{\Box}(\ensuremath{\exists}\textsuperscript{E} x, G x)\ensuremath{\rfloor} :=          -- from A5, God, NE and T2
    fun _ \ensuremath{\langle}x, _, hG\ensuremath{\rangle} => hG NE (A5 _) G (T2 x _ hG)
  intro w
  have \ensuremath{\langle}v, hv, hex\ensuremath{\rangle} := Coro w                          -- a God-like being is possible
  exact h2 w (h2 v hex w (Rsymm w v hv))               -- only symmetry of `\textsf{\bfseries r}` is needed

theorem MC (\ensuremath{\varphi} : \ensuremath{\sigma}) : \ensuremath{\lfloor}\ensuremath{\varphi} \ensuremath{\rightarrow}\textsuperscript{m} \ensuremath{\Box}\ensuremath{\varphi}\ensuremath{\rfloor} := by                   -- modal collapse
  intro w h\ensuremath{\varphi} v hv
  have \ensuremath{\langle}x, _, hGx\ensuremath{\rangle} := T3 v w (Rsymm w v hv)
  have \ensuremath{\langle}z, hzv, hGz\ensuremath{\rangle} := T3 w v hv
  exact (T2 x w hGx).2 (fun _ => \ensuremath{\varphi}) h\ensuremath{\varphi} v hv z hzv hGz

theorem PosProps : \ensuremath{\lfloor}P (fun _ : e => \ensuremath{\top}\textsuperscript{m}) \ensuremath{\wedge}\textsuperscript{m} P (fun x : e => x =\textsuperscript{m} x)\ensuremath{\rfloor} := fun w =>
  \ensuremath{\langle}Classical.byContradiction fun hn =>
     let \ensuremath{\langle}_, _, _, _, hx\ensuremath{\rangle} := T1 _ w ((A1 _ w).1 hn); hx trivial,
   Classical.byContradiction fun hn =>
     let \ensuremath{\langle}_, _, _, _, hx\ensuremath{\rangle} := T1 _ w ((A1 _ w).1 hn); hx rfl\ensuremath{\rangle}

theorem NegProps : \ensuremath{\lfloor}\ensuremath{\neg}\textsuperscript{m}P (fun _ : e => \ensuremath{\bot}\textsuperscript{m}) \ensuremath{\wedge}\textsuperscript{m} \ensuremath{\neg}\textsuperscript{m}P (fun x : e => x \ensuremath{\neq}\textsuperscript{m} x)\ensuremath{\rfloor} := fun w =>
  \ensuremath{\langle}fun hp => let \ensuremath{\langle}_, _, _, _, hx\ensuremath{\rangle} := T1 _ w hp; hx,
   fun hp => let \ensuremath{\langle}_, _, _, _, hx\ensuremath{\rangle} := T1 _ w hp; hx rfl\ensuremath{\rangle}

theorem UniqueEss1 (\ensuremath{\varphi} \ensuremath{\psi} : e \ensuremath{\rightarrow} \ensuremath{\sigma}) (x : e) : \ensuremath{\lfloor}Ess \ensuremath{\varphi} x \ensuremath{\wedge}\textsuperscript{m} Ess \ensuremath{\psi} x \ensuremath{\rightarrow}\textsuperscript{m} \ensuremath{\Box}(\ensuremath{\forall}\textsuperscript{E} y, \ensuremath{\varphi} y \ensuremath{\leftrightarrow}\textsuperscript{m} \ensuremath{\psi} y)\ensuremath{\rfloor} :=
  fun _ \ensuremath{\langle}\ensuremath{\langle}h\ensuremath{\varphi}x, h\ensuremath{\varphi}\ensuremath{\rangle}, \ensuremath{\langle}h\ensuremath{\psi}x, h\ensuremath{\psi}\ensuremath{\rangle}\ensuremath{\rangle} v hv y hy => \ensuremath{\langle}h\ensuremath{\varphi} \ensuremath{\psi} h\ensuremath{\psi}x v hv y hy, h\ensuremath{\psi} \ensuremath{\varphi} h\ensuremath{\varphi}x v hv y hy\ensuremath{\rangle}

-- UniqueEss2 : \ensuremath{\lfloor}Ess \ensuremath{\varphi} x \ensuremath{\wedge}\textsuperscript{m} Ess \ensuremath{\psi} x \ensuremath{\rightarrow}\textsuperscript{m} \ensuremath{\Box}(\ensuremath{\varphi} \ensuremath{\equiv}\textsuperscript{m} \ensuremath{\psi})\ensuremath{\rfloor}   -- `nitpick[card i=1]`: countermodel found
example (\ensuremath{\varphi} \ensuremath{\psi} : e \ensuremath{\rightarrow} \ensuremath{\sigma}) (x : e) : \ensuremath{\lfloor}Ess \ensuremath{\varphi} x \ensuremath{\wedge}\textsuperscript{m} Ess \ensuremath{\psi} x \ensuremath{\rightarrow}\textsuperscript{m} \ensuremath{\Box}(\ensuremath{\varphi} \ensuremath{\equiv}\textsuperscript{m} \ensuremath{\psi})\ensuremath{\rfloor} := by countermodel

theorem UniqueEss3 (\ensuremath{\varphi} : e \ensuremath{\rightarrow} \ensuremath{\sigma}) (x : e) : \ensuremath{\lfloor}Ess \ensuremath{\varphi} x \ensuremath{\rightarrow}\textsuperscript{m} \ensuremath{\Box}(\ensuremath{\forall}\textsuperscript{E} y, \ensuremath{\varphi} y \ensuremath{\rightarrow}\textsuperscript{m} y \ensuremath{\equiv}\textsuperscript{m} x)\ensuremath{\rfloor} :=
  fun _ \ensuremath{\langle}_, h\ensuremath{\rangle} => h (fun z => z \ensuremath{\equiv}\textsuperscript{m} x) (fun _ hp => hp)

theorem Monotheism (x y : e) : \ensuremath{\lfloor}G x \ensuremath{\wedge}\textsuperscript{m} G y \ensuremath{\rightarrow}\textsuperscript{m} x \ensuremath{\equiv}\textsuperscript{m} y\ensuremath{\rfloor} :=
  fun _ \ensuremath{\langle}hx, hy\ensuremath{\rangle} Q hQ => hy Q (PosOfGod hx hQ)

/-- Everything true of every actual God-like being at every successor world is positive: by
`Coro` some successor holds an actual God-like being, which has every positive property, and
positivity is necessary (`A4`).  No frame condition is used. -/
theorem Pos (\ensuremath{\chi} : e \ensuremath{\rightarrow} \ensuremath{\sigma}) (w : i) (h : \ensuremath{\forall} v, w \textsf{\bfseries r} v \ensuremath{\rightarrow} \ensuremath{\forall} x, x @\textsuperscript{m} v \ensuremath{\rightarrow} G x v \ensuremath{\rightarrow} \ensuremath{\chi} x v) : P \ensuremath{\chi} w :=
  Classical.byContradiction fun hn =>
    let \ensuremath{\langle}v, hv, x, hxa, hG\ensuremath{\rangle} := Coro w
    hG ~\textsuperscript{m}\ensuremath{\chi} (A4 ~\textsuperscript{m}\ensuremath{\chi} w ((A1 \ensuremath{\chi} w).1 hn) v hv) (h v hv x hxa hG)

theorem PisFilter : \ensuremath{\lfloor}Filter P\ensuremath{\rfloor} := fun w =>
  \ensuremath{\langle}\ensuremath{\langle}\ensuremath{\langle}(PosProps w).1, (NegProps w).1\ensuremath{\rangle},
    -- the inclusion holds at `w`; `MC` carries it to the successor worlds, then `A2` applies
    fun \ensuremath{\varphi} \ensuremath{\psi} \ensuremath{\langle}h\ensuremath{\varphi}, hsub\ensuremath{\rangle} => A2 \ensuremath{\varphi} \ensuremath{\psi} w \ensuremath{\langle}h\ensuremath{\varphi}, MC (\ensuremath{\forall}\textsuperscript{E} x, \ensuremath{\varphi} x \ensuremath{\rightarrow}\textsuperscript{m} \ensuremath{\psi} x) w hsub\ensuremath{\rangle}\ensuremath{\rangle},
   fun \ensuremath{\varphi} \ensuremath{\psi} \ensuremath{\langle}h\ensuremath{\varphi}, h\ensuremath{\psi}\ensuremath{\rangle} => Pos _ w (fun v hv _ _ hG => \ensuremath{\langle}hG \ensuremath{\varphi} (A4 \ensuremath{\varphi} w h\ensuremath{\varphi} v hv), hG \ensuremath{\psi} (A4 \ensuremath{\psi} w h\ensuremath{\psi} v hv)\ensuremath{\rangle})\ensuremath{\rangle}

theorem PisUFilter : \ensuremath{\lfloor}UFilter P\ensuremath{\rfloor} := fun w =>
  \ensuremath{\langle}PisFilter w, fun \ensuremath{\varphi} => (Classical.em (P \ensuremath{\varphi} w)).imp id (A1 \ensuremath{\varphi} w).1\ensuremath{\rangle}

-- `lemma True nitpick[satisfy,card=1,eval="\ensuremath{\lfloor}P (\ensuremath{\lambda}x.\ensuremath{\bot})\ensuremath{\rfloor}"] oops`
--   One model found of cardinality one (consistency check)
\end{Verbatim}
\subsubsection{HOMLinHOLonlyK.lean (slight variation of Figure 3 of \texorpdfstring{\cite{J75}}{the Notes})}
Shallow embedding of higher-order modal logic (HOML) in the classical higher-order logic (HOL) of Lean 4, utilizing the LogiKEy methodology.  Here logic K is introduced.

The global parameter settings of the Isabelle sources configure the model finder \texttt{nitpick} and the parser; they have no Lean counterpart, as Lean has no model finder.

\begin{Verbatim}[commandchars=\\\{\},fontsize=\scriptsize,xleftmargin=1em]
-- nitpick_params[user_axioms,expect=genuine,show_all,format=2,max_genuine=3]
-- declare[[syntax_ambiguity_warning=false]]
\end{Verbatim}
Lean's \texttt{Prop} is intuitionistic, HOL is classical; classical reasoning is enabled globally.

\begin{Verbatim}[commandchars=\\\{\},fontsize=\scriptsize,xleftmargin=1em]
attribute [instance] Classical.propDecidable

/-- Counterpart of Isabelle's `oops` after `nitpick` found a countermodel.  Lean has no model
finder, so such statements are only type-checked, never proved (and never added to the context). -/
macro "countermodel" : tactic => `(tactic| sorry)
/-- Counterpart of Isabelle's `oops` on a statement left genuinely open: neither a proof
nor a countermodel was obtained. -/
macro "openproblem" : tactic => `(tactic| sorry)
\end{Verbatim}
Type \texttt{i} is associated with possible worlds and type \texttt{e} with entities

\begin{Verbatim}[commandchars=\\\{\},fontsize=\scriptsize,xleftmargin=1em]
axiom i : Type                       -- Possible worlds
axiom e : Type                       -- Individuals/entities
@[instance] axiom iNonempty : Nonempty i
@[instance] axiom eNonempty : Nonempty e
abbrev \ensuremath{\sigma} := i \ensuremath{\rightarrow} Prop                 -- World-lifted propositions
abbrev \ensuremath{\tau} := e \ensuremath{\rightarrow} \ensuremath{\sigma}                    -- Modal properties

/-- Accessibility relation between worlds -/
axiom R : i \ensuremath{\rightarrow} i \ensuremath{\rightarrow} Prop
@[inherit_doc] infix:60 " \textsf{\bfseries r} " => R

-- No frame conditions are imposed: the base logic is K.
\end{Verbatim}
Logical connectives (operating on truth-sets)

\begin{Verbatim}[commandchars=\\\{\},fontsize=\scriptsize,xleftmargin=1em]
@[simp, grind] def Mbot : \ensuremath{\sigma} := fun _ => False
@[simp, grind] def Mtop : \ensuremath{\sigma} := fun _ => True
@[simp, grind] def Mneg (\ensuremath{\varphi} : \ensuremath{\sigma}) : \ensuremath{\sigma} := fun w => \ensuremath{\neg} \ensuremath{\varphi} w
@[simp, grind] def Mand (\ensuremath{\varphi} \ensuremath{\psi} : \ensuremath{\sigma}) : \ensuremath{\sigma} := fun w => \ensuremath{\varphi} w \ensuremath{\wedge} \ensuremath{\psi} w
@[simp, grind] def Mor (\ensuremath{\varphi} \ensuremath{\psi} : \ensuremath{\sigma}) : \ensuremath{\sigma} := fun w => \ensuremath{\varphi} w \ensuremath{\vee} \ensuremath{\psi} w
@[simp, grind] def Mimp (\ensuremath{\varphi} \ensuremath{\psi} : \ensuremath{\sigma}) : \ensuremath{\sigma} := fun w => \ensuremath{\varphi} w \ensuremath{\rightarrow} \ensuremath{\psi} w
@[simp, grind] def Mequiv (\ensuremath{\varphi} \ensuremath{\psi} : \ensuremath{\sigma}) : \ensuremath{\sigma} := fun w => \ensuremath{\varphi} w \ensuremath{\leftrightarrow} \ensuremath{\psi} w
@[simp, grind] def Mbox (\ensuremath{\varphi} : \ensuremath{\sigma}) : \ensuremath{\sigma} := fun w => \ensuremath{\forall} v, w \textsf{\bfseries r} v \ensuremath{\rightarrow} \ensuremath{\varphi} v
@[simp, grind] def Mdia (\ensuremath{\varphi} : \ensuremath{\sigma}) : \ensuremath{\sigma} := fun w => \ensuremath{\exists} v, w \textsf{\bfseries r} v \ensuremath{\wedge} \ensuremath{\varphi} v
@[simp, grind] def Mprimeq \{\ensuremath{\alpha}\} (x y : \ensuremath{\alpha}) : \ensuremath{\sigma} := fun _ => x = y
@[simp, grind] def Mprimneg \{\ensuremath{\alpha}\} (x y : \ensuremath{\alpha}) : \ensuremath{\sigma} := fun _ => x \ensuremath{\neq} y
@[simp, grind] def Mnegpred (\ensuremath{\Phi} : \ensuremath{\tau}) : \ensuremath{\tau} := fun x w => \ensuremath{\neg} \ensuremath{\Phi} x w
@[simp, grind] def Mconpred (\ensuremath{\Phi} \ensuremath{\Psi} : \ensuremath{\tau}) : \ensuremath{\tau} := fun x w => \ensuremath{\Phi} x w \ensuremath{\wedge} \ensuremath{\Psi} x w

notation:max "\ensuremath{\bot}\textsuperscript{m}" => Mbot
notation:max "\ensuremath{\top}\textsuperscript{m}" => Mtop
prefix:53 "\ensuremath{\neg}\textsuperscript{m}" => Mneg
infixl:50 " \ensuremath{\wedge}\textsuperscript{m} " => Mand
infixl:49 " \ensuremath{\vee}\textsuperscript{m} " => Mor
infixr:48 " \ensuremath{\rightarrow}\textsuperscript{m} " => Mimp
infixl:47 " \ensuremath{\leftrightarrow}\textsuperscript{m} " => Mequiv
prefix:55 "\ensuremath{\Box}" => Mbox
prefix:55 "\ensuremath{\Diamond}" => Mdia
infix:50 " =\textsuperscript{m} " => Mprimeq
infix:50 " \ensuremath{\neq}\textsuperscript{m} " => Mprimneg
prefix:max "~\textsuperscript{m}" => Mnegpred
infixl:50 " \ensuremath{\centerdot}\textsuperscript{m} " => Mconpred

@[simp, grind] def Mexclor (\ensuremath{\varphi} \ensuremath{\psi} : \ensuremath{\sigma}) : \ensuremath{\sigma} := (\ensuremath{\varphi} \ensuremath{\vee}\textsuperscript{m} \ensuremath{\psi}) \ensuremath{\wedge}\textsuperscript{m} \ensuremath{\neg}\textsuperscript{m}(\ensuremath{\varphi} \ensuremath{\wedge}\textsuperscript{m} \ensuremath{\psi})
infixl:49 " \ensuremath{\vee}\textsuperscript{e} " => Mexclor
\end{Verbatim}
Possibilist quantifiers (polymorphic)

\begin{Verbatim}[commandchars=\\\{\},fontsize=\scriptsize,xleftmargin=1em]
@[simp, grind] def Mallposs \{\ensuremath{\alpha}\} (\ensuremath{\Phi} : \ensuremath{\alpha} \ensuremath{\rightarrow} \ensuremath{\sigma}) : \ensuremath{\sigma} := fun w => \ensuremath{\forall} x, \ensuremath{\Phi} x w
@[simp, grind] def Mexiposs \{\ensuremath{\alpha}\} (\ensuremath{\Phi} : \ensuremath{\alpha} \ensuremath{\rightarrow} \ensuremath{\sigma}) : \ensuremath{\sigma} := fun w => \ensuremath{\exists} x, \ensuremath{\Phi} x w
\end{Verbatim}
Actualist quantifiers (for individuals/entities)

\begin{Verbatim}[commandchars=\\\{\},fontsize=\scriptsize,xleftmargin=1em]
/-- Existence (actuality) of an entity at a world -/
axiom existsAt : e \ensuremath{\rightarrow} \ensuremath{\sigma}
@[inherit_doc] infix:60 " @\textsuperscript{m} " => existsAt
@[simp, grind] def Mallact (\ensuremath{\Phi} : e \ensuremath{\rightarrow} \ensuremath{\sigma}) : \ensuremath{\sigma} := fun w => \ensuremath{\forall} x, x @\textsuperscript{m} w \ensuremath{\rightarrow} \ensuremath{\Phi} x w
@[simp, grind] def Mexiact (\ensuremath{\Phi} : e \ensuremath{\rightarrow} \ensuremath{\sigma}) : \ensuremath{\sigma} := fun w => \ensuremath{\exists} x, x @\textsuperscript{m} w \ensuremath{\wedge} \ensuremath{\Phi} x w

open Lean TSyntax.Compat in
macro "\ensuremath{\forall}\textsuperscript{m}" xs:explicitBinders ", " b:term : term => expandExplicitBinders ``Mallposs xs b
open Lean TSyntax.Compat in
macro "\ensuremath{\exists}\textsuperscript{m}" xs:explicitBinders ", " b:term : term => expandExplicitBinders ``Mexiposs xs b
open Lean TSyntax.Compat in
macro "\ensuremath{\forall}\textsuperscript{E}" xs:explicitBinders ", " b:term : term => expandExplicitBinders ``Mallact xs b
open Lean TSyntax.Compat in
macro "\ensuremath{\exists}\textsuperscript{E}" xs:explicitBinders ", " b:term : term => expandExplicitBinders ``Mexiact xs b
\end{Verbatim}
Leibniz equality (polymorphic)

\begin{Verbatim}[commandchars=\\\{\},fontsize=\scriptsize,xleftmargin=1em]
@[simp, grind] def Mleibeq \{\ensuremath{\alpha}\} (x y : \ensuremath{\alpha}) : \ensuremath{\sigma} := \ensuremath{\forall}\textsuperscript{m} (P : \ensuremath{\alpha} \ensuremath{\rightarrow} \ensuremath{\sigma}), P x \ensuremath{\rightarrow}\textsuperscript{m} P y
infix:50 " \ensuremath{\equiv}\textsuperscript{m} " => Mleibeq
\end{Verbatim}
Meta-logical predicate for global validity

\begin{Verbatim}[commandchars=\\\{\},fontsize=\scriptsize,xleftmargin=1em]
@[simp, grind] def Mvalid (\ensuremath{\psi} : \ensuremath{\sigma}) : Prop := \ensuremath{\forall} w, \ensuremath{\psi} w
notation:max "\ensuremath{\lfloor}" \ensuremath{\psi} "\ensuremath{\rfloor}" => Mvalid \ensuremath{\psi}
\end{Verbatim}
\subsubsection{ScottVariantHOMLinK.lean (Figure 13 of \texorpdfstring{\cite{J75}}{the Notes})}
Scott's variant of G\"odel's argument fails for base logic K (but only in the last step).
\begin{Verbatim}[commandchars=\\\{\},fontsize=\scriptsize,xleftmargin=1em]
import Notes.HOMLinHOLonlyK
\end{Verbatim}

\begin{Verbatim}[commandchars=\\\{\},fontsize=\scriptsize,xleftmargin=1em]
/-- Positive property -/
axiom P : (e \ensuremath{\rightarrow} \ensuremath{\sigma}) \ensuremath{\rightarrow} \ensuremath{\sigma}

axiom A1 (\ensuremath{\varphi} : e \ensuremath{\rightarrow} \ensuremath{\sigma}) : \ensuremath{\lfloor}\ensuremath{\neg}\textsuperscript{m}(P \ensuremath{\varphi}) \ensuremath{\leftrightarrow}\textsuperscript{m} P ~\textsuperscript{m}\ensuremath{\varphi}\ensuremath{\rfloor}

axiom A2 (\ensuremath{\varphi} \ensuremath{\psi} : e \ensuremath{\rightarrow} \ensuremath{\sigma}) : \ensuremath{\lfloor}P \ensuremath{\varphi} \ensuremath{\wedge}\textsuperscript{m} \ensuremath{\Box}(\ensuremath{\forall}\textsuperscript{E} y, \ensuremath{\varphi} y \ensuremath{\rightarrow}\textsuperscript{m} \ensuremath{\psi} y) \ensuremath{\rightarrow}\textsuperscript{m} P \ensuremath{\psi}\ensuremath{\rfloor}

theorem T1 (\ensuremath{\varphi} : e \ensuremath{\rightarrow} \ensuremath{\sigma}) : \ensuremath{\lfloor}P \ensuremath{\varphi} \ensuremath{\rightarrow}\textsuperscript{m} \ensuremath{\Diamond}(\ensuremath{\exists}\textsuperscript{E} x, \ensuremath{\varphi} x)\ensuremath{\rfloor} := fun w h =>
  Classical.byContradiction fun hn =>
    (A1 \ensuremath{\varphi} w).2 (A2 \ensuremath{\varphi} ~\textsuperscript{m}\ensuremath{\varphi} w \ensuremath{\langle}h, fun v hv y hy h\ensuremath{\varphi} _ => hn \ensuremath{\langle}v, hv, y, hy, h\ensuremath{\varphi}\ensuremath{\rangle}\ensuremath{\rangle}) h

/-- God-like -/
def G (x : e) : \ensuremath{\sigma} := \ensuremath{\forall}\textsuperscript{m} (\ensuremath{\varphi} : e \ensuremath{\rightarrow} \ensuremath{\sigma}), P \ensuremath{\varphi} \ensuremath{\rightarrow}\textsuperscript{m} \ensuremath{\varphi} x

axiom A3 : \ensuremath{\lfloor}P G\ensuremath{\rfloor}

-- `nitpick[satisfy,eval="G"]` inspects a model of `Coro` and the value of `G` in it.
theorem Coro : \ensuremath{\lfloor}\ensuremath{\Diamond}(\ensuremath{\exists}\textsuperscript{E} x, G x)\ensuremath{\rfloor} := fun w => T1 G w (A3 w)

axiom A4 (\ensuremath{\varphi} : e \ensuremath{\rightarrow} \ensuremath{\sigma}) : \ensuremath{\lfloor}P \ensuremath{\varphi} \ensuremath{\rightarrow}\textsuperscript{m} \ensuremath{\Box} P \ensuremath{\varphi}\ensuremath{\rfloor}

/-- `Ess \ensuremath{\varphi} x`: \ensuremath{\varphi} is an essence of x -/
def Ess (\ensuremath{\varphi} : e \ensuremath{\rightarrow} \ensuremath{\sigma}) (x : e) : \ensuremath{\sigma} := \ensuremath{\varphi} x \ensuremath{\wedge}\textsuperscript{m} (\ensuremath{\forall}\textsuperscript{m} (\ensuremath{\psi} : e \ensuremath{\rightarrow} \ensuremath{\sigma}), \ensuremath{\psi} x \ensuremath{\rightarrow}\textsuperscript{m} \ensuremath{\Box}(\ensuremath{\forall}\textsuperscript{E} y, \ensuremath{\varphi} y \ensuremath{\rightarrow}\textsuperscript{m} \ensuremath{\psi} y))

/-- A property exemplified by a God-like being is positive (a consequence of `A1`). -/
theorem PosOfGod \{x : e\} \{\ensuremath{\psi} : e \ensuremath{\rightarrow} \ensuremath{\sigma}\} \{w : i\} (hG : G x w) (h : \ensuremath{\psi} x w) : P \ensuremath{\psi} w :=
  Classical.byContradiction fun hn => hG ~\textsuperscript{m}\ensuremath{\psi} ((A1 \ensuremath{\psi} w).1 hn) h

theorem T2 (x : e) : \ensuremath{\lfloor}G x \ensuremath{\rightarrow}\textsuperscript{m} Ess G x\ensuremath{\rfloor} := fun w hG =>
  \ensuremath{\langle}hG, fun \ensuremath{\psi} h\ensuremath{\psi} v hv _y _ hGy => hGy \ensuremath{\psi} (A4 \ensuremath{\psi} w (PosOfGod hG h\ensuremath{\psi}) v hv)\ensuremath{\rangle}

/-- Necessary existence -/
def NE (x : e) : \ensuremath{\sigma} := \ensuremath{\forall}\textsuperscript{m} (\ensuremath{\varphi} : e \ensuremath{\rightarrow} \ensuremath{\sigma}), Ess \ensuremath{\varphi} x \ensuremath{\rightarrow}\textsuperscript{m} \ensuremath{\Box}(\ensuremath{\exists}\textsuperscript{E} x, \ensuremath{\varphi} x)

axiom A5 : \ensuremath{\lfloor}P NE\ensuremath{\rfloor}

-- `lemma True nitpick[satisfy,card=1,eval="\ensuremath{\lfloor}P (\ensuremath{\lambda}x.\ensuremath{\top})\ensuremath{\rfloor}"] oops`
--   One model found of cardinality one (consistency check)

-- T3 : \ensuremath{\lfloor}\ensuremath{\Box}(\ensuremath{\exists}\textsuperscript{E} x, G x)\ensuremath{\rfloor}   -- `nitpick[card e=1, card i=2, eval="G"]`: countermodel found
example : \ensuremath{\lfloor}\ensuremath{\Box}(\ensuremath{\exists}\textsuperscript{E} x, G x)\ensuremath{\rfloor} := by countermodel

-- MC : \ensuremath{\lfloor}\ensuremath{\varphi} \ensuremath{\rightarrow}\textsuperscript{m} \ensuremath{\Box}\ensuremath{\varphi}\ensuremath{\rfloor}   -- `nitpick[card e=1, card i=2, eval="G"]`: countermodel found
example (\ensuremath{\varphi} : \ensuremath{\sigma}) : \ensuremath{\lfloor}\ensuremath{\varphi} \ensuremath{\rightarrow}\textsuperscript{m} \ensuremath{\Box}\ensuremath{\varphi}\ensuremath{\rfloor} := by countermodel
\end{Verbatim}
\subsubsection{ScottVariantHOMLAndersonQuant.lean (Figure 15 of \texorpdfstring{\cite{J75}}{the Notes})}
Verification of Scott's variant of G\"odel's argument with a mixed use of actualist and possibilist quantifiers for entities; cf. Footnote 20 in \cite{J75}.
\begin{Verbatim}[commandchars=\\\{\},fontsize=\scriptsize,xleftmargin=1em]
import Notes.HOMLinHOL
import Notes.ModalFilter
\end{Verbatim}

\begin{Verbatim}[commandchars=\\\{\},fontsize=\scriptsize,xleftmargin=1em]
/-- Positive property -/
axiom P : (e \ensuremath{\rightarrow} \ensuremath{\sigma}) \ensuremath{\rightarrow} \ensuremath{\sigma}

axiom A1 (\ensuremath{\varphi} : e \ensuremath{\rightarrow} \ensuremath{\sigma}) : \ensuremath{\lfloor}\ensuremath{\neg}\textsuperscript{m}(P \ensuremath{\varphi}) \ensuremath{\leftrightarrow}\textsuperscript{m} P ~\textsuperscript{m}\ensuremath{\varphi}\ensuremath{\rfloor}

axiom A2 (\ensuremath{\varphi} \ensuremath{\psi} : e \ensuremath{\rightarrow} \ensuremath{\sigma}) : \ensuremath{\lfloor}P \ensuremath{\varphi} \ensuremath{\wedge}\textsuperscript{m} \ensuremath{\Box}(\ensuremath{\forall}\textsuperscript{m} y, \ensuremath{\varphi} y \ensuremath{\rightarrow}\textsuperscript{m} \ensuremath{\psi} y) \ensuremath{\rightarrow}\textsuperscript{m} P \ensuremath{\psi}\ensuremath{\rfloor}

theorem T1 (\ensuremath{\varphi} : e \ensuremath{\rightarrow} \ensuremath{\sigma}) : \ensuremath{\lfloor}P \ensuremath{\varphi} \ensuremath{\rightarrow}\textsuperscript{m} \ensuremath{\Diamond}(\ensuremath{\exists}\textsuperscript{m} x, \ensuremath{\varphi} x)\ensuremath{\rfloor} := fun w h =>
  Classical.byContradiction fun hn =>
    (A1 \ensuremath{\varphi} w).2 (A2 \ensuremath{\varphi} ~\textsuperscript{m}\ensuremath{\varphi} w \ensuremath{\langle}h, fun v hv y h\ensuremath{\varphi} _ => hn \ensuremath{\langle}v, hv, y, h\ensuremath{\varphi}\ensuremath{\rangle}\ensuremath{\rangle}) h

/-- God-like -/
def G (x : e) : \ensuremath{\sigma} := \ensuremath{\forall}\textsuperscript{m} (\ensuremath{\varphi} : e \ensuremath{\rightarrow} \ensuremath{\sigma}), P \ensuremath{\varphi} \ensuremath{\rightarrow}\textsuperscript{m} \ensuremath{\varphi} x

axiom A3 : \ensuremath{\lfloor}P G\ensuremath{\rfloor}

theorem Coro : \ensuremath{\lfloor}\ensuremath{\Diamond}(\ensuremath{\exists}\textsuperscript{m} x, G x)\ensuremath{\rfloor} := fun w => T1 G w (A3 w)

axiom A4 (\ensuremath{\varphi} : e \ensuremath{\rightarrow} \ensuremath{\sigma}) : \ensuremath{\lfloor}P \ensuremath{\varphi} \ensuremath{\rightarrow}\textsuperscript{m} \ensuremath{\Box} P \ensuremath{\varphi}\ensuremath{\rfloor}

/-- `Ess \ensuremath{\varphi} x`: \ensuremath{\varphi} is an essence of x -/
def Ess (\ensuremath{\varphi} : e \ensuremath{\rightarrow} \ensuremath{\sigma}) (x : e) : \ensuremath{\sigma} := \ensuremath{\varphi} x \ensuremath{\wedge}\textsuperscript{m} (\ensuremath{\forall}\textsuperscript{m} (\ensuremath{\psi} : e \ensuremath{\rightarrow} \ensuremath{\sigma}), \ensuremath{\psi} x \ensuremath{\rightarrow}\textsuperscript{m} \ensuremath{\Box}(\ensuremath{\forall}\textsuperscript{m} (y : e), \ensuremath{\varphi} y \ensuremath{\rightarrow}\textsuperscript{m} \ensuremath{\psi} y))

/-- A property exemplified by a God-like being is positive (a consequence of `A1`). -/
theorem PosOfGod \{x : e\} \{\ensuremath{\psi} : e \ensuremath{\rightarrow} \ensuremath{\sigma}\} \{w : i\} (hG : G x w) (h : \ensuremath{\psi} x w) : P \ensuremath{\psi} w :=
  Classical.byContradiction fun hn => hG ~\textsuperscript{m}\ensuremath{\psi} ((A1 \ensuremath{\psi} w).1 hn) h

theorem T2 (x : e) : \ensuremath{\lfloor}G x \ensuremath{\rightarrow}\textsuperscript{m} Ess G x\ensuremath{\rfloor} := fun w hG =>
  \ensuremath{\langle}hG, fun \ensuremath{\psi} h\ensuremath{\psi} v hv _y hGy => hGy \ensuremath{\psi} (A4 \ensuremath{\psi} w (PosOfGod hG h\ensuremath{\psi}) v hv)\ensuremath{\rangle}

/-- Necessary existence (with the actualist quantifier) -/
def NE (x : e) : \ensuremath{\sigma} := \ensuremath{\forall}\textsuperscript{m} (\ensuremath{\varphi} : e \ensuremath{\rightarrow} \ensuremath{\sigma}), Ess \ensuremath{\varphi} x \ensuremath{\rightarrow}\textsuperscript{m} \ensuremath{\Box}(\ensuremath{\exists}\textsuperscript{E} x, \ensuremath{\varphi} x)

axiom A5 : \ensuremath{\lfloor}P NE\ensuremath{\rfloor}

-- `lemma True nitpick[satisfy,card=1,eval="\ensuremath{\lfloor}P (\ensuremath{\lambda}x.\ensuremath{\top})\ensuremath{\rfloor}"] oops`
--   One model found of cardinality one (consistency check)

theorem T3 : \ensuremath{\lfloor}\ensuremath{\Box}(\ensuremath{\exists}\textsuperscript{E} x, G x)\ensuremath{\rfloor} := by
  have h2 : \ensuremath{\lfloor}(\ensuremath{\exists}\textsuperscript{m} x, G x) \ensuremath{\rightarrow}\textsuperscript{m} \ensuremath{\Box}(\ensuremath{\exists}\textsuperscript{E} x, G x)\ensuremath{\rfloor} :=
    fun _ \ensuremath{\langle}x, hG\ensuremath{\rangle} => hG NE (A5 _) G (T2 x _ hG)
  intro w
  have \ensuremath{\langle}v, hv, hex\ensuremath{\rangle} := Coro w
  have \ensuremath{\langle}x, _, hG\ensuremath{\rangle} := h2 v hex w (Rsymm w v hv)
  exact h2 w \ensuremath{\langle}x, hG\ensuremath{\rangle}

theorem MC (\ensuremath{\varphi} : \ensuremath{\sigma}) : \ensuremath{\lfloor}\ensuremath{\varphi} \ensuremath{\rightarrow}\textsuperscript{m} \ensuremath{\Box}\ensuremath{\varphi}\ensuremath{\rfloor} := by                   -- modal collapse
  intro w h\ensuremath{\varphi} v hv
  have \ensuremath{\langle}x, _, hGx\ensuremath{\rangle} := T3 v w (Rsymm w v hv)
  have \ensuremath{\langle}z, _, hGz\ensuremath{\rangle} := T3 w v hv
  exact (T2 x w hGx).2 (fun _ => \ensuremath{\varphi}) h\ensuremath{\varphi} v hv z hGz

theorem PosProps : \ensuremath{\lfloor}P (fun _ : e => \ensuremath{\top}\textsuperscript{m}) \ensuremath{\wedge}\textsuperscript{m} P (fun x : e => x =\textsuperscript{m} x)\ensuremath{\rfloor} := fun w =>
  \ensuremath{\langle}Classical.byContradiction fun hn => let \ensuremath{\langle}_, _, _, hx\ensuremath{\rangle} := T1 _ w ((A1 _ w).1 hn); hx trivial,
   Classical.byContradiction fun hn => let \ensuremath{\langle}_, _, _, hx\ensuremath{\rangle} := T1 _ w ((A1 _ w).1 hn); hx rfl\ensuremath{\rangle}

theorem NegProps : \ensuremath{\lfloor}\ensuremath{\neg}\textsuperscript{m}P (fun _ : e => \ensuremath{\bot}\textsuperscript{m}) \ensuremath{\wedge}\textsuperscript{m} \ensuremath{\neg}\textsuperscript{m}P (fun x : e => x \ensuremath{\neq}\textsuperscript{m} x)\ensuremath{\rfloor} := fun w =>
  \ensuremath{\langle}fun hp => let \ensuremath{\langle}_, _, _, hx\ensuremath{\rangle} := T1 _ w hp; hx,
   fun hp => let \ensuremath{\langle}_, _, _, hx\ensuremath{\rangle} := T1 _ w hp; hx rfl\ensuremath{\rangle}

theorem UniqueEss1 (\ensuremath{\varphi} \ensuremath{\psi} : e \ensuremath{\rightarrow} \ensuremath{\sigma}) (x : e) : \ensuremath{\lfloor}Ess \ensuremath{\varphi} x \ensuremath{\wedge}\textsuperscript{m} Ess \ensuremath{\psi} x \ensuremath{\rightarrow}\textsuperscript{m} \ensuremath{\Box}(\ensuremath{\forall}\textsuperscript{m} y, \ensuremath{\varphi} y \ensuremath{\leftrightarrow}\textsuperscript{m} \ensuremath{\psi} y)\ensuremath{\rfloor} :=
  fun _ \ensuremath{\langle}\ensuremath{\langle}h\ensuremath{\varphi}x, h\ensuremath{\varphi}\ensuremath{\rangle}, \ensuremath{\langle}h\ensuremath{\psi}x, h\ensuremath{\psi}\ensuremath{\rangle}\ensuremath{\rangle} v hv y => \ensuremath{\langle}h\ensuremath{\varphi} \ensuremath{\psi} h\ensuremath{\psi}x v hv y, h\ensuremath{\psi} \ensuremath{\varphi} h\ensuremath{\varphi}x v hv y\ensuremath{\rangle}

-- UniqueEss2 : \ensuremath{\lfloor}Ess \ensuremath{\varphi} x \ensuremath{\wedge}\textsuperscript{m} Ess \ensuremath{\psi} x \ensuremath{\rightarrow}\textsuperscript{m} \ensuremath{\Box}(\ensuremath{\varphi} =\textsuperscript{m} \ensuremath{\psi})\ensuremath{\rfloor}   -- `nitpick[card i=2]`: countermodel found
example (\ensuremath{\varphi} \ensuremath{\psi} : e \ensuremath{\rightarrow} \ensuremath{\sigma}) (x : e) : \ensuremath{\lfloor}Ess \ensuremath{\varphi} x \ensuremath{\wedge}\textsuperscript{m} Ess \ensuremath{\psi} x \ensuremath{\rightarrow}\textsuperscript{m} \ensuremath{\Box}(\ensuremath{\varphi} =\textsuperscript{m} \ensuremath{\psi})\ensuremath{\rfloor} := by countermodel

theorem UniqueEss3 (\ensuremath{\varphi} : e \ensuremath{\rightarrow} \ensuremath{\sigma}) (x : e) : \ensuremath{\lfloor}Ess \ensuremath{\varphi} x \ensuremath{\rightarrow}\textsuperscript{m} \ensuremath{\Box}(\ensuremath{\forall}\textsuperscript{m} y, \ensuremath{\varphi} y \ensuremath{\rightarrow}\textsuperscript{m} y \ensuremath{\equiv}\textsuperscript{m} x)\ensuremath{\rfloor} :=
  fun _ \ensuremath{\langle}_, h\ensuremath{\rangle} => h (fun z => z \ensuremath{\equiv}\textsuperscript{m} x) (fun _ hp => hp)

theorem Monotheism (x y : e) : \ensuremath{\lfloor}G x \ensuremath{\wedge}\textsuperscript{m} G y \ensuremath{\rightarrow}\textsuperscript{m} x \ensuremath{\equiv}\textsuperscript{m} y\ensuremath{\rfloor} :=
  fun _ \ensuremath{\langle}hx, hy\ensuremath{\rangle} Q hQ => hy Q (PosOfGod hx hQ)

/-- Everything true of every God-like being at every successor world is positive: by `Coro`
some successor holds a God-like being, which has every positive property, and positivity is
necessary (`A4`).  No frame condition is used. -/
theorem Pos (\ensuremath{\chi} : e \ensuremath{\rightarrow} \ensuremath{\sigma}) (w : i) (h : \ensuremath{\forall} v, w \textsf{\bfseries r} v \ensuremath{\rightarrow} \ensuremath{\forall} x, G x v \ensuremath{\rightarrow} \ensuremath{\chi} x v) : P \ensuremath{\chi} w :=
  Classical.byContradiction fun hn =>
    let \ensuremath{\langle}v, hv, x, hG\ensuremath{\rangle} := Coro w
    hG ~\textsuperscript{m}\ensuremath{\chi} (A4 ~\textsuperscript{m}\ensuremath{\chi} w ((A1 \ensuremath{\chi} w).1 hn) v hv) (h v hv x hG)

theorem PisFilter : \ensuremath{\lfloor}Filter P\ensuremath{\rfloor} := fun w =>
  \ensuremath{\langle}\ensuremath{\langle}\ensuremath{\langle}(PosProps w).1, (NegProps w).1\ensuremath{\rangle},
    -- `\ensuremath{\varphi} \ensuremath{\subseteq}\textsuperscript{E} \ensuremath{\psi}` is an actualist inclusion, `A2` needs a possibilist one; so argue directly: `MC`
    -- carries the inclusion to a successor `v` (one exists by `Coro`) holding an actual God-like
    -- being `x` (`T3`), `x` has every positive property, and positivity is necessary (`A4`)
    fun \ensuremath{\varphi} \ensuremath{\psi} \ensuremath{\langle}h\ensuremath{\varphi}, hsub\ensuremath{\rangle} => Classical.byContradiction fun hn =>
      let \ensuremath{\langle}v, hv, _\ensuremath{\rangle} := Coro w
      let \ensuremath{\langle}x, hxa, hGx\ensuremath{\rangle} := T3 w v hv
      hGx ~\textsuperscript{m}\ensuremath{\psi} (A4 ~\textsuperscript{m}\ensuremath{\psi} w ((A1 \ensuremath{\psi} w).1 hn) v hv)
        (MC (\ensuremath{\forall}\textsuperscript{E} x, \ensuremath{\varphi} x \ensuremath{\rightarrow}\textsuperscript{m} \ensuremath{\psi} x) w hsub v hv x hxa (hGx \ensuremath{\varphi} (A4 \ensuremath{\varphi} w h\ensuremath{\varphi} v hv)))\ensuremath{\rangle},
   fun \ensuremath{\varphi} \ensuremath{\psi} \ensuremath{\langle}h\ensuremath{\varphi}, h\ensuremath{\psi}\ensuremath{\rangle} => Pos _ w (fun v hv _ hG => \ensuremath{\langle}hG \ensuremath{\varphi} (A4 \ensuremath{\varphi} w h\ensuremath{\varphi} v hv), hG \ensuremath{\psi} (A4 \ensuremath{\psi} w h\ensuremath{\psi} v hv)\ensuremath{\rangle})\ensuremath{\rangle}

theorem PisUFilter : \ensuremath{\lfloor}UFilter P\ensuremath{\rfloor} := fun w =>
  \ensuremath{\langle}PisFilter w, fun \ensuremath{\varphi} => (Classical.em (P \ensuremath{\varphi} w)).imp id (A1 \ensuremath{\varphi} w).1\ensuremath{\rangle}

-- `lemma True nitpick[satisfy,card=1,eval="\ensuremath{\lfloor}P (\ensuremath{\lambda}x.\ensuremath{\bot})\ensuremath{\rfloor}"] oops`
--   One model found of cardinality one (consistency check)
\end{Verbatim}
\subsection{Appendix}
\subsubsection{GoedelVariantHOML1poss.lean (Figure 16 of \texorpdfstring{\cite{J75}}{the Notes})}
G\"odel's axioms and definitions, as presented in the 1970 manuscript, are inconsistent.  In contrast to Figure 6 we here use only possibilist quantifiers and still derive falsity.
\begin{Verbatim}[commandchars=\\\{\},fontsize=\scriptsize,xleftmargin=1em]
import Notes.HOMLinHOL
\end{Verbatim}

\begin{Verbatim}[commandchars=\\\{\},fontsize=\scriptsize,xleftmargin=1em]
/-- Positive property -/
axiom P : (e \ensuremath{\rightarrow} \ensuremath{\sigma}) \ensuremath{\rightarrow} \ensuremath{\sigma}

axiom Ax1 (\ensuremath{\varphi} \ensuremath{\psi} : e \ensuremath{\rightarrow} \ensuremath{\sigma}) : \ensuremath{\lfloor}P \ensuremath{\varphi} \ensuremath{\wedge}\textsuperscript{m} P \ensuremath{\psi} \ensuremath{\rightarrow}\textsuperscript{m} P (\ensuremath{\varphi} \ensuremath{\centerdot}\textsuperscript{m} \ensuremath{\psi})\ensuremath{\rfloor}

axiom Ax2a (\ensuremath{\varphi} : e \ensuremath{\rightarrow} \ensuremath{\sigma}) : \ensuremath{\lfloor}P \ensuremath{\varphi} \ensuremath{\vee}\textsuperscript{e} P ~\textsuperscript{m}\ensuremath{\varphi}\ensuremath{\rfloor}

/-- Auxiliary reformulation of `Ax2a` (Lean has no `sledgehammer`). -/
theorem Ax2a' (\ensuremath{\varphi} : e \ensuremath{\rightarrow} \ensuremath{\sigma}) : \ensuremath{\lfloor}\ensuremath{\neg}\textsuperscript{m}(P \ensuremath{\varphi}) \ensuremath{\leftrightarrow}\textsuperscript{m} P ~\textsuperscript{m}\ensuremath{\varphi}\ensuremath{\rfloor} :=
  fun w => \ensuremath{\langle}(Ax2a \ensuremath{\varphi} w).1.resolve_left, fun hp hq => (Ax2a \ensuremath{\varphi} w).2 \ensuremath{\langle}hq, hp\ensuremath{\rangle}\ensuremath{\rangle}

/-- God-like -/
def G (x : e) : \ensuremath{\sigma} := \ensuremath{\forall}\textsuperscript{m} (\ensuremath{\varphi} : e \ensuremath{\rightarrow} \ensuremath{\sigma}), P \ensuremath{\varphi} \ensuremath{\rightarrow}\textsuperscript{m} \ensuremath{\varphi} x

/-- Necessary property inclusion -/
@[simp, grind] def PInc (\ensuremath{\varphi} \ensuremath{\psi} : e \ensuremath{\rightarrow} \ensuremath{\sigma}) : \ensuremath{\sigma} := \ensuremath{\Box}(\ensuremath{\forall}\textsuperscript{m} (y : e), \ensuremath{\varphi} y \ensuremath{\rightarrow}\textsuperscript{m} \ensuremath{\psi} y)
infixr:48 " \ensuremath{\supset}\textsuperscript{N} " => PInc

/-- `Ess \ensuremath{\varphi} x`: \ensuremath{\varphi} is an essence of x -/
def Ess (\ensuremath{\varphi} : e \ensuremath{\rightarrow} \ensuremath{\sigma}) (x : e) : \ensuremath{\sigma} := \ensuremath{\forall}\textsuperscript{m} (\ensuremath{\psi} : e \ensuremath{\rightarrow} \ensuremath{\sigma}), \ensuremath{\psi} x \ensuremath{\rightarrow}\textsuperscript{m} (\ensuremath{\varphi} \ensuremath{\supset}\textsuperscript{N} \ensuremath{\psi})

axiom Ax2b (\ensuremath{\varphi} : e \ensuremath{\rightarrow} \ensuremath{\sigma}) : \ensuremath{\lfloor}P \ensuremath{\varphi} \ensuremath{\rightarrow}\textsuperscript{m} \ensuremath{\Box} P \ensuremath{\varphi}\ensuremath{\rfloor}

theorem Ax2b' (\ensuremath{\varphi} : e \ensuremath{\rightarrow} \ensuremath{\sigma}) : \ensuremath{\lfloor}\ensuremath{\neg}\textsuperscript{m}(P \ensuremath{\varphi}) \ensuremath{\rightarrow}\textsuperscript{m} \ensuremath{\Box}(\ensuremath{\neg}\textsuperscript{m}(P \ensuremath{\varphi}))\ensuremath{\rfloor} :=
  fun w hn v hv => (Ax2a' \ensuremath{\varphi} v).2 (Ax2b ~\textsuperscript{m}\ensuremath{\varphi} w ((Ax2a' \ensuremath{\varphi} w).1 hn) v hv)

/-- A property exemplified by a God-like being is positive (a consequence of `Ax2a`). -/
theorem PosOfGod \{x : e\} \{\ensuremath{\psi} : e \ensuremath{\rightarrow} \ensuremath{\sigma}\} \{w : i\} (hG : G x w) (h : \ensuremath{\psi} x w) : P \ensuremath{\psi} w :=
  Classical.byContradiction fun hn => hG ~\textsuperscript{m}\ensuremath{\psi} ((Ax2a' \ensuremath{\psi} w).1 hn) h

theorem Th1 (x : e) : \ensuremath{\lfloor}G x \ensuremath{\rightarrow}\textsuperscript{m} Ess G x\ensuremath{\rfloor} :=
  fun w hG \ensuremath{\psi} h\ensuremath{\psi} v hv _y hGy => hGy \ensuremath{\psi} (Ax2b \ensuremath{\psi} w (PosOfGod hG h\ensuremath{\psi}) v hv)

/-- Necessary existence -/
def E (x : e) : \ensuremath{\sigma} := \ensuremath{\forall}\textsuperscript{m} (\ensuremath{\varphi} : e \ensuremath{\rightarrow} \ensuremath{\sigma}), Ess \ensuremath{\varphi} x \ensuremath{\rightarrow}\textsuperscript{m} \ensuremath{\Box}(\ensuremath{\exists}\textsuperscript{m} x, \ensuremath{\varphi} x)

axiom Ax3 : \ensuremath{\lfloor}P E\ensuremath{\rfloor}

theorem Th2 (x : e) : \ensuremath{\lfloor}G x \ensuremath{\rightarrow}\textsuperscript{m} \ensuremath{\Box}(\ensuremath{\exists}\textsuperscript{m} y, G y)\ensuremath{\rfloor} := fun w hG => hG E (Ax3 w) G (Th1 x w hG)

theorem Th3 : \ensuremath{\lfloor}\ensuremath{\Diamond}(\ensuremath{\exists}\textsuperscript{m} x, G x) \ensuremath{\rightarrow}\textsuperscript{m} \ensuremath{\Box}(\ensuremath{\exists}\textsuperscript{m} y, G y)\ensuremath{\rfloor} := by
  have h1 : \ensuremath{\lfloor}(\ensuremath{\exists}\textsuperscript{m} x, G x) \ensuremath{\rightarrow}\textsuperscript{m} \ensuremath{\Box}(\ensuremath{\exists}\textsuperscript{m} y, G y)\ensuremath{\rfloor} := fun _ \ensuremath{\langle}x, hG\ensuremath{\rangle} => Th2 x _ hG
  intro w \ensuremath{\langle}v, hv, hex\ensuremath{\rangle}
  exact h1 w (h1 v hex w (Rsymm w v hv))               -- only symmetry of `\textsf{\bfseries r}` is needed

axiom Ax4 (\ensuremath{\varphi} \ensuremath{\psi} : e \ensuremath{\rightarrow} \ensuremath{\sigma}) : \ensuremath{\lfloor}P \ensuremath{\varphi} \ensuremath{\wedge}\textsuperscript{m} (\ensuremath{\varphi} \ensuremath{\supset}\textsuperscript{N} \ensuremath{\psi}) \ensuremath{\rightarrow}\textsuperscript{m} P \ensuremath{\psi}\ensuremath{\rfloor}

-- `lemma True nitpick[satisfy,expect=unknown] oops`
--   No model found (consistency check)

theorem EmptyEssL (x : e) : \ensuremath{\lfloor}Ess (fun _ : e => \ensuremath{\bot}\textsuperscript{m}) x\ensuremath{\rfloor} := fun _ _ _ _ _ _ h => h.elim

theorem Inconsistency : False := by
  have h1 : \ensuremath{\lfloor}\ensuremath{\neg}\textsuperscript{m}(P (fun _ : e => \ensuremath{\bot}\textsuperscript{m}))\ensuremath{\rfloor} := fun w hp =>
    (Ax2a _ w).2 \ensuremath{\langle}hp, Ax4 (fun _ : e => \ensuremath{\bot}\textsuperscript{m}) _ w \ensuremath{\langle}hp, fun _ _ _ h => h.elim\ensuremath{\rangle}\ensuremath{\rangle}
  have h2 : \ensuremath{\lfloor}P (fun x : e => Ess (fun _ : e => \ensuremath{\bot}\textsuperscript{m}) x \ensuremath{\rightarrow}\textsuperscript{m} \ensuremath{\Box}(\ensuremath{\exists}\textsuperscript{m} z, (fun _ : e => \ensuremath{\bot}\textsuperscript{m}) z))\ensuremath{\rfloor} :=
    fun w => Ax4 E _ w \ensuremath{\langle}Ax3 w, fun _ _ _ hE => hE (fun _ => \ensuremath{\bot}\textsuperscript{m})\ensuremath{\rangle}
  have h3 : \ensuremath{\lfloor}P (fun _ : e => \ensuremath{\Box}(\ensuremath{\exists}\textsuperscript{m} z, (fun _ : e => \ensuremath{\bot}\textsuperscript{m}) z))\ensuremath{\rfloor} :=
    fun w => Ax4 _ _ w \ensuremath{\langle}h2 w, fun v _ y h => h (EmptyEssL y v)\ensuremath{\rangle}
  have h4 : \ensuremath{\lfloor}P (fun _ : e => \ensuremath{\Box} \ensuremath{\bot}\textsuperscript{m})\ensuremath{\rfloor} := by
    have heq : (fun _ : e => \ensuremath{\Box}(\ensuremath{\exists}\textsuperscript{m} z, (fun _ : e => \ensuremath{\bot}\textsuperscript{m}) z)) = (fun _ : e => \ensuremath{\Box} \ensuremath{\bot}\textsuperscript{m}) := by
      -- an explicit term rather than `simp`: `simp` would draw in `eNonempty`, which the
      -- argument does not need (`#print axioms`)
      funext _ w; exact propext \ensuremath{\langle}fun h v hv => match h v hv with | \ensuremath{\langle}_, hf\ensuremath{\rangle} => hf, fun h v hv =>
          (h v hv).elim\ensuremath{\rangle}
    exact heq \ensuremath{\blacktriangleright} h3
  -- Isabelle discharges this with `smt` from `4 Ax2a Ax4`; no frame condition is needed.
  have h5 : \ensuremath{\lfloor}P (fun _ : e => \ensuremath{\bot}\textsuperscript{m})\ensuremath{\rfloor} := by
    intro w
    by_cases hdead : \ensuremath{\exists} v : i, \ensuremath{\forall} u, \ensuremath{\neg} v \textsf{\bfseries r} u
    \textperiodcentered{} -- some world is a dead end: there `Ax4`'s premise is vacuous, so `Ax2a` is violated
      have \ensuremath{\langle}v, hv\ensuremath{\rangle} := hdead
      exact ((Ax2a (fun _ : e => \ensuremath{\Box} \ensuremath{\bot}\textsuperscript{m}) v).2
        \ensuremath{\langle}h4 v, Ax4 _ _ v \ensuremath{\langle}h4 v, fun u hu => absurd hu (hv u)\ensuremath{\rangle}\ensuremath{\rangle}).elim
    \textperiodcentered{} -- otherwise every world has a successor, so `\ensuremath{\Box}\ensuremath{\bot}` is false and the properties coincide
      have ser : \ensuremath{\forall} v : i, \ensuremath{\exists} u, v \textsf{\bfseries r} u := fun v =>
        Classical.byContradiction fun h => hdead \ensuremath{\langle}v, fun u hu => h \ensuremath{\langle}u, hu\ensuremath{\rangle}\ensuremath{\rangle}
      have heq : (fun _ : e => \ensuremath{\Box} \ensuremath{\bot}\textsuperscript{m}) = (fun _ : e => \ensuremath{\bot}\textsuperscript{m}) := by
        funext _ v
        have \ensuremath{\langle}u, hu\ensuremath{\rangle} := ser v
        exact propext \ensuremath{\langle}fun h => h u hu, fun h => h.elim\ensuremath{\rangle}
      exact heq \ensuremath{\blacktriangleright} h4 w
  have w : i := Classical.ofNonempty
  exact h1 w (h5 w)
\end{Verbatim}
\subsubsection{GoedelVariantHOML2poss.lean (Figure 17 of \texorpdfstring{\cite{J75}}{the Notes})}
After an appropriate modification of the definition of essence, the inconsistency revealed in Figure 16 is avoided, and the argument can be successfully verified in modal logic S5 (indeed, as shown, only the modal schema B is actually needed).  In contrast to Figure 7 we here use only possibilist quantifiers to obtain these results.
\begin{Verbatim}[commandchars=\\\{\},fontsize=\scriptsize,xleftmargin=1em]
import Notes.HOMLinHOL
import Notes.ModalFilter
\end{Verbatim}

\begin{Verbatim}[commandchars=\\\{\},fontsize=\scriptsize,xleftmargin=1em]
/-- Positive property -/
axiom P : (e \ensuremath{\rightarrow} \ensuremath{\sigma}) \ensuremath{\rightarrow} \ensuremath{\sigma}

axiom Ax1 (\ensuremath{\varphi} \ensuremath{\psi} : e \ensuremath{\rightarrow} \ensuremath{\sigma}) : \ensuremath{\lfloor}P \ensuremath{\varphi} \ensuremath{\wedge}\textsuperscript{m} P \ensuremath{\psi} \ensuremath{\rightarrow}\textsuperscript{m} P (\ensuremath{\varphi} \ensuremath{\centerdot}\textsuperscript{m} \ensuremath{\psi})\ensuremath{\rfloor}

axiom Ax2a (\ensuremath{\varphi} : e \ensuremath{\rightarrow} \ensuremath{\sigma}) : \ensuremath{\lfloor}P \ensuremath{\varphi} \ensuremath{\vee}\textsuperscript{e} P ~\textsuperscript{m}\ensuremath{\varphi}\ensuremath{\rfloor}

/-- Auxiliary reformulation of `Ax2a` (Lean has no `sledgehammer`). -/
theorem Ax2a' (\ensuremath{\varphi} : e \ensuremath{\rightarrow} \ensuremath{\sigma}) : \ensuremath{\lfloor}\ensuremath{\neg}\textsuperscript{m}(P \ensuremath{\varphi}) \ensuremath{\leftrightarrow}\textsuperscript{m} P ~\textsuperscript{m}\ensuremath{\varphi}\ensuremath{\rfloor} :=
  fun w => \ensuremath{\langle}(Ax2a \ensuremath{\varphi} w).1.resolve_left, fun hp hq => (Ax2a \ensuremath{\varphi} w).2 \ensuremath{\langle}hq, hp\ensuremath{\rangle}\ensuremath{\rangle}

/-- God-like -/
def G (x : e) : \ensuremath{\sigma} := \ensuremath{\forall}\textsuperscript{m} (\ensuremath{\varphi} : e \ensuremath{\rightarrow} \ensuremath{\sigma}), P \ensuremath{\varphi} \ensuremath{\rightarrow}\textsuperscript{m} \ensuremath{\varphi} x

/-- Necessary property inclusion -/
@[simp, grind] def PInc (\ensuremath{\varphi} \ensuremath{\psi} : e \ensuremath{\rightarrow} \ensuremath{\sigma}) : \ensuremath{\sigma} := \ensuremath{\Box}(\ensuremath{\forall}\textsuperscript{m} (y : e), \ensuremath{\varphi} y \ensuremath{\rightarrow}\textsuperscript{m} \ensuremath{\psi} y)
infixr:48 " \ensuremath{\supset}\textsuperscript{N} " => PInc

/-- `Ess \ensuremath{\varphi} x`: \ensuremath{\varphi} is an essence of x (modified: \ensuremath{\varphi} must be exemplified by x) -/
def Ess (\ensuremath{\varphi} : e \ensuremath{\rightarrow} \ensuremath{\sigma}) (x : e) : \ensuremath{\sigma} := \ensuremath{\varphi} x \ensuremath{\wedge}\textsuperscript{m} (\ensuremath{\forall}\textsuperscript{m} (\ensuremath{\psi} : e \ensuremath{\rightarrow} \ensuremath{\sigma}), \ensuremath{\psi} x \ensuremath{\rightarrow}\textsuperscript{m} (\ensuremath{\varphi} \ensuremath{\supset}\textsuperscript{N} \ensuremath{\psi}))

axiom Ax2b (\ensuremath{\varphi} : e \ensuremath{\rightarrow} \ensuremath{\sigma}) : \ensuremath{\lfloor}P \ensuremath{\varphi} \ensuremath{\rightarrow}\textsuperscript{m} \ensuremath{\Box} P \ensuremath{\varphi}\ensuremath{\rfloor}

theorem Ax2b' (\ensuremath{\varphi} : e \ensuremath{\rightarrow} \ensuremath{\sigma}) : \ensuremath{\lfloor}\ensuremath{\neg}\textsuperscript{m}(P \ensuremath{\varphi}) \ensuremath{\rightarrow}\textsuperscript{m} \ensuremath{\Box}(\ensuremath{\neg}\textsuperscript{m}(P \ensuremath{\varphi}))\ensuremath{\rfloor} :=
  fun w hn v hv => (Ax2a' \ensuremath{\varphi} v).2 (Ax2b ~\textsuperscript{m}\ensuremath{\varphi} w ((Ax2a' \ensuremath{\varphi} w).1 hn) v hv)

/-- A property exemplified by a God-like being is positive (a consequence of `Ax2a`). -/
theorem PosOfGod \{x : e\} \{\ensuremath{\psi} : e \ensuremath{\rightarrow} \ensuremath{\sigma}\} \{w : i\} (hG : G x w) (h : \ensuremath{\psi} x w) : P \ensuremath{\psi} w :=
  Classical.byContradiction fun hn => hG ~\textsuperscript{m}\ensuremath{\psi} ((Ax2a' \ensuremath{\psi} w).1 hn) h

theorem Th1 (x : e) : \ensuremath{\lfloor}G x \ensuremath{\rightarrow}\textsuperscript{m} Ess G x\ensuremath{\rfloor} := fun w hG =>
  \ensuremath{\langle}hG, fun \ensuremath{\psi} h\ensuremath{\psi} v hv _y hGy => hGy \ensuremath{\psi} (Ax2b \ensuremath{\psi} w (PosOfGod hG h\ensuremath{\psi}) v hv)\ensuremath{\rangle}

/-- Necessary existence -/
def E (x : e) : \ensuremath{\sigma} := \ensuremath{\forall}\textsuperscript{m} (\ensuremath{\varphi} : e \ensuremath{\rightarrow} \ensuremath{\sigma}), Ess \ensuremath{\varphi} x \ensuremath{\rightarrow}\textsuperscript{m} \ensuremath{\Box}(\ensuremath{\exists}\textsuperscript{m} x, \ensuremath{\varphi} x)

axiom Ax3 : \ensuremath{\lfloor}P E\ensuremath{\rfloor}

theorem Th2 (x : e) : \ensuremath{\lfloor}G x \ensuremath{\rightarrow}\textsuperscript{m} \ensuremath{\Box}(\ensuremath{\exists}\textsuperscript{m} y, G y)\ensuremath{\rfloor} := fun w hG => hG E (Ax3 w) G (Th1 x w hG)

theorem Th3 : \ensuremath{\lfloor}\ensuremath{\Diamond}(\ensuremath{\exists}\textsuperscript{m} x, G x) \ensuremath{\rightarrow}\textsuperscript{m} \ensuremath{\Box}(\ensuremath{\exists}\textsuperscript{m} y, G y)\ensuremath{\rfloor} := by
  have h1 : \ensuremath{\lfloor}(\ensuremath{\exists}\textsuperscript{m} x, G x) \ensuremath{\rightarrow}\textsuperscript{m} \ensuremath{\Box}(\ensuremath{\exists}\textsuperscript{m} y, G y)\ensuremath{\rfloor} := fun _ \ensuremath{\langle}x, hG\ensuremath{\rangle} => Th2 x _ hG
  intro w \ensuremath{\langle}v, hv, hex\ensuremath{\rangle}
  exact h1 w (h1 v hex w (Rsymm w v hv))               -- only symmetry of `\textsf{\bfseries r}` is needed

axiom Ax4 (\ensuremath{\varphi} \ensuremath{\psi} : e \ensuremath{\rightarrow} \ensuremath{\sigma}) : \ensuremath{\lfloor}P \ensuremath{\varphi} \ensuremath{\wedge}\textsuperscript{m} (\ensuremath{\varphi} \ensuremath{\supset}\textsuperscript{N} \ensuremath{\psi}) \ensuremath{\rightarrow}\textsuperscript{m} P \ensuremath{\psi}\ensuremath{\rfloor}

-- `lemma True nitpick[satisfy,card=1,eval="\ensuremath{\lfloor}P (\ensuremath{\lambda}x.\ensuremath{\bot})\ensuremath{\rfloor}"] oops`
--   One model found of cardinality one (consistency check)

@[simp, grind] def PosProps (\ensuremath{\Phi} : (e \ensuremath{\rightarrow} \ensuremath{\sigma}) \ensuremath{\rightarrow} \ensuremath{\sigma}) : \ensuremath{\sigma} := \ensuremath{\forall}\textsuperscript{m} \ensuremath{\varphi}, \ensuremath{\Phi} \ensuremath{\varphi} \ensuremath{\rightarrow}\textsuperscript{m} P \ensuremath{\varphi}
@[simp, grind] def ConjOfPropsFrom (\ensuremath{\varphi} : e \ensuremath{\rightarrow} \ensuremath{\sigma}) (\ensuremath{\Phi} : (e \ensuremath{\rightarrow} \ensuremath{\sigma}) \ensuremath{\rightarrow} \ensuremath{\sigma}) : \ensuremath{\sigma} :=
  \ensuremath{\Box}(\ensuremath{\forall}\textsuperscript{m} z, \ensuremath{\varphi} z \ensuremath{\leftrightarrow}\textsuperscript{m} (\ensuremath{\forall}\textsuperscript{m} \ensuremath{\psi}, \ensuremath{\Phi} \ensuremath{\psi} \ensuremath{\rightarrow}\textsuperscript{m} \ensuremath{\psi} z))
axiom Ax1Gen (\ensuremath{\Phi} : (e \ensuremath{\rightarrow} \ensuremath{\sigma}) \ensuremath{\rightarrow} \ensuremath{\sigma}) (\ensuremath{\varphi} : e \ensuremath{\rightarrow} \ensuremath{\sigma}) :
  \ensuremath{\lfloor}(PosProps \ensuremath{\Phi} \ensuremath{\wedge}\textsuperscript{m} ConjOfPropsFrom \ensuremath{\varphi} \ensuremath{\Phi}) \ensuremath{\rightarrow}\textsuperscript{m} P \ensuremath{\varphi}\ensuremath{\rfloor}

theorem L : \ensuremath{\lfloor}P G\ensuremath{\rfloor} := fun w => Ax1Gen P G w \ensuremath{\langle}fun _ h => h, fun _ _ _ => Iff.rfl\ensuremath{\rangle}

/-- Possibly there is a God-like being.  The Isabelle proof cites `Ax2a`, `Ax4` and `L`; `Ax4` is
dispensable: if no God-like being were possible, `Ax1Gen` would make the empty property
positive, as the conjunction of the positive properties `\{G\}`, while it makes the universal
property positive as the conjunction of `\ensuremath{\emptyset}`, against the exclusivity in `Ax2a`.  Only `Ax1Gen`
and `Ax2a` are used, and no frame condition. -/
theorem Th4 : \ensuremath{\lfloor}\ensuremath{\Diamond}(\ensuremath{\exists}\textsuperscript{m} x, G x)\ensuremath{\rfloor} := fun w =>
  Classical.byContradiction fun hn =>
    have hbot : P (fun _ : e => \ensuremath{\bot}\textsuperscript{m}) w :=
      Ax1Gen (fun \ensuremath{\psi} => \ensuremath{\psi} =\textsuperscript{m} G) (fun _ => \ensuremath{\bot}\textsuperscript{m}) w
        \ensuremath{\langle}fun \ensuremath{\psi} h\ensuremath{\psi} => by simp only [Mprimeq] at h\ensuremath{\psi}; subst h\ensuremath{\psi}; exact L w,
         fun v hv z => \ensuremath{\langle}fun h => h.elim, fun h => hn \ensuremath{\langle}v, hv, z, h G rfl\ensuremath{\rangle}\ensuremath{\rangle}\ensuremath{\rangle}
    have htop : P (fun _ : e => \ensuremath{\top}\textsuperscript{m}) w :=
      Ax1Gen (fun _ => \ensuremath{\bot}\textsuperscript{m}) (fun _ => \ensuremath{\top}\textsuperscript{m}) w
        \ensuremath{\langle}fun _ h => h.elim, fun _ _ _ => \ensuremath{\langle}fun _ _ h => h.elim, fun _ => trivial\ensuremath{\rangle}\ensuremath{\rangle}
    have heq : (~\textsuperscript{m}(fun _ : e => \ensuremath{\bot}\textsuperscript{m})) = (fun _ : e => \ensuremath{\top}\textsuperscript{m}) := by funext x v; simp
    (Ax2a (fun _ => \ensuremath{\bot}\textsuperscript{m}) w).2 \ensuremath{\langle}hbot, heq \ensuremath{\blacktriangleright} htop\ensuremath{\rangle}

theorem Th5 : \ensuremath{\lfloor}\ensuremath{\Box}(\ensuremath{\exists}\textsuperscript{m} x, G x)\ensuremath{\rfloor} := fun w => Th3 w (Th4 w)

theorem MC (\ensuremath{\varphi} : \ensuremath{\sigma}) : \ensuremath{\lfloor}\ensuremath{\varphi} \ensuremath{\rightarrow}\textsuperscript{m} \ensuremath{\Box}\ensuremath{\varphi}\ensuremath{\rfloor} := by                   -- modal collapse
  intro w h\ensuremath{\varphi} v hv
  have \ensuremath{\langle}x, hGx\ensuremath{\rangle} := Th5 v w (Rsymm w v hv)
  have \ensuremath{\langle}z, hGz\ensuremath{\rangle} := Th5 w v hv
  exact (Th1 x w hGx).2 (fun _ => \ensuremath{\varphi}) h\ensuremath{\varphi} v hv z hGz

/-- The universal property is positive, from `Ax2a` and `Ax4` alone: were its complement
positive, `Ax4` would make every property positive, against the exclusivity in `Ax2a`. -/
theorem PosTop (w : i) : P (fun _ : e => \ensuremath{\top}\textsuperscript{m}) w :=
  (Ax2a (fun _ : e => \ensuremath{\top}\textsuperscript{m}) w).1.elim id fun hB =>
    ((Ax2a (fun _ : e => \ensuremath{\top}\textsuperscript{m}) w).2 \ensuremath{\langle}Ax4 _ _ w \ensuremath{\langle}hB, fun _ _ _ _ => trivial\ensuremath{\rangle}, hB\ensuremath{\rangle}).elim

theorem PosProps' : \ensuremath{\lfloor}P (fun _ : e => \ensuremath{\top}\textsuperscript{m}) \ensuremath{\wedge}\textsuperscript{m} P (fun x : e => x =\textsuperscript{m} x)\ensuremath{\rfloor} := fun w =>
  \ensuremath{\langle}PosTop w, Ax4 _ _ w \ensuremath{\langle}PosTop w, fun _ _ _ _ => rfl\ensuremath{\rangle}\ensuremath{\rangle}

theorem NegProps : \ensuremath{\lfloor}\ensuremath{\neg}\textsuperscript{m}P (fun _ : e => \ensuremath{\bot}\textsuperscript{m}) \ensuremath{\wedge}\textsuperscript{m} \ensuremath{\neg}\textsuperscript{m}P (fun x : e => x \ensuremath{\neq}\textsuperscript{m} x)\ensuremath{\rfloor} := fun w =>
  \ensuremath{\langle}fun hp => (Ax2a _ w).2 \ensuremath{\langle}hp, Ax4 _ _ w \ensuremath{\langle}hp, fun _ _ _ h => h.elim\ensuremath{\rangle}\ensuremath{\rangle},
   fun hp => (Ax2a _ w).2 \ensuremath{\langle}hp, Ax4 _ _ w \ensuremath{\langle}hp, fun _ _ _ h => (h rfl).elim\ensuremath{\rangle}\ensuremath{\rangle}\ensuremath{\rangle}

theorem UniqueEss1 (\ensuremath{\varphi} \ensuremath{\psi} : e \ensuremath{\rightarrow} \ensuremath{\sigma}) (x : e) : \ensuremath{\lfloor}Ess \ensuremath{\varphi} x \ensuremath{\wedge}\textsuperscript{m} Ess \ensuremath{\psi} x \ensuremath{\rightarrow}\textsuperscript{m} \ensuremath{\Box}(\ensuremath{\forall}\textsuperscript{m} y, \ensuremath{\varphi} y \ensuremath{\leftrightarrow}\textsuperscript{m} \ensuremath{\psi} y)\ensuremath{\rfloor} :=
  fun _ \ensuremath{\langle}\ensuremath{\langle}h\ensuremath{\varphi}x, h\ensuremath{\varphi}\ensuremath{\rangle}, \ensuremath{\langle}h\ensuremath{\psi}x, h\ensuremath{\psi}\ensuremath{\rangle}\ensuremath{\rangle} v hv y => \ensuremath{\langle}h\ensuremath{\varphi} \ensuremath{\psi} h\ensuremath{\psi}x v hv y, h\ensuremath{\psi} \ensuremath{\varphi} h\ensuremath{\varphi}x v hv y\ensuremath{\rangle}

-- UniqueEss2 : \ensuremath{\lfloor}Ess \ensuremath{\varphi} x \ensuremath{\wedge}\textsuperscript{m} Ess \ensuremath{\psi} x \ensuremath{\rightarrow}\textsuperscript{m} \ensuremath{\Box}(\ensuremath{\varphi} \ensuremath{\equiv}\textsuperscript{m} \ensuremath{\psi})\ensuremath{\rfloor}   -- `nitpick[card i=2]`: countermodel found
example (\ensuremath{\varphi} \ensuremath{\psi} : e \ensuremath{\rightarrow} \ensuremath{\sigma}) (x : e) : \ensuremath{\lfloor}Ess \ensuremath{\varphi} x \ensuremath{\wedge}\textsuperscript{m} Ess \ensuremath{\psi} x \ensuremath{\rightarrow}\textsuperscript{m} \ensuremath{\Box}(\ensuremath{\varphi} \ensuremath{\equiv}\textsuperscript{m} \ensuremath{\psi})\ensuremath{\rfloor} := by countermodel

theorem UniqueEss3 (\ensuremath{\varphi} : e \ensuremath{\rightarrow} \ensuremath{\sigma}) (x : e) : \ensuremath{\lfloor}Ess \ensuremath{\varphi} x \ensuremath{\rightarrow}\textsuperscript{m} \ensuremath{\Box}(\ensuremath{\forall}\textsuperscript{m} y, \ensuremath{\varphi} y \ensuremath{\rightarrow}\textsuperscript{m} y \ensuremath{\equiv}\textsuperscript{m} x)\ensuremath{\rfloor} :=
  fun _ \ensuremath{\langle}_, h\ensuremath{\rangle} => h (fun z => z \ensuremath{\equiv}\textsuperscript{m} x) (fun _ hp => hp)

theorem Monotheism (x y : e) : \ensuremath{\lfloor}G x \ensuremath{\wedge}\textsuperscript{m} G y \ensuremath{\rightarrow}\textsuperscript{m} x \ensuremath{\equiv}\textsuperscript{m} y\ensuremath{\rfloor} :=
  fun _ \ensuremath{\langle}hx, hy\ensuremath{\rangle} Q hQ => hy Q (PosOfGod hx hQ)

theorem PisFilterP : \ensuremath{\lfloor}FilterP P\ensuremath{\rfloor} := fun w =>
  \ensuremath{\langle}\ensuremath{\langle}\ensuremath{\langle}(PosProps' w).1, (NegProps w).1\ensuremath{\rangle},
    -- the inclusion holds at `w`; `MC` carries it to the successor worlds, then `Ax4` applies
    fun \ensuremath{\varphi} \ensuremath{\psi} \ensuremath{\langle}h\ensuremath{\varphi}, hsub\ensuremath{\rangle} => Ax4 \ensuremath{\varphi} \ensuremath{\psi} w \ensuremath{\langle}h\ensuremath{\varphi}, MC (\ensuremath{\forall}\textsuperscript{m} x, \ensuremath{\varphi} x \ensuremath{\rightarrow}\textsuperscript{m} \ensuremath{\psi} x) w hsub\ensuremath{\rangle}\ensuremath{\rangle},
   fun \ensuremath{\varphi} \ensuremath{\psi} \ensuremath{\langle}h\ensuremath{\varphi}, h\ensuremath{\psi}\ensuremath{\rangle} => Ax1 \ensuremath{\varphi} \ensuremath{\psi} w \ensuremath{\langle}h\ensuremath{\varphi}, h\ensuremath{\psi}\ensuremath{\rangle}\ensuremath{\rangle}

theorem PisUFilterP : \ensuremath{\lfloor}UFilterP P\ensuremath{\rfloor} := fun w =>
  \ensuremath{\langle}PisFilterP w, fun \ensuremath{\varphi} => (Classical.em (P \ensuremath{\varphi} w)).imp id (Ax2a' \ensuremath{\varphi} w).1\ensuremath{\rangle}

-- `lemma True nitpick[satisfy,card=1,eval="\ensuremath{\lfloor}P (\ensuremath{\lambda}x.\ensuremath{\bot})\ensuremath{\rfloor}"] oops`
--   One model found of cardinality one (consistency check)
\end{Verbatim}
\subsubsection{GoedelVariantHOML3poss.lean (Figure 18 of \texorpdfstring{\cite{J75}}{the Notes})}
After an appropriate modification of the definition of necessary property implication, the inconsistency shown in Figure 16 is avoided, and the argument can be successfully verified.  As shown here, this still holds when using possibilist quantifiers only.
\begin{Verbatim}[commandchars=\\\{\},fontsize=\scriptsize,xleftmargin=1em]
import Notes.HOMLinHOL
import Notes.ModalFilter
\end{Verbatim}

\begin{Verbatim}[commandchars=\\\{\},fontsize=\scriptsize,xleftmargin=1em]
/-- Positive property -/
axiom P : (e \ensuremath{\rightarrow} \ensuremath{\sigma}) \ensuremath{\rightarrow} \ensuremath{\sigma}

axiom Ax1 (\ensuremath{\varphi} \ensuremath{\psi} : e \ensuremath{\rightarrow} \ensuremath{\sigma}) : \ensuremath{\lfloor}P \ensuremath{\varphi} \ensuremath{\wedge}\textsuperscript{m} P \ensuremath{\psi} \ensuremath{\rightarrow}\textsuperscript{m} P (\ensuremath{\varphi} \ensuremath{\centerdot}\textsuperscript{m} \ensuremath{\psi})\ensuremath{\rfloor}

axiom Ax2a (\ensuremath{\varphi} : e \ensuremath{\rightarrow} \ensuremath{\sigma}) : \ensuremath{\lfloor}P \ensuremath{\varphi} \ensuremath{\vee}\textsuperscript{e} P ~\textsuperscript{m}\ensuremath{\varphi}\ensuremath{\rfloor}

/-- Auxiliary reformulation of `Ax2a` (Lean has no `sledgehammer`). -/
theorem Ax2a' (\ensuremath{\varphi} : e \ensuremath{\rightarrow} \ensuremath{\sigma}) : \ensuremath{\lfloor}\ensuremath{\neg}\textsuperscript{m}(P \ensuremath{\varphi}) \ensuremath{\leftrightarrow}\textsuperscript{m} P ~\textsuperscript{m}\ensuremath{\varphi}\ensuremath{\rfloor} :=
  fun w => \ensuremath{\langle}(Ax2a \ensuremath{\varphi} w).1.resolve_left, fun hp hq => (Ax2a \ensuremath{\varphi} w).2 \ensuremath{\langle}hq, hp\ensuremath{\rangle}\ensuremath{\rangle}

/-- God-like -/
def G (x : e) : \ensuremath{\sigma} := \ensuremath{\forall}\textsuperscript{m} (\ensuremath{\varphi} : e \ensuremath{\rightarrow} \ensuremath{\sigma}), P \ensuremath{\varphi} \ensuremath{\rightarrow}\textsuperscript{m} \ensuremath{\varphi} x

/-- Necessary property inclusion (modified: \ensuremath{\varphi} must be non-empty) -/
@[simp, grind] def PInc (\ensuremath{\varphi} \ensuremath{\psi} : e \ensuremath{\rightarrow} \ensuremath{\sigma}) : \ensuremath{\sigma} :=
  \ensuremath{\Box}((\ensuremath{\varphi} \ensuremath{\neq}\textsuperscript{m} (fun _ : e => \ensuremath{\bot}\textsuperscript{m})) \ensuremath{\wedge}\textsuperscript{m} (\ensuremath{\forall}\textsuperscript{m} (y : e), \ensuremath{\varphi} y \ensuremath{\rightarrow}\textsuperscript{m} \ensuremath{\psi} y))
infixr:48 " \ensuremath{\supset}\textsuperscript{N} " => PInc

/-- `Ess \ensuremath{\varphi} x`: \ensuremath{\varphi} is an essence of x -/
def Ess (\ensuremath{\varphi} : e \ensuremath{\rightarrow} \ensuremath{\sigma}) (x : e) : \ensuremath{\sigma} := \ensuremath{\forall}\textsuperscript{m} (\ensuremath{\psi} : e \ensuremath{\rightarrow} \ensuremath{\sigma}), \ensuremath{\psi} x \ensuremath{\rightarrow}\textsuperscript{m} (\ensuremath{\varphi} \ensuremath{\supset}\textsuperscript{N} \ensuremath{\psi})

axiom Ax2b (\ensuremath{\varphi} : e \ensuremath{\rightarrow} \ensuremath{\sigma}) : \ensuremath{\lfloor}P \ensuremath{\varphi} \ensuremath{\rightarrow}\textsuperscript{m} \ensuremath{\Box} P \ensuremath{\varphi}\ensuremath{\rfloor}

theorem Ax2b' (\ensuremath{\varphi} : e \ensuremath{\rightarrow} \ensuremath{\sigma}) : \ensuremath{\lfloor}\ensuremath{\neg}\textsuperscript{m}(P \ensuremath{\varphi}) \ensuremath{\rightarrow}\textsuperscript{m} \ensuremath{\Box}(\ensuremath{\neg}\textsuperscript{m}(P \ensuremath{\varphi}))\ensuremath{\rfloor} :=
  fun w hn v hv => (Ax2a' \ensuremath{\varphi} v).2 (Ax2b ~\textsuperscript{m}\ensuremath{\varphi} w ((Ax2a' \ensuremath{\varphi} w).1 hn) v hv)

/-- A property exemplified by a God-like being is positive (a consequence of `Ax2a`). -/
theorem PosOfGod \{x : e\} \{\ensuremath{\psi} : e \ensuremath{\rightarrow} \ensuremath{\sigma}\} \{w : i\} (hG : G x w) (h : \ensuremath{\psi} x w) : P \ensuremath{\psi} w :=
  Classical.byContradiction fun hn => hG ~\textsuperscript{m}\ensuremath{\psi} ((Ax2a' \ensuremath{\psi} w).1 hn) h

theorem Th1 (x : e) : \ensuremath{\lfloor}G x \ensuremath{\rightarrow}\textsuperscript{m} Ess G x\ensuremath{\rfloor} := by
  intro w hG \ensuremath{\psi} h\ensuremath{\psi} v hv
  refine \ensuremath{\langle}fun heq => ?_, fun y hGy => hGy \ensuremath{\psi} (Ax2b \ensuremath{\psi} w (PosOfGod hG h\ensuremath{\psi}) v hv)\ensuremath{\rangle}
  rw [heq] at hG; exact hG                             -- `G` is non-empty since `G x` holds

/-- Necessary existence -/
def E (x : e) : \ensuremath{\sigma} := \ensuremath{\forall}\textsuperscript{m} (\ensuremath{\varphi} : e \ensuremath{\rightarrow} \ensuremath{\sigma}), Ess \ensuremath{\varphi} x \ensuremath{\rightarrow}\textsuperscript{m} \ensuremath{\Box}(\ensuremath{\exists}\textsuperscript{m} x, \ensuremath{\varphi} x)

axiom Ax3 : \ensuremath{\lfloor}P E\ensuremath{\rfloor}

theorem Th2 (x : e) : \ensuremath{\lfloor}G x \ensuremath{\rightarrow}\textsuperscript{m} \ensuremath{\Box}(\ensuremath{\exists}\textsuperscript{m} y, G y)\ensuremath{\rfloor} := fun w hG => hG E (Ax3 w) G (Th1 x w hG)

theorem Th3 : \ensuremath{\lfloor}\ensuremath{\Diamond}(\ensuremath{\exists}\textsuperscript{m} x, G x) \ensuremath{\rightarrow}\textsuperscript{m} \ensuremath{\Box}(\ensuremath{\exists}\textsuperscript{m} y, G y)\ensuremath{\rfloor} := by
  have h1 : \ensuremath{\lfloor}(\ensuremath{\exists}\textsuperscript{m} x, G x) \ensuremath{\rightarrow}\textsuperscript{m} \ensuremath{\Box}(\ensuremath{\exists}\textsuperscript{m} y, G y)\ensuremath{\rfloor} := fun _ \ensuremath{\langle}x, hG\ensuremath{\rangle} => Th2 x _ hG
  intro w \ensuremath{\langle}v, hv, hex\ensuremath{\rangle}
  exact h1 w (h1 v hex w (Rsymm w v hv))               -- only symmetry of `\textsf{\bfseries r}` is needed

axiom Ax4 (\ensuremath{\varphi} \ensuremath{\psi} : e \ensuremath{\rightarrow} \ensuremath{\sigma}) : \ensuremath{\lfloor}P \ensuremath{\varphi} \ensuremath{\wedge}\textsuperscript{m} (\ensuremath{\varphi} \ensuremath{\supset}\textsuperscript{N} \ensuremath{\psi}) \ensuremath{\rightarrow}\textsuperscript{m} P \ensuremath{\psi}\ensuremath{\rfloor}

-- `lemma True nitpick[satisfy,card=1,eval="\ensuremath{\lfloor}P (\ensuremath{\lambda}x.\ensuremath{\bot})\ensuremath{\rfloor}"] oops`
--   One model found of cardinality one (consistency check)

@[simp, grind] def PosProps (\ensuremath{\Phi} : (e \ensuremath{\rightarrow} \ensuremath{\sigma}) \ensuremath{\rightarrow} \ensuremath{\sigma}) : \ensuremath{\sigma} := \ensuremath{\forall}\textsuperscript{m} \ensuremath{\varphi}, \ensuremath{\Phi} \ensuremath{\varphi} \ensuremath{\rightarrow}\textsuperscript{m} P \ensuremath{\varphi}
@[simp, grind] def ConjOfPropsFrom (\ensuremath{\varphi} : e \ensuremath{\rightarrow} \ensuremath{\sigma}) (\ensuremath{\Phi} : (e \ensuremath{\rightarrow} \ensuremath{\sigma}) \ensuremath{\rightarrow} \ensuremath{\sigma}) : \ensuremath{\sigma} :=
  \ensuremath{\Box}(\ensuremath{\forall}\textsuperscript{m} z, \ensuremath{\varphi} z \ensuremath{\leftrightarrow}\textsuperscript{m} (\ensuremath{\forall}\textsuperscript{m} \ensuremath{\psi}, \ensuremath{\Phi} \ensuremath{\psi} \ensuremath{\rightarrow}\textsuperscript{m} \ensuremath{\psi} z))
axiom Ax1Gen (\ensuremath{\Phi} : (e \ensuremath{\rightarrow} \ensuremath{\sigma}) \ensuremath{\rightarrow} \ensuremath{\sigma}) (\ensuremath{\varphi} : e \ensuremath{\rightarrow} \ensuremath{\sigma}) :
  \ensuremath{\lfloor}(PosProps \ensuremath{\Phi} \ensuremath{\wedge}\textsuperscript{m} ConjOfPropsFrom \ensuremath{\varphi} \ensuremath{\Phi}) \ensuremath{\rightarrow}\textsuperscript{m} P \ensuremath{\varphi}\ensuremath{\rfloor}

theorem L : \ensuremath{\lfloor}P G\ensuremath{\rfloor} := fun w => Ax1Gen P G w \ensuremath{\langle}fun _ h => h, fun _ _ _ => Iff.rfl\ensuremath{\rangle}

/-- The AFP sources leave `Th4` unreplayed (`oops`, then a postulate), although automated
provers report a proof from `Ax2a`, `L` and `Ax1Gen`.  Here is one: if no God-like being were
possible, `Ax1Gen` would make the empty property positive, as the conjunction of `\{G\}`; it also
makes the universal property positive, as the conjunction of `\ensuremath{\emptyset}`; the two contradict the
exclusivity in `Ax2a`.  Only `Ax1Gen` and `Ax2a` are used: the theorem holds in logic K. -/
theorem Th4 : \ensuremath{\lfloor}\ensuremath{\Diamond}(\ensuremath{\exists}\textsuperscript{m} x, G x)\ensuremath{\rfloor} := fun w =>
  Classical.byContradiction fun hn =>
    have hbot : P (fun _ : e => \ensuremath{\bot}\textsuperscript{m}) w :=
      Ax1Gen (fun \ensuremath{\psi} => \ensuremath{\psi} =\textsuperscript{m} G) (fun _ => \ensuremath{\bot}\textsuperscript{m}) w
        \ensuremath{\langle}fun \ensuremath{\psi} h\ensuremath{\psi} => by simp only [Mprimeq] at h\ensuremath{\psi}; subst h\ensuremath{\psi}; exact L w,
         fun v hv z => \ensuremath{\langle}fun h => h.elim, fun h => hn \ensuremath{\langle}v, hv, z, h G rfl\ensuremath{\rangle}\ensuremath{\rangle}\ensuremath{\rangle}
    have htop : P (fun _ : e => \ensuremath{\top}\textsuperscript{m}) w :=
      Ax1Gen (fun _ => \ensuremath{\bot}\textsuperscript{m}) (fun _ => \ensuremath{\top}\textsuperscript{m}) w
        \ensuremath{\langle}fun _ h => h.elim, fun _ _ _ => \ensuremath{\langle}fun _ _ h => h.elim, fun _ => trivial\ensuremath{\rangle}\ensuremath{\rangle}
    have heq : (~\textsuperscript{m}(fun _ : e => \ensuremath{\bot}\textsuperscript{m})) = (fun _ : e => \ensuremath{\top}\textsuperscript{m}) := by funext x v; simp
    (Ax2a (fun _ => \ensuremath{\bot}\textsuperscript{m}) w).2 \ensuremath{\langle}hbot, heq \ensuremath{\blacktriangleright} htop\ensuremath{\rangle}

theorem Th5 : \ensuremath{\lfloor}\ensuremath{\Box}(\ensuremath{\exists}\textsuperscript{m} x, G x)\ensuremath{\rfloor} := fun w => Th3 w (Th4 w)

theorem MC (\ensuremath{\varphi} : \ensuremath{\sigma}) : \ensuremath{\lfloor}\ensuremath{\varphi} \ensuremath{\rightarrow}\textsuperscript{m} \ensuremath{\Box}\ensuremath{\varphi}\ensuremath{\rfloor} := by                   -- modal collapse
  intro w h\ensuremath{\varphi} v hv
  have \ensuremath{\langle}x, hGx\ensuremath{\rangle} := Th5 v w (Rsymm w v hv)
  have \ensuremath{\langle}z, hGz\ensuremath{\rangle} := Th5 w v hv
  exact (Th1 x w hGx) (fun _ => \ensuremath{\varphi}) h\ensuremath{\varphi} v hv |>.2 z hGz

/-- `G` is not the empty property: by `Th4` it is exemplified at some successor of `w`. -/
theorem GNonempty (w : i) : G \ensuremath{\neq} (fun _ : e => \ensuremath{\bot}\textsuperscript{m}) := fun h =>
  let \ensuremath{\langle}_, _, _, hG\ensuremath{\rangle} := Th4 w; (h \ensuremath{\blacktriangleright} hG : \ensuremath{\bot}\textsuperscript{m} _)

/-- `Ax4` applied to `G`: everything true of every God-like being at every successor world is
positive.  Uses `L`, `Th4` and `Ax4`, hence `Ax1Gen`, `Ax2a`, `Ax4`; no frame condition. -/
theorem PosIncl (\ensuremath{\chi} : e \ensuremath{\rightarrow} \ensuremath{\sigma}) (w : i) (h : \ensuremath{\forall} v, w \textsf{\bfseries r} v \ensuremath{\rightarrow} \ensuremath{\forall} x, G x v \ensuremath{\rightarrow} \ensuremath{\chi} x v) : P \ensuremath{\chi} w :=
  Ax4 G \ensuremath{\chi} w \ensuremath{\langle}L w, fun v hv => \ensuremath{\langle}GNonempty w, h v hv\ensuremath{\rangle}\ensuremath{\rangle}

theorem PosProps' : \ensuremath{\lfloor}P (fun _ : e => \ensuremath{\top}\textsuperscript{m}) \ensuremath{\wedge}\textsuperscript{m} P (fun x : e => x =\textsuperscript{m} x)\ensuremath{\rfloor} := fun w =>
  \ensuremath{\langle}PosIncl _ w (fun _ _ _ _ => trivial), PosIncl _ w (fun _ _ _ _ => rfl)\ensuremath{\rangle}

theorem NegProps : \ensuremath{\lfloor}\ensuremath{\neg}\textsuperscript{m}P (fun _ : e => \ensuremath{\bot}\textsuperscript{m}) \ensuremath{\wedge}\textsuperscript{m} \ensuremath{\neg}\textsuperscript{m}P (fun x : e => x \ensuremath{\neq}\textsuperscript{m} x)\ensuremath{\rfloor} := fun w =>
  \ensuremath{\langle}fun hB => (Ax2a _ w).2 \ensuremath{\langle}hB, (show (~\textsuperscript{m}(fun _ : e => \ensuremath{\bot}\textsuperscript{m})) = (fun _ : e => \ensuremath{\top}\textsuperscript{m}) by
      funext x v; simp) \ensuremath{\blacktriangleright} (PosProps' w).1\ensuremath{\rangle},
   fun hN => (Ax2a _ w).2 \ensuremath{\langle}hN, (show (~\textsuperscript{m}(fun x : e => x \ensuremath{\neq}\textsuperscript{m} x)) = (fun _ : e => \ensuremath{\top}\textsuperscript{m}) by
      funext x v; simp) \ensuremath{\blacktriangleright} (PosProps' w).1\ensuremath{\rangle}\ensuremath{\rangle}

-- UniqueEss1 : \ensuremath{\lfloor}Ess \ensuremath{\varphi} x \ensuremath{\wedge}\textsuperscript{m} Ess \ensuremath{\psi} x \ensuremath{\rightarrow}\textsuperscript{m} \ensuremath{\Box}(\ensuremath{\forall}\textsuperscript{m} y, \ensuremath{\varphi} y \ensuremath{\leftrightarrow}\textsuperscript{m} \ensuremath{\psi} y)\ensuremath{\rfloor}      -- Unclear, open question
example (\ensuremath{\varphi} \ensuremath{\psi} : e \ensuremath{\rightarrow} \ensuremath{\sigma}) (x : e) : \ensuremath{\lfloor}Ess \ensuremath{\varphi} x \ensuremath{\wedge}\textsuperscript{m} Ess \ensuremath{\psi} x \ensuremath{\rightarrow}\textsuperscript{m} \ensuremath{\Box}(\ensuremath{\forall}\textsuperscript{m} y, \ensuremath{\varphi} y \ensuremath{\leftrightarrow}\textsuperscript{m} \ensuremath{\psi} y)\ensuremath{\rfloor} := by openproblem
-- UniqueEss2 : \ensuremath{\lfloor}Ess \ensuremath{\varphi} x \ensuremath{\wedge}\textsuperscript{m} Ess \ensuremath{\psi} x \ensuremath{\rightarrow}\textsuperscript{m} \ensuremath{\Box}(\ensuremath{\varphi} \ensuremath{\equiv}\textsuperscript{m} \ensuremath{\psi})\ensuremath{\rfloor}                -- Unclear, open question
example (\ensuremath{\varphi} \ensuremath{\psi} : e \ensuremath{\rightarrow} \ensuremath{\sigma}) (x : e) : \ensuremath{\lfloor}Ess \ensuremath{\varphi} x \ensuremath{\wedge}\textsuperscript{m} Ess \ensuremath{\psi} x \ensuremath{\rightarrow}\textsuperscript{m} \ensuremath{\Box}(\ensuremath{\varphi} \ensuremath{\equiv}\textsuperscript{m} \ensuremath{\psi})\ensuremath{\rfloor} := by openproblem

theorem UniqueEss3 (\ensuremath{\varphi} : e \ensuremath{\rightarrow} \ensuremath{\sigma}) (x : e) : \ensuremath{\lfloor}Ess \ensuremath{\varphi} x \ensuremath{\rightarrow}\textsuperscript{m} \ensuremath{\Box}(\ensuremath{\forall}\textsuperscript{m} y, \ensuremath{\varphi} y \ensuremath{\rightarrow}\textsuperscript{m} y \ensuremath{\equiv}\textsuperscript{m} x)\ensuremath{\rfloor} :=
  fun _ h v hv => (h (fun z => z \ensuremath{\equiv}\textsuperscript{m} x) (fun _ hp => hp) v hv).2

theorem Monotheism (x y : e) : \ensuremath{\lfloor}G x \ensuremath{\wedge}\textsuperscript{m} G y \ensuremath{\rightarrow}\textsuperscript{m} x \ensuremath{\equiv}\textsuperscript{m} y\ensuremath{\rfloor} :=
  fun _ \ensuremath{\langle}hx, hy\ensuremath{\rangle} Q hQ => hy Q (PosOfGod hx hQ)

theorem PisFilterP : \ensuremath{\lfloor}FilterP P\ensuremath{\rfloor} := fun w =>
  \ensuremath{\langle}\ensuremath{\langle}\ensuremath{\langle}(PosProps' w).1, (NegProps w).1\ensuremath{\rangle},
    -- the inclusion holds at `w`; `MC` carries it to the successor worlds, then `Ax4` applies
    fun \ensuremath{\varphi} \ensuremath{\psi} \ensuremath{\langle}h\ensuremath{\varphi}, hsub\ensuremath{\rangle} => Ax4 \ensuremath{\varphi} \ensuremath{\psi} w \ensuremath{\langle}h\ensuremath{\varphi}, fun v hv =>
      \ensuremath{\langle}fun h => (NegProps w).1 (h \ensuremath{\blacktriangleright} h\ensuremath{\varphi}), MC (\ensuremath{\forall}\textsuperscript{m} x, \ensuremath{\varphi} x \ensuremath{\rightarrow}\textsuperscript{m} \ensuremath{\psi} x) w hsub v hv\ensuremath{\rangle}\ensuremath{\rangle}\ensuremath{\rangle},
   fun \ensuremath{\varphi} \ensuremath{\psi} \ensuremath{\langle}h\ensuremath{\varphi}, h\ensuremath{\psi}\ensuremath{\rangle} => Ax1 \ensuremath{\varphi} \ensuremath{\psi} w \ensuremath{\langle}h\ensuremath{\varphi}, h\ensuremath{\psi}\ensuremath{\rangle}\ensuremath{\rangle}

theorem PisUFilterP : \ensuremath{\lfloor}UFilterP P\ensuremath{\rfloor} := fun w =>
  \ensuremath{\langle}PisFilterP w, fun \ensuremath{\varphi} => (Classical.em (P \ensuremath{\varphi} w)).imp id (Ax2a' \ensuremath{\varphi} w).1\ensuremath{\rangle}

-- `lemma True nitpick[satisfy,card=1,eval="\ensuremath{\lfloor}P (\ensuremath{\lambda}x.\ensuremath{\top})\ensuremath{\rfloor}"] oops`
--   One model found of cardinality one (consistency check)
\end{Verbatim}
\subsubsection{ScottVariantHOMLposs.lean (Figure 19 of \texorpdfstring{\cite{J75}}{the Notes})}
Scott's variant of G\"odel's ontological proof is still valid when using possibilist quantifiers only.
\begin{Verbatim}[commandchars=\\\{\},fontsize=\scriptsize,xleftmargin=1em]
import Notes.HOMLinHOL
import Notes.ModalFilter
\end{Verbatim}

\begin{Verbatim}[commandchars=\\\{\},fontsize=\scriptsize,xleftmargin=1em]
/-- Positive property -/
axiom P : (e \ensuremath{\rightarrow} \ensuremath{\sigma}) \ensuremath{\rightarrow} \ensuremath{\sigma}

axiom A1 (\ensuremath{\varphi} : e \ensuremath{\rightarrow} \ensuremath{\sigma}) : \ensuremath{\lfloor}\ensuremath{\neg}\textsuperscript{m}(P \ensuremath{\varphi}) \ensuremath{\leftrightarrow}\textsuperscript{m} P ~\textsuperscript{m}\ensuremath{\varphi}\ensuremath{\rfloor}

axiom A2 (\ensuremath{\varphi} \ensuremath{\psi} : e \ensuremath{\rightarrow} \ensuremath{\sigma}) : \ensuremath{\lfloor}P \ensuremath{\varphi} \ensuremath{\wedge}\textsuperscript{m} \ensuremath{\Box}(\ensuremath{\forall}\textsuperscript{m} y, \ensuremath{\varphi} y \ensuremath{\rightarrow}\textsuperscript{m} \ensuremath{\psi} y) \ensuremath{\rightarrow}\textsuperscript{m} P \ensuremath{\psi}\ensuremath{\rfloor}

theorem T1 (\ensuremath{\varphi} : e \ensuremath{\rightarrow} \ensuremath{\sigma}) : \ensuremath{\lfloor}P \ensuremath{\varphi} \ensuremath{\rightarrow}\textsuperscript{m} \ensuremath{\Diamond}(\ensuremath{\exists}\textsuperscript{m} x, \ensuremath{\varphi} x)\ensuremath{\rfloor} := fun w h =>
  Classical.byContradiction fun hn =>
    (A1 \ensuremath{\varphi} w).2 (A2 \ensuremath{\varphi} ~\textsuperscript{m}\ensuremath{\varphi} w \ensuremath{\langle}h, fun v hv y h\ensuremath{\varphi} _ => hn \ensuremath{\langle}v, hv, y, h\ensuremath{\varphi}\ensuremath{\rangle}\ensuremath{\rangle}) h

/-- God-like -/
def G (x : e) : \ensuremath{\sigma} := \ensuremath{\forall}\textsuperscript{m} (\ensuremath{\varphi} : e \ensuremath{\rightarrow} \ensuremath{\sigma}), P \ensuremath{\varphi} \ensuremath{\rightarrow}\textsuperscript{m} \ensuremath{\varphi} x

axiom A3 : \ensuremath{\lfloor}P G\ensuremath{\rfloor}

theorem Coro : \ensuremath{\lfloor}\ensuremath{\Diamond}(\ensuremath{\exists}\textsuperscript{m} x, G x)\ensuremath{\rfloor} := fun w => T1 G w (A3 w)

axiom A4 (\ensuremath{\varphi} : e \ensuremath{\rightarrow} \ensuremath{\sigma}) : \ensuremath{\lfloor}P \ensuremath{\varphi} \ensuremath{\rightarrow}\textsuperscript{m} \ensuremath{\Box} P \ensuremath{\varphi}\ensuremath{\rfloor}

/-- `Ess \ensuremath{\varphi} x`: \ensuremath{\varphi} is an essence of x -/
def Ess (\ensuremath{\varphi} : e \ensuremath{\rightarrow} \ensuremath{\sigma}) (x : e) : \ensuremath{\sigma} := \ensuremath{\varphi} x \ensuremath{\wedge}\textsuperscript{m} (\ensuremath{\forall}\textsuperscript{m} (\ensuremath{\psi} : e \ensuremath{\rightarrow} \ensuremath{\sigma}), \ensuremath{\psi} x \ensuremath{\rightarrow}\textsuperscript{m} \ensuremath{\Box}(\ensuremath{\forall}\textsuperscript{m} (y : e), \ensuremath{\varphi} y \ensuremath{\rightarrow}\textsuperscript{m} \ensuremath{\psi} y))

/-- A property exemplified by a God-like being is positive (a consequence of `A1`). -/
theorem PosOfGod \{x : e\} \{\ensuremath{\psi} : e \ensuremath{\rightarrow} \ensuremath{\sigma}\} \{w : i\} (hG : G x w) (h : \ensuremath{\psi} x w) : P \ensuremath{\psi} w :=
  Classical.byContradiction fun hn => hG ~\textsuperscript{m}\ensuremath{\psi} ((A1 \ensuremath{\psi} w).1 hn) h

theorem T2 (x : e) : \ensuremath{\lfloor}G x \ensuremath{\rightarrow}\textsuperscript{m} Ess G x\ensuremath{\rfloor} := fun w hG =>
  \ensuremath{\langle}hG, fun \ensuremath{\psi} h\ensuremath{\psi} v hv _y hGy => hGy \ensuremath{\psi} (A4 \ensuremath{\psi} w (PosOfGod hG h\ensuremath{\psi}) v hv)\ensuremath{\rangle}

/-- Necessary existence -/
def NE (x : e) : \ensuremath{\sigma} := \ensuremath{\forall}\textsuperscript{m} (\ensuremath{\varphi} : e \ensuremath{\rightarrow} \ensuremath{\sigma}), Ess \ensuremath{\varphi} x \ensuremath{\rightarrow}\textsuperscript{m} \ensuremath{\Box}(\ensuremath{\exists}\textsuperscript{m} x, \ensuremath{\varphi} x)

axiom A5 : \ensuremath{\lfloor}P NE\ensuremath{\rfloor}

-- `lemma True nitpick[satisfy,card=1,eval="\ensuremath{\lfloor}P (\ensuremath{\lambda}x.\ensuremath{\bot})\ensuremath{\rfloor}"] oops`
--   One model found of cardinality one (consistency check)

theorem T3 : \ensuremath{\lfloor}\ensuremath{\Box}(\ensuremath{\exists}\textsuperscript{m} x, G x)\ensuremath{\rfloor} := by
  have h2 : \ensuremath{\lfloor}(\ensuremath{\exists}\textsuperscript{m} x, G x) \ensuremath{\rightarrow}\textsuperscript{m} \ensuremath{\Box}(\ensuremath{\exists}\textsuperscript{m} x, G x)\ensuremath{\rfloor} :=
    fun _ \ensuremath{\langle}x, hG\ensuremath{\rangle} => hG NE (A5 _) G (T2 x _ hG)
  intro w
  have \ensuremath{\langle}v, hv, hex\ensuremath{\rangle} := Coro w
  exact h2 w (h2 v hex w (Rsymm w v hv))

theorem MC (\ensuremath{\varphi} : \ensuremath{\sigma}) : \ensuremath{\lfloor}\ensuremath{\varphi} \ensuremath{\rightarrow}\textsuperscript{m} \ensuremath{\Box}\ensuremath{\varphi}\ensuremath{\rfloor} := by                   -- modal collapse
  intro w h\ensuremath{\varphi} v hv
  have \ensuremath{\langle}x, hGx\ensuremath{\rangle} := T3 v w (Rsymm w v hv)
  have \ensuremath{\langle}z, hGz\ensuremath{\rangle} := T3 w v hv
  exact (T2 x w hGx).2 (fun _ => \ensuremath{\varphi}) h\ensuremath{\varphi} v hv z hGz

theorem PosProps : \ensuremath{\lfloor}P (fun _ : e => \ensuremath{\top}\textsuperscript{m}) \ensuremath{\wedge}\textsuperscript{m} P (fun x : e => x =\textsuperscript{m} x)\ensuremath{\rfloor} := fun w =>
  \ensuremath{\langle}Classical.byContradiction fun hn => let \ensuremath{\langle}_, _, _, hx\ensuremath{\rangle} := T1 _ w ((A1 _ w).1 hn); hx trivial,
   Classical.byContradiction fun hn => let \ensuremath{\langle}_, _, _, hx\ensuremath{\rangle} := T1 _ w ((A1 _ w).1 hn); hx rfl\ensuremath{\rangle}

theorem NegProps : \ensuremath{\lfloor}\ensuremath{\neg}\textsuperscript{m}P (fun _ : e => \ensuremath{\bot}\textsuperscript{m}) \ensuremath{\wedge}\textsuperscript{m} \ensuremath{\neg}\textsuperscript{m}P (fun x : e => x \ensuremath{\neq}\textsuperscript{m} x)\ensuremath{\rfloor} := fun w =>
  \ensuremath{\langle}fun hp => let \ensuremath{\langle}_, _, _, hx\ensuremath{\rangle} := T1 _ w hp; hx,
   fun hp => let \ensuremath{\langle}_, _, _, hx\ensuremath{\rangle} := T1 _ w hp; hx rfl\ensuremath{\rangle}

theorem UniqueEss1 (\ensuremath{\varphi} \ensuremath{\psi} : e \ensuremath{\rightarrow} \ensuremath{\sigma}) (x : e) : \ensuremath{\lfloor}Ess \ensuremath{\varphi} x \ensuremath{\wedge}\textsuperscript{m} Ess \ensuremath{\psi} x \ensuremath{\rightarrow}\textsuperscript{m} \ensuremath{\Box}(\ensuremath{\forall}\textsuperscript{m} y, \ensuremath{\varphi} y \ensuremath{\leftrightarrow}\textsuperscript{m} \ensuremath{\psi} y)\ensuremath{\rfloor} :=
  fun _ \ensuremath{\langle}\ensuremath{\langle}h\ensuremath{\varphi}x, h\ensuremath{\varphi}\ensuremath{\rangle}, \ensuremath{\langle}h\ensuremath{\psi}x, h\ensuremath{\psi}\ensuremath{\rangle}\ensuremath{\rangle} v hv y => \ensuremath{\langle}h\ensuremath{\varphi} \ensuremath{\psi} h\ensuremath{\psi}x v hv y, h\ensuremath{\psi} \ensuremath{\varphi} h\ensuremath{\varphi}x v hv y\ensuremath{\rangle}

-- UniqueEss2 : \ensuremath{\lfloor}Ess \ensuremath{\varphi} x \ensuremath{\wedge}\textsuperscript{m} Ess \ensuremath{\psi} x \ensuremath{\rightarrow}\textsuperscript{m} \ensuremath{\Box}(\ensuremath{\varphi} \ensuremath{\equiv}\textsuperscript{m} \ensuremath{\psi})\ensuremath{\rfloor}   -- `nitpick[card i=2]`: countermodel found
example (\ensuremath{\varphi} \ensuremath{\psi} : e \ensuremath{\rightarrow} \ensuremath{\sigma}) (x : e) : \ensuremath{\lfloor}Ess \ensuremath{\varphi} x \ensuremath{\wedge}\textsuperscript{m} Ess \ensuremath{\psi} x \ensuremath{\rightarrow}\textsuperscript{m} \ensuremath{\Box}(\ensuremath{\varphi} \ensuremath{\equiv}\textsuperscript{m} \ensuremath{\psi})\ensuremath{\rfloor} := by countermodel

theorem UniqueEss3 (\ensuremath{\varphi} : e \ensuremath{\rightarrow} \ensuremath{\sigma}) (x : e) : \ensuremath{\lfloor}Ess \ensuremath{\varphi} x \ensuremath{\rightarrow}\textsuperscript{m} \ensuremath{\Box}(\ensuremath{\forall}\textsuperscript{m} y, \ensuremath{\varphi} y \ensuremath{\rightarrow}\textsuperscript{m} y \ensuremath{\equiv}\textsuperscript{m} x)\ensuremath{\rfloor} :=
  fun _ \ensuremath{\langle}_, h\ensuremath{\rangle} => h (fun z => z \ensuremath{\equiv}\textsuperscript{m} x) (fun _ hp => hp)

theorem Monotheism (x y : e) : \ensuremath{\lfloor}G x \ensuremath{\wedge}\textsuperscript{m} G y \ensuremath{\rightarrow}\textsuperscript{m} x \ensuremath{\equiv}\textsuperscript{m} y\ensuremath{\rfloor} :=
  fun _ \ensuremath{\langle}hx, hy\ensuremath{\rangle} Q hQ => hy Q (PosOfGod hx hQ)

/-- Everything true of every God-like being at every successor world is positive: by `Coro`
some successor holds a God-like being, which has every positive property, and positivity is
necessary (`A4`).  No frame condition is used. -/
theorem Pos (\ensuremath{\chi} : e \ensuremath{\rightarrow} \ensuremath{\sigma}) (w : i) (h : \ensuremath{\forall} v, w \textsf{\bfseries r} v \ensuremath{\rightarrow} \ensuremath{\forall} x, G x v \ensuremath{\rightarrow} \ensuremath{\chi} x v) : P \ensuremath{\chi} w :=
  Classical.byContradiction fun hn =>
    let \ensuremath{\langle}v, hv, x, hG\ensuremath{\rangle} := Coro w
    hG ~\textsuperscript{m}\ensuremath{\chi} (A4 ~\textsuperscript{m}\ensuremath{\chi} w ((A1 \ensuremath{\chi} w).1 hn) v hv) (h v hv x hG)

theorem PisFilterP : \ensuremath{\lfloor}FilterP P\ensuremath{\rfloor} := fun w =>
  \ensuremath{\langle}\ensuremath{\langle}\ensuremath{\langle}(PosProps w).1, (NegProps w).1\ensuremath{\rangle},
    -- the inclusion holds at `w`; `MC` carries it to the successor worlds, then `A2` applies
    fun \ensuremath{\varphi} \ensuremath{\psi} \ensuremath{\langle}h\ensuremath{\varphi}, hsub\ensuremath{\rangle} => A2 \ensuremath{\varphi} \ensuremath{\psi} w \ensuremath{\langle}h\ensuremath{\varphi}, MC (\ensuremath{\forall}\textsuperscript{m} x, \ensuremath{\varphi} x \ensuremath{\rightarrow}\textsuperscript{m} \ensuremath{\psi} x) w hsub\ensuremath{\rangle}\ensuremath{\rangle},
   fun \ensuremath{\varphi} \ensuremath{\psi} \ensuremath{\langle}h\ensuremath{\varphi}, h\ensuremath{\psi}\ensuremath{\rangle} => Pos _ w (fun v hv _ hG => \ensuremath{\langle}hG \ensuremath{\varphi} (A4 \ensuremath{\varphi} w h\ensuremath{\varphi} v hv), hG \ensuremath{\psi} (A4 \ensuremath{\psi} w h\ensuremath{\psi} v hv)\ensuremath{\rangle})\ensuremath{\rangle}

theorem PisUFilterP : \ensuremath{\lfloor}UFilterP P\ensuremath{\rfloor} := fun w =>
  \ensuremath{\langle}PisFilterP w, fun \ensuremath{\varphi} => (Classical.em (P \ensuremath{\varphi} w)).imp id (A1 \ensuremath{\varphi} w).1\ensuremath{\rangle}

-- `lemma True nitpick[satisfy,card=1,eval="\ensuremath{\lfloor}P (\ensuremath{\lambda}x.\ensuremath{\bot})\ensuremath{\rfloor}"] oops`
--   One model found of cardinality one (consistency check)
\end{Verbatim}
\subsubsection{EvilDerivable.lean (Figure 20 of \texorpdfstring{\cite{J75}}{the Notes})}
The necessary existence of an Evil-like entity proved from (controversially) modified assumptions. By rejecting G\"odel's assumptions and instead postulating corresponding negative versions of them, the necessary existence of Evil becomes derivable.  The non-positive properties of this Evil-like entity are however identical to the positive properties of G\"odel's God-like entity.
\begin{Verbatim}[commandchars=\\\{\},fontsize=\scriptsize,xleftmargin=1em]
import Notes.HOMLinHOL
import Notes.ModalFilter
\end{Verbatim}

\begin{Verbatim}[commandchars=\\\{\},fontsize=\scriptsize,xleftmargin=1em]
/-- Positive property -/
axiom P : (e \ensuremath{\rightarrow} \ensuremath{\sigma}) \ensuremath{\rightarrow} \ensuremath{\sigma}

/-- An entity possessing all non-positive properties -/
def Evil (x : e) : \ensuremath{\sigma} := \ensuremath{\forall}\textsuperscript{m} (\ensuremath{\varphi} : e \ensuremath{\rightarrow} \ensuremath{\sigma}), \ensuremath{\neg}\textsuperscript{m}(P \ensuremath{\varphi}) \ensuremath{\rightarrow}\textsuperscript{m} \ensuremath{\varphi} x

/-- `Ess \ensuremath{\varphi} x`: \ensuremath{\varphi} is an essence of x -/
def Ess (\ensuremath{\varphi} : e \ensuremath{\rightarrow} \ensuremath{\sigma}) (x : e) : \ensuremath{\sigma} := \ensuremath{\varphi} x \ensuremath{\wedge}\textsuperscript{m} (\ensuremath{\forall}\textsuperscript{m} (\ensuremath{\psi} : e \ensuremath{\rightarrow} \ensuremath{\sigma}), \ensuremath{\psi} x \ensuremath{\rightarrow}\textsuperscript{m} \ensuremath{\Box}(\ensuremath{\forall}\textsuperscript{E} y, \ensuremath{\varphi} y \ensuremath{\rightarrow}\textsuperscript{m} \ensuremath{\psi} y))

/-- Necessary existence -/
def E (x : e) : \ensuremath{\sigma} := \ensuremath{\forall}\textsuperscript{m} (\ensuremath{\varphi} : e \ensuremath{\rightarrow} \ensuremath{\sigma}), Ess \ensuremath{\varphi} x \ensuremath{\rightarrow}\textsuperscript{m} \ensuremath{\Box}(\ensuremath{\exists}\textsuperscript{E} x, \ensuremath{\varphi} x)

axiom A1 (\ensuremath{\varphi} : e \ensuremath{\rightarrow} \ensuremath{\sigma}) : \ensuremath{\lfloor}\ensuremath{\neg}\textsuperscript{m}(P \ensuremath{\varphi}) \ensuremath{\leftrightarrow}\textsuperscript{m} P ~\textsuperscript{m}\ensuremath{\varphi}\ensuremath{\rfloor}

axiom A2 (\ensuremath{\varphi} \ensuremath{\psi} : e \ensuremath{\rightarrow} \ensuremath{\sigma}) : \ensuremath{\lfloor}\ensuremath{\neg}\textsuperscript{m}(P \ensuremath{\varphi}) \ensuremath{\wedge}\textsuperscript{m} \ensuremath{\Box}(\ensuremath{\forall}\textsuperscript{E} y, \ensuremath{\varphi} y \ensuremath{\rightarrow}\textsuperscript{m} \ensuremath{\psi} y) \ensuremath{\rightarrow}\textsuperscript{m} \ensuremath{\neg}\textsuperscript{m}(P \ensuremath{\psi})\ensuremath{\rfloor}

axiom A4 : \ensuremath{\lfloor}\ensuremath{\neg}\textsuperscript{m}(P Evil)\ensuremath{\rfloor}

axiom A3 (\ensuremath{\varphi} : e \ensuremath{\rightarrow} \ensuremath{\sigma}) : \ensuremath{\lfloor}\ensuremath{\neg}\textsuperscript{m}(P \ensuremath{\varphi}) \ensuremath{\rightarrow}\textsuperscript{m} \ensuremath{\Box}(\ensuremath{\neg}\textsuperscript{m}(P \ensuremath{\varphi}))\ensuremath{\rfloor}

axiom A5 : \ensuremath{\lfloor}\ensuremath{\neg}\textsuperscript{m}(P E)\ensuremath{\rfloor}

-- `lemma True nitpick[satisfy,card i=1,eval="\ensuremath{\lfloor}P (\ensuremath{\lambda}x.\ensuremath{\bot})\ensuremath{\rfloor}",eval="\ensuremath{\lfloor}P (\ensuremath{\lambda}x.\ensuremath{\top})\ensuremath{\rfloor}"] oops`
--   Model found (consistency check)

theorem T1 (\ensuremath{\varphi} : e \ensuremath{\rightarrow} \ensuremath{\sigma}) : \ensuremath{\lfloor}\ensuremath{\neg}\textsuperscript{m}(P \ensuremath{\varphi}) \ensuremath{\rightarrow}\textsuperscript{m} \ensuremath{\Diamond}(\ensuremath{\exists}\textsuperscript{E} x, \ensuremath{\varphi} x)\ensuremath{\rfloor} := fun w h =>
  Classical.byContradiction fun hn =>
    A2 \ensuremath{\varphi} ~\textsuperscript{m}\ensuremath{\varphi} w \ensuremath{\langle}h, fun v hv y hy h\ensuremath{\varphi} _ => hn \ensuremath{\langle}v, hv, y, hy, h\ensuremath{\varphi}\ensuremath{\rangle}\ensuremath{\rangle} ((A1 \ensuremath{\varphi} w).1 h)

theorem T2 : \ensuremath{\lfloor}\ensuremath{\Diamond}(\ensuremath{\exists}\textsuperscript{E} x, Evil x)\ensuremath{\rfloor} := fun w => T1 Evil w (A4 w)

/-- A property exemplified by an Evil-like being is non-positive (a consequence of `A1`). -/
theorem NegOfEvil \{x : e\} \{\ensuremath{\psi} : e \ensuremath{\rightarrow} \ensuremath{\sigma}\} \{w : i\} (hE : Evil x w) (h : \ensuremath{\psi} x w) : (\ensuremath{\neg}\textsuperscript{m}(P \ensuremath{\psi})) w :=
  (A1 \ensuremath{\psi} w).2 (Classical.byContradiction fun hn => hE ~\textsuperscript{m}\ensuremath{\psi} hn h)

theorem T3 (x : e) : \ensuremath{\lfloor}Evil x \ensuremath{\rightarrow}\textsuperscript{m} Ess Evil x\ensuremath{\rfloor} := fun w hE =>
  \ensuremath{\langle}hE, fun \ensuremath{\psi} h\ensuremath{\psi} v hv _y _ hEy => hEy \ensuremath{\psi} (A3 \ensuremath{\psi} w (NegOfEvil hE h\ensuremath{\psi}) v hv)\ensuremath{\rangle}

theorem T4 : \ensuremath{\lfloor}\ensuremath{\Diamond}(\ensuremath{\exists}\textsuperscript{E} x, Evil x) \ensuremath{\rightarrow}\textsuperscript{m} \ensuremath{\Box}(\ensuremath{\exists}\textsuperscript{E} y, Evil y)\ensuremath{\rfloor} := by
  have h1 : \ensuremath{\lfloor}(\ensuremath{\exists}\textsuperscript{E} x, Evil x) \ensuremath{\rightarrow}\textsuperscript{m} \ensuremath{\Box}(\ensuremath{\exists}\textsuperscript{E} y, Evil y)\ensuremath{\rfloor} :=
    fun _ \ensuremath{\langle}x, _, hE\ensuremath{\rangle} => hE E (A5 _) Evil (T3 x _ hE)
  intro w \ensuremath{\langle}v, hv, hex\ensuremath{\rangle}
  exact h1 w (h1 v hex w (Rsymm w v hv))

theorem T5 : \ensuremath{\lfloor}\ensuremath{\Box}(\ensuremath{\exists}\textsuperscript{E} x, Evil x)\ensuremath{\rfloor} := fun w => T4 w (T2 w)

/-- Modal collapse.  The AFP sources leave it unreplayed (`oops`), although sledgehammer
reports a proof from `A1`, `A3`, `T5`, `Evil_def` and `Rsymm`.  It is the modal-collapse proof
of the God-like variants with `Evil` in place of `G`: an actual Evil-like being exists at `w`
(by `T5` and symmetry) and at every successor `v` (by `T5`); the former has the essence `Evil`
(`T3`), whose necessary implication of `\ensuremath{\varphi}` then delivers `\ensuremath{\varphi}` at `v`. -/
theorem MC (\ensuremath{\varphi} : \ensuremath{\sigma}) : \ensuremath{\lfloor}\ensuremath{\varphi} \ensuremath{\rightarrow}\textsuperscript{m} \ensuremath{\Box}\ensuremath{\varphi}\ensuremath{\rfloor} := by
  intro w h\ensuremath{\varphi} v hv
  have \ensuremath{\langle}x, _, hEx\ensuremath{\rangle} := T5 v w (Rsymm w v hv)
  have \ensuremath{\langle}z, hzv, hEz\ensuremath{\rangle} := T5 w v hv
  exact (T3 x w hEx).2 (fun _ => \ensuremath{\varphi}) h\ensuremath{\varphi} v hv z hzv hEz

theorem PosProps : \ensuremath{\lfloor}P (fun _ : e => \ensuremath{\bot}\textsuperscript{m}) \ensuremath{\wedge}\textsuperscript{m} P (fun x : e => x \ensuremath{\neq}\textsuperscript{m} x)\ensuremath{\rfloor} := fun w =>
  \ensuremath{\langle}Classical.byContradiction fun hn => let \ensuremath{\langle}_, _, _, _, h\ensuremath{\rangle} := T1 _ w hn; h,
   Classical.byContradiction fun hn => let \ensuremath{\langle}_, _, _, _, h\ensuremath{\rangle} := T1 _ w hn; h rfl\ensuremath{\rangle}

theorem NegProps : \ensuremath{\lfloor}\ensuremath{\neg}\textsuperscript{m}P (fun _ : e => \ensuremath{\top}\textsuperscript{m}) \ensuremath{\wedge}\textsuperscript{m} \ensuremath{\neg}\textsuperscript{m}P (fun x : e => x =\textsuperscript{m} x)\ensuremath{\rfloor} := fun w =>
  \ensuremath{\langle}(A1 _ w).2 (Classical.byContradiction fun hn => let \ensuremath{\langle}_, _, _, _, h\ensuremath{\rangle} := T1 _ w hn; h trivial),
   (A1 _ w).2 (Classical.byContradiction fun hn => let \ensuremath{\langle}_, _, _, _, h\ensuremath{\rangle} := T1 _ w hn; h rfl)\ensuremath{\rangle}

theorem UniqueEss1 (\ensuremath{\varphi} \ensuremath{\psi} : e \ensuremath{\rightarrow} \ensuremath{\sigma}) (x : e) : \ensuremath{\lfloor}Ess \ensuremath{\varphi} x \ensuremath{\wedge}\textsuperscript{m} Ess \ensuremath{\psi} x \ensuremath{\rightarrow}\textsuperscript{m} \ensuremath{\Box}(\ensuremath{\forall}\textsuperscript{E} y, \ensuremath{\varphi} y \ensuremath{\leftrightarrow}\textsuperscript{m} \ensuremath{\psi} y)\ensuremath{\rfloor} :=
  fun _ \ensuremath{\langle}\ensuremath{\langle}h\ensuremath{\varphi}x, h\ensuremath{\varphi}\ensuremath{\rangle}, \ensuremath{\langle}h\ensuremath{\psi}x, h\ensuremath{\psi}\ensuremath{\rangle}\ensuremath{\rangle} v hv y hy => \ensuremath{\langle}h\ensuremath{\varphi} \ensuremath{\psi} h\ensuremath{\psi}x v hv y hy, h\ensuremath{\psi} \ensuremath{\varphi} h\ensuremath{\varphi}x v hv y hy\ensuremath{\rangle}

-- UniqueEss2 : \ensuremath{\lfloor}Ess \ensuremath{\varphi} x \ensuremath{\wedge}\textsuperscript{m} Ess \ensuremath{\psi} x \ensuremath{\rightarrow}\textsuperscript{m} \ensuremath{\Box}(\ensuremath{\varphi} =\textsuperscript{m} \ensuremath{\psi})\ensuremath{\rfloor}   -- `nitpick[card i=2]`: countermodel found
example (\ensuremath{\varphi} \ensuremath{\psi} : e \ensuremath{\rightarrow} \ensuremath{\sigma}) (x : e) : \ensuremath{\lfloor}Ess \ensuremath{\varphi} x \ensuremath{\wedge}\textsuperscript{m} Ess \ensuremath{\psi} x \ensuremath{\rightarrow}\textsuperscript{m} \ensuremath{\Box}(\ensuremath{\varphi} =\textsuperscript{m} \ensuremath{\psi})\ensuremath{\rfloor} := by countermodel

theorem Monoevilism (x y : e) : \ensuremath{\lfloor}Evil x \ensuremath{\wedge}\textsuperscript{m} Evil y \ensuremath{\rightarrow}\textsuperscript{m} x \ensuremath{\equiv}\textsuperscript{m} y\ensuremath{\rfloor} :=
  fun _ \ensuremath{\langle}hx, hy\ensuremath{\rangle} Q hQ => hy Q (NegOfEvil hx hQ)

/-- Everything true of every actual Evil-like being at every successor world is non-positive:
by `T2` some successor holds an actual Evil-like being, which has every non-positive property,
and non-positivity is necessary (`A3`).  No frame condition is used. -/
theorem NegP (\ensuremath{\chi} : e \ensuremath{\rightarrow} \ensuremath{\sigma}) (w : i) (h : \ensuremath{\forall} v, w \textsf{\bfseries r} v \ensuremath{\rightarrow} \ensuremath{\forall} x, x @\textsuperscript{m} v \ensuremath{\rightarrow} Evil x v \ensuremath{\rightarrow} \ensuremath{\chi} x v) :
    (\ensuremath{\neg}\textsuperscript{m}(P \ensuremath{\chi})) w := fun hp =>
  let \ensuremath{\langle}v, hv, x, hxa, hE\ensuremath{\rangle} := T2 w
  hE ~\textsuperscript{m}\ensuremath{\chi} (A3 ~\textsuperscript{m}\ensuremath{\chi} w (fun hn => (A1 \ensuremath{\chi} w).2 hn hp) v hv) (h v hv x hxa hE)

theorem NisFilter : \ensuremath{\lfloor}Filter (fun \ensuremath{\varphi} => \ensuremath{\neg}\textsuperscript{m}(P \ensuremath{\varphi}))\ensuremath{\rfloor} := fun w =>
  \ensuremath{\langle}\ensuremath{\langle}\ensuremath{\langle}(NegProps w).1, fun hn => hn (PosProps w).1\ensuremath{\rangle},
    -- the inclusion holds at `w`; `MC` carries it to the successor worlds, then `A2` applies
    fun \ensuremath{\varphi} \ensuremath{\psi} \ensuremath{\langle}h\ensuremath{\varphi}, hsub\ensuremath{\rangle} => A2 \ensuremath{\varphi} \ensuremath{\psi} w \ensuremath{\langle}h\ensuremath{\varphi}, MC (\ensuremath{\forall}\textsuperscript{E} x, \ensuremath{\varphi} x \ensuremath{\rightarrow}\textsuperscript{m} \ensuremath{\psi} x) w hsub\ensuremath{\rangle}\ensuremath{\rangle},
   fun \ensuremath{\varphi} \ensuremath{\psi} \ensuremath{\langle}h\ensuremath{\varphi}, h\ensuremath{\psi}\ensuremath{\rangle} => NegP _ w (fun v hv _ _ hE => \ensuremath{\langle}hE \ensuremath{\varphi} (A3 \ensuremath{\varphi} w h\ensuremath{\varphi} v hv), hE \ensuremath{\psi} (A3 \ensuremath{\psi} w h\ensuremath{\psi} v hv)\ensuremath{\rangle})\ensuremath{\rangle}

theorem NisUFilter : \ensuremath{\lfloor}UFilter (fun \ensuremath{\varphi} => \ensuremath{\neg}\textsuperscript{m}(P \ensuremath{\varphi}))\ensuremath{\rfloor} := fun w =>
  \ensuremath{\langle}NisFilter w, fun \ensuremath{\varphi} => (Classical.em ((\ensuremath{\neg}\textsuperscript{m}(P \ensuremath{\varphi})) w)).imp id
    (fun hnn hp => (A1 \ensuremath{\varphi} w).2 hp (Classical.byContradiction hnn))\ensuremath{\rangle}
\end{Verbatim}
\subsection{Further Appendices}
\subsubsection{GoedelVariantHOML1AndersonQuant.lean}
The same as GoedelVariantHOML1, but now for a mixed use of actualist and possibilist quantifiers for entities; cf. Footnote 20 in \cite{J75}.
\begin{Verbatim}[commandchars=\\\{\},fontsize=\scriptsize,xleftmargin=1em]
import Notes.HOMLinHOL
\end{Verbatim}

\begin{Verbatim}[commandchars=\\\{\},fontsize=\scriptsize,xleftmargin=1em]
/-- Positive property -/
axiom P : (e \ensuremath{\rightarrow} \ensuremath{\sigma}) \ensuremath{\rightarrow} \ensuremath{\sigma}

axiom Ax1 (\ensuremath{\varphi} \ensuremath{\psi} : e \ensuremath{\rightarrow} \ensuremath{\sigma}) : \ensuremath{\lfloor}P \ensuremath{\varphi} \ensuremath{\wedge}\textsuperscript{m} P \ensuremath{\psi} \ensuremath{\rightarrow}\textsuperscript{m} P (\ensuremath{\varphi} \ensuremath{\centerdot}\textsuperscript{m} \ensuremath{\psi})\ensuremath{\rfloor}

axiom Ax2a (\ensuremath{\varphi} : e \ensuremath{\rightarrow} \ensuremath{\sigma}) : \ensuremath{\lfloor}P \ensuremath{\varphi} \ensuremath{\vee}\textsuperscript{e} P ~\textsuperscript{m}\ensuremath{\varphi}\ensuremath{\rfloor}

/-- Auxiliary reformulation of `Ax2a` (Lean has no `sledgehammer`). -/
theorem Ax2a' (\ensuremath{\varphi} : e \ensuremath{\rightarrow} \ensuremath{\sigma}) : \ensuremath{\lfloor}\ensuremath{\neg}\textsuperscript{m}(P \ensuremath{\varphi}) \ensuremath{\leftrightarrow}\textsuperscript{m} P ~\textsuperscript{m}\ensuremath{\varphi}\ensuremath{\rfloor} :=
  fun w => \ensuremath{\langle}(Ax2a \ensuremath{\varphi} w).1.resolve_left, fun hp hq => (Ax2a \ensuremath{\varphi} w).2 \ensuremath{\langle}hq, hp\ensuremath{\rangle}\ensuremath{\rangle}

/-- God-like -/
def G (x : e) : \ensuremath{\sigma} := \ensuremath{\forall}\textsuperscript{m} (\ensuremath{\varphi} : e \ensuremath{\rightarrow} \ensuremath{\sigma}), P \ensuremath{\varphi} \ensuremath{\rightarrow}\textsuperscript{m} \ensuremath{\varphi} x

/-- Necessary property inclusion -/
@[simp, grind] def PInc (\ensuremath{\varphi} \ensuremath{\psi} : e \ensuremath{\rightarrow} \ensuremath{\sigma}) : \ensuremath{\sigma} := \ensuremath{\Box}(\ensuremath{\forall}\textsuperscript{m} (y : e), \ensuremath{\varphi} y \ensuremath{\rightarrow}\textsuperscript{m} \ensuremath{\psi} y)
infixr:48 " \ensuremath{\supset}\textsuperscript{N} " => PInc

/-- `Ess \ensuremath{\varphi} x`: \ensuremath{\varphi} is an essence of x -/
def Ess (\ensuremath{\varphi} : e \ensuremath{\rightarrow} \ensuremath{\sigma}) (x : e) : \ensuremath{\sigma} := \ensuremath{\forall}\textsuperscript{m} (\ensuremath{\psi} : e \ensuremath{\rightarrow} \ensuremath{\sigma}), \ensuremath{\psi} x \ensuremath{\rightarrow}\textsuperscript{m} (\ensuremath{\varphi} \ensuremath{\supset}\textsuperscript{N} \ensuremath{\psi})

axiom Ax2b (\ensuremath{\varphi} : e \ensuremath{\rightarrow} \ensuremath{\sigma}) : \ensuremath{\lfloor}P \ensuremath{\varphi} \ensuremath{\rightarrow}\textsuperscript{m} \ensuremath{\Box} P \ensuremath{\varphi}\ensuremath{\rfloor}

theorem Ax2b' (\ensuremath{\varphi} : e \ensuremath{\rightarrow} \ensuremath{\sigma}) : \ensuremath{\lfloor}\ensuremath{\neg}\textsuperscript{m}(P \ensuremath{\varphi}) \ensuremath{\rightarrow}\textsuperscript{m} \ensuremath{\Box}(\ensuremath{\neg}\textsuperscript{m}(P \ensuremath{\varphi}))\ensuremath{\rfloor} :=
  fun w hn v hv => (Ax2a' \ensuremath{\varphi} v).2 (Ax2b ~\textsuperscript{m}\ensuremath{\varphi} w ((Ax2a' \ensuremath{\varphi} w).1 hn) v hv)

/-- A property exemplified by a God-like being is positive (a consequence of `Ax2a`). -/
theorem PosOfGod \{x : e\} \{\ensuremath{\psi} : e \ensuremath{\rightarrow} \ensuremath{\sigma}\} \{w : i\} (hG : G x w) (h : \ensuremath{\psi} x w) : P \ensuremath{\psi} w :=
  Classical.byContradiction fun hn => hG ~\textsuperscript{m}\ensuremath{\psi} ((Ax2a' \ensuremath{\psi} w).1 hn) h

theorem Th1 (x : e) : \ensuremath{\lfloor}G x \ensuremath{\rightarrow}\textsuperscript{m} Ess G x\ensuremath{\rfloor} :=
  fun w hG \ensuremath{\psi} h\ensuremath{\psi} v hv _y hGy => hGy \ensuremath{\psi} (Ax2b \ensuremath{\psi} w (PosOfGod hG h\ensuremath{\psi}) v hv)

/-- Necessary existence -/
def E (x : e) : \ensuremath{\sigma} := \ensuremath{\forall}\textsuperscript{m} (\ensuremath{\varphi} : e \ensuremath{\rightarrow} \ensuremath{\sigma}), Ess \ensuremath{\varphi} x \ensuremath{\rightarrow}\textsuperscript{m} \ensuremath{\Box}(\ensuremath{\exists}\textsuperscript{E} x, \ensuremath{\varphi} x)

axiom Ax3 : \ensuremath{\lfloor}P E\ensuremath{\rfloor}

theorem Th2 (x : e) : \ensuremath{\lfloor}G x \ensuremath{\rightarrow}\textsuperscript{m} \ensuremath{\Box}(\ensuremath{\exists}\textsuperscript{E} y, G y)\ensuremath{\rfloor} := fun w hG => hG E (Ax3 w) G (Th1 x w hG)

theorem Th3 : \ensuremath{\lfloor}\ensuremath{\Diamond}(\ensuremath{\exists}\textsuperscript{E} x, G x) \ensuremath{\rightarrow}\textsuperscript{m} \ensuremath{\Box}(\ensuremath{\exists}\textsuperscript{E} y, G y)\ensuremath{\rfloor} := by
  have h1 : \ensuremath{\lfloor}(\ensuremath{\exists}\textsuperscript{E} x, G x) \ensuremath{\rightarrow}\textsuperscript{m} \ensuremath{\Box}(\ensuremath{\exists}\textsuperscript{E} y, G y)\ensuremath{\rfloor} := fun _ \ensuremath{\langle}x, _, hG\ensuremath{\rangle} => Th2 x _ hG
  intro w \ensuremath{\langle}v, hv, hex\ensuremath{\rangle}
  exact h1 w (h1 v hex w (Rsymm w v hv))               -- only symmetry of `\textsf{\bfseries r}` is needed

axiom Ax4 (\ensuremath{\varphi} \ensuremath{\psi} : e \ensuremath{\rightarrow} \ensuremath{\sigma}) : \ensuremath{\lfloor}P \ensuremath{\varphi} \ensuremath{\wedge}\textsuperscript{m} (\ensuremath{\varphi} \ensuremath{\supset}\textsuperscript{N} \ensuremath{\psi}) \ensuremath{\rightarrow}\textsuperscript{m} P \ensuremath{\psi}\ensuremath{\rfloor}

-- `lemma True nitpick[satisfy,expect=unknown] oops`
--   No model found (consistency check)

theorem EmptyEssL (x : e) : \ensuremath{\lfloor}Ess (fun _ : e => \ensuremath{\bot}\textsuperscript{m}) x\ensuremath{\rfloor} := fun _ _ _ _ _ _ h => h.elim

theorem Inconsistency : False := by
  have h1 : \ensuremath{\lfloor}\ensuremath{\neg}\textsuperscript{m}(P (fun _ : e => \ensuremath{\bot}\textsuperscript{m}))\ensuremath{\rfloor} := fun w hp =>
    (Ax2a _ w).2 \ensuremath{\langle}hp, Ax4 (fun _ : e => \ensuremath{\bot}\textsuperscript{m}) _ w \ensuremath{\langle}hp, fun _ _ _ h => h.elim\ensuremath{\rangle}\ensuremath{\rangle}
  have h2 : \ensuremath{\lfloor}P (fun x : e => Ess (fun _ : e => \ensuremath{\bot}\textsuperscript{m}) x \ensuremath{\rightarrow}\textsuperscript{m} \ensuremath{\Box}(\ensuremath{\exists}\textsuperscript{E} z, (fun _ : e => \ensuremath{\bot}\textsuperscript{m}) z))\ensuremath{\rfloor} :=
    fun w => Ax4 E _ w \ensuremath{\langle}Ax3 w, fun _ _ _ hE => hE (fun _ => \ensuremath{\bot}\textsuperscript{m})\ensuremath{\rangle}
  have h3 : \ensuremath{\lfloor}P (fun _ : e => \ensuremath{\Box}(\ensuremath{\exists}\textsuperscript{E} z, (fun _ : e => \ensuremath{\bot}\textsuperscript{m}) z))\ensuremath{\rfloor} :=
    fun w => Ax4 _ _ w \ensuremath{\langle}h2 w, fun v _ y h => h (EmptyEssL y v)\ensuremath{\rangle}
  have h4 : \ensuremath{\lfloor}P (fun _ : e => \ensuremath{\Box} \ensuremath{\bot}\textsuperscript{m})\ensuremath{\rfloor} := by
    have heq : (fun _ : e => \ensuremath{\Box}(\ensuremath{\exists}\textsuperscript{E} z, (fun _ : e => \ensuremath{\bot}\textsuperscript{m}) z)) = (fun _ : e => \ensuremath{\Box} \ensuremath{\bot}\textsuperscript{m}) := by
      -- an explicit term rather than `simp`: `simp` would draw in `eNonempty`, which the
      -- argument does not need (`#print axioms`)
      funext _ w; exact propext \ensuremath{\langle}fun h v hv => match h v hv with | \ensuremath{\langle}_, _, hf\ensuremath{\rangle} => hf, fun h v hv =>
          (h v hv).elim\ensuremath{\rangle}
    exact heq \ensuremath{\blacktriangleright} h3
  -- Isabelle discharges this with `smt` from `4 Ax2a Ax4`; no frame condition is needed.
  have h5 : \ensuremath{\lfloor}P (fun _ : e => \ensuremath{\bot}\textsuperscript{m})\ensuremath{\rfloor} := by
    intro w
    by_cases hdead : \ensuremath{\exists} v : i, \ensuremath{\forall} u, \ensuremath{\neg} v \textsf{\bfseries r} u
    \textperiodcentered{} -- some world is a dead end: there `Ax4`'s premise is vacuous, so `Ax2a` is violated
      have \ensuremath{\langle}v, hv\ensuremath{\rangle} := hdead
      exact ((Ax2a (fun _ : e => \ensuremath{\Box} \ensuremath{\bot}\textsuperscript{m}) v).2
        \ensuremath{\langle}h4 v, Ax4 _ _ v \ensuremath{\langle}h4 v, fun u hu => absurd hu (hv u)\ensuremath{\rangle}\ensuremath{\rangle}).elim
    \textperiodcentered{} -- otherwise every world has a successor, so `\ensuremath{\Box}\ensuremath{\bot}` is false and the properties coincide
      have ser : \ensuremath{\forall} v : i, \ensuremath{\exists} u, v \textsf{\bfseries r} u := fun v =>
        Classical.byContradiction fun h => hdead \ensuremath{\langle}v, fun u hu => h \ensuremath{\langle}u, hu\ensuremath{\rangle}\ensuremath{\rangle}
      have heq : (fun _ : e => \ensuremath{\Box} \ensuremath{\bot}\textsuperscript{m}) = (fun _ : e => \ensuremath{\bot}\textsuperscript{m}) := by
        funext _ v
        have \ensuremath{\langle}u, hu\ensuremath{\rangle} := ser v
        exact propext \ensuremath{\langle}fun h => h u hu, fun h => h.elim\ensuremath{\rangle}
      exact heq \ensuremath{\blacktriangleright} h4 w
  have w : i := Classical.ofNonempty
  exact h1 w (h5 w)
\end{Verbatim}
\subsubsection{GoedelVariantHOML2AndersonQuant.lean}
The same as GoedelVariantHOML2, but now for a mixed use of actualist and possibilist quantifiers for entities; cf. Footnote 20 in \cite{J75}.
\begin{Verbatim}[commandchars=\\\{\},fontsize=\scriptsize,xleftmargin=1em]
import Notes.HOMLinHOL
import Notes.ModalFilter
\end{Verbatim}

\begin{Verbatim}[commandchars=\\\{\},fontsize=\scriptsize,xleftmargin=1em]
/-- Positive property -/
axiom P : (e \ensuremath{\rightarrow} \ensuremath{\sigma}) \ensuremath{\rightarrow} \ensuremath{\sigma}

axiom Ax1 (\ensuremath{\varphi} \ensuremath{\psi} : e \ensuremath{\rightarrow} \ensuremath{\sigma}) : \ensuremath{\lfloor}P \ensuremath{\varphi} \ensuremath{\wedge}\textsuperscript{m} P \ensuremath{\psi} \ensuremath{\rightarrow}\textsuperscript{m} P (\ensuremath{\varphi} \ensuremath{\centerdot}\textsuperscript{m} \ensuremath{\psi})\ensuremath{\rfloor}

axiom Ax2a (\ensuremath{\varphi} : e \ensuremath{\rightarrow} \ensuremath{\sigma}) : \ensuremath{\lfloor}P \ensuremath{\varphi} \ensuremath{\vee}\textsuperscript{e} P ~\textsuperscript{m}\ensuremath{\varphi}\ensuremath{\rfloor}

/-- Auxiliary reformulation of `Ax2a` (Lean has no `sledgehammer`). -/
theorem Ax2a' (\ensuremath{\varphi} : e \ensuremath{\rightarrow} \ensuremath{\sigma}) : \ensuremath{\lfloor}\ensuremath{\neg}\textsuperscript{m}(P \ensuremath{\varphi}) \ensuremath{\leftrightarrow}\textsuperscript{m} P ~\textsuperscript{m}\ensuremath{\varphi}\ensuremath{\rfloor} :=
  fun w => \ensuremath{\langle}(Ax2a \ensuremath{\varphi} w).1.resolve_left, fun hp hq => (Ax2a \ensuremath{\varphi} w).2 \ensuremath{\langle}hq, hp\ensuremath{\rangle}\ensuremath{\rangle}

/-- God-like -/
def G (x : e) : \ensuremath{\sigma} := \ensuremath{\forall}\textsuperscript{m} (\ensuremath{\varphi} : e \ensuremath{\rightarrow} \ensuremath{\sigma}), P \ensuremath{\varphi} \ensuremath{\rightarrow}\textsuperscript{m} \ensuremath{\varphi} x

/-- Necessary property inclusion -/
@[simp, grind] def PInc (\ensuremath{\varphi} \ensuremath{\psi} : e \ensuremath{\rightarrow} \ensuremath{\sigma}) : \ensuremath{\sigma} := \ensuremath{\Box}(\ensuremath{\forall}\textsuperscript{m} (y : e), \ensuremath{\varphi} y \ensuremath{\rightarrow}\textsuperscript{m} \ensuremath{\psi} y)
infixr:48 " \ensuremath{\supset}\textsuperscript{N} " => PInc

/-- `Ess \ensuremath{\varphi} x`: \ensuremath{\varphi} is an essence of x (modified: \ensuremath{\varphi} must be exemplified by x) -/
def Ess (\ensuremath{\varphi} : e \ensuremath{\rightarrow} \ensuremath{\sigma}) (x : e) : \ensuremath{\sigma} := \ensuremath{\varphi} x \ensuremath{\wedge}\textsuperscript{m} (\ensuremath{\forall}\textsuperscript{m} (\ensuremath{\psi} : e \ensuremath{\rightarrow} \ensuremath{\sigma}), \ensuremath{\psi} x \ensuremath{\rightarrow}\textsuperscript{m} (\ensuremath{\varphi} \ensuremath{\supset}\textsuperscript{N} \ensuremath{\psi}))

axiom Ax2b (\ensuremath{\varphi} : e \ensuremath{\rightarrow} \ensuremath{\sigma}) : \ensuremath{\lfloor}P \ensuremath{\varphi} \ensuremath{\rightarrow}\textsuperscript{m} \ensuremath{\Box} P \ensuremath{\varphi}\ensuremath{\rfloor}

theorem Ax2b' (\ensuremath{\varphi} : e \ensuremath{\rightarrow} \ensuremath{\sigma}) : \ensuremath{\lfloor}\ensuremath{\neg}\textsuperscript{m}(P \ensuremath{\varphi}) \ensuremath{\rightarrow}\textsuperscript{m} \ensuremath{\Box}(\ensuremath{\neg}\textsuperscript{m}(P \ensuremath{\varphi}))\ensuremath{\rfloor} :=
  fun w hn v hv => (Ax2a' \ensuremath{\varphi} v).2 (Ax2b ~\textsuperscript{m}\ensuremath{\varphi} w ((Ax2a' \ensuremath{\varphi} w).1 hn) v hv)

/-- A property exemplified by a God-like being is positive (a consequence of `Ax2a`). -/
theorem PosOfGod \{x : e\} \{\ensuremath{\psi} : e \ensuremath{\rightarrow} \ensuremath{\sigma}\} \{w : i\} (hG : G x w) (h : \ensuremath{\psi} x w) : P \ensuremath{\psi} w :=
  Classical.byContradiction fun hn => hG ~\textsuperscript{m}\ensuremath{\psi} ((Ax2a' \ensuremath{\psi} w).1 hn) h

theorem Th1 (x : e) : \ensuremath{\lfloor}G x \ensuremath{\rightarrow}\textsuperscript{m} Ess G x\ensuremath{\rfloor} := fun w hG =>
  \ensuremath{\langle}hG, fun \ensuremath{\psi} h\ensuremath{\psi} v hv _y hGy => hGy \ensuremath{\psi} (Ax2b \ensuremath{\psi} w (PosOfGod hG h\ensuremath{\psi}) v hv)\ensuremath{\rangle}

/-- Necessary existence -/
def E (x : e) : \ensuremath{\sigma} := \ensuremath{\forall}\textsuperscript{m} (\ensuremath{\varphi} : e \ensuremath{\rightarrow} \ensuremath{\sigma}), Ess \ensuremath{\varphi} x \ensuremath{\rightarrow}\textsuperscript{m} \ensuremath{\Box}(\ensuremath{\exists}\textsuperscript{E} x, \ensuremath{\varphi} x)

axiom Ax3 : \ensuremath{\lfloor}P E\ensuremath{\rfloor}

theorem Th2 (x : e) : \ensuremath{\lfloor}G x \ensuremath{\rightarrow}\textsuperscript{m} \ensuremath{\Box}(\ensuremath{\exists}\textsuperscript{E} y, G y)\ensuremath{\rfloor} := fun w hG => hG E (Ax3 w) G (Th1 x w hG)

theorem Th3 : \ensuremath{\lfloor}\ensuremath{\Diamond}(\ensuremath{\exists}\textsuperscript{E} x, G x) \ensuremath{\rightarrow}\textsuperscript{m} \ensuremath{\Box}(\ensuremath{\exists}\textsuperscript{E} y, G y)\ensuremath{\rfloor} := by
  have h1 : \ensuremath{\lfloor}(\ensuremath{\exists}\textsuperscript{E} x, G x) \ensuremath{\rightarrow}\textsuperscript{m} \ensuremath{\Box}(\ensuremath{\exists}\textsuperscript{E} y, G y)\ensuremath{\rfloor} := fun _ \ensuremath{\langle}x, _, hG\ensuremath{\rangle} => Th2 x _ hG
  intro w \ensuremath{\langle}v, hv, hex\ensuremath{\rangle}
  exact h1 w (h1 v hex w (Rsymm w v hv))               -- only symmetry of `\textsf{\bfseries r}` is needed

axiom Ax4 (\ensuremath{\varphi} \ensuremath{\psi} : e \ensuremath{\rightarrow} \ensuremath{\sigma}) : \ensuremath{\lfloor}P \ensuremath{\varphi} \ensuremath{\wedge}\textsuperscript{m} (\ensuremath{\varphi} \ensuremath{\supset}\textsuperscript{N} \ensuremath{\psi}) \ensuremath{\rightarrow}\textsuperscript{m} P \ensuremath{\psi}\ensuremath{\rfloor}

-- `lemma True nitpick[satisfy,card=1,eval="\ensuremath{\lfloor}P (\ensuremath{\lambda}x.\ensuremath{\bot})\ensuremath{\rfloor}"] oops`
--   One model found of cardinality one (consistency check)

@[simp, grind] def PosProps (\ensuremath{\Phi} : (e \ensuremath{\rightarrow} \ensuremath{\sigma}) \ensuremath{\rightarrow} \ensuremath{\sigma}) : \ensuremath{\sigma} := \ensuremath{\forall}\textsuperscript{m} \ensuremath{\varphi}, \ensuremath{\Phi} \ensuremath{\varphi} \ensuremath{\rightarrow}\textsuperscript{m} P \ensuremath{\varphi}
@[simp, grind] def ConjOfPropsFrom (\ensuremath{\varphi} : e \ensuremath{\rightarrow} \ensuremath{\sigma}) (\ensuremath{\Phi} : (e \ensuremath{\rightarrow} \ensuremath{\sigma}) \ensuremath{\rightarrow} \ensuremath{\sigma}) : \ensuremath{\sigma} :=
  \ensuremath{\Box}(\ensuremath{\forall}\textsuperscript{m} z, \ensuremath{\varphi} z \ensuremath{\leftrightarrow}\textsuperscript{m} (\ensuremath{\forall}\textsuperscript{m} \ensuremath{\psi}, \ensuremath{\Phi} \ensuremath{\psi} \ensuremath{\rightarrow}\textsuperscript{m} \ensuremath{\psi} z))
axiom Ax1Gen (\ensuremath{\Phi} : (e \ensuremath{\rightarrow} \ensuremath{\sigma}) \ensuremath{\rightarrow} \ensuremath{\sigma}) (\ensuremath{\varphi} : e \ensuremath{\rightarrow} \ensuremath{\sigma}) :
  \ensuremath{\lfloor}(PosProps \ensuremath{\Phi} \ensuremath{\wedge}\textsuperscript{m} ConjOfPropsFrom \ensuremath{\varphi} \ensuremath{\Phi}) \ensuremath{\rightarrow}\textsuperscript{m} P \ensuremath{\varphi}\ensuremath{\rfloor}

theorem L : \ensuremath{\lfloor}P G\ensuremath{\rfloor} := fun w => Ax1Gen P G w \ensuremath{\langle}fun _ h => h, fun _ _ _ => Iff.rfl\ensuremath{\rangle}

/-- Possibly there is an actual God-like being.  Here the argument of the actualist and
possibilist variants does not go through: `Ax1Gen` speaks of all beings at the successor worlds,
`Th4` only of the actual ones.  Instead, `Ax4` with `G \ensuremath{\supset}\textsuperscript{N} ~\textsuperscript{m}G`: a possibilist God-like being `y`
at a successor `v` yields, by `Th2`, actual God-like beings at every successor of `v`; symmetry
brings this back to `w`, and `Th2` once more forward to `v`.  This is the Isabelle proof
(`Ax2a Ax4 L Rsymm Th2`); cf. Footnote 25 of [J75].  Reflexivity would serve equally. -/
theorem Th4 : \ensuremath{\lfloor}\ensuremath{\Diamond}(\ensuremath{\exists}\textsuperscript{E} x, G x)\ensuremath{\rfloor} := fun w =>
  Classical.byContradiction fun hn =>
    (Ax2a' G w).2 (Ax4 G ~\textsuperscript{m}G w \ensuremath{\langle}L w, fun v hv y hg _ =>
      let \ensuremath{\langle}x, _, hGx\ensuremath{\rangle} := Th2 y v hg w (Rsymm w v hv)
      hn \ensuremath{\langle}v, hv, Th2 x w hGx v hv\ensuremath{\rangle}\ensuremath{\rangle}) (L w)

theorem Th5 : \ensuremath{\lfloor}\ensuremath{\Box}(\ensuremath{\exists}\textsuperscript{E} x, G x)\ensuremath{\rfloor} := fun w => Th3 w (Th4 w)

theorem MC (\ensuremath{\varphi} : \ensuremath{\sigma}) : \ensuremath{\lfloor}\ensuremath{\varphi} \ensuremath{\rightarrow}\textsuperscript{m} \ensuremath{\Box}\ensuremath{\varphi}\ensuremath{\rfloor} := by                   -- modal collapse
  intro w h\ensuremath{\varphi} v hv
  have \ensuremath{\langle}x, _, hGx\ensuremath{\rangle} := Th5 v w (Rsymm w v hv)
  have \ensuremath{\langle}z, _, hGz\ensuremath{\rangle} := Th5 w v hv
  exact (Th1 x w hGx).2 (fun _ => \ensuremath{\varphi}) h\ensuremath{\varphi} v hv z hGz

/-- The universal property is positive, from `Ax2a` and `Ax4` alone: were its complement
positive, `Ax4` would make every property positive, against the exclusivity in `Ax2a`. -/
theorem PosTop (w : i) : P (fun _ : e => \ensuremath{\top}\textsuperscript{m}) w :=
  (Ax2a (fun _ : e => \ensuremath{\top}\textsuperscript{m}) w).1.elim id fun hB =>
    ((Ax2a (fun _ : e => \ensuremath{\top}\textsuperscript{m}) w).2 \ensuremath{\langle}Ax4 _ _ w \ensuremath{\langle}hB, fun _ _ _ _ => trivial\ensuremath{\rangle}, hB\ensuremath{\rangle}).elim

theorem PosProps' : \ensuremath{\lfloor}P (fun _ : e => \ensuremath{\top}\textsuperscript{m}) \ensuremath{\wedge}\textsuperscript{m} P (fun x : e => x =\textsuperscript{m} x)\ensuremath{\rfloor} := fun w =>
  \ensuremath{\langle}PosTop w, Ax4 _ _ w \ensuremath{\langle}PosTop w, fun _ _ _ _ => rfl\ensuremath{\rangle}\ensuremath{\rangle}

theorem NegProps : \ensuremath{\lfloor}\ensuremath{\neg}\textsuperscript{m}P (fun _ : e => \ensuremath{\bot}\textsuperscript{m}) \ensuremath{\wedge}\textsuperscript{m} \ensuremath{\neg}\textsuperscript{m}P (fun x : e => x \ensuremath{\neq}\textsuperscript{m} x)\ensuremath{\rfloor} := fun w =>
  \ensuremath{\langle}fun hp => (Ax2a _ w).2 \ensuremath{\langle}hp, Ax4 _ _ w \ensuremath{\langle}hp, fun _ _ _ h => h.elim\ensuremath{\rangle}\ensuremath{\rangle},
   fun hp => (Ax2a _ w).2 \ensuremath{\langle}hp, Ax4 _ _ w \ensuremath{\langle}hp, fun _ _ _ h => (h rfl).elim\ensuremath{\rangle}\ensuremath{\rangle}\ensuremath{\rangle}

theorem UniqueEss1 (\ensuremath{\varphi} \ensuremath{\psi} : e \ensuremath{\rightarrow} \ensuremath{\sigma}) (x : e) : \ensuremath{\lfloor}Ess \ensuremath{\varphi} x \ensuremath{\wedge}\textsuperscript{m} Ess \ensuremath{\psi} x \ensuremath{\rightarrow}\textsuperscript{m} \ensuremath{\Box}(\ensuremath{\forall}\textsuperscript{m} y, \ensuremath{\varphi} y \ensuremath{\leftrightarrow}\textsuperscript{m} \ensuremath{\psi} y)\ensuremath{\rfloor} :=
  fun _ \ensuremath{\langle}\ensuremath{\langle}h\ensuremath{\varphi}x, h\ensuremath{\varphi}\ensuremath{\rangle}, \ensuremath{\langle}h\ensuremath{\psi}x, h\ensuremath{\psi}\ensuremath{\rangle}\ensuremath{\rangle} v hv y => \ensuremath{\langle}h\ensuremath{\varphi} \ensuremath{\psi} h\ensuremath{\psi}x v hv y, h\ensuremath{\psi} \ensuremath{\varphi} h\ensuremath{\varphi}x v hv y\ensuremath{\rangle}

-- UniqueEss2 : \ensuremath{\lfloor}Ess \ensuremath{\varphi} x \ensuremath{\wedge}\textsuperscript{m} Ess \ensuremath{\psi} x \ensuremath{\rightarrow}\textsuperscript{m} \ensuremath{\Box}(\ensuremath{\varphi} \ensuremath{\equiv}\textsuperscript{m} \ensuremath{\psi})\ensuremath{\rfloor}   -- `nitpick[card i=2]`: countermodel found
example (\ensuremath{\varphi} \ensuremath{\psi} : e \ensuremath{\rightarrow} \ensuremath{\sigma}) (x : e) : \ensuremath{\lfloor}Ess \ensuremath{\varphi} x \ensuremath{\wedge}\textsuperscript{m} Ess \ensuremath{\psi} x \ensuremath{\rightarrow}\textsuperscript{m} \ensuremath{\Box}(\ensuremath{\varphi} \ensuremath{\equiv}\textsuperscript{m} \ensuremath{\psi})\ensuremath{\rfloor} := by countermodel

theorem UniqueEss3 (\ensuremath{\varphi} : e \ensuremath{\rightarrow} \ensuremath{\sigma}) (x : e) : \ensuremath{\lfloor}Ess \ensuremath{\varphi} x \ensuremath{\rightarrow}\textsuperscript{m} \ensuremath{\Box}(\ensuremath{\forall}\textsuperscript{m} y, \ensuremath{\varphi} y \ensuremath{\rightarrow}\textsuperscript{m} y \ensuremath{\equiv}\textsuperscript{m} x)\ensuremath{\rfloor} :=
  fun _ \ensuremath{\langle}_, h\ensuremath{\rangle} => h (fun z => z \ensuremath{\equiv}\textsuperscript{m} x) (fun _ hp => hp)

theorem Monotheism (x y : e) : \ensuremath{\lfloor}G x \ensuremath{\wedge}\textsuperscript{m} G y \ensuremath{\rightarrow}\textsuperscript{m} x \ensuremath{\equiv}\textsuperscript{m} y\ensuremath{\rfloor} :=
  fun _ \ensuremath{\langle}hx, hy\ensuremath{\rangle} Q hQ => hy Q (PosOfGod hx hQ)

theorem PisFilter : \ensuremath{\lfloor}Filter P\ensuremath{\rfloor} := fun w =>
  \ensuremath{\langle}\ensuremath{\langle}\ensuremath{\langle}(PosProps' w).1, (NegProps w).1\ensuremath{\rangle},
    -- `\ensuremath{\varphi} \ensuremath{\subseteq}\textsuperscript{E} \ensuremath{\psi}` is an actualist inclusion, `Ax4` needs a possibilist one; so argue directly:
    -- `MC` carries the inclusion to a successor `v` holding an actual God-like being `x` (`Th4`),
    -- `x` has every positive property, and positivity is necessary (`Ax2b`)
    fun \ensuremath{\varphi} \ensuremath{\psi} \ensuremath{\langle}h\ensuremath{\varphi}, hsub\ensuremath{\rangle} => Classical.byContradiction fun hn =>
      let \ensuremath{\langle}v, hv, x, hxa, hGx\ensuremath{\rangle} := Th4 w
      hGx ~\textsuperscript{m}\ensuremath{\psi} (Ax2b ~\textsuperscript{m}\ensuremath{\psi} w ((Ax2a' \ensuremath{\psi} w).1 hn) v hv)
        (MC (\ensuremath{\forall}\textsuperscript{E} x, \ensuremath{\varphi} x \ensuremath{\rightarrow}\textsuperscript{m} \ensuremath{\psi} x) w hsub v hv x hxa (hGx \ensuremath{\varphi} (Ax2b \ensuremath{\varphi} w h\ensuremath{\varphi} v hv)))\ensuremath{\rangle},
   fun \ensuremath{\varphi} \ensuremath{\psi} \ensuremath{\langle}h\ensuremath{\varphi}, h\ensuremath{\psi}\ensuremath{\rangle} => Ax1 \ensuremath{\varphi} \ensuremath{\psi} w \ensuremath{\langle}h\ensuremath{\varphi}, h\ensuremath{\psi}\ensuremath{\rangle}\ensuremath{\rangle}

theorem PisUFilter : \ensuremath{\lfloor}UFilter P\ensuremath{\rfloor} := fun w =>
  \ensuremath{\langle}PisFilter w, fun \ensuremath{\varphi} => (Classical.em (P \ensuremath{\varphi} w)).imp id (Ax2a' \ensuremath{\varphi} w).1\ensuremath{\rangle}

-- `lemma True nitpick[satisfy,card=1,eval="\ensuremath{\lfloor}P (\ensuremath{\lambda}x.\ensuremath{\bot})\ensuremath{\rfloor}"] oops`
--   One model found of cardinality one (consistency check)
\end{Verbatim}
\subsubsection{GoedelVariantHOML3AndersonQuant.lean}
The same as GoedelVariantHOML3, but now for a mixed use of actualist and possibilist quantifiers for entities; cf. Footnote 20 in \cite{J75}.
\begin{Verbatim}[commandchars=\\\{\},fontsize=\scriptsize,xleftmargin=1em]
import Notes.HOMLinHOL
import Notes.ModalFilter
\end{Verbatim}

\begin{Verbatim}[commandchars=\\\{\},fontsize=\scriptsize,xleftmargin=1em]
/-- Positive property -/
axiom P : (e \ensuremath{\rightarrow} \ensuremath{\sigma}) \ensuremath{\rightarrow} \ensuremath{\sigma}

axiom Ax1 (\ensuremath{\varphi} \ensuremath{\psi} : e \ensuremath{\rightarrow} \ensuremath{\sigma}) : \ensuremath{\lfloor}P \ensuremath{\varphi} \ensuremath{\wedge}\textsuperscript{m} P \ensuremath{\psi} \ensuremath{\rightarrow}\textsuperscript{m} P (\ensuremath{\varphi} \ensuremath{\centerdot}\textsuperscript{m} \ensuremath{\psi})\ensuremath{\rfloor}

axiom Ax2a (\ensuremath{\varphi} : e \ensuremath{\rightarrow} \ensuremath{\sigma}) : \ensuremath{\lfloor}P \ensuremath{\varphi} \ensuremath{\vee}\textsuperscript{e} P ~\textsuperscript{m}\ensuremath{\varphi}\ensuremath{\rfloor}

/-- Auxiliary reformulation of `Ax2a` (Lean has no `sledgehammer`). -/
theorem Ax2a' (\ensuremath{\varphi} : e \ensuremath{\rightarrow} \ensuremath{\sigma}) : \ensuremath{\lfloor}\ensuremath{\neg}\textsuperscript{m}(P \ensuremath{\varphi}) \ensuremath{\leftrightarrow}\textsuperscript{m} P ~\textsuperscript{m}\ensuremath{\varphi}\ensuremath{\rfloor} :=
  fun w => \ensuremath{\langle}(Ax2a \ensuremath{\varphi} w).1.resolve_left, fun hp hq => (Ax2a \ensuremath{\varphi} w).2 \ensuremath{\langle}hq, hp\ensuremath{\rangle}\ensuremath{\rangle}

/-- God-like -/
def G (x : e) : \ensuremath{\sigma} := \ensuremath{\forall}\textsuperscript{m} (\ensuremath{\varphi} : e \ensuremath{\rightarrow} \ensuremath{\sigma}), P \ensuremath{\varphi} \ensuremath{\rightarrow}\textsuperscript{m} \ensuremath{\varphi} x

/-- Necessary property inclusion (modified: \ensuremath{\varphi} must be non-empty) -/
@[simp, grind] def PInc (\ensuremath{\varphi} \ensuremath{\psi} : e \ensuremath{\rightarrow} \ensuremath{\sigma}) : \ensuremath{\sigma} :=
  \ensuremath{\Box}((\ensuremath{\varphi} \ensuremath{\neq}\textsuperscript{m} (fun _ : e => \ensuremath{\bot}\textsuperscript{m})) \ensuremath{\wedge}\textsuperscript{m} (\ensuremath{\forall}\textsuperscript{m} (y : e), \ensuremath{\varphi} y \ensuremath{\rightarrow}\textsuperscript{m} \ensuremath{\psi} y))
infixr:48 " \ensuremath{\supset}\textsuperscript{N} " => PInc

/-- `Ess \ensuremath{\varphi} x`: \ensuremath{\varphi} is an essence of x -/
def Ess (\ensuremath{\varphi} : e \ensuremath{\rightarrow} \ensuremath{\sigma}) (x : e) : \ensuremath{\sigma} := \ensuremath{\forall}\textsuperscript{m} (\ensuremath{\psi} : e \ensuremath{\rightarrow} \ensuremath{\sigma}), \ensuremath{\psi} x \ensuremath{\rightarrow}\textsuperscript{m} (\ensuremath{\varphi} \ensuremath{\supset}\textsuperscript{N} \ensuremath{\psi})

axiom Ax2b (\ensuremath{\varphi} : e \ensuremath{\rightarrow} \ensuremath{\sigma}) : \ensuremath{\lfloor}P \ensuremath{\varphi} \ensuremath{\rightarrow}\textsuperscript{m} \ensuremath{\Box} P \ensuremath{\varphi}\ensuremath{\rfloor}

theorem Ax2b' (\ensuremath{\varphi} : e \ensuremath{\rightarrow} \ensuremath{\sigma}) : \ensuremath{\lfloor}\ensuremath{\neg}\textsuperscript{m}(P \ensuremath{\varphi}) \ensuremath{\rightarrow}\textsuperscript{m} \ensuremath{\Box}(\ensuremath{\neg}\textsuperscript{m}(P \ensuremath{\varphi}))\ensuremath{\rfloor} :=
  fun w hn v hv => (Ax2a' \ensuremath{\varphi} v).2 (Ax2b ~\textsuperscript{m}\ensuremath{\varphi} w ((Ax2a' \ensuremath{\varphi} w).1 hn) v hv)

/-- A property exemplified by a God-like being is positive (a consequence of `Ax2a`). -/
theorem PosOfGod \{x : e\} \{\ensuremath{\psi} : e \ensuremath{\rightarrow} \ensuremath{\sigma}\} \{w : i\} (hG : G x w) (h : \ensuremath{\psi} x w) : P \ensuremath{\psi} w :=
  Classical.byContradiction fun hn => hG ~\textsuperscript{m}\ensuremath{\psi} ((Ax2a' \ensuremath{\psi} w).1 hn) h

theorem Th1 (x : e) : \ensuremath{\lfloor}G x \ensuremath{\rightarrow}\textsuperscript{m} Ess G x\ensuremath{\rfloor} := by
  intro w hG \ensuremath{\psi} h\ensuremath{\psi} v hv
  refine \ensuremath{\langle}fun heq => ?_, fun y hGy => hGy \ensuremath{\psi} (Ax2b \ensuremath{\psi} w (PosOfGod hG h\ensuremath{\psi}) v hv)\ensuremath{\rangle}
  rw [heq] at hG; exact hG                             -- `G` is non-empty since `G x` holds

/-- Necessary existence -/
def E (x : e) : \ensuremath{\sigma} := \ensuremath{\forall}\textsuperscript{m} (\ensuremath{\varphi} : e \ensuremath{\rightarrow} \ensuremath{\sigma}), Ess \ensuremath{\varphi} x \ensuremath{\rightarrow}\textsuperscript{m} \ensuremath{\Box}(\ensuremath{\exists}\textsuperscript{E} x, \ensuremath{\varphi} x)

axiom Ax3 : \ensuremath{\lfloor}P E\ensuremath{\rfloor}

theorem Th2 (x : e) : \ensuremath{\lfloor}G x \ensuremath{\rightarrow}\textsuperscript{m} \ensuremath{\Box}(\ensuremath{\exists}\textsuperscript{E} y, G y)\ensuremath{\rfloor} := fun w hG => hG E (Ax3 w) G (Th1 x w hG)

theorem Th3 : \ensuremath{\lfloor}\ensuremath{\Diamond}(\ensuremath{\exists}\textsuperscript{E} x, G x) \ensuremath{\rightarrow}\textsuperscript{m} \ensuremath{\Box}(\ensuremath{\exists}\textsuperscript{E} y, G y)\ensuremath{\rfloor} := by
  have h1 : \ensuremath{\lfloor}(\ensuremath{\exists}\textsuperscript{E} x, G x) \ensuremath{\rightarrow}\textsuperscript{m} \ensuremath{\Box}(\ensuremath{\exists}\textsuperscript{E} y, G y)\ensuremath{\rfloor} := fun _ \ensuremath{\langle}x, _, hG\ensuremath{\rangle} => Th2 x _ hG
  intro w \ensuremath{\langle}v, hv, hex\ensuremath{\rangle}
  exact h1 w (h1 v hex w (Rsymm w v hv))               -- only symmetry of `\textsf{\bfseries r}` is needed

axiom Ax4 (\ensuremath{\varphi} \ensuremath{\psi} : e \ensuremath{\rightarrow} \ensuremath{\sigma}) : \ensuremath{\lfloor}P \ensuremath{\varphi} \ensuremath{\wedge}\textsuperscript{m} (\ensuremath{\varphi} \ensuremath{\supset}\textsuperscript{N} \ensuremath{\psi}) \ensuremath{\rightarrow}\textsuperscript{m} P \ensuremath{\psi}\ensuremath{\rfloor}

-- `lemma True nitpick[satisfy,card=1,eval="\ensuremath{\lfloor}P (\ensuremath{\lambda}x.\ensuremath{\bot})\ensuremath{\rfloor}"] oops`
--   Two models found of cardinality one (consistency check)

@[simp, grind] def PosProps (\ensuremath{\Phi} : (e \ensuremath{\rightarrow} \ensuremath{\sigma}) \ensuremath{\rightarrow} \ensuremath{\sigma}) : \ensuremath{\sigma} := \ensuremath{\forall}\textsuperscript{m} \ensuremath{\varphi}, \ensuremath{\Phi} \ensuremath{\varphi} \ensuremath{\rightarrow}\textsuperscript{m} P \ensuremath{\varphi}
@[simp, grind] def ConjOfPropsFrom (\ensuremath{\varphi} : e \ensuremath{\rightarrow} \ensuremath{\sigma}) (\ensuremath{\Phi} : (e \ensuremath{\rightarrow} \ensuremath{\sigma}) \ensuremath{\rightarrow} \ensuremath{\sigma}) : \ensuremath{\sigma} :=
  \ensuremath{\Box}(\ensuremath{\forall}\textsuperscript{E} z, \ensuremath{\varphi} z \ensuremath{\leftrightarrow}\textsuperscript{m} (\ensuremath{\forall}\textsuperscript{m} \ensuremath{\psi}, \ensuremath{\Phi} \ensuremath{\psi} \ensuremath{\rightarrow}\textsuperscript{m} \ensuremath{\psi} z))
axiom Ax1Gen (\ensuremath{\Phi} : (e \ensuremath{\rightarrow} \ensuremath{\sigma}) \ensuremath{\rightarrow} \ensuremath{\sigma}) (\ensuremath{\varphi} : e \ensuremath{\rightarrow} \ensuremath{\sigma}) :
  \ensuremath{\lfloor}(PosProps \ensuremath{\Phi} \ensuremath{\wedge}\textsuperscript{m} ConjOfPropsFrom \ensuremath{\varphi} \ensuremath{\Phi}) \ensuremath{\rightarrow}\textsuperscript{m} P \ensuremath{\varphi}\ensuremath{\rfloor}

theorem L : \ensuremath{\lfloor}P G\ensuremath{\rfloor} := fun w => Ax1Gen P G w \ensuremath{\langle}fun _ h => h, fun _ _ _ _ => Iff.rfl\ensuremath{\rangle}

/-- The AFP sources leave `Th4` unreplayed (`oops`, then a postulate), although automated
provers report a proof from `Ax2a`, `L` and `Ax1Gen`.  Here is one: if no God-like being were
possible, `Ax1Gen` would make the empty property positive, as the conjunction of `\{G\}`; it also
makes the universal property positive, as the conjunction of `\ensuremath{\emptyset}`; the two contradict the
exclusivity in `Ax2a`.  Only `Ax1Gen` and `Ax2a` are used: the theorem holds in logic K.  (In
this mixed-quantifier variant `ConjOfPropsFrom` quantifies actualistically, so the argument is
the same as in the actualist variant; cf. Footnote 29 of [J75].) -/
theorem Th4 : \ensuremath{\lfloor}\ensuremath{\Diamond}(\ensuremath{\exists}\textsuperscript{E} x, G x)\ensuremath{\rfloor} := fun w =>
  Classical.byContradiction fun hn =>
    have hbot : P (fun _ : e => \ensuremath{\bot}\textsuperscript{m}) w :=
      Ax1Gen (fun \ensuremath{\psi} => \ensuremath{\psi} =\textsuperscript{m} G) (fun _ => \ensuremath{\bot}\textsuperscript{m}) w
        \ensuremath{\langle}fun \ensuremath{\psi} h\ensuremath{\psi} => by simp only [Mprimeq] at h\ensuremath{\psi}; subst h\ensuremath{\psi}; exact L w,
         fun v hv z hz => \ensuremath{\langle}fun h => h.elim, fun h => hn \ensuremath{\langle}v, hv, z, hz, h G rfl\ensuremath{\rangle}\ensuremath{\rangle}\ensuremath{\rangle}
    have htop : P (fun _ : e => \ensuremath{\top}\textsuperscript{m}) w :=
      Ax1Gen (fun _ => \ensuremath{\bot}\textsuperscript{m}) (fun _ => \ensuremath{\top}\textsuperscript{m}) w
        \ensuremath{\langle}fun _ h => h.elim, fun _ _ _ _ => \ensuremath{\langle}fun _ _ h => h.elim, fun _ => trivial\ensuremath{\rangle}\ensuremath{\rangle}
    have heq : (~\textsuperscript{m}(fun _ : e => \ensuremath{\bot}\textsuperscript{m})) = (fun _ : e => \ensuremath{\top}\textsuperscript{m}) := by funext x v; simp
    (Ax2a (fun _ => \ensuremath{\bot}\textsuperscript{m}) w).2 \ensuremath{\langle}hbot, heq \ensuremath{\blacktriangleright} htop\ensuremath{\rangle}

theorem Th5 : \ensuremath{\lfloor}\ensuremath{\Box}(\ensuremath{\exists}\textsuperscript{E} x, G x)\ensuremath{\rfloor} := fun w => Th3 w (Th4 w)

theorem MC (\ensuremath{\varphi} : \ensuremath{\sigma}) : \ensuremath{\lfloor}\ensuremath{\varphi} \ensuremath{\rightarrow}\textsuperscript{m} \ensuremath{\Box}\ensuremath{\varphi}\ensuremath{\rfloor} := by                   -- modal collapse
  intro w h\ensuremath{\varphi} v hv
  have \ensuremath{\langle}x, _, hGx\ensuremath{\rangle} := Th5 v w (Rsymm w v hv)
  have \ensuremath{\langle}z, _, hGz\ensuremath{\rangle} := Th5 w v hv
  exact (Th1 x w hGx) (fun _ => \ensuremath{\varphi}) h\ensuremath{\varphi} v hv |>.2 z hGz

/-- `G` is not the empty property: by `Th4` it is exemplified at some successor of `w`. -/
theorem GNonempty (w : i) : G \ensuremath{\neq} (fun _ : e => \ensuremath{\bot}\textsuperscript{m}) := fun h =>
  let \ensuremath{\langle}_, _, _, _, hG\ensuremath{\rangle} := Th4 w; (h \ensuremath{\blacktriangleright} hG : \ensuremath{\bot}\textsuperscript{m} _)

/-- `Ax4` applied to `G`: everything true of every God-like being at every successor world is
positive.  Uses `L`, `Th4` and `Ax4`. -/
theorem PosIncl (\ensuremath{\chi} : e \ensuremath{\rightarrow} \ensuremath{\sigma}) (w : i) (h : \ensuremath{\forall} v, w \textsf{\bfseries r} v \ensuremath{\rightarrow} \ensuremath{\forall} x, G x v \ensuremath{\rightarrow} \ensuremath{\chi} x v) : P \ensuremath{\chi} w :=
  Ax4 G \ensuremath{\chi} w \ensuremath{\langle}L w, fun v hv => \ensuremath{\langle}GNonempty w, h v hv\ensuremath{\rangle}\ensuremath{\rangle}

theorem PosProps' : \ensuremath{\lfloor}P (fun _ : e => \ensuremath{\top}\textsuperscript{m}) \ensuremath{\wedge}\textsuperscript{m} P (fun x : e => x =\textsuperscript{m} x)\ensuremath{\rfloor} := fun w =>
  \ensuremath{\langle}PosIncl _ w (fun _ _ _ _ => trivial), PosIncl _ w (fun _ _ _ _ => rfl)\ensuremath{\rangle}

theorem NegProps : \ensuremath{\lfloor}\ensuremath{\neg}\textsuperscript{m}P (fun _ : e => \ensuremath{\bot}\textsuperscript{m}) \ensuremath{\wedge}\textsuperscript{m} \ensuremath{\neg}\textsuperscript{m}P (fun x : e => x \ensuremath{\neq}\textsuperscript{m} x)\ensuremath{\rfloor} := fun w =>
  \ensuremath{\langle}fun hB => (Ax2a _ w).2 \ensuremath{\langle}hB, (show (~\textsuperscript{m}(fun _ : e => \ensuremath{\bot}\textsuperscript{m})) = (fun _ : e => \ensuremath{\top}\textsuperscript{m}) by
      funext x v; simp) \ensuremath{\blacktriangleright} (PosProps' w).1\ensuremath{\rangle},
   fun hN => (Ax2a _ w).2 \ensuremath{\langle}hN, (show (~\textsuperscript{m}(fun x : e => x \ensuremath{\neq}\textsuperscript{m} x)) = (fun _ : e => \ensuremath{\top}\textsuperscript{m}) by
      funext x v; simp) \ensuremath{\blacktriangleright} (PosProps' w).1\ensuremath{\rangle}\ensuremath{\rangle}

-- UniqueEss1 : \ensuremath{\lfloor}Ess \ensuremath{\varphi} x \ensuremath{\wedge}\textsuperscript{m} Ess \ensuremath{\psi} x \ensuremath{\rightarrow}\textsuperscript{m} \ensuremath{\Box}(\ensuremath{\forall}\textsuperscript{m} y, \ensuremath{\varphi} y \ensuremath{\leftrightarrow}\textsuperscript{m} \ensuremath{\psi} y)\ensuremath{\rfloor}      -- Unclear, open question
example (\ensuremath{\varphi} \ensuremath{\psi} : e \ensuremath{\rightarrow} \ensuremath{\sigma}) (x : e) : \ensuremath{\lfloor}Ess \ensuremath{\varphi} x \ensuremath{\wedge}\textsuperscript{m} Ess \ensuremath{\psi} x \ensuremath{\rightarrow}\textsuperscript{m} \ensuremath{\Box}(\ensuremath{\forall}\textsuperscript{m} y, \ensuremath{\varphi} y \ensuremath{\leftrightarrow}\textsuperscript{m} \ensuremath{\psi} y)\ensuremath{\rfloor} := by openproblem
-- UniqueEss2 : \ensuremath{\lfloor}Ess \ensuremath{\varphi} x \ensuremath{\wedge}\textsuperscript{m} Ess \ensuremath{\psi} x \ensuremath{\rightarrow}\textsuperscript{m} \ensuremath{\Box}(\ensuremath{\varphi} \ensuremath{\equiv}\textsuperscript{m} \ensuremath{\psi})\ensuremath{\rfloor}                -- Unclear, open question
example (\ensuremath{\varphi} \ensuremath{\psi} : e \ensuremath{\rightarrow} \ensuremath{\sigma}) (x : e) : \ensuremath{\lfloor}Ess \ensuremath{\varphi} x \ensuremath{\wedge}\textsuperscript{m} Ess \ensuremath{\psi} x \ensuremath{\rightarrow}\textsuperscript{m} \ensuremath{\Box}(\ensuremath{\varphi} \ensuremath{\equiv}\textsuperscript{m} \ensuremath{\psi})\ensuremath{\rfloor} := by openproblem

theorem UniqueEss3 (\ensuremath{\varphi} : e \ensuremath{\rightarrow} \ensuremath{\sigma}) (x : e) : \ensuremath{\lfloor}Ess \ensuremath{\varphi} x \ensuremath{\rightarrow}\textsuperscript{m} \ensuremath{\Box}(\ensuremath{\forall}\textsuperscript{m} y, \ensuremath{\varphi} y \ensuremath{\rightarrow}\textsuperscript{m} y \ensuremath{\equiv}\textsuperscript{m} x)\ensuremath{\rfloor} :=
  fun _ h v hv => (h (fun z => z \ensuremath{\equiv}\textsuperscript{m} x) (fun _ hp => hp) v hv).2

theorem Monotheism (x y : e) : \ensuremath{\lfloor}G x \ensuremath{\wedge}\textsuperscript{m} G y \ensuremath{\rightarrow}\textsuperscript{m} x \ensuremath{\equiv}\textsuperscript{m} y\ensuremath{\rfloor} :=
  fun _ \ensuremath{\langle}hx, hy\ensuremath{\rangle} Q hQ => hy Q (PosOfGod hx hQ)

theorem PisFilter : \ensuremath{\lfloor}Filter P\ensuremath{\rfloor} := fun w =>
  \ensuremath{\langle}\ensuremath{\langle}\ensuremath{\langle}(PosProps' w).1, (NegProps w).1\ensuremath{\rangle},
    -- `\ensuremath{\varphi} \ensuremath{\subseteq}\textsuperscript{E} \ensuremath{\psi}` is an actualist inclusion, `Ax4` needs a possibilist one; so argue directly:
    -- `MC` carries the inclusion to a successor `v` holding an actual God-like being `x` (`Th4`),
    -- `x` has every positive property, and positivity is necessary (`Ax2b`)
    fun \ensuremath{\varphi} \ensuremath{\psi} \ensuremath{\langle}h\ensuremath{\varphi}, hsub\ensuremath{\rangle} => Classical.byContradiction fun hn =>
      let \ensuremath{\langle}v, hv, x, hxa, hGx\ensuremath{\rangle} := Th4 w
      hGx ~\textsuperscript{m}\ensuremath{\psi} (Ax2b ~\textsuperscript{m}\ensuremath{\psi} w ((Ax2a' \ensuremath{\psi} w).1 hn) v hv)
        (MC (\ensuremath{\forall}\textsuperscript{E} x, \ensuremath{\varphi} x \ensuremath{\rightarrow}\textsuperscript{m} \ensuremath{\psi} x) w hsub v hv x hxa (hGx \ensuremath{\varphi} (Ax2b \ensuremath{\varphi} w h\ensuremath{\varphi} v hv)))\ensuremath{\rangle},
   fun \ensuremath{\varphi} \ensuremath{\psi} \ensuremath{\langle}h\ensuremath{\varphi}, h\ensuremath{\psi}\ensuremath{\rangle} => Ax1 \ensuremath{\varphi} \ensuremath{\psi} w \ensuremath{\langle}h\ensuremath{\varphi}, h\ensuremath{\psi}\ensuremath{\rangle}\ensuremath{\rangle}

theorem PisUFilter : \ensuremath{\lfloor}UFilter P\ensuremath{\rfloor} := fun w =>
  \ensuremath{\langle}PisFilter w, fun \ensuremath{\varphi} => (Classical.em (P \ensuremath{\varphi} w)).imp id (Ax2a' \ensuremath{\varphi} w).1\ensuremath{\rangle}

-- `lemma True nitpick[satisfy,card=1,eval="\ensuremath{\lfloor}P (\ensuremath{\lambda}x.\ensuremath{\top})\ensuremath{\rfloor}"] oops`
--   One model found of cardinality one (consistency check)
\end{Verbatim}
\subsubsection{HOMLinHOLonlyS4.lean (slight variation of Figure 3 of \texorpdfstring{\cite{J75}}{the Notes})}
Shallow embedding of higher-order modal logic (HOML) in the classical higher-order logic (HOL) of Lean 4, utilizing the LogiKEy methodology.  Here logic S4 is introduced.

The global parameter settings of the Isabelle sources configure the model finder \texttt{nitpick} and the parser; they have no Lean counterpart, as Lean has no model finder.

\begin{Verbatim}[commandchars=\\\{\},fontsize=\scriptsize,xleftmargin=1em]
-- nitpick_params[user_axioms,expect=genuine,show_all,format=2,max_genuine=3]
-- declare[[syntax_ambiguity_warning=false]]
\end{Verbatim}
Lean's \texttt{Prop} is intuitionistic, HOL is classical; classical reasoning is enabled globally.

\begin{Verbatim}[commandchars=\\\{\},fontsize=\scriptsize,xleftmargin=1em]
attribute [instance] Classical.propDecidable

/-- Counterpart of Isabelle's `oops` after `nitpick` found a countermodel.  Lean has no model
finder, so such statements are only type-checked, never proved (and never added to the context). -/
macro "countermodel" : tactic => `(tactic| sorry)
/-- Counterpart of Isabelle's `oops` on a statement left genuinely open: neither a proof
nor a countermodel was obtained. -/
macro "openproblem" : tactic => `(tactic| sorry)
\end{Verbatim}
Type \texttt{i} is associated with possible worlds and type \texttt{e} with entities

\begin{Verbatim}[commandchars=\\\{\},fontsize=\scriptsize,xleftmargin=1em]
axiom i : Type                       -- Possible worlds
axiom e : Type                       -- Individuals/entities
@[instance] axiom iNonempty : Nonempty i
@[instance] axiom eNonempty : Nonempty e
abbrev \ensuremath{\sigma} := i \ensuremath{\rightarrow} Prop                 -- World-lifted propositions
abbrev \ensuremath{\tau} := e \ensuremath{\rightarrow} \ensuremath{\sigma}                    -- Modal properties

/-- Accessibility relation between worlds -/
axiom R : i \ensuremath{\rightarrow} i \ensuremath{\rightarrow} Prop
@[inherit_doc] infix:60 " \textsf{\bfseries r} " => R

axiom Rrefl  : \ensuremath{\forall} x, x \textsf{\bfseries r} x
axiom Rtrans : \ensuremath{\forall} x y z, x \textsf{\bfseries r} y \ensuremath{\wedge} y \textsf{\bfseries r} z \ensuremath{\rightarrow} x \textsf{\bfseries r} z
\end{Verbatim}
Logical connectives (operating on truth-sets)

\begin{Verbatim}[commandchars=\\\{\},fontsize=\scriptsize,xleftmargin=1em]
@[simp, grind] def Mbot : \ensuremath{\sigma} := fun _ => False
@[simp, grind] def Mtop : \ensuremath{\sigma} := fun _ => True
@[simp, grind] def Mneg (\ensuremath{\varphi} : \ensuremath{\sigma}) : \ensuremath{\sigma} := fun w => \ensuremath{\neg} \ensuremath{\varphi} w
@[simp, grind] def Mand (\ensuremath{\varphi} \ensuremath{\psi} : \ensuremath{\sigma}) : \ensuremath{\sigma} := fun w => \ensuremath{\varphi} w \ensuremath{\wedge} \ensuremath{\psi} w
@[simp, grind] def Mor (\ensuremath{\varphi} \ensuremath{\psi} : \ensuremath{\sigma}) : \ensuremath{\sigma} := fun w => \ensuremath{\varphi} w \ensuremath{\vee} \ensuremath{\psi} w
@[simp, grind] def Mimp (\ensuremath{\varphi} \ensuremath{\psi} : \ensuremath{\sigma}) : \ensuremath{\sigma} := fun w => \ensuremath{\varphi} w \ensuremath{\rightarrow} \ensuremath{\psi} w
@[simp, grind] def Mequiv (\ensuremath{\varphi} \ensuremath{\psi} : \ensuremath{\sigma}) : \ensuremath{\sigma} := fun w => \ensuremath{\varphi} w \ensuremath{\leftrightarrow} \ensuremath{\psi} w
@[simp, grind] def Mbox (\ensuremath{\varphi} : \ensuremath{\sigma}) : \ensuremath{\sigma} := fun w => \ensuremath{\forall} v, w \textsf{\bfseries r} v \ensuremath{\rightarrow} \ensuremath{\varphi} v
@[simp, grind] def Mdia (\ensuremath{\varphi} : \ensuremath{\sigma}) : \ensuremath{\sigma} := fun w => \ensuremath{\exists} v, w \textsf{\bfseries r} v \ensuremath{\wedge} \ensuremath{\varphi} v
@[simp, grind] def Mprimeq \{\ensuremath{\alpha}\} (x y : \ensuremath{\alpha}) : \ensuremath{\sigma} := fun _ => x = y
@[simp, grind] def Mprimneg \{\ensuremath{\alpha}\} (x y : \ensuremath{\alpha}) : \ensuremath{\sigma} := fun _ => x \ensuremath{\neq} y
@[simp, grind] def Mnegpred (\ensuremath{\Phi} : \ensuremath{\tau}) : \ensuremath{\tau} := fun x w => \ensuremath{\neg} \ensuremath{\Phi} x w
@[simp, grind] def Mconpred (\ensuremath{\Phi} \ensuremath{\Psi} : \ensuremath{\tau}) : \ensuremath{\tau} := fun x w => \ensuremath{\Phi} x w \ensuremath{\wedge} \ensuremath{\Psi} x w

notation:max "\ensuremath{\bot}\textsuperscript{m}" => Mbot
notation:max "\ensuremath{\top}\textsuperscript{m}" => Mtop
prefix:53 "\ensuremath{\neg}\textsuperscript{m}" => Mneg
infixl:50 " \ensuremath{\wedge}\textsuperscript{m} " => Mand
infixl:49 " \ensuremath{\vee}\textsuperscript{m} " => Mor
infixr:48 " \ensuremath{\rightarrow}\textsuperscript{m} " => Mimp
infixl:47 " \ensuremath{\leftrightarrow}\textsuperscript{m} " => Mequiv
prefix:55 "\ensuremath{\Box}" => Mbox
prefix:55 "\ensuremath{\Diamond}" => Mdia
infix:50 " =\textsuperscript{m} " => Mprimeq
infix:50 " \ensuremath{\neq}\textsuperscript{m} " => Mprimneg
prefix:max "~\textsuperscript{m}" => Mnegpred
infixl:50 " \ensuremath{\centerdot}\textsuperscript{m} " => Mconpred

@[simp, grind] def Mexclor (\ensuremath{\varphi} \ensuremath{\psi} : \ensuremath{\sigma}) : \ensuremath{\sigma} := (\ensuremath{\varphi} \ensuremath{\vee}\textsuperscript{m} \ensuremath{\psi}) \ensuremath{\wedge}\textsuperscript{m} \ensuremath{\neg}\textsuperscript{m}(\ensuremath{\varphi} \ensuremath{\wedge}\textsuperscript{m} \ensuremath{\psi})
infixl:49 " \ensuremath{\vee}\textsuperscript{e} " => Mexclor
\end{Verbatim}
Possibilist quantifiers (polymorphic)

\begin{Verbatim}[commandchars=\\\{\},fontsize=\scriptsize,xleftmargin=1em]
@[simp, grind] def Mallposs \{\ensuremath{\alpha}\} (\ensuremath{\Phi} : \ensuremath{\alpha} \ensuremath{\rightarrow} \ensuremath{\sigma}) : \ensuremath{\sigma} := fun w => \ensuremath{\forall} x, \ensuremath{\Phi} x w
@[simp, grind] def Mexiposs \{\ensuremath{\alpha}\} (\ensuremath{\Phi} : \ensuremath{\alpha} \ensuremath{\rightarrow} \ensuremath{\sigma}) : \ensuremath{\sigma} := fun w => \ensuremath{\exists} x, \ensuremath{\Phi} x w
\end{Verbatim}
Actualist quantifiers (for individuals/entities)

\begin{Verbatim}[commandchars=\\\{\},fontsize=\scriptsize,xleftmargin=1em]
/-- Existence (actuality) of an entity at a world -/
axiom existsAt : e \ensuremath{\rightarrow} \ensuremath{\sigma}
@[inherit_doc] infix:60 " @\textsuperscript{m} " => existsAt
@[simp, grind] def Mallact (\ensuremath{\Phi} : e \ensuremath{\rightarrow} \ensuremath{\sigma}) : \ensuremath{\sigma} := fun w => \ensuremath{\forall} x, x @\textsuperscript{m} w \ensuremath{\rightarrow} \ensuremath{\Phi} x w
@[simp, grind] def Mexiact (\ensuremath{\Phi} : e \ensuremath{\rightarrow} \ensuremath{\sigma}) : \ensuremath{\sigma} := fun w => \ensuremath{\exists} x, x @\textsuperscript{m} w \ensuremath{\wedge} \ensuremath{\Phi} x w

open Lean TSyntax.Compat in
macro "\ensuremath{\forall}\textsuperscript{m}" xs:explicitBinders ", " b:term : term => expandExplicitBinders ``Mallposs xs b
open Lean TSyntax.Compat in
macro "\ensuremath{\exists}\textsuperscript{m}" xs:explicitBinders ", " b:term : term => expandExplicitBinders ``Mexiposs xs b
open Lean TSyntax.Compat in
macro "\ensuremath{\forall}\textsuperscript{E}" xs:explicitBinders ", " b:term : term => expandExplicitBinders ``Mallact xs b
open Lean TSyntax.Compat in
macro "\ensuremath{\exists}\textsuperscript{E}" xs:explicitBinders ", " b:term : term => expandExplicitBinders ``Mexiact xs b
\end{Verbatim}
Leibniz equality (polymorphic)

\begin{Verbatim}[commandchars=\\\{\},fontsize=\scriptsize,xleftmargin=1em]
@[simp, grind] def Mleibeq \{\ensuremath{\alpha}\} (x y : \ensuremath{\alpha}) : \ensuremath{\sigma} := \ensuremath{\forall}\textsuperscript{m} (P : \ensuremath{\alpha} \ensuremath{\rightarrow} \ensuremath{\sigma}), P x \ensuremath{\rightarrow}\textsuperscript{m} P y
infix:50 " \ensuremath{\equiv}\textsuperscript{m} " => Mleibeq
\end{Verbatim}
Meta-logical predicate for global validity

\begin{Verbatim}[commandchars=\\\{\},fontsize=\scriptsize,xleftmargin=1em]
@[simp, grind] def Mvalid (\ensuremath{\psi} : \ensuremath{\sigma}) : Prop := \ensuremath{\forall} w, \ensuremath{\psi} w
notation:max "\ensuremath{\lfloor}" \ensuremath{\psi} "\ensuremath{\rfloor}" => Mvalid \ensuremath{\psi}
\end{Verbatim}
\subsubsection{TestsHOMLinS4.lean}
Tests and verifications of properties for the embedding of HOML (S4) in HOL.
\begin{Verbatim}[commandchars=\\\{\},fontsize=\scriptsize,xleftmargin=1em]
import Notes.HOMLinHOLonlyS4
\end{Verbatim}

\begin{Verbatim}[commandchars=\\\{\},fontsize=\scriptsize,xleftmargin=1em]
variable \{\ensuremath{\alpha} : Type\} \{A B C \ensuremath{\varphi} \ensuremath{\psi} : \ensuremath{\sigma}\} \{p q : \ensuremath{\alpha} \ensuremath{\rightarrow} \ensuremath{\sigma}\} \{P Q : e \ensuremath{\rightarrow} \ensuremath{\sigma}\} \{x y z t : \ensuremath{\alpha}\}
\end{Verbatim}
Test for S4 modal logic

\begin{Verbatim}[commandchars=\\\{\},fontsize=\scriptsize,xleftmargin=1em]
theorem axM : \ensuremath{\lfloor}\ensuremath{\Box}\ensuremath{\varphi} \ensuremath{\rightarrow}\textsuperscript{m} \ensuremath{\varphi}\ensuremath{\rfloor} := fun w h => h w (Rrefl w)
theorem axD : \ensuremath{\lfloor}\ensuremath{\Box}\ensuremath{\varphi} \ensuremath{\rightarrow}\textsuperscript{m} \ensuremath{\Diamond}\ensuremath{\varphi}\ensuremath{\rfloor} := fun w h => \ensuremath{\langle}w, Rrefl w, h w (Rrefl w)\ensuremath{\rangle}
-- axB   -- `nitpick[expect=genuine]`: countermodel found
example : \ensuremath{\lfloor}\ensuremath{\varphi} \ensuremath{\rightarrow}\textsuperscript{m} \ensuremath{\Box}\ensuremath{\Diamond}\ensuremath{\varphi}\ensuremath{\rfloor} := by countermodel
theorem ax4 : \ensuremath{\lfloor}\ensuremath{\Box}\ensuremath{\varphi} \ensuremath{\rightarrow}\textsuperscript{m} \ensuremath{\Box}\ensuremath{\Box}\ensuremath{\varphi}\ensuremath{\rfloor} := fun w h v hv u hu => h u (Rtrans w v u \ensuremath{\langle}hv, hu\ensuremath{\rangle})
-- ax5   -- `nitpick[expect=genuine]`: countermodel found
example : \ensuremath{\lfloor}\ensuremath{\Diamond}\ensuremath{\varphi} \ensuremath{\rightarrow}\textsuperscript{m} \ensuremath{\Box}\ensuremath{\Diamond}\ensuremath{\varphi}\ensuremath{\rfloor} := by countermodel
\end{Verbatim}
Test for Barcan and converse Barcan formula

\begin{Verbatim}[commandchars=\\\{\},fontsize=\scriptsize,xleftmargin=1em]
-- BarcanAct   -- `nitpick[expect=genuine]`: countermodel found
example : \ensuremath{\lfloor}(\ensuremath{\forall}\textsuperscript{E} x, \ensuremath{\Box}(P x)) \ensuremath{\rightarrow}\textsuperscript{m} \ensuremath{\Box}(\ensuremath{\forall}\textsuperscript{E} x, P x)\ensuremath{\rfloor} := by countermodel
-- ConvBarcanAct   -- `nitpick[expect=genuine]`: countermodel found
example : \ensuremath{\lfloor}\ensuremath{\Box}(\ensuremath{\forall}\textsuperscript{E} x, P x) \ensuremath{\rightarrow}\textsuperscript{m} (\ensuremath{\forall}\textsuperscript{E} x, \ensuremath{\Box}(P x))\ensuremath{\rfloor} := by countermodel
theorem BarcanPoss : \ensuremath{\lfloor}(\ensuremath{\forall}\textsuperscript{m} x, \ensuremath{\Box}(p x)) \ensuremath{\rightarrow}\textsuperscript{m} \ensuremath{\Box}(\ensuremath{\forall}\textsuperscript{m} x, p x)\ensuremath{\rfloor} := fun _ h v hv x => h x v hv
theorem ConvBarcanPoss : \ensuremath{\lfloor}\ensuremath{\Box}(\ensuremath{\forall}\textsuperscript{m} x, p x) \ensuremath{\rightarrow}\textsuperscript{m} (\ensuremath{\forall}\textsuperscript{m} x, \ensuremath{\Box}(p x))\ensuremath{\rfloor} := fun _ h x v hv => h v hv x
\end{Verbatim}
A simple Hilbert system for classical propositional logic is derived

\begin{Verbatim}[commandchars=\\\{\},fontsize=\scriptsize,xleftmargin=1em]
theorem Hilbert_A1 : \ensuremath{\lfloor}A \ensuremath{\rightarrow}\textsuperscript{m} (B \ensuremath{\rightarrow}\textsuperscript{m} A)\ensuremath{\rfloor} := by grind
theorem Hilbert_A2 : \ensuremath{\lfloor}(A \ensuremath{\rightarrow}\textsuperscript{m} (B \ensuremath{\rightarrow}\textsuperscript{m} C)) \ensuremath{\rightarrow}\textsuperscript{m} ((A \ensuremath{\rightarrow}\textsuperscript{m} B) \ensuremath{\rightarrow}\textsuperscript{m} (A \ensuremath{\rightarrow}\textsuperscript{m} C))\ensuremath{\rfloor} := by grind
theorem Hilbert_MP (h1 : \ensuremath{\lfloor}A\ensuremath{\rfloor}) (h2 : \ensuremath{\lfloor}A \ensuremath{\rightarrow}\textsuperscript{m} B\ensuremath{\rfloor}) : \ensuremath{\lfloor}B\ensuremath{\rfloor} := fun w => h2 w (h1 w)
\end{Verbatim}
We have a polymorphic possibilist quantifier for which existential import holds

\begin{Verbatim}[commandchars=\\\{\},fontsize=\scriptsize,xleftmargin=1em]
theorem Quant_1 (h : \ensuremath{\lfloor}A\ensuremath{\rfloor}) : \ensuremath{\lfloor}\ensuremath{\forall}\textsuperscript{m} (_ : \ensuremath{\alpha}), A\ensuremath{\rfloor} := by grind
\end{Verbatim}
Existential import holds for possibilist quantifiers

\begin{Verbatim}[commandchars=\\\{\},fontsize=\scriptsize,xleftmargin=1em]
theorem ExImPossibilist1 : \ensuremath{\lfloor}\ensuremath{\exists}\textsuperscript{m} (x : e), x =\textsuperscript{m} x\ensuremath{\rfloor} := fun _ => \ensuremath{\langle}Classical.ofNonempty, rfl\ensuremath{\rangle}
theorem ExImPossibilist2 : \ensuremath{\lfloor}\ensuremath{\exists}\textsuperscript{m} (x : e), x \ensuremath{\equiv}\textsuperscript{m} x\ensuremath{\rfloor} := fun _ => \ensuremath{\langle}Classical.ofNonempty, fun _ h => h\ensuremath{\rangle}
theorem ExImPossibilist3 \{t : e\} : \ensuremath{\lfloor}\ensuremath{\exists}\textsuperscript{m} (x : e), x =\textsuperscript{m} t\ensuremath{\rfloor} := fun _ => \ensuremath{\langle}t, rfl\ensuremath{\rangle}
theorem ExImPossibilist4 : \ensuremath{\lfloor}\ensuremath{\exists}\textsuperscript{m} (x : \ensuremath{\alpha}), x \ensuremath{\equiv}\textsuperscript{m} t\ensuremath{\rfloor} := fun _ => \ensuremath{\langle}t, fun _ h => h\ensuremath{\rangle}
theorem ExImPossibilist [Nonempty \ensuremath{\alpha}] : \ensuremath{\lfloor}\ensuremath{\exists}\textsuperscript{m} (_ : \ensuremath{\alpha}), \ensuremath{\top}\textsuperscript{m}\ensuremath{\rfloor} := fun _ => \ensuremath{\langle}Classical.ofNonempty, trivial\ensuremath{\rangle}
\end{Verbatim}
We have an actualist quantifier for individuals for which existential import does not hold

\begin{Verbatim}[commandchars=\\\{\},fontsize=\scriptsize,xleftmargin=1em]
theorem Quant_2 (h : \ensuremath{\lfloor}A\ensuremath{\rfloor}) : \ensuremath{\lfloor}\ensuremath{\forall}\textsuperscript{E} (_ : e), A\ensuremath{\rfloor} := by grind
\end{Verbatim}
Existential import does not hold for our actualist quantifiers (for individuals)

\begin{Verbatim}[commandchars=\\\{\},fontsize=\scriptsize,xleftmargin=1em]
-- ExImActualist1   -- `nitpick[card=1,expect=genuine]`: countermodel found
example : \ensuremath{\lfloor}\ensuremath{\exists}\textsuperscript{E} (x : e), x =\textsuperscript{m} x\ensuremath{\rfloor} := by countermodel
-- ExImActualist2   -- `nitpick[card=1,expect=genuine]`: countermodel found
example : \ensuremath{\lfloor}\ensuremath{\exists}\textsuperscript{E} (x : e), x \ensuremath{\equiv}\textsuperscript{m} x\ensuremath{\rfloor} := by countermodel
-- ExImActualist3   -- `nitpick[card=1,expect=genuine]`: countermodel found
example \{t : e\} : \ensuremath{\lfloor}\ensuremath{\exists}\textsuperscript{E} (x : e), x =\textsuperscript{m} t\ensuremath{\rfloor} := by countermodel
-- ExImActualist   -- `nitpick[card=1,expect=genuine]`: countermodel found
example : \ensuremath{\lfloor}\ensuremath{\exists}\textsuperscript{E} (_ : e), \ensuremath{\top}\textsuperscript{m}\ensuremath{\rfloor} := by countermodel
\end{Verbatim}
Properties of the embedded primitive equality, which coincides with Leibniz equality

\begin{Verbatim}[commandchars=\\\{\},fontsize=\scriptsize,xleftmargin=1em]
theorem EqRefl : \ensuremath{\lfloor}x =\textsuperscript{m} x\ensuremath{\rfloor} := by grind
theorem EqSym : \ensuremath{\lfloor}(x =\textsuperscript{m} y) \ensuremath{\leftrightarrow}\textsuperscript{m} (y =\textsuperscript{m} x)\ensuremath{\rfloor} := by grind
theorem EqTrans : \ensuremath{\lfloor}((x =\textsuperscript{m} y) \ensuremath{\wedge}\textsuperscript{m} (y =\textsuperscript{m} z)) \ensuremath{\rightarrow}\textsuperscript{m} (x =\textsuperscript{m} z)\ensuremath{\rfloor} := by grind
theorem EQCong \{\ensuremath{\beta}\} \{f : \ensuremath{\alpha} \ensuremath{\rightarrow} \ensuremath{\beta}\} : \ensuremath{\lfloor}(x =\textsuperscript{m} y) \ensuremath{\rightarrow}\textsuperscript{m} ((f x) =\textsuperscript{m} (f y))\ensuremath{\rfloor} := by grind
theorem EQFuncExt : \ensuremath{\lfloor}(p =\textsuperscript{m} q) \ensuremath{\rightarrow}\textsuperscript{m} (\ensuremath{\forall}\textsuperscript{m} x, ((p x) =\textsuperscript{m} (q x)))\ensuremath{\rfloor} := by grind
theorem EQBoolExt1 : \ensuremath{\lfloor}(\ensuremath{\varphi} =\textsuperscript{m} \ensuremath{\psi}) \ensuremath{\rightarrow}\textsuperscript{m} (\ensuremath{\varphi} \ensuremath{\leftrightarrow}\textsuperscript{m} \ensuremath{\psi})\ensuremath{\rfloor} := by grind
-- EQBoolExt2   -- `nitpick[card=2]`: countermodel found
example : \ensuremath{\lfloor}(\ensuremath{\varphi} \ensuremath{\leftrightarrow}\textsuperscript{m} \ensuremath{\psi}) \ensuremath{\rightarrow}\textsuperscript{m} (\ensuremath{\varphi} =\textsuperscript{m} \ensuremath{\psi})\ensuremath{\rfloor} := by countermodel
theorem EQBoolExt3 : \ensuremath{\lfloor}\ensuremath{\varphi} \ensuremath{\leftrightarrow}\textsuperscript{m} \ensuremath{\psi}\ensuremath{\rfloor} \ensuremath{\rightarrow} \ensuremath{\lfloor}\ensuremath{\varphi} =\textsuperscript{m} \ensuremath{\psi}\ensuremath{\rfloor} := fun h _ => funext fun v => propext (h v)
theorem EqPrimLeib : \ensuremath{\lfloor}(x =\textsuperscript{m} y) \ensuremath{\leftrightarrow}\textsuperscript{m} (x \ensuremath{\equiv}\textsuperscript{m} y)\ensuremath{\rfloor} :=
  fun _ => \ensuremath{\langle}fun h _ hp => h \ensuremath{\blacktriangleright} hp, fun h => h (fun z _ => x = z) rfl\ensuremath{\rangle}
\end{Verbatim}
Comprehension is natively supported in HOL (due to lambda-abstraction)

\begin{Verbatim}[commandchars=\\\{\},fontsize=\scriptsize,xleftmargin=1em]
theorem Comprehension1 : \ensuremath{\lfloor}\ensuremath{\exists}\textsuperscript{m} (f : \ensuremath{\alpha} \ensuremath{\rightarrow} \ensuremath{\sigma}), \ensuremath{\forall}\textsuperscript{m} x, (f x) \ensuremath{\leftrightarrow}\textsuperscript{m} A\ensuremath{\rfloor} :=
  fun _ => \ensuremath{\langle}fun _ => A, fun _ => Iff.rfl\ensuremath{\rangle}
theorem Comprehension2 \{A1 : \ensuremath{\alpha} \ensuremath{\rightarrow} \ensuremath{\sigma}\} : \ensuremath{\lfloor}\ensuremath{\exists}\textsuperscript{m} (f : \ensuremath{\alpha} \ensuremath{\rightarrow} \ensuremath{\sigma}), \ensuremath{\forall}\textsuperscript{m} x, (f x) \ensuremath{\leftrightarrow}\textsuperscript{m} (A1 x)\ensuremath{\rfloor} :=
  fun _ => \ensuremath{\langle}A1, fun _ => Iff.rfl\ensuremath{\rangle}
theorem Comprehension3 \{\ensuremath{\beta}\} \{A2 : \ensuremath{\alpha} \ensuremath{\rightarrow} \ensuremath{\beta} \ensuremath{\rightarrow} \ensuremath{\sigma}\} :
    \ensuremath{\lfloor}\ensuremath{\exists}\textsuperscript{m} (f : \ensuremath{\alpha} \ensuremath{\rightarrow} \ensuremath{\beta} \ensuremath{\rightarrow} \ensuremath{\sigma}), \ensuremath{\forall}\textsuperscript{m} x, \ensuremath{\forall}\textsuperscript{m} y, (f x y) \ensuremath{\leftrightarrow}\textsuperscript{m} (A2 x y)\ensuremath{\rfloor} := fun _ => \ensuremath{\langle}A2, fun _ _ => Iff.rfl\ensuremath{\rangle}
\end{Verbatim}
Modal collapse does not hold

\begin{Verbatim}[commandchars=\\\{\},fontsize=\scriptsize,xleftmargin=1em]
-- ModalCollapse   -- `nitpick[card=2,expect=genuine]`: countermodel found
example : \ensuremath{\lfloor}\ensuremath{\forall}\textsuperscript{m} (f : \ensuremath{\sigma}), f \ensuremath{\rightarrow}\textsuperscript{m} \ensuremath{\Box}f\ensuremath{\rfloor} := by countermodel
\end{Verbatim}
Empty property and self-difference

\begin{Verbatim}[commandchars=\\\{\},fontsize=\scriptsize,xleftmargin=1em]
theorem TruePropertyAndSelfIdentity : \ensuremath{\lfloor}(fun _ : e => \ensuremath{\top}\textsuperscript{m}) =\textsuperscript{m} (fun x : e => x =\textsuperscript{m} x)\ensuremath{\rfloor} := by
  intro _; funext x v; simp
theorem EmptyPropertyAndSelfDifference : \ensuremath{\lfloor}(fun _ : e => \ensuremath{\bot}\textsuperscript{m}) =\textsuperscript{m} (fun x : e => x \ensuremath{\neq}\textsuperscript{m} x)\ensuremath{\rfloor} := by
  intro _; funext x v; simp
theorem EmptyProperty2 : \ensuremath{\lfloor}\ensuremath{\exists}\textsuperscript{m} x, P x\ensuremath{\rfloor} \ensuremath{\rightarrow} \ensuremath{\lfloor}P \ensuremath{\neq}\textsuperscript{m} (fun _ : e => \ensuremath{\bot}\textsuperscript{m})\ensuremath{\rfloor} := by
  intro h w hEq; have \ensuremath{\langle}x, hx\ensuremath{\rangle} := h w; rw [hEq] at hx; exact hx
theorem EmptyProperty3 : \ensuremath{\lfloor}\ensuremath{\exists}\textsuperscript{E} x, P x\ensuremath{\rfloor} \ensuremath{\rightarrow} \ensuremath{\lfloor}P \ensuremath{\neq}\textsuperscript{m} (fun _ : e => \ensuremath{\bot}\textsuperscript{m})\ensuremath{\rfloor} := by
  intro h w hEq; have \ensuremath{\langle}x, _, hx\ensuremath{\rangle} := h w; rw [hEq] at hx; exact hx
-- EmptyProperty4   -- `nitpick[expect=genuine]`: countermodel found
example : \ensuremath{\lfloor}P \ensuremath{\neq}\textsuperscript{m} (fun _ : e => \ensuremath{\bot}\textsuperscript{m})\ensuremath{\rfloor} \ensuremath{\rightarrow} \ensuremath{\lfloor}\ensuremath{\exists}\textsuperscript{m} x, P x\ensuremath{\rfloor} := by countermodel
-- EmptyProperty5   -- `nitpick[expect=genuine]`: countermodel found
example : \ensuremath{\lfloor}P \ensuremath{\neq}\textsuperscript{m} (fun _ : e => \ensuremath{\bot}\textsuperscript{m})\ensuremath{\rfloor} \ensuremath{\rightarrow} \ensuremath{\lfloor}\ensuremath{\exists}\textsuperscript{E} x, P x\ensuremath{\rfloor} := by countermodel
\end{Verbatim}
\subsubsection{GoedelVariantHOML1inS4.lean}
The same as GoedelVariantHOML1, but now in logic S4.
\begin{Verbatim}[commandchars=\\\{\},fontsize=\scriptsize,xleftmargin=1em]
import Notes.HOMLinHOLonlyS4
\end{Verbatim}

\begin{Verbatim}[commandchars=\\\{\},fontsize=\scriptsize,xleftmargin=1em]
/-- Positive property -/
axiom P : (e \ensuremath{\rightarrow} \ensuremath{\sigma}) \ensuremath{\rightarrow} \ensuremath{\sigma}

axiom Ax1 (\ensuremath{\varphi} \ensuremath{\psi} : e \ensuremath{\rightarrow} \ensuremath{\sigma}) : \ensuremath{\lfloor}P \ensuremath{\varphi} \ensuremath{\wedge}\textsuperscript{m} P \ensuremath{\psi} \ensuremath{\rightarrow}\textsuperscript{m} P (\ensuremath{\varphi} \ensuremath{\centerdot}\textsuperscript{m} \ensuremath{\psi})\ensuremath{\rfloor}

axiom Ax2a (\ensuremath{\varphi} : e \ensuremath{\rightarrow} \ensuremath{\sigma}) : \ensuremath{\lfloor}P \ensuremath{\varphi} \ensuremath{\vee}\textsuperscript{e} P ~\textsuperscript{m}\ensuremath{\varphi}\ensuremath{\rfloor}

/-- Auxiliary reformulation of `Ax2a` (Lean has no `sledgehammer`). -/
theorem Ax2a' (\ensuremath{\varphi} : e \ensuremath{\rightarrow} \ensuremath{\sigma}) : \ensuremath{\lfloor}\ensuremath{\neg}\textsuperscript{m}(P \ensuremath{\varphi}) \ensuremath{\leftrightarrow}\textsuperscript{m} P ~\textsuperscript{m}\ensuremath{\varphi}\ensuremath{\rfloor} :=
  fun w => \ensuremath{\langle}(Ax2a \ensuremath{\varphi} w).1.resolve_left, fun hp hq => (Ax2a \ensuremath{\varphi} w).2 \ensuremath{\langle}hq, hp\ensuremath{\rangle}\ensuremath{\rangle}

/-- God-like -/
def G (x : e) : \ensuremath{\sigma} := \ensuremath{\forall}\textsuperscript{m} (\ensuremath{\varphi} : e \ensuremath{\rightarrow} \ensuremath{\sigma}), P \ensuremath{\varphi} \ensuremath{\rightarrow}\textsuperscript{m} \ensuremath{\varphi} x

/-- Necessary property inclusion -/
@[simp, grind] def PInc (\ensuremath{\varphi} \ensuremath{\psi} : e \ensuremath{\rightarrow} \ensuremath{\sigma}) : \ensuremath{\sigma} := \ensuremath{\Box}(\ensuremath{\forall}\textsuperscript{E} y, \ensuremath{\varphi} y \ensuremath{\rightarrow}\textsuperscript{m} \ensuremath{\psi} y)
infixr:48 " \ensuremath{\supset}\textsuperscript{N} " => PInc

/-- `Ess \ensuremath{\varphi} x`: \ensuremath{\varphi} is an essence of x -/
def Ess (\ensuremath{\varphi} : e \ensuremath{\rightarrow} \ensuremath{\sigma}) (x : e) : \ensuremath{\sigma} := \ensuremath{\forall}\textsuperscript{m} (\ensuremath{\psi} : e \ensuremath{\rightarrow} \ensuremath{\sigma}), \ensuremath{\psi} x \ensuremath{\rightarrow}\textsuperscript{m} (\ensuremath{\varphi} \ensuremath{\supset}\textsuperscript{N} \ensuremath{\psi})

axiom Ax2b (\ensuremath{\varphi} : e \ensuremath{\rightarrow} \ensuremath{\sigma}) : \ensuremath{\lfloor}P \ensuremath{\varphi} \ensuremath{\rightarrow}\textsuperscript{m} \ensuremath{\Box} P \ensuremath{\varphi}\ensuremath{\rfloor}

theorem Ax2b' (\ensuremath{\varphi} : e \ensuremath{\rightarrow} \ensuremath{\sigma}) : \ensuremath{\lfloor}\ensuremath{\neg}\textsuperscript{m}(P \ensuremath{\varphi}) \ensuremath{\rightarrow}\textsuperscript{m} \ensuremath{\Box}(\ensuremath{\neg}\textsuperscript{m}(P \ensuremath{\varphi}))\ensuremath{\rfloor} :=
  fun w hn v hv => (Ax2a' \ensuremath{\varphi} v).2 (Ax2b ~\textsuperscript{m}\ensuremath{\varphi} w ((Ax2a' \ensuremath{\varphi} w).1 hn) v hv)

/-- A property exemplified by a God-like being is positive (a consequence of `Ax2a`). -/
theorem PosOfGod \{x : e\} \{\ensuremath{\psi} : e \ensuremath{\rightarrow} \ensuremath{\sigma}\} \{w : i\} (hG : G x w) (h : \ensuremath{\psi} x w) : P \ensuremath{\psi} w :=
  Classical.byContradiction fun hn => hG ~\textsuperscript{m}\ensuremath{\psi} ((Ax2a' \ensuremath{\psi} w).1 hn) h

theorem Th1 (x : e) : \ensuremath{\lfloor}G x \ensuremath{\rightarrow}\textsuperscript{m} Ess G x\ensuremath{\rfloor} :=
  fun w hG \ensuremath{\psi} h\ensuremath{\psi} v hv _y _ hGy => hGy \ensuremath{\psi} (Ax2b \ensuremath{\psi} w (PosOfGod hG h\ensuremath{\psi}) v hv)

/-- Necessary existence -/
def E (x : e) : \ensuremath{\sigma} := \ensuremath{\forall}\textsuperscript{m} (\ensuremath{\varphi} : e \ensuremath{\rightarrow} \ensuremath{\sigma}), Ess \ensuremath{\varphi} x \ensuremath{\rightarrow}\textsuperscript{m} \ensuremath{\Box}(\ensuremath{\exists}\textsuperscript{E} x, \ensuremath{\varphi} x)

axiom Ax3 : \ensuremath{\lfloor}P E\ensuremath{\rfloor}

theorem Th2 (x : e) : \ensuremath{\lfloor}G x \ensuremath{\rightarrow}\textsuperscript{m} \ensuremath{\Box}(\ensuremath{\exists}\textsuperscript{E} y, G y)\ensuremath{\rfloor} := fun w hG => hG E (Ax3 w) G (Th1 x w hG)

axiom Ax4 (\ensuremath{\varphi} \ensuremath{\psi} : e \ensuremath{\rightarrow} \ensuremath{\sigma}) : \ensuremath{\lfloor}P \ensuremath{\varphi} \ensuremath{\wedge}\textsuperscript{m} (\ensuremath{\varphi} \ensuremath{\supset}\textsuperscript{N} \ensuremath{\psi}) \ensuremath{\rightarrow}\textsuperscript{m} P \ensuremath{\psi}\ensuremath{\rfloor}

/-- Auxiliary (as in Figure 6): the empty property is an essence of anything. -/
theorem EmptyEssL (x : e) : \ensuremath{\lfloor}Ess (fun _ : e => \ensuremath{\bot}\textsuperscript{m}) x\ensuremath{\rfloor} := fun _ _ _ _ _ _ _ h => h.elim

theorem Th3 : \ensuremath{\lfloor}\ensuremath{\Diamond}(\ensuremath{\exists}\textsuperscript{E} x, G x) \ensuremath{\rightarrow}\textsuperscript{m} \ensuremath{\Box}(\ensuremath{\exists}\textsuperscript{E} y, G y)\ensuremath{\rfloor} := by
  intro _ \ensuremath{\langle}v, _, x, _, hG\ensuremath{\rangle}                             -- a God-like `x` at the reachable world `v`
  -- `E x v` applied to the empty essence says `v` has no successor, contradicting reflexivity
  have \ensuremath{\langle}_, _, h\ensuremath{\rangle} := hG E (Ax3 v) (fun _ : e => \ensuremath{\bot}\textsuperscript{m}) (EmptyEssL x v) v (Rrefl v)
  exact h.elim
\end{Verbatim}
\subsubsection{GoedelVariantHOML2inS4.lean}
The same as GoedelVariantHOML2, but now in logic S4, where the proof of theorem Th3 fails.
\begin{Verbatim}[commandchars=\\\{\},fontsize=\scriptsize,xleftmargin=1em]
import Notes.HOMLinHOLonlyS4
\end{Verbatim}

\begin{Verbatim}[commandchars=\\\{\},fontsize=\scriptsize,xleftmargin=1em]
/-- Positive property -/
axiom P : (e \ensuremath{\rightarrow} \ensuremath{\sigma}) \ensuremath{\rightarrow} \ensuremath{\sigma}

axiom Ax1 (\ensuremath{\varphi} \ensuremath{\psi} : e \ensuremath{\rightarrow} \ensuremath{\sigma}) : \ensuremath{\lfloor}P \ensuremath{\varphi} \ensuremath{\wedge}\textsuperscript{m} P \ensuremath{\psi} \ensuremath{\rightarrow}\textsuperscript{m} P (\ensuremath{\varphi} \ensuremath{\centerdot}\textsuperscript{m} \ensuremath{\psi})\ensuremath{\rfloor}

@[simp, grind] def PosProps (\ensuremath{\Phi} : (e \ensuremath{\rightarrow} \ensuremath{\sigma}) \ensuremath{\rightarrow} \ensuremath{\sigma}) : \ensuremath{\sigma} := \ensuremath{\forall}\textsuperscript{m} \ensuremath{\varphi}, \ensuremath{\Phi} \ensuremath{\varphi} \ensuremath{\rightarrow}\textsuperscript{m} P \ensuremath{\varphi}
@[simp, grind] def ConjOfPropsFrom (\ensuremath{\varphi} : e \ensuremath{\rightarrow} \ensuremath{\sigma}) (\ensuremath{\Phi} : (e \ensuremath{\rightarrow} \ensuremath{\sigma}) \ensuremath{\rightarrow} \ensuremath{\sigma}) : \ensuremath{\sigma} :=
  \ensuremath{\Box}(\ensuremath{\forall}\textsuperscript{E} z, \ensuremath{\varphi} z \ensuremath{\leftrightarrow}\textsuperscript{m} (\ensuremath{\forall}\textsuperscript{m} \ensuremath{\psi}, \ensuremath{\Phi} \ensuremath{\psi} \ensuremath{\rightarrow}\textsuperscript{m} \ensuremath{\psi} z))
axiom Ax1Gen (\ensuremath{\Phi} : (e \ensuremath{\rightarrow} \ensuremath{\sigma}) \ensuremath{\rightarrow} \ensuremath{\sigma}) (\ensuremath{\varphi} : e \ensuremath{\rightarrow} \ensuremath{\sigma}) :
  \ensuremath{\lfloor}(PosProps \ensuremath{\Phi} \ensuremath{\wedge}\textsuperscript{m} ConjOfPropsFrom \ensuremath{\varphi} \ensuremath{\Phi}) \ensuremath{\rightarrow}\textsuperscript{m} P \ensuremath{\varphi}\ensuremath{\rfloor}

axiom Ax2a (\ensuremath{\varphi} : e \ensuremath{\rightarrow} \ensuremath{\sigma}) : \ensuremath{\lfloor}P \ensuremath{\varphi} \ensuremath{\vee}\textsuperscript{e} P ~\textsuperscript{m}\ensuremath{\varphi}\ensuremath{\rfloor}

/-- Auxiliary reformulation of `Ax2a` (Lean has no `sledgehammer`). -/
theorem Ax2a' (\ensuremath{\varphi} : e \ensuremath{\rightarrow} \ensuremath{\sigma}) : \ensuremath{\lfloor}\ensuremath{\neg}\textsuperscript{m}(P \ensuremath{\varphi}) \ensuremath{\leftrightarrow}\textsuperscript{m} P ~\textsuperscript{m}\ensuremath{\varphi}\ensuremath{\rfloor} :=
  fun w => \ensuremath{\langle}(Ax2a \ensuremath{\varphi} w).1.resolve_left, fun hp hq => (Ax2a \ensuremath{\varphi} w).2 \ensuremath{\langle}hq, hp\ensuremath{\rangle}\ensuremath{\rangle}

/-- God-like -/
def G (x : e) : \ensuremath{\sigma} := \ensuremath{\forall}\textsuperscript{m} (\ensuremath{\varphi} : e \ensuremath{\rightarrow} \ensuremath{\sigma}), P \ensuremath{\varphi} \ensuremath{\rightarrow}\textsuperscript{m} \ensuremath{\varphi} x

/-- Necessary property inclusion -/
@[simp, grind] def PInc (\ensuremath{\varphi} \ensuremath{\psi} : e \ensuremath{\rightarrow} \ensuremath{\sigma}) : \ensuremath{\sigma} := \ensuremath{\Box}(\ensuremath{\forall}\textsuperscript{E} y, \ensuremath{\varphi} y \ensuremath{\rightarrow}\textsuperscript{m} \ensuremath{\psi} y)
infixr:48 " \ensuremath{\supset}\textsuperscript{N} " => PInc

/-- `Ess \ensuremath{\varphi} x`: \ensuremath{\varphi} is an essence of x (modified: \ensuremath{\varphi} must be exemplified by x) -/
def Ess (\ensuremath{\varphi} : e \ensuremath{\rightarrow} \ensuremath{\sigma}) (x : e) : \ensuremath{\sigma} := \ensuremath{\varphi} x \ensuremath{\wedge}\textsuperscript{m} (\ensuremath{\forall}\textsuperscript{m} (\ensuremath{\psi} : e \ensuremath{\rightarrow} \ensuremath{\sigma}), \ensuremath{\psi} x \ensuremath{\rightarrow}\textsuperscript{m} (\ensuremath{\varphi} \ensuremath{\supset}\textsuperscript{N} \ensuremath{\psi}))

axiom Ax2b (\ensuremath{\varphi} : e \ensuremath{\rightarrow} \ensuremath{\sigma}) : \ensuremath{\lfloor}P \ensuremath{\varphi} \ensuremath{\rightarrow}\textsuperscript{m} \ensuremath{\Box} P \ensuremath{\varphi}\ensuremath{\rfloor}

theorem Ax2b' (\ensuremath{\varphi} : e \ensuremath{\rightarrow} \ensuremath{\sigma}) : \ensuremath{\lfloor}\ensuremath{\neg}\textsuperscript{m}(P \ensuremath{\varphi}) \ensuremath{\rightarrow}\textsuperscript{m} \ensuremath{\Box}(\ensuremath{\neg}\textsuperscript{m}(P \ensuremath{\varphi}))\ensuremath{\rfloor} :=
  fun w hn v hv => (Ax2a' \ensuremath{\varphi} v).2 (Ax2b ~\textsuperscript{m}\ensuremath{\varphi} w ((Ax2a' \ensuremath{\varphi} w).1 hn) v hv)

/-- A property exemplified by a God-like being is positive (a consequence of `Ax2a`). -/
theorem PosOfGod \{x : e\} \{\ensuremath{\psi} : e \ensuremath{\rightarrow} \ensuremath{\sigma}\} \{w : i\} (hG : G x w) (h : \ensuremath{\psi} x w) : P \ensuremath{\psi} w :=
  Classical.byContradiction fun hn => hG ~\textsuperscript{m}\ensuremath{\psi} ((Ax2a' \ensuremath{\psi} w).1 hn) h

theorem Th1 (x : e) : \ensuremath{\lfloor}G x \ensuremath{\rightarrow}\textsuperscript{m} Ess G x\ensuremath{\rfloor} := fun w hG =>
  \ensuremath{\langle}hG, fun \ensuremath{\psi} h\ensuremath{\psi} v hv _y _ hGy => hGy \ensuremath{\psi} (Ax2b \ensuremath{\psi} w (PosOfGod hG h\ensuremath{\psi}) v hv)\ensuremath{\rangle}

/-- Necessary existence -/
def E (x : e) : \ensuremath{\sigma} := \ensuremath{\forall}\textsuperscript{m} (\ensuremath{\varphi} : e \ensuremath{\rightarrow} \ensuremath{\sigma}), Ess \ensuremath{\varphi} x \ensuremath{\rightarrow}\textsuperscript{m} \ensuremath{\Box}(\ensuremath{\exists}\textsuperscript{E} x, \ensuremath{\varphi} x)

axiom Ax3 : \ensuremath{\lfloor}P E\ensuremath{\rfloor}

theorem Th2 (x : e) : \ensuremath{\lfloor}G x \ensuremath{\rightarrow}\textsuperscript{m} \ensuremath{\Box}(\ensuremath{\exists}\textsuperscript{E} y, G y)\ensuremath{\rfloor} := fun w hG => hG E (Ax3 w) G (Th1 x w hG)

axiom Ax4 (\ensuremath{\varphi} \ensuremath{\psi} : e \ensuremath{\rightarrow} \ensuremath{\sigma}) : \ensuremath{\lfloor}P \ensuremath{\varphi} \ensuremath{\wedge}\textsuperscript{m} (\ensuremath{\varphi} \ensuremath{\supset}\textsuperscript{N} \ensuremath{\psi}) \ensuremath{\rightarrow}\textsuperscript{m} P \ensuremath{\psi}\ensuremath{\rfloor}

-- Th3 : \ensuremath{\lfloor}\ensuremath{\Diamond}(\ensuremath{\exists}\textsuperscript{E} x, G x) \ensuremath{\rightarrow}\textsuperscript{m} \ensuremath{\Box}(\ensuremath{\exists}\textsuperscript{E} y, G y)\ensuremath{\rfloor}   -- `nitpick`: open problem
example : \ensuremath{\lfloor}\ensuremath{\Diamond}(\ensuremath{\exists}\textsuperscript{E} x, G x) \ensuremath{\rightarrow}\textsuperscript{m} \ensuremath{\Box}(\ensuremath{\exists}\textsuperscript{E} y, G y)\ensuremath{\rfloor} := by openproblem
\end{Verbatim}
\subsubsection{GoedelVariantHOML2possInS4.lean}
The same as GoedelVariantHOML2poss, but now in logic S4, where the proof of theorem Th3 fails.
\begin{Verbatim}[commandchars=\\\{\},fontsize=\scriptsize,xleftmargin=1em]
import Notes.HOMLinHOLonlyS4
\end{Verbatim}

\begin{Verbatim}[commandchars=\\\{\},fontsize=\scriptsize,xleftmargin=1em]
/-- Positive property -/
axiom P : (e \ensuremath{\rightarrow} \ensuremath{\sigma}) \ensuremath{\rightarrow} \ensuremath{\sigma}

axiom Ax1 (\ensuremath{\varphi} \ensuremath{\psi} : e \ensuremath{\rightarrow} \ensuremath{\sigma}) : \ensuremath{\lfloor}P \ensuremath{\varphi} \ensuremath{\wedge}\textsuperscript{m} P \ensuremath{\psi} \ensuremath{\rightarrow}\textsuperscript{m} P (\ensuremath{\varphi} \ensuremath{\centerdot}\textsuperscript{m} \ensuremath{\psi})\ensuremath{\rfloor}

@[simp, grind] def PosProps (\ensuremath{\Phi} : (e \ensuremath{\rightarrow} \ensuremath{\sigma}) \ensuremath{\rightarrow} \ensuremath{\sigma}) : \ensuremath{\sigma} := \ensuremath{\forall}\textsuperscript{m} \ensuremath{\varphi}, \ensuremath{\Phi} \ensuremath{\varphi} \ensuremath{\rightarrow}\textsuperscript{m} P \ensuremath{\varphi}
@[simp, grind] def ConjOfPropsFrom (\ensuremath{\varphi} : e \ensuremath{\rightarrow} \ensuremath{\sigma}) (\ensuremath{\Phi} : (e \ensuremath{\rightarrow} \ensuremath{\sigma}) \ensuremath{\rightarrow} \ensuremath{\sigma}) : \ensuremath{\sigma} :=
  \ensuremath{\Box}(\ensuremath{\forall}\textsuperscript{m} z, \ensuremath{\varphi} z \ensuremath{\leftrightarrow}\textsuperscript{m} (\ensuremath{\forall}\textsuperscript{m} \ensuremath{\psi}, \ensuremath{\Phi} \ensuremath{\psi} \ensuremath{\rightarrow}\textsuperscript{m} \ensuremath{\psi} z))
axiom Ax1Gen (\ensuremath{\Phi} : (e \ensuremath{\rightarrow} \ensuremath{\sigma}) \ensuremath{\rightarrow} \ensuremath{\sigma}) (\ensuremath{\varphi} : e \ensuremath{\rightarrow} \ensuremath{\sigma}) :
  \ensuremath{\lfloor}(PosProps \ensuremath{\Phi} \ensuremath{\wedge}\textsuperscript{m} ConjOfPropsFrom \ensuremath{\varphi} \ensuremath{\Phi}) \ensuremath{\rightarrow}\textsuperscript{m} P \ensuremath{\varphi}\ensuremath{\rfloor}

axiom Ax2a (\ensuremath{\varphi} : e \ensuremath{\rightarrow} \ensuremath{\sigma}) : \ensuremath{\lfloor}P \ensuremath{\varphi} \ensuremath{\vee}\textsuperscript{e} P ~\textsuperscript{m}\ensuremath{\varphi}\ensuremath{\rfloor}

/-- Auxiliary reformulation of `Ax2a` (Lean has no `sledgehammer`). -/
theorem Ax2a' (\ensuremath{\varphi} : e \ensuremath{\rightarrow} \ensuremath{\sigma}) : \ensuremath{\lfloor}\ensuremath{\neg}\textsuperscript{m}(P \ensuremath{\varphi}) \ensuremath{\leftrightarrow}\textsuperscript{m} P ~\textsuperscript{m}\ensuremath{\varphi}\ensuremath{\rfloor} :=
  fun w => \ensuremath{\langle}(Ax2a \ensuremath{\varphi} w).1.resolve_left, fun hp hq => (Ax2a \ensuremath{\varphi} w).2 \ensuremath{\langle}hq, hp\ensuremath{\rangle}\ensuremath{\rangle}

/-- God-like -/
def G (x : e) : \ensuremath{\sigma} := \ensuremath{\forall}\textsuperscript{m} (\ensuremath{\varphi} : e \ensuremath{\rightarrow} \ensuremath{\sigma}), P \ensuremath{\varphi} \ensuremath{\rightarrow}\textsuperscript{m} \ensuremath{\varphi} x

/-- Necessary property inclusion -/
@[simp, grind] def PInc (\ensuremath{\varphi} \ensuremath{\psi} : e \ensuremath{\rightarrow} \ensuremath{\sigma}) : \ensuremath{\sigma} := \ensuremath{\Box}(\ensuremath{\forall}\textsuperscript{m} (y : e), \ensuremath{\varphi} y \ensuremath{\rightarrow}\textsuperscript{m} \ensuremath{\psi} y)
infixr:48 " \ensuremath{\supset}\textsuperscript{N} " => PInc

/-- `Ess \ensuremath{\varphi} x`: \ensuremath{\varphi} is an essence of x (modified: \ensuremath{\varphi} must be exemplified by x) -/
def Ess (\ensuremath{\varphi} : e \ensuremath{\rightarrow} \ensuremath{\sigma}) (x : e) : \ensuremath{\sigma} := \ensuremath{\varphi} x \ensuremath{\wedge}\textsuperscript{m} (\ensuremath{\forall}\textsuperscript{m} (\ensuremath{\psi} : e \ensuremath{\rightarrow} \ensuremath{\sigma}), \ensuremath{\psi} x \ensuremath{\rightarrow}\textsuperscript{m} (\ensuremath{\varphi} \ensuremath{\supset}\textsuperscript{N} \ensuremath{\psi}))

axiom Ax2b (\ensuremath{\varphi} : e \ensuremath{\rightarrow} \ensuremath{\sigma}) : \ensuremath{\lfloor}P \ensuremath{\varphi} \ensuremath{\rightarrow}\textsuperscript{m} \ensuremath{\Box} P \ensuremath{\varphi}\ensuremath{\rfloor}

theorem Ax2b' (\ensuremath{\varphi} : e \ensuremath{\rightarrow} \ensuremath{\sigma}) : \ensuremath{\lfloor}\ensuremath{\neg}\textsuperscript{m}(P \ensuremath{\varphi}) \ensuremath{\rightarrow}\textsuperscript{m} \ensuremath{\Box}(\ensuremath{\neg}\textsuperscript{m}(P \ensuremath{\varphi}))\ensuremath{\rfloor} :=
  fun w hn v hv => (Ax2a' \ensuremath{\varphi} v).2 (Ax2b ~\textsuperscript{m}\ensuremath{\varphi} w ((Ax2a' \ensuremath{\varphi} w).1 hn) v hv)

/-- A property exemplified by a God-like being is positive (a consequence of `Ax2a`). -/
theorem PosOfGod \{x : e\} \{\ensuremath{\psi} : e \ensuremath{\rightarrow} \ensuremath{\sigma}\} \{w : i\} (hG : G x w) (h : \ensuremath{\psi} x w) : P \ensuremath{\psi} w :=
  Classical.byContradiction fun hn => hG ~\textsuperscript{m}\ensuremath{\psi} ((Ax2a' \ensuremath{\psi} w).1 hn) h

theorem Th1 (x : e) : \ensuremath{\lfloor}G x \ensuremath{\rightarrow}\textsuperscript{m} Ess G x\ensuremath{\rfloor} := fun w hG =>
  \ensuremath{\langle}hG, fun \ensuremath{\psi} h\ensuremath{\psi} v hv _y hGy => hGy \ensuremath{\psi} (Ax2b \ensuremath{\psi} w (PosOfGod hG h\ensuremath{\psi}) v hv)\ensuremath{\rangle}

/-- Necessary existence -/
def E (x : e) : \ensuremath{\sigma} := \ensuremath{\forall}\textsuperscript{m} (\ensuremath{\varphi} : e \ensuremath{\rightarrow} \ensuremath{\sigma}), Ess \ensuremath{\varphi} x \ensuremath{\rightarrow}\textsuperscript{m} \ensuremath{\Box}(\ensuremath{\exists}\textsuperscript{m} x, \ensuremath{\varphi} x)

axiom Ax3 : \ensuremath{\lfloor}P E\ensuremath{\rfloor}

theorem Th2 (x : e) : \ensuremath{\lfloor}G x \ensuremath{\rightarrow}\textsuperscript{m} \ensuremath{\Box}(\ensuremath{\exists}\textsuperscript{m} y, G y)\ensuremath{\rfloor} := fun w hG => hG E (Ax3 w) G (Th1 x w hG)

axiom Ax4 (\ensuremath{\varphi} \ensuremath{\psi} : e \ensuremath{\rightarrow} \ensuremath{\sigma}) : \ensuremath{\lfloor}P \ensuremath{\varphi} \ensuremath{\wedge}\textsuperscript{m} (\ensuremath{\varphi} \ensuremath{\supset}\textsuperscript{N} \ensuremath{\psi}) \ensuremath{\rightarrow}\textsuperscript{m} P \ensuremath{\psi}\ensuremath{\rfloor}

-- Th3 : \ensuremath{\lfloor}\ensuremath{\Diamond}(\ensuremath{\exists}\textsuperscript{m} x, G x) \ensuremath{\rightarrow}\textsuperscript{m} \ensuremath{\Box}(\ensuremath{\exists}\textsuperscript{m} y, G y)\ensuremath{\rfloor}   -- `nitpick`: open problem
example : \ensuremath{\lfloor}\ensuremath{\Diamond}(\ensuremath{\exists}\textsuperscript{m} x, G x) \ensuremath{\rightarrow}\textsuperscript{m} \ensuremath{\Box}(\ensuremath{\exists}\textsuperscript{m} y, G y)\ensuremath{\rfloor} := by openproblem
\end{Verbatim}
\subsubsection{GoedelVariantHOML3inS4.lean}
The same as GoedelVariantHOML3, but now in logic S4, where the proof of theorem Th3 fails.
\begin{Verbatim}[commandchars=\\\{\},fontsize=\scriptsize,xleftmargin=1em]
import Notes.HOMLinHOLonlyS4
\end{Verbatim}

\begin{Verbatim}[commandchars=\\\{\},fontsize=\scriptsize,xleftmargin=1em]
/-- Positive property -/
axiom P : (e \ensuremath{\rightarrow} \ensuremath{\sigma}) \ensuremath{\rightarrow} \ensuremath{\sigma}

axiom Ax1 (\ensuremath{\varphi} \ensuremath{\psi} : e \ensuremath{\rightarrow} \ensuremath{\sigma}) : \ensuremath{\lfloor}P \ensuremath{\varphi} \ensuremath{\wedge}\textsuperscript{m} P \ensuremath{\psi} \ensuremath{\rightarrow}\textsuperscript{m} P (\ensuremath{\varphi} \ensuremath{\centerdot}\textsuperscript{m} \ensuremath{\psi})\ensuremath{\rfloor}

@[simp, grind] def PosProps (\ensuremath{\Phi} : (e \ensuremath{\rightarrow} \ensuremath{\sigma}) \ensuremath{\rightarrow} \ensuremath{\sigma}) : \ensuremath{\sigma} := \ensuremath{\forall}\textsuperscript{m} \ensuremath{\varphi}, \ensuremath{\Phi} \ensuremath{\varphi} \ensuremath{\rightarrow}\textsuperscript{m} P \ensuremath{\varphi}
@[simp, grind] def ConjOfPropsFrom (\ensuremath{\varphi} : e \ensuremath{\rightarrow} \ensuremath{\sigma}) (\ensuremath{\Phi} : (e \ensuremath{\rightarrow} \ensuremath{\sigma}) \ensuremath{\rightarrow} \ensuremath{\sigma}) : \ensuremath{\sigma} :=
  \ensuremath{\Box}(\ensuremath{\forall}\textsuperscript{E} z, \ensuremath{\varphi} z \ensuremath{\leftrightarrow}\textsuperscript{m} (\ensuremath{\forall}\textsuperscript{m} \ensuremath{\psi}, \ensuremath{\Phi} \ensuremath{\psi} \ensuremath{\rightarrow}\textsuperscript{m} \ensuremath{\psi} z))
axiom Ax1Gen (\ensuremath{\Phi} : (e \ensuremath{\rightarrow} \ensuremath{\sigma}) \ensuremath{\rightarrow} \ensuremath{\sigma}) (\ensuremath{\varphi} : e \ensuremath{\rightarrow} \ensuremath{\sigma}) :
  \ensuremath{\lfloor}(PosProps \ensuremath{\Phi} \ensuremath{\wedge}\textsuperscript{m} ConjOfPropsFrom \ensuremath{\varphi} \ensuremath{\Phi}) \ensuremath{\rightarrow}\textsuperscript{m} P \ensuremath{\varphi}\ensuremath{\rfloor}

axiom Ax2a (\ensuremath{\varphi} : e \ensuremath{\rightarrow} \ensuremath{\sigma}) : \ensuremath{\lfloor}P \ensuremath{\varphi} \ensuremath{\vee}\textsuperscript{e} P ~\textsuperscript{m}\ensuremath{\varphi}\ensuremath{\rfloor}

/-- Auxiliary reformulation of `Ax2a` (Lean has no `sledgehammer`). -/
theorem Ax2a' (\ensuremath{\varphi} : e \ensuremath{\rightarrow} \ensuremath{\sigma}) : \ensuremath{\lfloor}\ensuremath{\neg}\textsuperscript{m}(P \ensuremath{\varphi}) \ensuremath{\leftrightarrow}\textsuperscript{m} P ~\textsuperscript{m}\ensuremath{\varphi}\ensuremath{\rfloor} :=
  fun w => \ensuremath{\langle}(Ax2a \ensuremath{\varphi} w).1.resolve_left, fun hp hq => (Ax2a \ensuremath{\varphi} w).2 \ensuremath{\langle}hq, hp\ensuremath{\rangle}\ensuremath{\rangle}

/-- God-like -/
def G (x : e) : \ensuremath{\sigma} := \ensuremath{\forall}\textsuperscript{m} (\ensuremath{\varphi} : e \ensuremath{\rightarrow} \ensuremath{\sigma}), P \ensuremath{\varphi} \ensuremath{\rightarrow}\textsuperscript{m} \ensuremath{\varphi} x

/-- Necessary property inclusion (modified: \ensuremath{\varphi} must be non-empty) -/
@[simp, grind] def PInc (\ensuremath{\varphi} \ensuremath{\psi} : e \ensuremath{\rightarrow} \ensuremath{\sigma}) : \ensuremath{\sigma} :=
  \ensuremath{\Box}((\ensuremath{\varphi} \ensuremath{\neq}\textsuperscript{m} (fun _ : e => \ensuremath{\bot}\textsuperscript{m})) \ensuremath{\wedge}\textsuperscript{m} (\ensuremath{\forall}\textsuperscript{E} y, \ensuremath{\varphi} y \ensuremath{\rightarrow}\textsuperscript{m} \ensuremath{\psi} y))
infixr:48 " \ensuremath{\supset}\textsuperscript{N} " => PInc

/-- `Ess \ensuremath{\varphi} x`: \ensuremath{\varphi} is an essence of x -/
def Ess (\ensuremath{\varphi} : e \ensuremath{\rightarrow} \ensuremath{\sigma}) (x : e) : \ensuremath{\sigma} := \ensuremath{\forall}\textsuperscript{m} (\ensuremath{\psi} : e \ensuremath{\rightarrow} \ensuremath{\sigma}), \ensuremath{\psi} x \ensuremath{\rightarrow}\textsuperscript{m} (\ensuremath{\varphi} \ensuremath{\supset}\textsuperscript{N} \ensuremath{\psi})

axiom Ax2b (\ensuremath{\varphi} : e \ensuremath{\rightarrow} \ensuremath{\sigma}) : \ensuremath{\lfloor}P \ensuremath{\varphi} \ensuremath{\rightarrow}\textsuperscript{m} \ensuremath{\Box} P \ensuremath{\varphi}\ensuremath{\rfloor}

theorem Ax2b' (\ensuremath{\varphi} : e \ensuremath{\rightarrow} \ensuremath{\sigma}) : \ensuremath{\lfloor}\ensuremath{\neg}\textsuperscript{m}(P \ensuremath{\varphi}) \ensuremath{\rightarrow}\textsuperscript{m} \ensuremath{\Box}(\ensuremath{\neg}\textsuperscript{m}(P \ensuremath{\varphi}))\ensuremath{\rfloor} :=
  fun w hn v hv => (Ax2a' \ensuremath{\varphi} v).2 (Ax2b ~\textsuperscript{m}\ensuremath{\varphi} w ((Ax2a' \ensuremath{\varphi} w).1 hn) v hv)

/-- A property exemplified by a God-like being is positive (a consequence of `Ax2a`). -/
theorem PosOfGod \{x : e\} \{\ensuremath{\psi} : e \ensuremath{\rightarrow} \ensuremath{\sigma}\} \{w : i\} (hG : G x w) (h : \ensuremath{\psi} x w) : P \ensuremath{\psi} w :=
  Classical.byContradiction fun hn => hG ~\textsuperscript{m}\ensuremath{\psi} ((Ax2a' \ensuremath{\psi} w).1 hn) h

theorem Th1 (x : e) : \ensuremath{\lfloor}G x \ensuremath{\rightarrow}\textsuperscript{m} Ess G x\ensuremath{\rfloor} := by
  intro w hG \ensuremath{\psi} h\ensuremath{\psi} v hv
  refine \ensuremath{\langle}fun heq => ?_, fun y _ hGy => hGy \ensuremath{\psi} (Ax2b \ensuremath{\psi} w (PosOfGod hG h\ensuremath{\psi}) v hv)\ensuremath{\rangle}
  rw [heq] at hG; exact hG                             -- `G` is non-empty since `G x` holds

/-- Necessary existence -/
def E (x : e) : \ensuremath{\sigma} := \ensuremath{\forall}\textsuperscript{m} (\ensuremath{\varphi} : e \ensuremath{\rightarrow} \ensuremath{\sigma}), Ess \ensuremath{\varphi} x \ensuremath{\rightarrow}\textsuperscript{m} \ensuremath{\Box}(\ensuremath{\exists}\textsuperscript{E} x, \ensuremath{\varphi} x)

axiom Ax3 : \ensuremath{\lfloor}P E\ensuremath{\rfloor}

theorem Th2 (x : e) : \ensuremath{\lfloor}G x \ensuremath{\rightarrow}\textsuperscript{m} \ensuremath{\Box}(\ensuremath{\exists}\textsuperscript{E} y, G y)\ensuremath{\rfloor} := fun w hG => hG E (Ax3 w) G (Th1 x w hG)

axiom Ax4 (\ensuremath{\varphi} \ensuremath{\psi} : e \ensuremath{\rightarrow} \ensuremath{\sigma}) : \ensuremath{\lfloor}P \ensuremath{\varphi} \ensuremath{\wedge}\textsuperscript{m} (\ensuremath{\varphi} \ensuremath{\supset}\textsuperscript{N} \ensuremath{\psi}) \ensuremath{\rightarrow}\textsuperscript{m} P \ensuremath{\psi}\ensuremath{\rfloor}

-- Th3 : \ensuremath{\lfloor}\ensuremath{\Diamond}(\ensuremath{\exists}\textsuperscript{E} x, G x) \ensuremath{\rightarrow}\textsuperscript{m} \ensuremath{\Box}(\ensuremath{\exists}\textsuperscript{E} y, G y)\ensuremath{\rfloor}   -- `nitpick`: open problem
example : \ensuremath{\lfloor}\ensuremath{\Diamond}(\ensuremath{\exists}\textsuperscript{E} x, G x) \ensuremath{\rightarrow}\textsuperscript{m} \ensuremath{\Box}(\ensuremath{\exists}\textsuperscript{E} y, G y)\ensuremath{\rfloor} := by openproblem
\end{Verbatim}
\subsubsection{GoedelVariantHOML3possInS4.lean}
The same as GoedelVariantHOML3poss, but now in logic S4, where the proof of theorem Th3 fails.
\begin{Verbatim}[commandchars=\\\{\},fontsize=\scriptsize,xleftmargin=1em]
import Notes.HOMLinHOLonlyS4
\end{Verbatim}

\begin{Verbatim}[commandchars=\\\{\},fontsize=\scriptsize,xleftmargin=1em]
/-- Positive property -/
axiom P : (e \ensuremath{\rightarrow} \ensuremath{\sigma}) \ensuremath{\rightarrow} \ensuremath{\sigma}

axiom Ax1 (\ensuremath{\varphi} \ensuremath{\psi} : e \ensuremath{\rightarrow} \ensuremath{\sigma}) : \ensuremath{\lfloor}P \ensuremath{\varphi} \ensuremath{\wedge}\textsuperscript{m} P \ensuremath{\psi} \ensuremath{\rightarrow}\textsuperscript{m} P (\ensuremath{\varphi} \ensuremath{\centerdot}\textsuperscript{m} \ensuremath{\psi})\ensuremath{\rfloor}

@[simp, grind] def PosProps (\ensuremath{\Phi} : (e \ensuremath{\rightarrow} \ensuremath{\sigma}) \ensuremath{\rightarrow} \ensuremath{\sigma}) : \ensuremath{\sigma} := \ensuremath{\forall}\textsuperscript{m} \ensuremath{\varphi}, \ensuremath{\Phi} \ensuremath{\varphi} \ensuremath{\rightarrow}\textsuperscript{m} P \ensuremath{\varphi}
@[simp, grind] def ConjOfPropsFrom (\ensuremath{\varphi} : e \ensuremath{\rightarrow} \ensuremath{\sigma}) (\ensuremath{\Phi} : (e \ensuremath{\rightarrow} \ensuremath{\sigma}) \ensuremath{\rightarrow} \ensuremath{\sigma}) : \ensuremath{\sigma} :=
  \ensuremath{\Box}(\ensuremath{\forall}\textsuperscript{m} z, \ensuremath{\varphi} z \ensuremath{\leftrightarrow}\textsuperscript{m} (\ensuremath{\forall}\textsuperscript{m} \ensuremath{\psi}, \ensuremath{\Phi} \ensuremath{\psi} \ensuremath{\rightarrow}\textsuperscript{m} \ensuremath{\psi} z))
axiom Ax1Gen (\ensuremath{\Phi} : (e \ensuremath{\rightarrow} \ensuremath{\sigma}) \ensuremath{\rightarrow} \ensuremath{\sigma}) (\ensuremath{\varphi} : e \ensuremath{\rightarrow} \ensuremath{\sigma}) :
  \ensuremath{\lfloor}(PosProps \ensuremath{\Phi} \ensuremath{\wedge}\textsuperscript{m} ConjOfPropsFrom \ensuremath{\varphi} \ensuremath{\Phi}) \ensuremath{\rightarrow}\textsuperscript{m} P \ensuremath{\varphi}\ensuremath{\rfloor}

axiom Ax2a (\ensuremath{\varphi} : e \ensuremath{\rightarrow} \ensuremath{\sigma}) : \ensuremath{\lfloor}P \ensuremath{\varphi} \ensuremath{\vee}\textsuperscript{e} P ~\textsuperscript{m}\ensuremath{\varphi}\ensuremath{\rfloor}

/-- Auxiliary reformulation of `Ax2a` (Lean has no `sledgehammer`). -/
theorem Ax2a' (\ensuremath{\varphi} : e \ensuremath{\rightarrow} \ensuremath{\sigma}) : \ensuremath{\lfloor}\ensuremath{\neg}\textsuperscript{m}(P \ensuremath{\varphi}) \ensuremath{\leftrightarrow}\textsuperscript{m} P ~\textsuperscript{m}\ensuremath{\varphi}\ensuremath{\rfloor} :=
  fun w => \ensuremath{\langle}(Ax2a \ensuremath{\varphi} w).1.resolve_left, fun hp hq => (Ax2a \ensuremath{\varphi} w).2 \ensuremath{\langle}hq, hp\ensuremath{\rangle}\ensuremath{\rangle}

/-- God-like -/
def G (x : e) : \ensuremath{\sigma} := \ensuremath{\forall}\textsuperscript{m} (\ensuremath{\varphi} : e \ensuremath{\rightarrow} \ensuremath{\sigma}), P \ensuremath{\varphi} \ensuremath{\rightarrow}\textsuperscript{m} \ensuremath{\varphi} x

/-- Necessary property inclusion (modified: \ensuremath{\varphi} must be non-empty) -/
@[simp, grind] def PInc (\ensuremath{\varphi} \ensuremath{\psi} : e \ensuremath{\rightarrow} \ensuremath{\sigma}) : \ensuremath{\sigma} :=
  \ensuremath{\Box}((\ensuremath{\varphi} \ensuremath{\neq}\textsuperscript{m} (fun _ : e => \ensuremath{\bot}\textsuperscript{m})) \ensuremath{\wedge}\textsuperscript{m} (\ensuremath{\forall}\textsuperscript{m} (y : e), \ensuremath{\varphi} y \ensuremath{\rightarrow}\textsuperscript{m} \ensuremath{\psi} y))
infixr:48 " \ensuremath{\supset}\textsuperscript{N} " => PInc

/-- `Ess \ensuremath{\varphi} x`: \ensuremath{\varphi} is an essence of x -/
def Ess (\ensuremath{\varphi} : e \ensuremath{\rightarrow} \ensuremath{\sigma}) (x : e) : \ensuremath{\sigma} := \ensuremath{\forall}\textsuperscript{m} (\ensuremath{\psi} : e \ensuremath{\rightarrow} \ensuremath{\sigma}), \ensuremath{\psi} x \ensuremath{\rightarrow}\textsuperscript{m} (\ensuremath{\varphi} \ensuremath{\supset}\textsuperscript{N} \ensuremath{\psi})

axiom Ax2b (\ensuremath{\varphi} : e \ensuremath{\rightarrow} \ensuremath{\sigma}) : \ensuremath{\lfloor}P \ensuremath{\varphi} \ensuremath{\rightarrow}\textsuperscript{m} \ensuremath{\Box} P \ensuremath{\varphi}\ensuremath{\rfloor}

theorem Ax2b' (\ensuremath{\varphi} : e \ensuremath{\rightarrow} \ensuremath{\sigma}) : \ensuremath{\lfloor}\ensuremath{\neg}\textsuperscript{m}(P \ensuremath{\varphi}) \ensuremath{\rightarrow}\textsuperscript{m} \ensuremath{\Box}(\ensuremath{\neg}\textsuperscript{m}(P \ensuremath{\varphi}))\ensuremath{\rfloor} :=
  fun w hn v hv => (Ax2a' \ensuremath{\varphi} v).2 (Ax2b ~\textsuperscript{m}\ensuremath{\varphi} w ((Ax2a' \ensuremath{\varphi} w).1 hn) v hv)

/-- A property exemplified by a God-like being is positive (a consequence of `Ax2a`). -/
theorem PosOfGod \{x : e\} \{\ensuremath{\psi} : e \ensuremath{\rightarrow} \ensuremath{\sigma}\} \{w : i\} (hG : G x w) (h : \ensuremath{\psi} x w) : P \ensuremath{\psi} w :=
  Classical.byContradiction fun hn => hG ~\textsuperscript{m}\ensuremath{\psi} ((Ax2a' \ensuremath{\psi} w).1 hn) h

theorem Th1 (x : e) : \ensuremath{\lfloor}G x \ensuremath{\rightarrow}\textsuperscript{m} Ess G x\ensuremath{\rfloor} := by
  intro w hG \ensuremath{\psi} h\ensuremath{\psi} v hv
  refine \ensuremath{\langle}fun heq => ?_, fun y hGy => hGy \ensuremath{\psi} (Ax2b \ensuremath{\psi} w (PosOfGod hG h\ensuremath{\psi}) v hv)\ensuremath{\rangle}
  rw [heq] at hG; exact hG                             -- `G` is non-empty since `G x` holds

/-- Necessary existence -/
def E (x : e) : \ensuremath{\sigma} := \ensuremath{\forall}\textsuperscript{m} (\ensuremath{\varphi} : e \ensuremath{\rightarrow} \ensuremath{\sigma}), Ess \ensuremath{\varphi} x \ensuremath{\rightarrow}\textsuperscript{m} \ensuremath{\Box}(\ensuremath{\exists}\textsuperscript{m} x, \ensuremath{\varphi} x)

axiom Ax3 : \ensuremath{\lfloor}P E\ensuremath{\rfloor}

theorem Th2 (x : e) : \ensuremath{\lfloor}G x \ensuremath{\rightarrow}\textsuperscript{m} \ensuremath{\Box}(\ensuremath{\exists}\textsuperscript{m} y, G y)\ensuremath{\rfloor} := fun w hG => hG E (Ax3 w) G (Th1 x w hG)

axiom Ax4 (\ensuremath{\varphi} \ensuremath{\psi} : e \ensuremath{\rightarrow} \ensuremath{\sigma}) : \ensuremath{\lfloor}P \ensuremath{\varphi} \ensuremath{\wedge}\textsuperscript{m} (\ensuremath{\varphi} \ensuremath{\supset}\textsuperscript{N} \ensuremath{\psi}) \ensuremath{\rightarrow}\textsuperscript{m} P \ensuremath{\psi}\ensuremath{\rfloor}

-- Th3 : \ensuremath{\lfloor}\ensuremath{\Diamond}(\ensuremath{\exists}\textsuperscript{m} x, G x) \ensuremath{\rightarrow}\textsuperscript{m} \ensuremath{\Box}(\ensuremath{\exists}\textsuperscript{m} y, G y)\ensuremath{\rfloor}   -- `nitpick`: open problem
example : \ensuremath{\lfloor}\ensuremath{\Diamond}(\ensuremath{\exists}\textsuperscript{m} x, G x) \ensuremath{\rightarrow}\textsuperscript{m} \ensuremath{\Box}(\ensuremath{\exists}\textsuperscript{m} y, G y)\ensuremath{\rfloor} := by openproblem
\end{Verbatim}
\subsubsection{ScottVariantHOMLinS4.lean}
The same as ScottVariantHOML, but now in logic S4, where \texttt{T3} and \texttt{MC} fail.
\begin{Verbatim}[commandchars=\\\{\},fontsize=\scriptsize,xleftmargin=1em]
import Notes.HOMLinHOLonlyS4
\end{Verbatim}

\begin{Verbatim}[commandchars=\\\{\},fontsize=\scriptsize,xleftmargin=1em]
/-- Positive property -/
axiom P : (e \ensuremath{\rightarrow} \ensuremath{\sigma}) \ensuremath{\rightarrow} \ensuremath{\sigma}

axiom A1 (\ensuremath{\varphi} : e \ensuremath{\rightarrow} \ensuremath{\sigma}) : \ensuremath{\lfloor}\ensuremath{\neg}\textsuperscript{m}(P \ensuremath{\varphi}) \ensuremath{\leftrightarrow}\textsuperscript{m} P ~\textsuperscript{m}\ensuremath{\varphi}\ensuremath{\rfloor}

axiom A2 (\ensuremath{\varphi} \ensuremath{\psi} : e \ensuremath{\rightarrow} \ensuremath{\sigma}) : \ensuremath{\lfloor}P \ensuremath{\varphi} \ensuremath{\wedge}\textsuperscript{m} \ensuremath{\Box}(\ensuremath{\forall}\textsuperscript{E} y, \ensuremath{\varphi} y \ensuremath{\rightarrow}\textsuperscript{m} \ensuremath{\psi} y) \ensuremath{\rightarrow}\textsuperscript{m} P \ensuremath{\psi}\ensuremath{\rfloor}

theorem T1 (\ensuremath{\varphi} : e \ensuremath{\rightarrow} \ensuremath{\sigma}) : \ensuremath{\lfloor}P \ensuremath{\varphi} \ensuremath{\rightarrow}\textsuperscript{m} \ensuremath{\Diamond}(\ensuremath{\exists}\textsuperscript{E} x, \ensuremath{\varphi} x)\ensuremath{\rfloor} := fun w h =>
  Classical.byContradiction fun hn =>
    (A1 \ensuremath{\varphi} w).2 (A2 \ensuremath{\varphi} ~\textsuperscript{m}\ensuremath{\varphi} w \ensuremath{\langle}h, fun v hv y hy h\ensuremath{\varphi} _ => hn \ensuremath{\langle}v, hv, y, hy, h\ensuremath{\varphi}\ensuremath{\rangle}\ensuremath{\rangle}) h

/-- God-like -/
def G (x : e) : \ensuremath{\sigma} := \ensuremath{\forall}\textsuperscript{m} (\ensuremath{\varphi} : e \ensuremath{\rightarrow} \ensuremath{\sigma}), P \ensuremath{\varphi} \ensuremath{\rightarrow}\textsuperscript{m} \ensuremath{\varphi} x

axiom A3 : \ensuremath{\lfloor}P G\ensuremath{\rfloor}

theorem Coro : \ensuremath{\lfloor}\ensuremath{\Diamond}(\ensuremath{\exists}\textsuperscript{E} x, G x)\ensuremath{\rfloor} := fun w => T1 G w (A3 w)

axiom A4 (\ensuremath{\varphi} : e \ensuremath{\rightarrow} \ensuremath{\sigma}) : \ensuremath{\lfloor}P \ensuremath{\varphi} \ensuremath{\rightarrow}\textsuperscript{m} \ensuremath{\Box} P \ensuremath{\varphi}\ensuremath{\rfloor}

/-- `Ess \ensuremath{\varphi} x`: \ensuremath{\varphi} is an essence of x -/
def Ess (\ensuremath{\varphi} : e \ensuremath{\rightarrow} \ensuremath{\sigma}) (x : e) : \ensuremath{\sigma} := \ensuremath{\varphi} x \ensuremath{\wedge}\textsuperscript{m} (\ensuremath{\forall}\textsuperscript{m} (\ensuremath{\psi} : e \ensuremath{\rightarrow} \ensuremath{\sigma}), \ensuremath{\psi} x \ensuremath{\rightarrow}\textsuperscript{m} \ensuremath{\Box}(\ensuremath{\forall}\textsuperscript{E} y, \ensuremath{\varphi} y \ensuremath{\rightarrow}\textsuperscript{m} \ensuremath{\psi} y))

/-- A property exemplified by a God-like being is positive (a consequence of `A1`). -/
theorem PosOfGod \{x : e\} \{\ensuremath{\psi} : e \ensuremath{\rightarrow} \ensuremath{\sigma}\} \{w : i\} (hG : G x w) (h : \ensuremath{\psi} x w) : P \ensuremath{\psi} w :=
  Classical.byContradiction fun hn => hG ~\textsuperscript{m}\ensuremath{\psi} ((A1 \ensuremath{\psi} w).1 hn) h

theorem T2 (x : e) : \ensuremath{\lfloor}G x \ensuremath{\rightarrow}\textsuperscript{m} Ess G x\ensuremath{\rfloor} := fun w hG =>
  \ensuremath{\langle}hG, fun \ensuremath{\psi} h\ensuremath{\psi} v hv _y _ hGy => hGy \ensuremath{\psi} (A4 \ensuremath{\psi} w (PosOfGod hG h\ensuremath{\psi}) v hv)\ensuremath{\rangle}

/-- Necessary existence -/
def NE (x : e) : \ensuremath{\sigma} := \ensuremath{\forall}\textsuperscript{m} (\ensuremath{\varphi} : e \ensuremath{\rightarrow} \ensuremath{\sigma}), Ess \ensuremath{\varphi} x \ensuremath{\rightarrow}\textsuperscript{m} \ensuremath{\Box}(\ensuremath{\exists}\textsuperscript{E} x, \ensuremath{\varphi} x)

axiom A5 : \ensuremath{\lfloor}P NE\ensuremath{\rfloor}

-- `lemma True nitpick[satisfy,card=1,eval="\ensuremath{\lfloor}P (\ensuremath{\lambda}x.\ensuremath{\top})\ensuremath{\rfloor}"] oops`
--   One model found of cardinality one (consistency check)

-- T3 : \ensuremath{\lfloor}\ensuremath{\Box}(\ensuremath{\exists}\textsuperscript{E} x, G x)\ensuremath{\rfloor}   -- `nitpick[card e=1, card i=2]`: countermodel found
example : \ensuremath{\lfloor}\ensuremath{\Box}(\ensuremath{\exists}\textsuperscript{E} x, G x)\ensuremath{\rfloor} := by countermodel

-- MC : \ensuremath{\lfloor}\ensuremath{\varphi} \ensuremath{\rightarrow}\textsuperscript{m} \ensuremath{\Box}\ensuremath{\varphi}\ensuremath{\rfloor}   -- `nitpick[card e=1, card i=2]`: countermodel found
example (\ensuremath{\varphi} : \ensuremath{\sigma}) : \ensuremath{\lfloor}\ensuremath{\varphi} \ensuremath{\rightarrow}\textsuperscript{m} \ensuremath{\Box}\ensuremath{\varphi}\ensuremath{\rfloor} := by countermodel
\end{Verbatim}
\subsubsection{ScottVariantHOMLpossInS4.lean}
The same as ScottVariantHOMLposs, but now in logic S4, where \texttt{T3} and \texttt{MC} fail.
\begin{Verbatim}[commandchars=\\\{\},fontsize=\scriptsize,xleftmargin=1em]
import Notes.HOMLinHOLonlyS4
\end{Verbatim}

\begin{Verbatim}[commandchars=\\\{\},fontsize=\scriptsize,xleftmargin=1em]
/-- Positive property -/
axiom P : (e \ensuremath{\rightarrow} \ensuremath{\sigma}) \ensuremath{\rightarrow} \ensuremath{\sigma}

axiom A1 (\ensuremath{\varphi} : e \ensuremath{\rightarrow} \ensuremath{\sigma}) : \ensuremath{\lfloor}\ensuremath{\neg}\textsuperscript{m}(P \ensuremath{\varphi}) \ensuremath{\leftrightarrow}\textsuperscript{m} P ~\textsuperscript{m}\ensuremath{\varphi}\ensuremath{\rfloor}

axiom A2 (\ensuremath{\varphi} \ensuremath{\psi} : e \ensuremath{\rightarrow} \ensuremath{\sigma}) : \ensuremath{\lfloor}P \ensuremath{\varphi} \ensuremath{\wedge}\textsuperscript{m} \ensuremath{\Box}(\ensuremath{\forall}\textsuperscript{m} y, \ensuremath{\varphi} y \ensuremath{\rightarrow}\textsuperscript{m} \ensuremath{\psi} y) \ensuremath{\rightarrow}\textsuperscript{m} P \ensuremath{\psi}\ensuremath{\rfloor}

theorem T1 (\ensuremath{\varphi} : e \ensuremath{\rightarrow} \ensuremath{\sigma}) : \ensuremath{\lfloor}P \ensuremath{\varphi} \ensuremath{\rightarrow}\textsuperscript{m} \ensuremath{\Diamond}(\ensuremath{\exists}\textsuperscript{m} x, \ensuremath{\varphi} x)\ensuremath{\rfloor} := fun w h =>
  Classical.byContradiction fun hn =>
    (A1 \ensuremath{\varphi} w).2 (A2 \ensuremath{\varphi} ~\textsuperscript{m}\ensuremath{\varphi} w \ensuremath{\langle}h, fun v hv y h\ensuremath{\varphi} _ => hn \ensuremath{\langle}v, hv, y, h\ensuremath{\varphi}\ensuremath{\rangle}\ensuremath{\rangle}) h

/-- God-like -/
def G (x : e) : \ensuremath{\sigma} := \ensuremath{\forall}\textsuperscript{m} (\ensuremath{\varphi} : e \ensuremath{\rightarrow} \ensuremath{\sigma}), P \ensuremath{\varphi} \ensuremath{\rightarrow}\textsuperscript{m} \ensuremath{\varphi} x

axiom A3 : \ensuremath{\lfloor}P G\ensuremath{\rfloor}

theorem Coro : \ensuremath{\lfloor}\ensuremath{\Diamond}(\ensuremath{\exists}\textsuperscript{m} x, G x)\ensuremath{\rfloor} := fun w => T1 G w (A3 w)

axiom A4 (\ensuremath{\varphi} : e \ensuremath{\rightarrow} \ensuremath{\sigma}) : \ensuremath{\lfloor}P \ensuremath{\varphi} \ensuremath{\rightarrow}\textsuperscript{m} \ensuremath{\Box} P \ensuremath{\varphi}\ensuremath{\rfloor}

/-- `Ess \ensuremath{\varphi} x`: \ensuremath{\varphi} is an essence of x -/
def Ess (\ensuremath{\varphi} : e \ensuremath{\rightarrow} \ensuremath{\sigma}) (x : e) : \ensuremath{\sigma} := \ensuremath{\varphi} x \ensuremath{\wedge}\textsuperscript{m} (\ensuremath{\forall}\textsuperscript{m} (\ensuremath{\psi} : e \ensuremath{\rightarrow} \ensuremath{\sigma}), \ensuremath{\psi} x \ensuremath{\rightarrow}\textsuperscript{m} \ensuremath{\Box}(\ensuremath{\forall}\textsuperscript{m} (y : e), \ensuremath{\varphi} y \ensuremath{\rightarrow}\textsuperscript{m} \ensuremath{\psi} y))

/-- A property exemplified by a God-like being is positive (a consequence of `A1`). -/
theorem PosOfGod \{x : e\} \{\ensuremath{\psi} : e \ensuremath{\rightarrow} \ensuremath{\sigma}\} \{w : i\} (hG : G x w) (h : \ensuremath{\psi} x w) : P \ensuremath{\psi} w :=
  Classical.byContradiction fun hn => hG ~\textsuperscript{m}\ensuremath{\psi} ((A1 \ensuremath{\psi} w).1 hn) h

theorem T2 (x : e) : \ensuremath{\lfloor}G x \ensuremath{\rightarrow}\textsuperscript{m} Ess G x\ensuremath{\rfloor} := fun w hG =>
  \ensuremath{\langle}hG, fun \ensuremath{\psi} h\ensuremath{\psi} v hv _y hGy => hGy \ensuremath{\psi} (A4 \ensuremath{\psi} w (PosOfGod hG h\ensuremath{\psi}) v hv)\ensuremath{\rangle}

/-- Necessary existence -/
def NE (x : e) : \ensuremath{\sigma} := \ensuremath{\forall}\textsuperscript{m} (\ensuremath{\varphi} : e \ensuremath{\rightarrow} \ensuremath{\sigma}), Ess \ensuremath{\varphi} x \ensuremath{\rightarrow}\textsuperscript{m} \ensuremath{\Box}(\ensuremath{\exists}\textsuperscript{m} x, \ensuremath{\varphi} x)

axiom A5 : \ensuremath{\lfloor}P NE\ensuremath{\rfloor}

-- `lemma True nitpick[satisfy,card=1,eval="\ensuremath{\lfloor}P (\ensuremath{\lambda}x.\ensuremath{\bot})\ensuremath{\rfloor}"] oops`
--   One model found of cardinality one (consistency check)

-- T3 : \ensuremath{\lfloor}\ensuremath{\Box}(\ensuremath{\exists}\textsuperscript{m} x, G x)\ensuremath{\rfloor}   -- `nitpick[card e=1, card i=2]`: countermodel found
example : \ensuremath{\lfloor}\ensuremath{\Box}(\ensuremath{\exists}\textsuperscript{m} x, G x)\ensuremath{\rfloor} := by countermodel

-- MC : \ensuremath{\lfloor}\ensuremath{\varphi} \ensuremath{\rightarrow}\textsuperscript{m} \ensuremath{\Box}\ensuremath{\varphi}\ensuremath{\rfloor}   -- `nitpick[card e=1, card i=2]`: countermodel found
example (\ensuremath{\varphi} : \ensuremath{\sigma}) : \ensuremath{\lfloor}\ensuremath{\varphi} \ensuremath{\rightarrow}\textsuperscript{m} \ensuremath{\Box}\ensuremath{\varphi}\ensuremath{\rfloor} := by countermodel
\end{Verbatim}